%% file: phys-rep.tex
\documentclass{elsart}

\newcommand{\Expect}[1]{{\mathbf E}\left[ #1 \right]}
\newcommand{\vev}[1]{\langle #1 \rangle}
\newcommand{\ket}[1]{| #1 \rangle}
\newcommand{\bra}[1]{\langle #1 |}

\newcommand{\statav}[1]{\langle #1 \rangle}
\newtheorem{ex}{Example}

\usepackage[final]{graphicx}
\usepackage{amssymb} 
\usepackage{oldgerm} 

\begin{document}

\begin{frontmatter}



\title{2D growth processes: \\  
SLE and Loewner~chains}


\author{Michel Bauer}
\ead{michel.bauer@cea.fr}
and
\author{Denis Bernard\corauthref{cor1}}
\ead{denis.bernard@cea.fr}
\corauth[cor1]{Member of CNRS}


\address{Service de Physique Th\'eorique de Saclay\thanksref{label1}\\
CE-Saclay, 91191 Gif-sur-Yvette, France}
\thanks[label1]{CEA/DSM/SPhT and URA 2306 du CNRS}

\address{and}

\address{Laboratoire de Physique Th\'eorique \thanksref{label2}\\
Ecole Normale Sup\'erieure, 24 rue Lhomond, 75005 Paris, France}
\thanks[label2]{UMR 8549 du CNRS}

\begin{abstract}
  This review provides an introduction to two dimensional growth
  processes. Although it covers a variety processes such as diffusion
  limited aggregation, it is mostly devoted to a detailed presentation
  of stochastic Schramm-Loewner evolutions (SLE) which are Markov
  processes describing interfaces in 2D critical systems. It starts
  with an informal discussion, using numerical simulations, of various
  examples of 2D growth processes and their connections with
  statistical mechanics. SLE is then introduced and Schramm's argument
  mapping conformally invariant interfaces to SLE is explained. A
  substantial part of the review is devoted to reveal the deep
  connections between statistical mechanics and processes, and more
  specifically to the present context, between 2D critical systems and
  SLE. Some of the SLE remarkable properties are explained, as well as
  the tools for computing with SLE. This review has been written with
  the aim of filling the gap between the mathematical and the physical
  literatures on the subject.
\end{abstract}

\end{frontmatter}

\newpage
\tableofcontents
\newpage 

{\bf Notations:}\\
${\bf P}[\cdots]=$ probability,\\
${\bf E}[\cdots]=$ expectation.\\
${\bf P}[\cdots\vert C]=$ probability conditioned on $C$,\\
${\bf E}[\cdots\vert C]=$ expectation conditioned on $C$.\\
$\mathcal{F}_t=$ filtration by $\sigma$-algebras.\\
$B_t=$ normalized Brownian motion with ${\bf E}[B_tB_s]={\rm
  min}(t,s)$.\\
$\xi_t=\sqrt{\kappa}\,B_t$, with covariance
${\bf E}[\xi_t\xi_s]=\kappa\ {\rm min}(t,s)$.\\
$\mathbb{U},\ \mathbb{D}=$ (planar) domain, ie. connected and simply
connected open subset of the complex plane $\mathbb{C}$.\\
$\mathbb{K}=$ hulls, ie. connected compact subset of a domain
$\mathbb{D}$ such that $\mathbb{D}\setminus\mathbb{K}$ is a domain.\\
$g_\mathbb{K}=$ holomorphic map uniformizing
$\mathbb{D}\setminus\mathbb{K}$ into $\mathbb{D}$, \\
$f_\mathbb{K}:\mathbb{D}\to\mathbb{D}\setminus\mathbb{K}$ its inverse.\\
$\gamma_{[0,t]}=$ the SLE curve with tip $\gamma_t$ at time $t$.\\
$\mathbb{K}_t=$ the SLE hull at time $t$.\\
$\mathbb{D}_t\equiv \mathbb{D}\setminus\mathbb{K}_t$, the domain 
$\mathbb{D}$ with the hull $\mathbb{K}_t$ removed.\\
$g_t:\mathbb{D}_t\to \mathbb{D}$, the SLE Loewner map and
$f_t:\mathbb{D}\to\mathbb{D}_t$, its inverse.\\
$U_t=g_t(\gamma_t)$= image of the tip of the SLE curve.\\
$h_t: \mathbb{D}_t\to\mathbb{D}$, mapping the tip
of the curve back to its starting point.\\
$\mathfrak{vir}=$ the Virasoro algebra.\\
$\mathfrak{g}_h$= a group element associated to a map $h$.\\
$G_h=$ representation of $\mathfrak{g}_h$ in CFT Hilbert spaces.\\
$h_{r;s}=[(r\kappa-4s)^2-(\kappa-4)^2]/16\kappa$ for
$c=1-6(\kappa-4)^2/4\kappa$.\\
$\ket{\psi_{r;s}}=$ highest weight vector with dimension $h_{r;s}$.\\
$\varphi_\delta(x)=$ boundary primary field with dimension $\delta$.\\
$\psi_{r;s}(x)=$ degenerate boundary primary field with dimension $h_{r;s}$.\\
$\Phi(z,\bar z)=$ bulk primary fields.\\
$\Phi_{r;s}(z,\bar z)=$ degenerate bulk primary field with dimension $2h_{r;s}$.\\
$\vev{\cdots}_\mathbb{D}=$ CFT correlation functions in a domain
$\mathbb{D}$.\\
$Z_\mathbb{D}=$ statistical partition function in a domain $\mathbb{D}$.\\
$\prec\cdots\succ_\mathbb{D}=$ statistical average in a domain
$\mathbb{D}$.\\ 

\vfill \eject

\section{Introduction}
\label{sec:intro}

\input{intro.tex}

\section{Constructive examples}
\label{sec:examples}

\input{chap1.tex}

\section{Loewner chains}
\label{sec:anal}

\input{chap2.tex}

\section{Stochastic Schramm-Loewner evolution}
\label{sec:sle}

\input{chap3.tex}

\section{Statistical mechanics and processes}
\label{sec:statmech}

\input{chap4.tex}

\section{SLE delicatessen}
\label{sec:delicat}

\input{chap5.tex}

\section{Computing with SLEs}
\label{sec:comput}

\input{chap6.tex}

\section{Other growth processes}
\label{sec:othergrowth}

\input{chap7.tex}


\vskip 1.0 truecm

{\bf Acknowledgements:}

We benefited from discussions with many of our colleagues.
We would like to thank especially Vincent Beffara, John Cardy,
Antti Kupiainen, Kalle Kyt\"ol\"a, Greg Lawler, Vincent Pasquier,
Wendelin Werner and Paul Wiegmann.

Work supported in part by EC contract number
HPRN-CT-2002-00325 of the EUCLID research training network, 
and in part by EC contract number MRTN-CT-2004-5652
 of the ENIGMA research training network.


\appendix

\section{Probabilistic  background}
\label{app:proba}

\input{app1.tex}

\vskip 1.5 truecm

\section{CFT background}
\label{app:cft}

\input{app2.tex}


\end{document}

%% file: intro.tex
The main subject of this report is stochastic Loewner evolutions, and
its interplay with statistical mechanics and conformal field theory. 

Stochastic Loewner evolutions are growth processes, and as such they
fall in the more general category of growth phenomena. These are
ubiquitous in the physical world at many scales, from crystals to
plants to dunes and larger. They can be studied in many frameworks,
deterministic of probabilistic, in discrete or continuous space and
time. Understanding growth is usually a very difficult task.  This is
true even in two dimensions, the case we concentrate on in these
notes. Yet two dimensions is a highly favorable situation because it allows
to make use of the power of complex analysis in one variable. In many
interesting cases, the growing object in two dimensions can be seen as
a domain, i.e. a contractile open subset of the Riemann sphere (the
complex plane with a point at infinity added) leading to a description
by so-called Loewner chains.

Stochastic Loewner evolution is a simple but particularly interesting
example of growth process for which the growth is local and continuous
so that the resulting set is a curve without branching.  Of course
other examples have been studied in connection with 2d physical
systems. The motivations are sometimes very practical. For instance,
is it efficient to put a pump in the center of oil film at the surface
of the ocean to fight against pollution? The answer has to do with the
Laplacian growth or Hele-Shaw problem. The names diffusion limited
aggregation and dielectric breakdown speak for themselves. Various
models have been invented, sometimes with less physical motivation,
but in order to find more manageable growth processes. These include
various models of iterated conformal maps, etc. As mentioned above, in
most cases the shape of the growing domains is encoded in a
uniformizing conformal map whose evolution describes the evolution of
the domain.  The dynamics can be either discrete or continuous in
time, it can be either deterministic or stochastic. But the growth
process is always described by a Loewner chain.

So we shall also give a pedagogical introduction to the beautiful
subject of general Loewner chains.  We wanted to show that it leads to
many basic mathematical structures whose appearance in the growth
context is not so easy to foresee, like integrable
systems and anomalies to mention just a few. We have also tried to
stress that some growth processes have rules which are easy to
simulate on the computer. A few minutes of CPU are enough to get an
idea of the shape of the growing patterns, to be convinced that
something interesting and non trivial is going on, and even sometimes
to get an idea of fractal dimensions. This is of course not to be
compared with serious large scale simulations, but it is a good
illustration of the big contrast between simple rules, complex
patterns and involved mathematical structures. However, other growth
models, and among those some have been conjectured to be equivalent to
simple ones, have resisted until recently to precise numerical
calculations due to instabilities.

To avoid any confusion, let us stress that being able to describe a
growth process using tools from complex analysis and conformal geometry
does not mean that the growth process itself is conformally invariant
at all. Conformal invariance of the growth process itself puts rather
drastic conditions on the density that appears in the Loewner chain
and lead to stochastic Loewner evolutions.

\vspace{.2cm}

Why do we think the emergence of stochastic Loewner evolutions is so
important ? This question has several answers at various levels. 

A first obvious answer is that stochastic Loewner evolutions are among
the very few growth processes that can be studied analytically in
great detail. The other growth processes we shall present in these
notes are still very poorly understood, and many basic qualitative
question like universality classes are still debated.

A second obvious answer is that stochastic Loewner evolutions solve a
problem that had remained open for two decades despite the fact that
the importance of conformal invariance had been fully recognized : the
description of conformally invariant extended objects. This
obvious answer is in fact best incorporated into a deeper one which is
rooted in history. 

There is a natural flow in the life of scientific discoveries, and   
conformal field theory was no exception to the rule. 

Starting in 1984, conformal field theory has been an object of study
for itself during a decade or so, revealing a fascinating richness.
At a critical point and for short range interactions, statistical
mechanics systems are expected to be conformally invariant. The
argument for that was given two decades ago in the seminal paper on
conformal field theory \cite{BPZ}. The rough idea is the following. At
a critical point, a system becomes scale invariant.  If the
interactions on the lattice are short range, the model is described in
the continuum limit by a local field theory and scale invariance
implies that the stress tensor is traceless. In two dimensions this is
enough to ensure that the theory transforms simply --no dynamics is
involved, only pure kinematics-- when the domain where it is defined
is changed by a conformal transformation. The local fields are
classified by representations of the infinite dimensional Virasoro
algebra and this dictates the way correlation functions transform.
This has led to a tremendous accumulation of exact results.

 From the start, conformal field theory was also seriously directed
towards applications, and this is even more true now that it has
reached technical maturity. During the last twenty years or so,
conformal field theory has become a standard tool, and a very powerful
one indeed, to tackle a variety of problems. Significant progresses in
condensed matter theory owe a lot to conformal field theory :
computation of universal amplitudes for the Kondo problem, various
aspects of the (fractional) quantum Hall effect, Luttinger liquid
theory are just a few examples. String theory has sowed conformal
field theory but also collected a lot.

This shift from goal to tool does not mean that everything is
understood. In fact nothing could be less exact. A situation that is
well under control is that of Virasoro unitary minimal models. The
Hilbert space of the system splits as a finite sum of representations
of the Virasoro algebra, each associated to a (local) primary field,
and the corresponding correlation functions can be described rather
explicitly. However, the initial hope of classifying all critical
phenomena in two dimensions has vanished.  Work has concentrated on
special, manageable, classes of theories generalizing the Virasoro
unitary minimal models. The most user-friendly theories are minimal
for algebras extending the Virasoro algebra. For these a finite number
of representations suffices to describe many physical properties of
the underlying model. Even the classification of minimal theories is a
formidable task and it is far from obvious that the goal will be
achieved ever. Surprisingly maybe, adding unitarity on top of
minimality does not help much.

On the other hand, many (most of the ?) important applications of
conformal field theories, emerging for instance from string theory or
disordered systems, involve non unitary and non minimal models. The
presence of an infinite number of fields/representations makes their
study extremely complex, and no unifying principle has emerged so far.
Great ingeniosity has been devoted obtaining a core of deep and
interesting but partial, scattered and sometimes controversial
results.

Concerning interfaces --for instance domain boundaries-- of critical
systems in two dimensions, the situation was until recently also quite
unsatisfactory. The few significant results obtained using conformal
field theory before the emergence of stochastic Loewner evolutions
were the outcome of highly clever craftsmanship and had nothing to do
with systematic techniques.  It should be stressed however that
formul\ae\ like Cardy's percolation probability distribution had not
escaped the notice of mathematicians, and have been a source of
motivation for them that has finally lead to Schramm's breakthrough.

Analysis of the interplay between conformal field theory and
stochastic Loewner evolutions leads to a very exciting and positive
message. The conformal field theories needed to understand interfaces
have many nasty features, non minimality, non unitarity, etc. However
for the first time physicists have a rigorous mathematical parapet,
they can check their predictions and learn how to tame the pathologies
that have prevented systematic progress until now. We are a long way from
such an horizon, but in the long run this might be the main impact of
stochastic Loewner evolutions in physics.

The Swiss army knife of axiomatic and/or constructive quantum field
theory contains in particular algebra and representation theory,
complex variables (for the analyticity of correlation functions and
the S matrix in axiomatic field theory) and measure theory (in
constructive quantum field theory). It is a happy accident, without
deep significance, that these tools are also at the heart of the
understanding of two dimensional critical interfaces that has emerged
at the turn of the millenium.

\vspace{.2cm}

Non local objects like interfaces are not classified by
representations of the Virasoro algebra but the reasoning that led O.
Schramm to the crucial breakthrough \cite{Schramm00}, i.e. the
definition of stochastic Loewner evolutions, rests on a fairly obvious
but cleverly exploited statement of what conformal invariance means
for an interface. Surprisingly it allows to turn this problem into
growth problem of Markovian character. From a na\"ive viewpoint, this
is one of the most surprising features of stochastic Loewner
evolutions. Maybe this is one of the reasons why they were not discovered
by the impressive army of conformal field theorists. After all, in a
statistical mechanics system with appropriate boundary conditions, a
complete domain boundary is present in each sample, any dynamics
building it piece after piece seems rather artificial, and
correlations between the pieces at not short range. The discrete
geometric random curves on which the interest of mathematicians has
focused also do not give a clue. While percolation and some of its
cousins and descendants can be very naturally viewed as growth
processes, this is more the exception than the rule. The case of self
avoiding walks is a significant example. The literature on the
subject repeatedly stresses that changing the length of a self
avoiding walk by one changes the measure globally in a complicated.

For some years, probabilistic techniques have been applied to
interfaces, leading to a systematic understanding that was lacking on
the conformal field theory side.  There is now a satisfactory
understanding of interfaces in the continuum limit.  However, from a
mathematical viewpoint, giving proofs that a discrete interface on the
lattice has a conformally invariant limit remains a hard challenge and
only a handful of cases has been settled up to now.

\vspace{.2cm}

The organization of these notes is as follows.

\vspace{.2cm}

Section \ref{sec:examples} is an informal presentation of discrete lattice
models, first of geometric random curves -- starting with the most
growth process like, percolation, and ending with the self avoiding
walk--, then of statistical mechanics domain boundaries --the Ising
model, the $O(n)$ models and $Q$-state Potts model--, ending with a
few growth processes that are not expected to be conformally invariant
in the continuum limit, like diffusion limited aggregation.

The first goal is to get some familiarity with the basic objects that are
studied in the rest of this report. In particular we show that
geometric random curves are easy to simulate and produce beautiful and
complicated patterns. We emphasize that many variants of these
geometric random curves are still to be discovered and studied. We
also recall that appropriate statistical mechanics models domain
boundaries are described by geometric random curves.

\vspace{.2cm}

Section \ref{sec:anal} introduces Loewner chains which are one of the
basic tools to describe growth process in two dimensions. Riemann's
mapping theorem states that two domains (= connected and simply
connected open sets different from $\mathbb C$ itself) are conformally
equivalent. This allows to use a fixed simple reference domain, which
is usually taken as the upper-half plane or the in/out side of the
unit disk. This conformal equivalence is unique once an appropriate
normalization, which may depend on the growth problem at hand, has
been chosen. Cauchy's theorem allows to write down an integral
representation for the conformal map as an integral along the boundary
of the reference domain, involving a (positive because of growth)
density which is time dependent. A nice way to specify the growth rule
is often directly on this density. The time derivative of the
conformal map has an analogous representation, leading to an equation
called a Loewner chain. Local growth is when the density is a finite
sum of Dirac peaks. The positions of these peaks are functions of time
and serve as of the Loewner evolution.  This case is the most
important for the ensuing study.

\vspace{.2cm}

Schramm-Loewner evolutions (also called stochastic Loewner
evolutions), the object of section \ref{sec:sle} occur when the
Loewner evolution measure is a single delta peak and the associated
parameter is a Brownian motion. We reproduce Schramm's argument that
this is exactly the setting that describes conformally invariant
measures on random curves. SLE has a number of avatars, depending on
whether the random curves go from one boundary point to another to
another boundary point --chordal SLE--, to a point in the bulk
--radial SLE-- or to an interval on the boundary --dipolar SLE--.  The
diffusion coefficient, i.e. the normalization of the Brownian motion
$\kappa$ is the only  parameter, and qualitative and
quantitative features of SLE$_{\kappa}$ samples depend on it.
SLE$_{\kappa}$ can be generalized to SLE$_{\kappa,\rho}$ which we
review briefly.  The group theoretic formulation of the various SLEs
as a random processes on groups is also presented

\vspace{.2cm}

Section \ref{sec:statmech} makes contact with statistical mechanics
and the interplay between the measure on domain boundaries and the
full initial measure on configuration. Roughly speaking, to check that
a measure on random curves is inherited from a statistical mechanics
model, one has to check that a correlation function with fixed domain
boundary, when averaged over the random curve measure supposed to
describe the domain boundary, yields back the original correlation
function. We rephrase this statement in terms of martingales.  These
observations, which are in general of little use --not only because
nobody has a measure on domains boundaries to offer but also because
the computation of correlation functions with fixed domain boundaries
is well out of reach-- becomes very efficient when conformal
invariance is imposed. Indeed conformal field theory is able to reduce
kinematically correlation functions in any domain to correlation
function in a reference domain, and the measure on domain boundaries
is an SLE. Hence It\^o calculus becomes an efficient tool.  This
strategy is made explicit in the operator formalism for the variants
of SLE introduced before. Its predictive power is illustrated on how
it leads naturally to multiple SLEs.

\vspace{.2cm}

Section \ref{sec:delicat} is concerned with geometric structures and
properties of SLE samples. The locality property of SLE$_6$ (related
to percolation) and the restriction property of SLE$_{8/3}$ (related
to self avoiding walks) are presented. The application to the fractal
dimension of the exterior perimeter of Brownian excursion is
explained. Duplantier's predictions concerning the fractal spectrum of
harmonic measures of conformally invariant hulls are also presented.
The section ends with a friendly introduction to the Brownian loop
soup.

\vspace{.2cm}

Section \ref{sec:comput} illustrates how to compute explicit
significant properties of  SLE using tools from stochastic calculus
and/or conformal field theories. Boundary hitting probabilities,
crossing formul\ae\, fractal dimensions, etc are computed. The last
part is devoted to a list of references to other important results.

\vspace{.2cm}

Section \ref{sec:othergrowth} is an introduction to the study of more
general growth processes via discrete and continuous time Loewner
chains. The relationship between Laplacian growth and integrability is
presented.

\vspace{.2cm}

For the sake of completeness, we have included two appendices. While
appendix \ref{app:cft} on conformal field theory basics is rather
short, appendix \ref{app:proba} is a more substantial --but of course
very limited-- introduction to probabilistic methods and stochastic
processes. This appendix contains enough material to help understand
the probabilistic tools used systematically in the rest of the report:
martingales, Brownian motion, It\^o calculus. It seemed to us that
these subject are not so familiar to physicists and that systematic
reference to the probabilistic literature (excellent as it can be)
would be awkward. This has not prevented us from giving a list of
books that have proved valuable for us.

%% file: chap1.tex
 
Before we embark on more formal aspects, it is good to give a few
explicit examples of the kind of structures that we aim to study, i.e.
conformally invariant random curves in two dimensions. 

SLE gives a description of such objects directly in the continuum, but
the starting point is usually a discrete model of random curves on a
lattice. It is a tough job, only achieved for a handful of cases at
the time of this writing, to start from such a definition and show
that in the continuum limit one recovers a conformally invariant
probability distribution. The variety of examples will amply show that
a general heuristic criterion to decide whether or not a given
discrete interface distribution has a conformally invariant continuum
limit is not so easy to exhibit. In quantum field theory, it is not
easy to exhibit local field theories which are scale invariant but not
conformally invariant \cite{CardyRiva}, and there is a heuristic
argument based on locality\footnote{With the quantum field theory
  meaning.} to explain why it is so. But a similar heuristic argument
for SLE does not exist. We shall make a few remarks on this in the sequel.

Another feature of SLE is to present the random curves as growth
processes: SLE gives a recipe to accumulate (infinitesimal) pieces
on top of each other, with a form of Markov property to be elucidated
below. For discrete models, a natural growth process definition is
more the exception than the rule.

Let us also note that the favorite examples in the mathematics and
physics community are not the same. Physicists are used to start from
lattice models where each lattice site carries a degree of freedom,
and the random distribution of these degrees of freedom is derived
from a Boltzmann weight, i.e. an unnormalized probability
distribution. In the presence of appropriate boundary conditions, some
one dimensional defects appear. The weight of a defect of given shape
can (in principle) be obtained by summing the Boltzmann weights over
all configurations exhibiting this defect. On the other hand,
mathematicians have often concentrated on interfaces with a more
algorithmic and direct definition. For the cost of numerical
simulations, this makes a real difference. At a more fundamental level
however, the distinction is artificial because it is usually possible
to cook up Boltzmann weights (for local degrees of freedom and with
local interactions) that do the job of reproducing an interface
distribution defined by more direct means or at least an interface
distribution which is in the same universality class.

The model whose definition fits best with the image of a growth
process is percolation, and we shall start with it. The growth aspect
of the two next examples, the harmonic navigator (the GPL version of
Schramm's harmonic explorer) and loop-erased random walks, is only
slightly less apparent. But self avoiding random walks to be
introduced right after are of a quite different nature. We shall
illustrate these cases with baby numerical simulations, referring the
interested readers to the specialized literature for careful and
clever large scale studies \cite{Wilson96} and
\cite{TKen02a,TKen02b,TKen02c}. Our aim is mainly to give
some concrete pictures of these remarkably beautiful objects. We
shall also see on concrete examples that the landscape of algorithms
used to produce the curves is rather varied and largely unexplored,
sheltering fundamental problems.

We shall then consider interfaces defined via lattice models in the
cases of the Potts and $O(n)$ models, with some pictures for the Ising
model.

We shall finally define diffusion limited aggregation (DLA), a growth
process which is expected to have a scaling but no conformally
invariant continuum limit. DLA, together with its cousins and
descendants, will reappear at the end of these notes because many of
those can be defined via Loewner chains. 

We start with some basic definitions. 

In the sequel we shall often need the notion of a lattice domain.  

A square lattice domain $\mathbb{D}$ is a domain in the usual sense,
which can be decomposed as a disjoint union of open squares with side
length 1 (faces), open segments of length 1 (edges) and points
(vertices), in such a way that each open segment belongs to the
boundary of two open squares and each vertex belongs to the boundary
of four open segments. Unless stated explicitly, we assume that the
number of faces is finite. 

An admissible boundary condition is a couple of distinct points
$(a,b)$, $a,b\notin \mathbb{D}$ such that there is a path from $a$ to
$b$ in $\mathbb{D}$ i.e. a number $n\geq 1$ and a sequence
$s_1,\cdots,s_{2n+1}$ where $a=s_1,b=s_{2n+1}$, the $s_{2m+1}$, $1
\leq m <n$, (if any) are distinct vertices of the decomposition of
$\mathbb{D}$ and the $s_{2m}$, $1 \leq m <n$, are distinct edges of
the decomposition of $\mathbb{D}$ with boundary
$\{s_{2m-1},s_{2m+1}\}$.  Any such path splits $\mathbb{D}$ into a
left and a right piece.

If $s_1,\cdots,s_{2n+1}$ is a path from $a$ to $b$ in
$\mathbb{D}$ and $0 \leq m <n$, the set $\mathbb{D}'$ obtained by
removing from $\mathbb{D}$ the sets $s_l$, $1 < l \leq s_{2m+1}$ is
still a domain, and $(s_{2m+1},b)$ is an admissible boundary
condition for $\mathbb{D}'$.

\begin{figure}[htbp]
\begin{center}
\includegraphics[width=\textwidth]{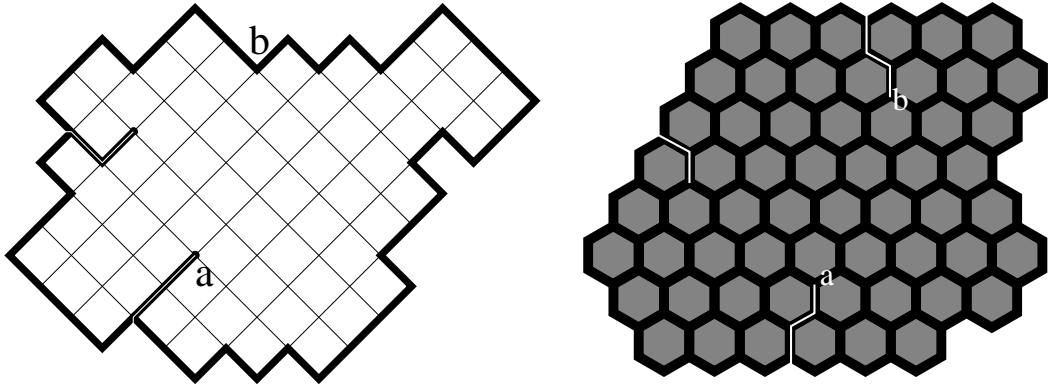}
\caption{Example of square and hexagonal lattice domains.}
     \label{fig:latdoms}
 \end{center}
\end{figure}

Similar definitions and properties would hold for an hexagonal lattice
domain, regular hexagons with (say) side of length 1 replacing the
squares, and three replacing four. The two examples in
fig.\ref{fig:latdoms} will probably make obvious what kind of domain
we have in mind. 

Our main interest in the next subsections will be in measures on paths
from $a$ to $b$ in $\mathbb{D}$ when $\mathbb{D}$ is a lattice domain
and $(a,b)$ an admissible boundary condition.

\begin{figure}[htbp]
\begin{center}
\includegraphics[width=\textwidth]{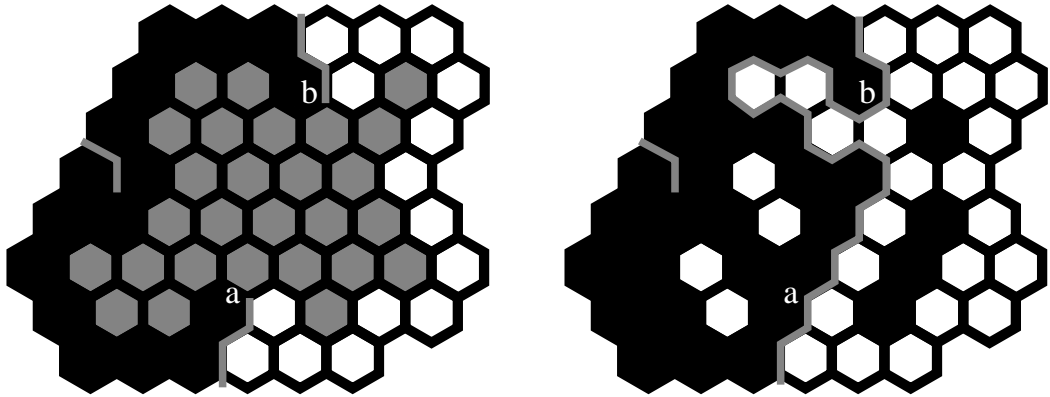}
\caption{Left : the boundary of an hexagonal lattice domain with
  boundary conditions. Right : the interface associated to a configuration.}
     \label{fig:hexample}
 \end{center}
\end{figure}

Hexagonal lattice domains have useful special properties.  Suppose
$(\mathbb{D},a,b)$ is an hexagonal lattice domain with admissible
boundary condition. The right (resp. left) hexagons are by definition
those which are on the right (resp. left) of every path from $a$ to $b$
in $\mathbb{D}$. Left and right hexagons are called boundary hexagons.
The other hexagons of $\mathbb{D}$ are called inner
hexagons\footnote{Note that being a boundary or an inner hexagon
  depends on $(a,b)$.}. Color the left hexagons in black (say) and the
right hexagons in white as in Fig.\ref{fig:hexample} on the left.
If one colors the inner hexagons arbitrarily in black or white, then there is
a single path from $a$ to $b$ in $\mathbb{D}$ such that the hexagon on
the left (resp. right) of any of its edges is black (resp. white).
This is illustrated in fig.\ref{fig:hexample} on the right. This path
can be defined recursively because $a$ is on the boundary of at least
one left and at least one right hexagon: as $a$ is not in
$\mathbb{D}$, in any coloring there is exactly one edge in
$\mathbb{D}$ with $a$ on its boundary and bounding two hexagons of
different colors. Start the path with this edge and go on.

All the examples of interfaces we shall deal with in the sequel can be
defined on arbitrary hexagonal lattice domain with admissible boundary
condition, though sometimes we shall use square lattice domains.
Geometrical examples will define directly a law for the interface or a
probabilistic algorithm to construct samples. Examples from
statistical mechanics will give a weight for each coloring of the
inner hexagons, and the law for the interface will be derived (at
least in principle) from this more fundamental weight. The model of
interface can depend on some parameters, called collectively $p$ (for
instance, temperature can be one of those).

Because arbitrary domains can be used, the statement of conformal
invariance is non trivial and can be checked numerically. Fix an
interface model and take a sequence of lattice domains
$(\mathbb{D}_n,a_n,b_n)$ and of positive scales $s_n \rightarrow 0^+$
such that (in an obvious notation) $s_n(\mathbb{D}_n,a_n,b_n)$
converges to a domain with two boundary points marked,
$(\mathbb{D},a,b)$. A continuum limit exists when there is a (domain
independent) function $p(s)$ such that the distribution of interfaces
in $s_n(\mathbb{D}_n,a_n,b_n)$ with parameters $p(s_n)$ converges to
some limit. Then, different domains can be compared and conformal
invariance can be checked on good lattice approximations of these
domains.

\subsection{Geometrical examples}
\label{sec:geoexamp}

\subsubsection{Percolation}
\label{sec:perco}

Let $(\mathbb{D},a,b)$ be an hexagonal lattice domain with admissible
boundary condition. Color the left hexagons in black (say) and the
right hexagons in white. A configuration is a choice of color (black
or white) for the inner hexagons. Give each configuration the same
probability. Equivalently, the colors of the inner hexagons are
independent random variables taking each color with probability $1/2$.
We could also introduce some asymmetry between the colors, but our main
interest will be in the symmetric case, because it has a continuum
limit, without adjusting any parameters. 

\begin{figure}[htbp]
\begin{center}
\includegraphics[width=0.7\textwidth]{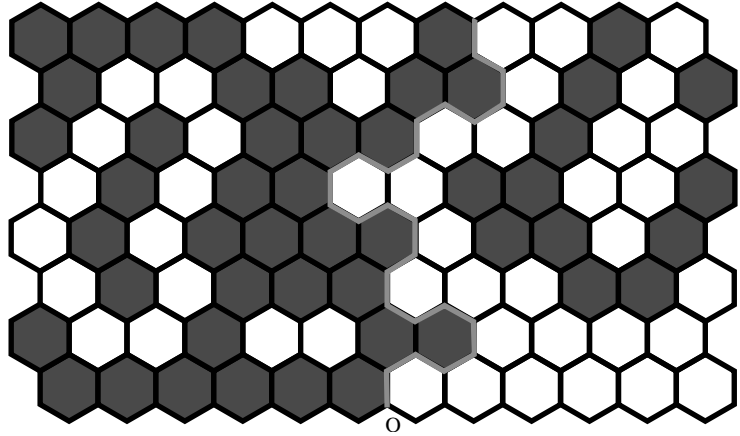}
\caption{The definition of the percolation interface.}
     \label{fig:percodef}
 \end{center}
\end{figure}

As recalled above, each configuration defines an interface, i.e. the
unique path from $a$ to $b$ in $\mathbb{D}$ such that the hexagon on
the left (resp. right) of any of its edges is black (resp. white), see
fig.\ref{fig:percodef}. Hence the probability distribution on
configurations induces a probability distribution on paths from $a$ to
$b$ in $\mathbb{D}$.  This is called the (symmetric) percolation
probability distribution.

Because inner hexagon colors are independent, it is easy to compute
the probability of a percolation path from $a$ to $b$ in
$\mathbb{D}$~: if a path has an edge in common with $l$ distinct inner
faces of $\mathbb{D}$, its probability is $2^{-l}$. The weight is
given by a purely local rule. If $(\mathbb{D}',a,b)$ is another
hexagonal lattice domain with admissible boundary condition, a path
common to $(\mathbb{D},a,b)$ and $(\mathbb{D}',a,b)$ touching the same
boundary and inner hexagons in both domains has the same probability
in both domains : the percolation interface does not depend on the
distribution of black and white sites away from itself. This is called
\textit{locality}, a property that singles out percolation. 

\begin{figure}[htbp]
\begin{center}
\includegraphics[width=\textwidth]{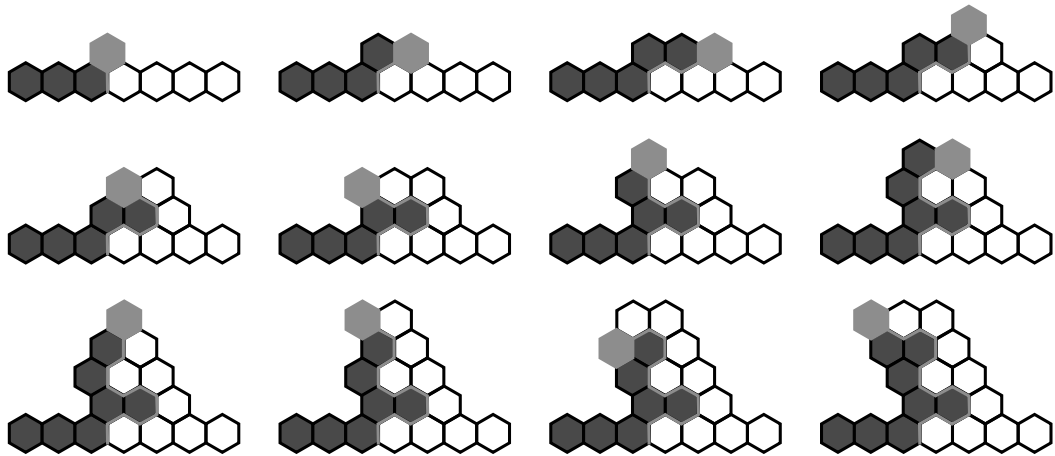}
\caption{The percolation interface as a growth process.}
     \label{fig:percolgrowth}
 \end{center}
\end{figure}

In particular, locality allows to view percolation as a simple growth
process, defined as follows. If $a$ is incident to
no inner hexagon of $\mathbb{D}$, there is no choice in the first step
of a path from $a$ to $b$ in $\mathbb{D}$. Else, $a$ is incident to
exactly one inner hexagon of $\mathbb{D}$. Color it black or white
using a fair coin, and make a step along the edge of $\mathbb{D}$
adjacent at $a$ whose adjacent faces have different colors. Then
remove from $\mathbb{D}$ the edge corresponding to the first step and
its other end point, call it $\dot{a}$ to get a new domain
$\dot{\mathbb{D}}$. If $\dot{a}=b$ stop. Else
$(\dot{\mathbb{D}},\dot{a},b)$ is a new hexagonal domain with admissible
boundary condition and one can iterate as shown
on the fig.\ref{fig:percolgrowth}.

There is exactly one coin toss for each inner face of $\mathbb{D}$
touching an edge of the path : this toss takes place the first time
the inner face is touched by the tip of the path. In the rest of the
process, this face becomes a boundary hexagon. Hence this growth
process gives the percolation measure.

A geometry which is of frequent use is to pave the upper-half plane
with regular hexagons and impose that the left (resp. right) hexagon
be those intersecting the negative (resp. positive) real axis.  This
is an example with an infinite number of faces. No limiting procedure
(taking larger and larger finite approximations of the upper-half
plane) is necessary to get the correct weight for the initial steps of
the percolation interfaces, again because of locality.

\begin{figure}[htbp]
\begin{center}
\includegraphics[width=0.85\textwidth]{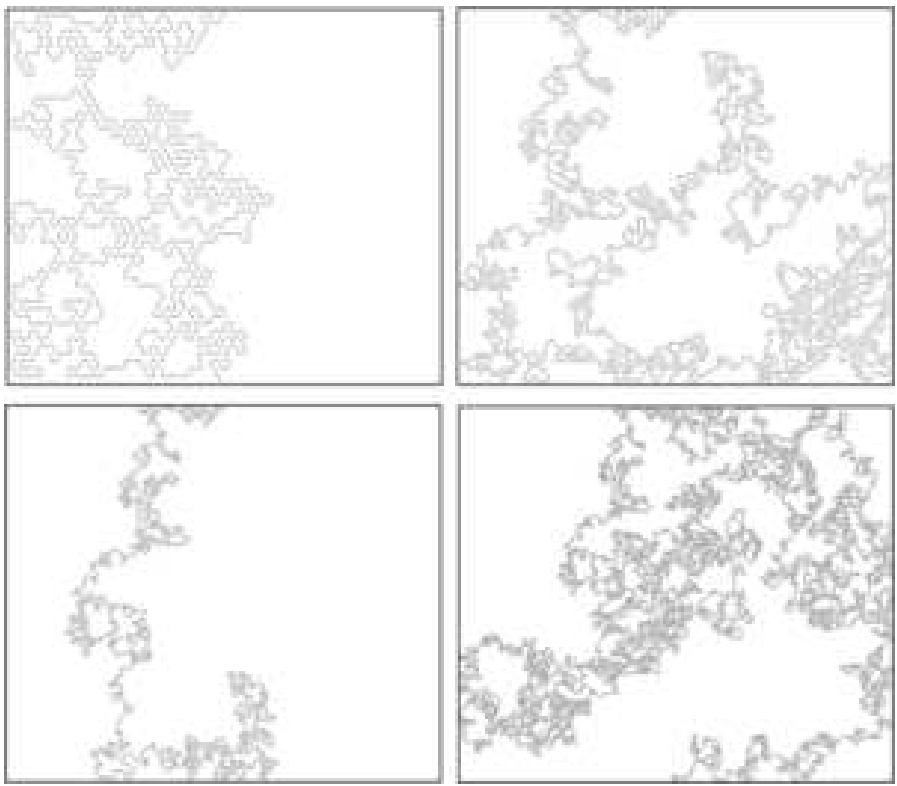}
\caption{Samples of the percolation interface for increasing
  sizes.}
      \label{fig:percosamples}
\end{center}\end{figure}

Fig.\ref{fig:percosamples} shows a few samples.
They join the middle horizontal sides of similar rectangles of
increasing size. The pseudo random sequence is the same for the four samples. 

Even for small samples, the percolation interface makes many twists
and turns.  By construction, the percolation interface is a simple
curve, but with the resolution of the figure, the percolation
interface for large samples does not look like a simple curve at all!

To estimate the (Hausdorff, fractal) dimension of the percolation
interface, we have generated samples in similar rectangular domains of
different sizes and made the statistics of the number of steps $S$ of
the interface as a function of the size $L$ of the rectangle domain.
One observes that $S \propto L^{\delta}$ and a modest numerical effort
(a few hours of CPU) leads to $\delta=1.75\pm.01$.

The percolation interface is build by applying local rules involving
only a few nearby sites, and we could wave our hands to argue that its
scale invariance should imply its conformal invariance in the
continuum limit. But the percolation process is one among the few
systems that has been rigorously proved to have a conformally
invariant distribution in the continuum limit, the fractal dimension
being exactly $7/4$. As suggested by numerical simulations, the
continuum limit does not describe simple curves but curves with a
dense set of double points, and in fact the --to be defined later--
SLE$_6$ process describes not only the percolation interface but also
the percolation hull, which is the complement of the set of points
that can be joined to infinity by a continuous path that does not
intersect the percolation interface.  As we shall see later, among
SLE$_\kappa$'s, SLE$_6$ is the only one that satisfies locality, so
what is really to prove in this case is conformal invariance in the
continuum limit (a nontrivial task), and the value of $\kappa$ is for
free.

\subsubsection{Harmonic navigator}
\label{sec:navig}

\begin{figure}[htbp]
\begin{center}
\includegraphics[width=0.85\textwidth]{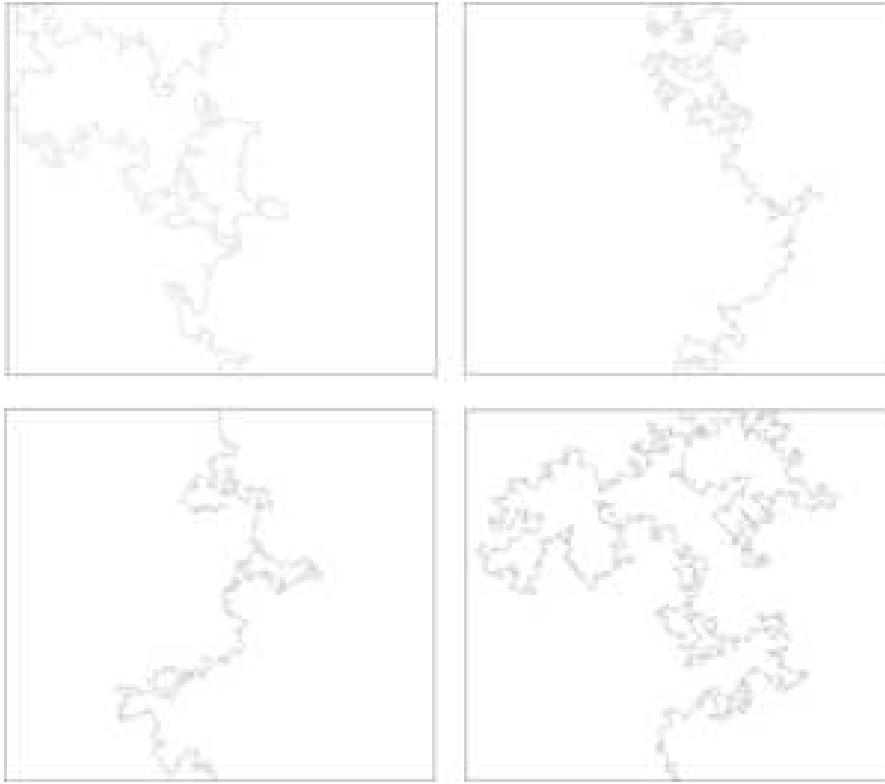}
\caption{Samples of the harmonic navigator for increasing sizes.}
      \label{fig:navisamples}
\end{center}\end{figure}

The harmonic navigator\footnote{We prefer the name ``navigator'' to
  the more standard ``explorer'' used by Schramm to avoid any
  Microsoft licence problem.} is a simple extension of the percolation
process. The only difference is in the way randomness enters whenever
a color choice for an hexagon has to be made. For percolation, one
simply tosses a fair coin. For the harmonic navigator, the choice
involves the spatial distribution of the boundary hexagons. Note that
not only the initial boundary hexagons, but also the ones colored
during the beginning of the process are considered as boundary
hexagons. Explicitly, a symmetric random walk is started at the
hexagon to be colored. The walk is stopped when it hits the boundary
for the first time. The color of the starting point is chosen to be
the color of the end point.  To put this differently, the boundary
splits into two pieces of different colors, and one tosses a coin
biased by the discrete harmonic measure of the two boundary pieces
seen from the hexagon to be colored. Fig.\ref{fig:navisamples} shows
a few samples in domains of increasing size.

We have also estimated the fractal dimension of the harmonic
navigator. One finds a number close to $\delta=1.50\pm.01$. Again an
accuracy of two significant digits can be achieved in a few hours of
CPU.  The computation time is of much longer than for percolation, and
the ratio of the two does grows slowly when the size of the
rectangular domain is changed. This is related to familiar properties
of random walks : quite often, the random walk finds the boundary
quickly, and hits it at a point nearby its starting point, most often
at an hexagon bounding the growing interface. However, a look at the
samples, obtained via the same pseudo random sequence but sharing only
a modest initial portion, gives convincing evidence that from time to
time, the walk hits the boundary far away from the interface. We shall
come back to this later.

The study of the convergence, in the
continuum limit, of the harmonic explorer to level lines in Gaussian
(free) field theory and to SLE$_4$ (whose fractal dimension is
exactly $3/2$) has seen important recent developments
\cite{SchrammSheffield03,SchrammSheffield05}.

The definition of the harmonic navigator can be extended in many
directions. 

\begin{figure}[htbp]
\begin{center}
\includegraphics[width=0.85\textwidth]{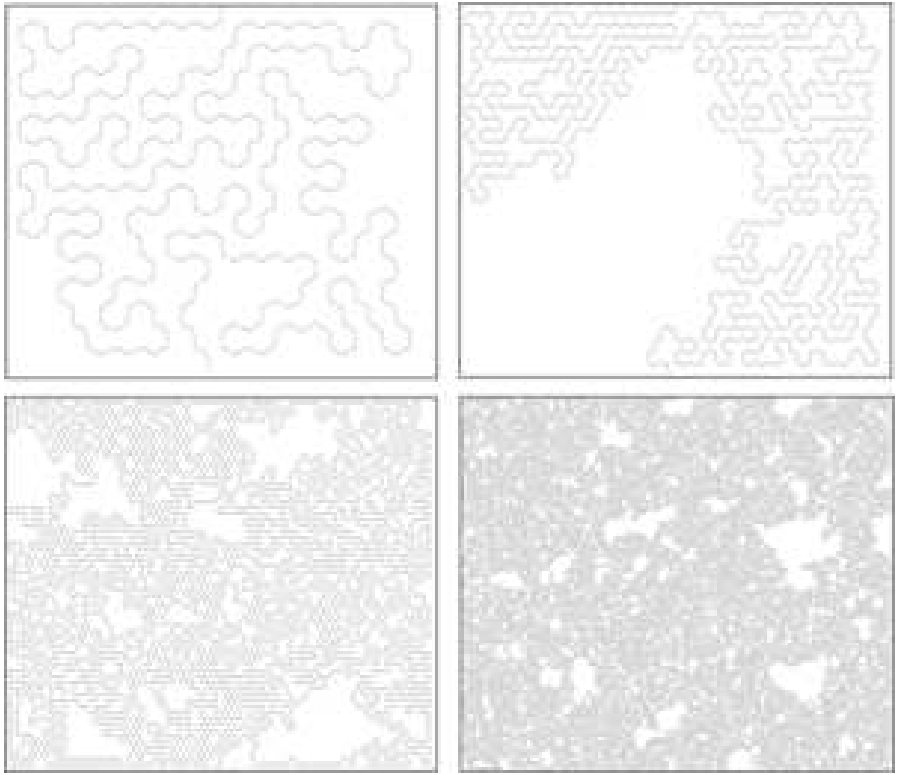}
\caption{Samples of the harmonic anti-navigator for increasing sizes.}
      \label{fig:antinavisamples}
\end{center}\end{figure}

\underline{The harmonic anti-navigator}. Observe that if the
neighborhood of an hexagon to be colored contains much more hexagons
of one color than of the other, then with high probability it will
get colored by the most abundant color. This means a repulsive force
or excluded volume that tends to prevent the path from coming too
close to another piece of itself. What if one decided to make the
opposite color choice at each step? Then the resulting object would be
much more dense, as confirmed by Fig.\ref{fig:antinavisamples} which
shows a few samples in domains of increasing size.

But does the harmonic anti-navigator
have an interesting continuum limit? Is it related to conformal
invariance? 

\begin{figure}[htbp]
\begin{center}
\includegraphics[width=0.85\textwidth]{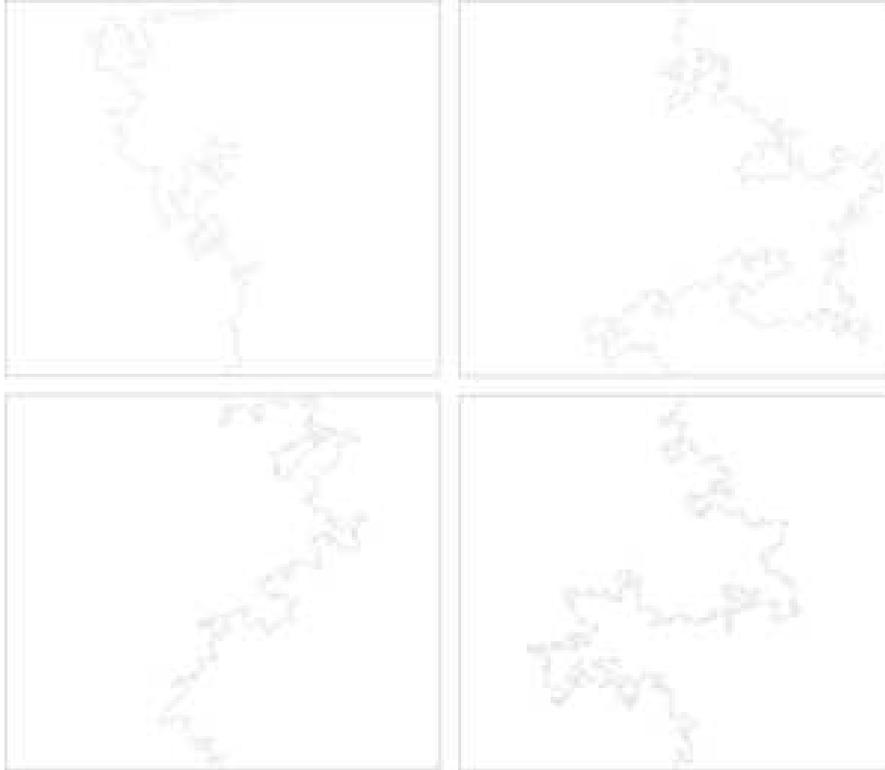}
\caption{Samples of the percolation navigator for increasing sizes.}
      \label{fig:perconavisamples}
\end{center}\end{figure}

\underline{The percolation navigator}. What if we would replace the
random walk by other processes that hit the boundary with probability
1~? This means replacing the harmonic measure by another measure. For
instance, we could start a percolation process at the tip of the
growing interface, see the color of the boundary at the first hitting
point and use this color for the new hexagon. It seems that nothing is
known about this process. The samples in Fig.\ref{fig:perconavisamples}
lead to expect nice fractals in the continuum limit. The fractal
dimension can be estimated to be $\delta\sim 1.42$ and does not look
like a simple number.

\begin{figure}[htbp]
\begin{center}
\includegraphics[width=0.85\textwidth]{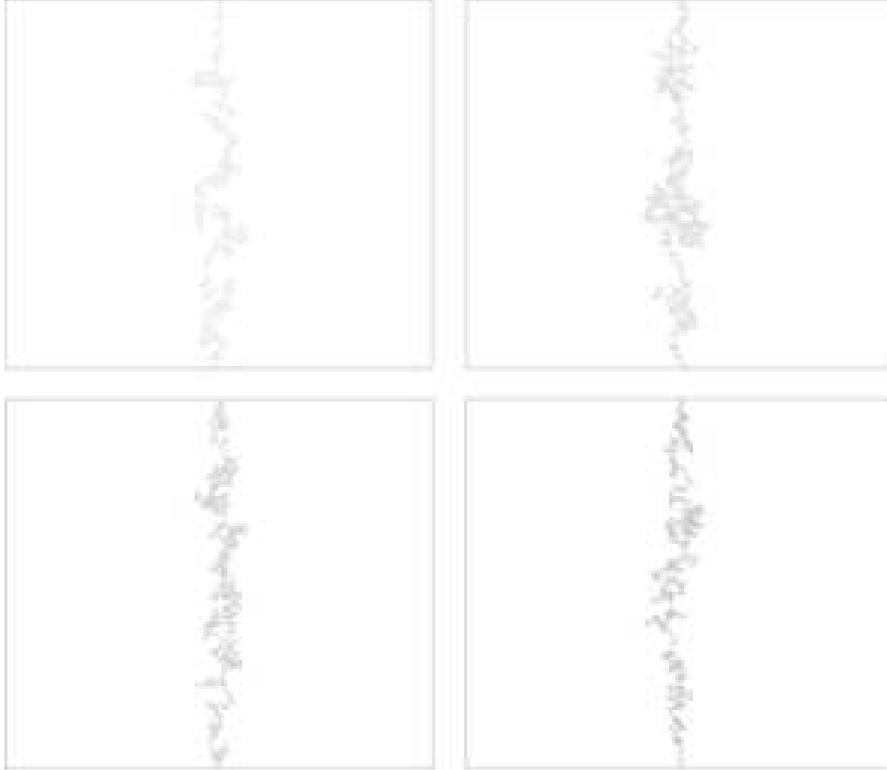}
\caption{Samples of the boundary harmonic navigator for increasing sizes.}
      \label{fig:naviboundarysamples}
\end{center}\end{figure}

\underline{The boundary harmonic navigator}. Yet another deformation
of the harmonic navigator would be to keep only the initial boundary
to compute the measure, i.e. let the interface be transparent to the
random walk. In that case, the probability to color some hexagon in
black or white depends only on the position of the hexagon, but not on
the beginning of the interface. In fact one can color each inner
hexagon by tossing a coin biased by the harmonic measure of the left
and right boundaries seen from the hexagon. This leads to a
statistical mechanics model with independent sites, and the
probability of a given interface is just the product of the
probabilities for all inner hexagons that have at least one edge on
the interface. Hence, this process is similar to inhomogeneous
percolation. The effect of the bias is a repulsive force away from the
boundary of the initial domain and in the long range, the interfaces
has a tendency to remain in regions where the bias is small and
explore only a small part of the available space. On the other hand,
in regions where the bias is small, at small scales the interface will
look like percolation i.e. make many twists and turn. This is indeed
the case, as shown on Fig.\ref{fig:naviboundarysamples}. Due to the
competition between small and large scales, conformal invariance is
not expected. The CPU time needed to draw a sample is now much larger
and grows faster when the size increases because the random walk has
to explore space until it hits the initial boundary.

This is the first process that we meet for which removing the
beginning of the path from the domain and starting the process for the
cut domain at the tip is not the same as continuing the process in the
initial domain. Thus this process does not have the so-called domain Markov
property, an important feature of conformally invariant interfaces to
which we shall come back later.

In fact all these variations --and many others-- can be mixed.
Deciding which one leads to a conformally invariant continuum limit is
not so obvious. This illustrates that the landscape of plausible
algorithms is vast and largely unexplored. There is room for numerical
experiments and a lot of theoretical work.

\subsubsection{Loop-erased random walks}
\label{sec:lerw}

This example still keeps some aspects of a growth process, in that new
pieces of the process can be added recursively. A loop-erased random
walk is a random walk with loops erased along as they appear. More
formally, if $X_0,X_1,\cdots,X_n$ is a finite sequence of abstract
objects, we define the associated loop-erased sequence by the
following recursive algorithm.

\vspace{.2cm}

\noindent Until all terms in the sequence are distinct,\\ 
\textbf{Step 1} Find the couple $(l,m)$ with
$0\leq l <m$ such that the terms with indexes from $0$ to $m-1
$ are all distinct but the terms with indexes $m$ and $l$ coincide. \\
\textbf{Step 2} Remove the terms with indexes from $l+1$ to $m$, and shift
the indexes larger than $m$ by $l-m$ to get a new sequence.

\vspace{.2cm}

Let us look at two examples. \\
For the ``month'' sequence \textit{j,f,m,a,m,j,j,a,s,o,n,d}, the first loop
is \textit{m,a,m}, whose removal leads to \textit{j,f,m,j,j,a,s,o,n,d}, then
\textit{j,f,m,j}, leading to \textit{j,j,a,s,o,n,d}, then \textit{j,j} leading to
\textit{j,a,s,o,n,d} where all terms are distinct. \\ 
For the ``reverse month'' sequence \textit{d,n,o,s,a,j,j,m,a,m,f,j}, the
first loop is \textit{j,j}, leading to \textit{d,n,o,s,a,j,m,a,m,f,j}, then
\textit{a,j,m,a} leading to \textit{d,n,o,s,a,m,f,j}. 

This shows that the procedure is not ``time-reversal'' invariant.
Moreover, terms that are within a loop can survive: in the second
example \textit{m,f}, which stands in the \textit{j,m,a,m,f,j} loop,
survives because the first \textit{j} is inside the loop
\textit{a,j,m,a} which is removed first.

\begin{figure}[htbp]
\begin{center}
\includegraphics[width=0.6\textwidth]{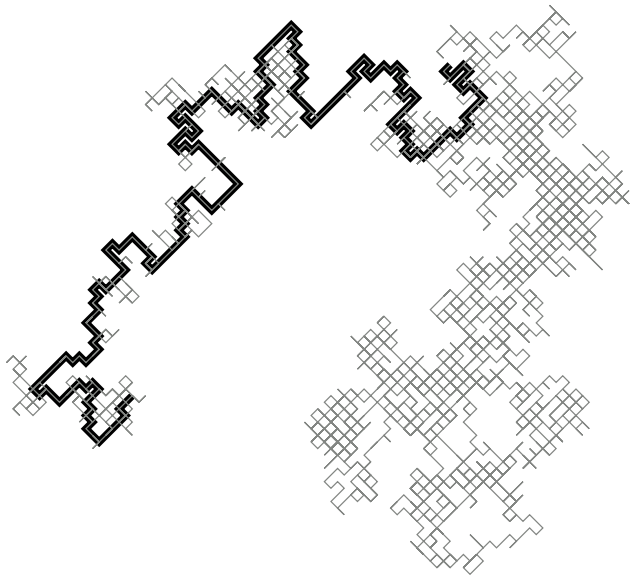} 
\end{center}
\caption{A loop-erased random walk with its shadow.}
     \label{fig:lerwtrace}
\end{figure}

A loop-erased random walk is when this procedure is applied to a (two
dimensional for our main interest) random walk. In the full plane this
is very easy to do. Fig.\ref{fig:lerwtrace} represents a loop-erased
walk of 200 steps obtained by removing the loops of a 4006 steps
random walk on the square lattice. The thin grey lines build the
shadow of the random walk (where shadow means that we do not keep
track of the order and multiplicity of the visits) and the thick line
is the corresponding loop-erased walk. The time asymmetry is clearly
visible and allows to assert with little uncertainty that the walk
starts on the top right corner.

The same procedure can be applied to walks in the upper half plane.
There are a few options for the choice of boundary conditions. \\
A first choice is to consider reflecting boundary conditions on the
real axis for the random walk. \\
Another choice is annihilating boundary conditions: if the random walk
hits the real axis, one forgets everything and starts anew at the
origin. Why this is the natural boundary condition has to wait until
Section \ref{sec:dmp}.

\begin{figure}[htbp]
\begin{center}
\includegraphics[width=\textwidth]{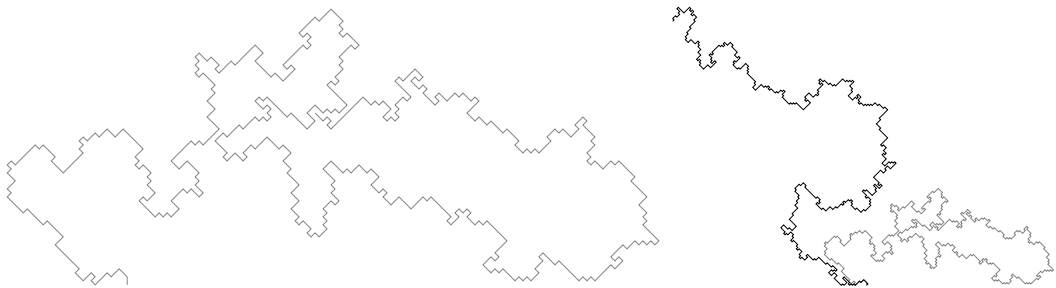}
\end{center}
\caption{On the left: a large loop is about to be created. 
On the right: the massive rearrangement to go from 633 to 634
steps.}
      \label{fig:lerw633-634}
\end{figure}

Due to the fact that on a two-dimensional lattice a random walk is
recurrent (with probability one it visits any site infinitely many
times), massive rearrangement occur with probability one. This is
already apparent on the small sample Fig.\ref{fig:lerwtrace} and means
that if one looks at the loop-erased random walk associated to a given
random walk, it does not have a limit in any sense when the size of
the random walk goes to infinity. Let us illustrate this point. The
samples in Fig.\ref{fig:lerw633-634} were obtained with reflecting
boundary conditions. It takes 12697 random walk steps to build a
loop-erased walk of length 633, but step 12698 of the random walk
closes a long loop, and then the first occurrence of a loop-erased
walk of length 634 is after 34066 random walk steps.  Observe that in
the mean time most of the initial steps of the loop-erased walk have
been reorganized.

However, simulations are possible because when the length of the
random walk tends to infinity, so does the maximal length of the
corresponding loop-erased walk with probability one: there are times
at which the loop-erased walk associated to a random walk will reach
any number of steps $S$ ascribed in advance. If one stops the
procedure the first time this happens, the random walk measure induces
a measure on non-intersecting walks of $S$ steps which can be taken as
a definition of the loop-erased random walk measure.

\vspace{.3cm}

In a square lattice domain with admissible boundary condition
$(\mathbb{D},a,b)$ we make the annihilating choice to define the
loop-erased random walk measure. Consider all walks from $a$ to $b$
that do not touch the boundary except at $a$ before the first step and
at $b$ after the last step and give each such walk of length $l$ a
weight $4^{-l}$.  Then erase the loops to get a probability
distribution for loop-erased random walks from $a$ to $b$ in the
domain. Observe that this choice is exactly the annihilating boundary
condition. The probability for the simple symmetric random walk to hit
the boundary for the first time at $b$ starting from $a$ can be
interpreted as the partition function for loop-erased walks.  A simple
but expansive way to make simulations is to simulate simple random
walks starting at $a$ and throw away those which hit the boundary
before they leave at $b$.

Though annihilating boundary conditions lead to remove even more parts
of the random walk than the reflecting ones, the corresponding process
in the upper half plane can be arranged (conditioned in probabilistic
jargon) to solve the problem of convergence as follows.

Instead of stopping the process when the loop-erased walk has reached
a given length, one can stop it when it reaches a certain altitude,
say $n$, along the $y$-axis. Whatever the corresponding random walk
has been, the only thing that matters is the last part of it,
connecting the origin to altitude $n$ without returning to altitude
$0$. Moreover, the first time the loop-erased walk reaches altitude
$n$ is exactly the first time the random walk reaches altitude $n$.
Now a small miracle happens: if a 1d symmetric random walk is
conditioned to reach altitude $n$ before it hits the origin again, the
resulting walk still has the Markov property. It is a discrete
equivalent to the 3d Bessel process (a Bessel process describes the
norm of a Brownian motion, however no knowledge of Bessel processes is
needed here, we just borrow the name). When at site m, $0 < m <n$, the
probability to go to $m\pm1$ is $(1\pm 1/m)/2$, independently of all
previous steps. Observe that there is no $n$ dependence so that we can
forget about $n$, i.e. let it go to infinity.  The discrete 3d Bessel
process is not recurrent and tends to infinity with probability one:
for any altitude $l$ there is with probability one a time after which
the discrete 3d Bessel process remains above $l$ for ever. Henceforth,
we choose to simulate a symmetric simple random walk along the $x$
axis and the discrete 3d Bessel process along the $y$-axis and we
erase the loops of this new process. This leads to the convergence of
the loop-erased walk and numerically to a more economical simulation.

\begin{figure}[htbp]
\begin{center}
\includegraphics[width=.8\textwidth]{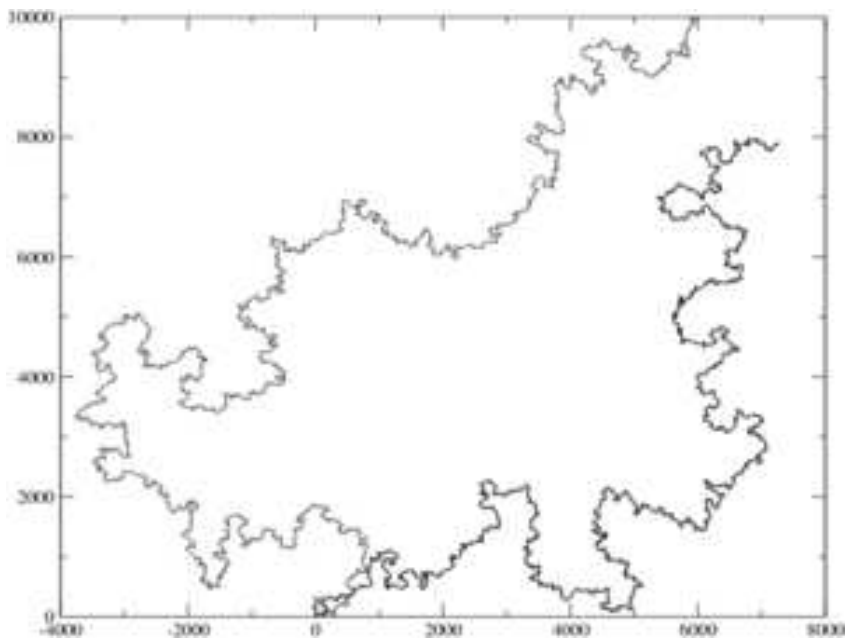}
\end{center}
\caption{A sample of the loop-erased random walk for the two boundary
conditions.}
      \label{fig:samplesLERW-BW}
\end{figure}

Fig.\ref{fig:samplesLERW-BW} is a simulation of about $10^5$ steps,
both for reflecting and annihilating boundary conditions. At first
glance, one observes in both cases similar simple (no multiple points)
but irregular curves with a likely fractal shape. The intuitive
explanation why a loop-erased random walk has a tendency not to come
back too close to itself is that if it would do so, then with large
probability a few more steps of the random walk would close a loop.

To estimate the Hausdorff dimensions in both cases, we have
generated samples of random walks, erased the loops and made the
statistics of the number of steps $S$ of the resulting walks compared
to a typical length $L$ (end-to-end distance for reflecting boundary
conditions, maximal altitude for annihilating boundary conditions). In
both cases, one observes that $S \propto L^{\delta}$ and again a modest
numerical effort (a few hours of CPU) leads to $\delta=1.25\pm.01$.
This is an indication that the boundary conditions do not change the
universality class. 

\begin{figure}[htbp]
\begin{center}
\includegraphics[width=0.5\textwidth]{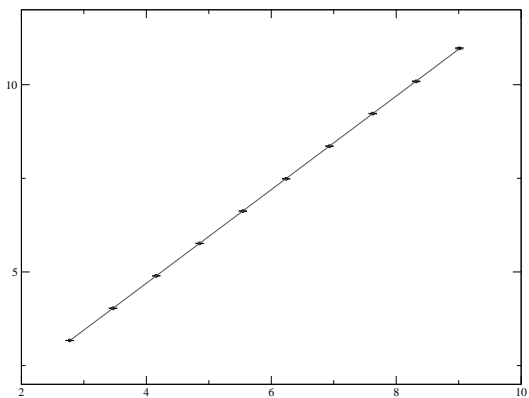}
\end{center}
\caption{The logarithm of the average length of the loop-erased random walk
  versus logarithm of the maximum altitude. The numerical results are
  the circles, the straight line is the linear regression, the error
  bars are shown.}
      \label{fig:moyenne-BW}
\end{figure}

To get an idea of how small the finite size corrections are, observe
Fig.\ref{fig:moyenne-BW}. The altitude was sampled from $2^4$ to
$2^{13}$. The best fit gives a slope $1.2496$
and the first two points already give $1.2403$.

As recalled in the introduction, it is believed on the basis of
intuitive arguments that in two dimensions scale invariance is almost
enough for conformal invariance, providing there are no long range
interactions.  What does this absence of long range interactions mean
for loop-erased random walks?  Clearly along the loop-erased walk
there are long range correlations, if only because a loop-erased
random walk cannot cross itself. A possibly more relevant feature is
that, in the underlying 2d physical space, interactions are indeed
short range. At each time step, the increment of the underlying random
walk is independent of the rest of the walk, and the formation of a
loop to be removed is known from data at the present position of the
random walk.

\vspace{.3cm}

 From the analytical viewpoint, the loop-erased random walk is one of
the few systems that has been proved to have a conformally invariant
distribution in the continuum limit, the fractal dimension being
exactly $5/4$. A naive idea to get directly a continuum limit
representation of loop-erased walks would be to remove the loops from
a Brownian motion.  This turns out to be impossible due to the
proliferation of overlapping loops of small scale. However, the
SLE$_2$ process, to be defined later, gives a direct definition.  In
fact, it is the consideration of loop-erased random walks that led
Schramm \cite{Schramm00} to propose SLE as a description of interfaces.

\subsubsection{SAW}
\label{sec:saw}

The self avoiding walk is one of the most important examples, and it
is known to lead to notoriously difficult questions. One of the
reasons is perhaps that a recursive definition is not known. And it is
likely that before the discovery of SLE few people would have bet that
the continuum limit of self avoiding walks would be described most
naturally as a (Markovian!) growth process. 

The statistical ensemble of self avoiding walks of $S$ steps can be
defined on an arbitrary simple graph. The probability space consists
of sequences of $S+1$ distinct adjacent vertices, and if not empty, it
is endowed with the uniform probability measure. Conditioning on the
initial and/or the end point leads to other ensembles, again with
uniform probability distribution. We are interested mainly in the case
when the graph is a simply connected piece of a 2-d lattice. One of
the difficulty is that if $S=S'+S''$ the $S'$ first steps obviously
build a self avoiding walks of length $S'$ but the number of possible
complements of length $S''$ depends on the first $S'$ steps, so that
the induced probability measure on the $S'$ first steps obtained by
summing over the last $S''$ steps is not uniform. So it is tricky to
produce samples of self avoiding walks by a recursive procedure. In
fact the most efficient way known at present to simulate self avoiding
walks is via a dynamical Monte Carlo algorithm. 

Let us pause for a second to recall the basic idea. To produce samples
of a finite probability space $\Omega=\{a,b,c,\cdots\}$ (which we can
assume to give a positive probability to each of its points), the
starting point of a dynamical Monte Carlo algorithm is to view the
points in the probability space as vertices of an abstract graph. The
task is then to define enough edges to make a connected graph and cook
up for each edge $\{a,b\}$ two oriented weights $z_{ab}$ to go from
point $a$ to point $b$ and $z_{ba}$ to go from $b$ to $a$ in such
a way that $p_a z_{ab}=p_b z_{ba}$ (detailed balance). Then a random
walk on the graph using the weights $z_{ab}$, with arbitrary initial
conditions, leads at large times to a stationary distribution which is
exactly the probability distribution one started with. The art is in a
clever choice of edges, also called elementary moves. The complete
graph is most of the time not an option, not only for size questions.
The point is that quite often $\Omega$ is hard to describe even if the
probability law itself is simple (even uniform) because $\Omega$ lacks
structure. But even in that situation, one can often guess simple
choices of elementary moves and show that they are enough to ensure
connectivity. This can be much easier than an enumeration of
$\Omega$.  

The simulation of self avoiding walks is a famous example of this
strategy.  On a regular lattice, a convenient choice of moves is given
by so called ``pivots'' which we describe briefly,
\cite{sokal,TKen02a}. To have a finite sample space of
non-intersecting walks, fix their length and initial point. Let us
describe a time step. Starting from any non-intersecting walk, at each
step choose a vertex (called the pivot) on the walk and a lattice
symmetry fixing the pivot, both with the uniform probability.  Keep
the part of the walk before the pivot, but apply the symmetry to the
part of the walk after the pivot. If the resulting walk intersects
itself, do nothing. Else move to the new walk. Decide that two
non-intersecting walks are connected if one can go from one to the
other in a time step. It is not too difficult to show that the
resulting graph on non-intersecting walks is connected and that
detailed balance holds for the uniform probability distribution on
non-intersecting walks.  Hence the stationary long time measure for
the pivot Monte Carlo algorithm is the self avoiding walk
measure\footnote{As a side remark, note that if the cases when the
  move is not possible are not counted as time steps, detailed balance
  does not hold anymore, but of course convergence to the right
  measure is preserved.}.

\begin{figure}[htbp]
\begin{center}
\includegraphics[width=0.7\textwidth]{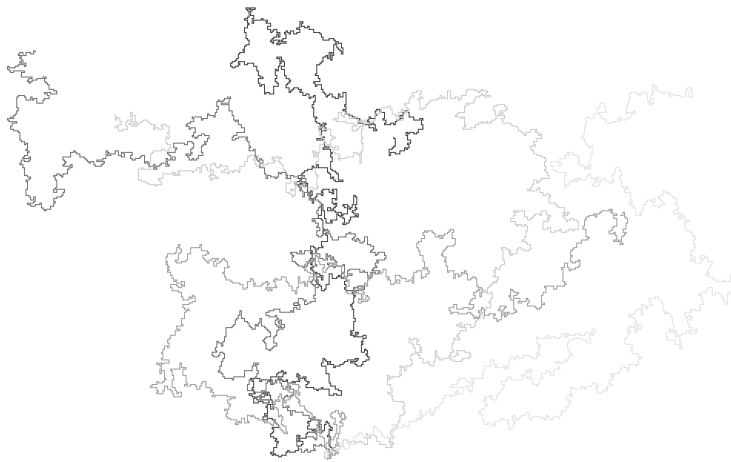}
\end{center}
\caption{A few self avoiding walks.}
      \label{fig:sawBW}
\end{figure}

Fig.\ref{fig:sawBW} shows a few samples. Producing a single clean
sample of reasonable size starting from a walk far from equilibrium
(like a straight segment) takes many Monte Carlo iterations. In fact
it takes roughly the time needed to compute the fractal dimension with
1 percent error for our previous examples. However, once the large
time regime is reached, one estimates that only a fraction of the
number of iterations needed to thermalize is enough to get a new
(almost) independent sample, so that a good numerical estimate of the
fractal dimension of the self avoiding walk can still be obtained via
a modest numerical effort. Thinking about the way samples are build,
it may seem hard to believe that the self avoiding walk can be viewed
as a growth process in a natural way, which is what SLE does.

In some respect the self avoiding walk is in a position similar to the
one of percolation because it has a compelling characteristic
property. Percolation has locality, and the self avoiding walk has the
restriction property.  If a sample space is endowed with the uniform
probability measure and one concentrates on a subspace (or, in
probabilistic language, conditions on a subspace) the measure induced
on the subspace is obviously still uniform. Hence the self avoiding
walk on a graph conditioned not to leave a certain subgraph is the
self avoiding walk on the subgraph. This is called restriction. As we
shall see later, among SLE$_\kappa$'s, SLE$_{8/3}$ is the only one
that satisfies restriction. So if the continuum limit of the self
avoiding walk exists and is conformally invariant --two facts which
are still conjectural at the moment despite hard efforts of gifted
people-- it has to be SLE$_{8/3}$ and the value of its fractal
dimension, $4/3$, comes for free.

It is also useful to consider ensembles of self avoiding walks of
variable length. In the full plane, the logarithm of the number of
self avoiding walks of $S$ steps is $\sim S \log \mu$ for large $S$
where $\mu$ is lattice dependant. To get a continuum limit made of
long fluctuating walks, it is thus necessary to weight each self
avoiding walk with weight $\mu^{-S}$.

We hope that these examples have convincingly supported our assessment
in the introduction that the world of interfaces and of algorithms to
explore it is incredibly rich and wide, harvesting many beautiful and
fragile objects.

\subsection{Examples from statistical mechanics}
\label{sec:statmechexamp}

\subsubsection{Ising model}
\label{sec:ising}

\begin{figure}[htbp]
\begin{center}
\includegraphics[width=.5\textwidth]{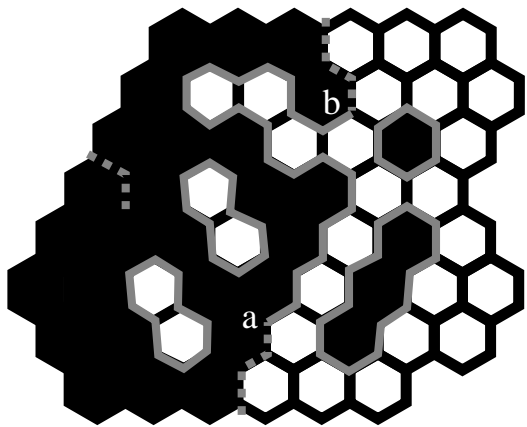}
\caption{A configuration of the Ising model.}
     \label{fig:isample}
 \end{center}
\end{figure}

Our first example from statistical mechanics is the celebrated Ising
model, where we choose to put the spin variables on the faces of an
hexagonal lattice domain with admissible boundary conditions
$(\mathbb{D},a,b)$ and we use the low temperature expansion.  The spins
are fixed to be $up$ on the left and $down$ on the right faces. The
energy of a configuration is proportional to the length of the curves
separating up and down islands. There is one interface from $a$ to $b$
and a number of loops, see Fig.\ref{fig:isample}. 

\begin{figure}[htbp]
\begin{center}
\includegraphics[width=\textwidth]{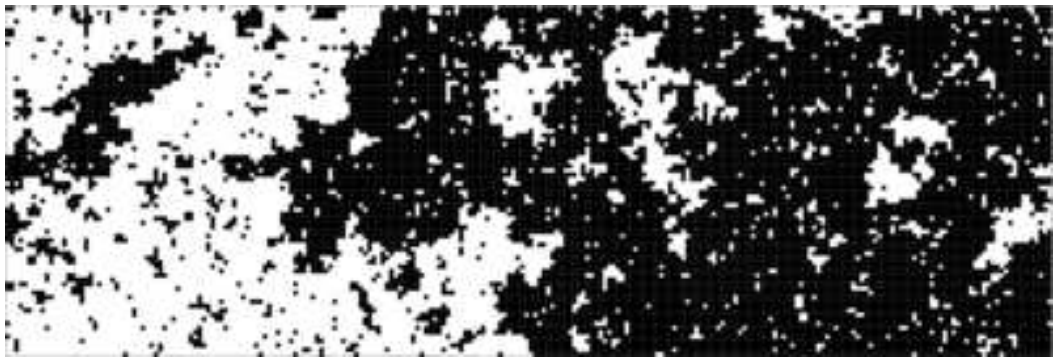}
\caption{A sample for the critical Ising model. The bottom line, where
  the spins are frozen --black on the right, white on the left-- is
  not represented. Courtesy of J. Houdayer.}
     \label{fig:isingsample}
 \end{center}
\end{figure}

The proportionality constant in the configuration energy has to be
adjusted carefully to lead to a critical system with long range
correlations. This time, making accurate simulations is much more
demanding. On the square lattice, the definition of the interface
suffers from ambiguities, but these become less relevant for larger
sample sizes. Fig.\ref{fig:isingsample} is an illustration.

Although there is no question that the fractal dimension of the Ising
interface with the above boundary conditions is $11/8$ and is
described by the --to be defined later-- SLE$_3$ ensemble, a
mathematical proof that a continuum limit distribution for the
interface exists and is conformally invariant is still out of reach.

\subsubsection{Potts models}
\label{sec:potts}

The $Q$-state Potts model can be defined on an arbitrary simple graph
$G=(V,E)$ with vertices $v \in V$ and edges $\{v,v'\}\in E \subset
\mathrm{sym}(V^2) $, the collection of two-elements subsets of $V$.
The parameter $Q$ is a positive integer to start with. Each vertex $v
\in V$ carries a variable $S_v \in \{1,\cdots,Q\}$. The Boltzmann
weight of a configuration is by definition
$$
\prod_{\{v,v'\}\in E}e^{\beta (\delta_{S_v,S_{v'}}-1)},
$$
where $\beta >0$ is the temperature. Write $e^{\beta
  (\delta_{S_v,S_{v'}}-1)}=p\delta_{S_v,S_{v'}}+(1-p)$ where $p\equiv
1-e^{-\beta}\in [0,1]$, view the first term, $p\delta_{S_v,S_{v'}}$,
as ``the edge $\{v,v'\}$ is occupied'', the second term as ``the edge
$\{v,v'\}$ is not occupied'' and expand the Boltzmann weight as a sum
of $2^{|E|}$ terms. Each term is associated to a subgraph $C$ of $G$
with the same vertex set $V$, but edges in $E_C$, the subset of $E$
made of the occupied edges. The partition function $Z$ is obtained by
summing each of the $2^{|E|}$ terms over the $Q^{|V|}$ spin
configurations. Each connected component of $C$ gives a non vanishing
factor only if all spins in it are the same. Hence, each cluster
(=connected component) of $C$ gives a factor $Q$ (isolated points
count as clusters) and the partition function can be rewritten,
following Fortuin-Kastelyn \cite{FortuinKastelyn}, as a sum over
cluster configurations
$$
Z= \sum_C\, p^{|E_C|}\,(1-p)^{|E|-|E_C|}\, Q^{N_C}
$$ 
where $N_C$ the number of clusters in the configuration $C$.
This formula makes sense for arbitrary $Q$ now. 

\begin{figure}[htbp]
  \begin{center}
    \includegraphics[width=0.7\textwidth]{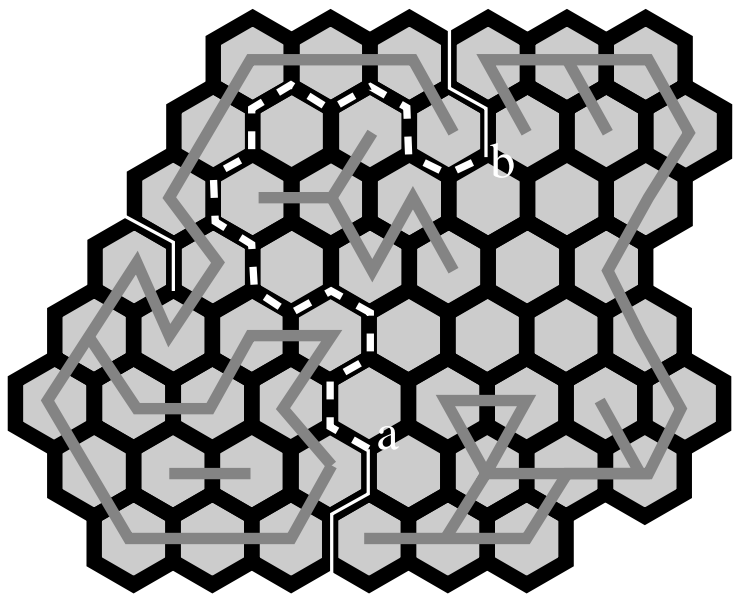}
      \caption{FK-clusters configuration in the Potts models and the
        corresponding interface.}
      \label{fig:pottsex}
  \end{center}
\end{figure}

To introduce interfaces, one can consider for instance that the
vertices of the graph on which the Potts model is defined are the
faces of an hexagonal lattice domain. Freeze the left faces to a given
color, so that a left cluster containing all left faces (plus possibly
some other) can be defined and either freeze the right vertices at a
different value, see Fig.\ref{fig:pottsex} for an illustration, or
condition on configurations such that the left cluster does not
contain right faces. There is a single simple lattice path bounded on
the left by the left cluster, and it defines an interface. If the
hexagons of the left cluster are colored black and the other ones
white, the interface separates the two colors.

For $Q\leq 4$ the parameter $p$ can be adjusted so that a continuum
scale invariant limit exists. The interface is conjectured to be
conformally invariant and statistically equivalent to an SLE trace
\cite{Schramm00,RohdeSchramm01}.

For $Q=2$, the Potts model Boltzmann weight is proportional to the
Ising model weight, and for general $Q$, again up to a constant, the
energy is given by the length of the curves separating islands of
identical spins.  However, when $Q > 2$, these curves are complicated
and not very manageable. This is related to the following fact. The
reader will have noticed that we always choose situations when the
lattice interface is a simple curve. This is needed to be in the SLE
framework, but this is not a generic situation. For instance the
physical interface separating clusters of different colors in the
$Q=3$ Potts model do exhibit points where three lines meet, loops et
c\ae tera.

\subsubsection{$O(n)$ models}
\label{sec:o(n)}

The $O(n)$ model can also be defined on an arbitrary simple graph
$G=(V,E)$ with vertices $v \in V$ and edges $\{v,v'\}\in E \subset
\mathrm{sym}(V^2) $. This time each vertex $v \in V$ carries a
variable $\sigma_v =(\sigma_v^1,\cdots,\sigma_v^n) \in S^{n-1}$, the
sphere in $n$ dimensions with radius $\sqrt{n}$. The measure $d\mu$ is
the rotation invariant measure of unit mass on that sphere, so that
$$
 \int d\mu(\sigma)=1 \quad \int d\mu(\sigma) \sigma^{\alpha}
\sigma^{\beta} = \delta^{\alpha\beta},
$$
while the integrals of odd functions of $\sigma$ vanish.

The Boltzmann weight  of a configuration is 
$$
\prod_{\{v,v'\}\in E} w(\sigma_v \cdot \sigma_{v'}),
$$
where $\sigma_v \cdot \sigma_{v'} \equiv \sigma_v^1 \sigma_{v'}^1 +
\cdots +\sigma_v^n \sigma_{v'}^n$ is the scalar product. In the
original version of the model, $w(x)=e^{\beta x}$, but for certain
classes of graphs, there is a more convenient choice to which we shall
come in a moment.

We start by defining the graph associated to an hexagonal lattice
domain $\mathbb{D}$. We forget the open hexagons and only keep the
edges and vertices in $\mathbb{D}$. Then we add the vertices needed to
get a closed set in the plane, yielding the desired (planar) graph
$G_{\mathbb{D}}$. Note that $\mathbb{D}$ can be recovered from  
$G_{\mathbb{D}}$ by adding the open hexagons needed to have each edge
bounded on both sides, and then taking the topological interior to
remove the unwanted vertices.

One good property of this class of graphs is that it is a subclass
(that we do not try to characterize) of the class of graphs
with vertices of valence at most three. A boundary vertex is by
definition a vertex of valence $<3$. On such graphs, it is convenient
to choose $w(x)=1+tx$ where $t$ is a parameter. The Boltzmann weight
is $\prod_{\{v,v'\}\in E} (1+t\, \sigma_v\cdot \sigma_{v'})$.

To get a graphical representation of the partition function, expand
the Boltzmann weight as a sum of monomials in the $\sigma_v\cdot
\sigma_{v'}$'s. Each monomial corresponds to a subgraph of $G$. Then
integrate each monomial against $\prod_{v\in V}d\mu(\sigma_v)$. Each
$\sigma_v$ appears at most three times in a monomial, so that the
trivial integrals listed above allow to compute everything. A monomial
gives a nonzero contribution if and only if the subgraph it describes
is a union of disjoint cycles, also called loops. Call such a subgraph
a loop subgraph of $G$. Then
$$
Z=\sum_{F\in \mathcal{F}} n^{l(F)}t^{b(F)},
$$
where $F$ runs over all
loop subgraphs of $G$, $l(F)$ is the number of loops of $F$ and $b(F)$
is the number of bonds (i.e. edges) in $F$. So we are summing over a
loop gas. The temperature-like parameter $t$ can be reinterpreted as a
bond fugacity.

Interfaces appear in a natural way via correlation functions. There
are several options and we shall use the simplest: choose a component
number, say $1$, and insert $\sigma^1$'s at boundary vertices
$v_1,\cdots,v_{2k}$. The insertion of an odd number of $\sigma^1$'s
gives $0$. Up to now, we have mostly considered the case $k=1$ when
only one interface is present. Again,
$Z_{v_1,\cdots,v_{2k}} \equiv Z\vev{\sigma_{v_1}^1 \cdots
  \sigma_{v_{2k}}^1}$ has a graphical expansion as a sum over
$\mathcal{F}_{v_1,\cdots,v_{2k}}$, the collection of subgraphs of $G$
consisting on the one hand of $k$ connected component which are 
(simple) lines pairing the insertion points and on the other hand of an
arbitrary number of connected component which are loops. Again, each
loop gives a factor $n$, but the lines give a factor $1$. Explicitly,
$$
Z_{v_1,\cdots,v_{2k}}=\sum_{F\in \mathcal{F}_{v_1,\cdots,v_{2k}}}
n^{l(F)}t^{b(F)}.
$$

Alternatively we could choose several component numbers (if $n$ is
large enough). Then each component number has to appear an even number
of times to give a non-vanishing result, and then different kinds of
lines appear, pairing insertion points with the same component
numbers. Note that this can be seen as a conditioning of the previous
situation.

We could also look at correlators which are scalar products,
yielding slightly different rules to weight the lines, depending
whether they connect two insertions which build a scalar product or
not. 

Up to now, we have seen the graphical expansion as a trick to study
the original spin model, which could be formulated only for integral
$n$. However, the graphical expansion gives a meaning when $n$ is a
formal parameter, in particular a real or complex number. The general
model is interesting for its own sake. For instance, one can introduce
conditioning. One can restrict the sums over subgraphs which
contain all vertices of $G$, leading to so-called fully packed models.
One can also impose say that a given bulk lattice point belongs to an
interface, and we would like to interpret the corresponding
partition function as a correlator with a certain field inserted at
that point. The price to pay for such extensions is that the original
local Boltzmann weight is replaced by nonlocal weights. We shall see
later that nevertheless the $O(n)$ model for general $n$ still has a
very important property, the domain Markov property.

Take an hexagonal lattice domain $\mathbb{D}$ and choose a ``loops and
lines'' configuration for $G_{\mathbb{D}}$. If one associates a $+$
sign to an arbitrary hexagon of $\mathbb{D}$ there is a single way to
extend this assignment to all hexagons of $\mathbb{D}$ by continuity,
flipping the sign only when a loop or a line is crossed. So there is
another version of the configuration space using Ising like variables.
A ``loops and lines'' configuration can be seen as the frontier
between island of opposite signs.

For $n=1$, we recover that Kramers-Wannier duality between the
low temperature expansion of the Ising model for spins on the faces of
$\mathbb{D}$ that we studied before and the high temperature expansion
of the Ising model for spins on the vertices of $G_{\mathbb{D}}$.

Note also that for $n=0$ one recovers the correct weight for
self-avoiding walks as introduced before. This is another illustration
that the physical approach via statistical mechanics and the
mathematical approach via combinatorics are in fact closely related. 

Considering the previous superficial remarks, it is probably not
surprising that the phase structure of $O(n)$ models is rather
complicated and interesting. when $n \in [-2,2]$, one can adjust $t$
so that a continuum scale invariant limit exists. The interface is
again conjectured to be conformally invariant and statistically
equivalent to an SLE trace.

\subsection{The domain Markov property}
\label{sec:dmp}

We have already insisted that the models of interfaces should be
defined on lattice domains of arbitrary shapes. Let us however note
that the possibility to have a natural definition on arbitrary lattice
domains is not so obvious. For models of geometric interfaces, there
is no general recipe, and for specific cases we have taken a definition
which may look arbitrary, as illustrated by the loop erased random
walk example. For statistical mechanics, the models we have introduced
have a natural definition on any domain because they are based on
nearest neighbor interactions and need only an abstract graph
structure.\\
Suppose that $(\mathbb{D},a,b)$ is a lattice domain with admissible
boundary condition and $\gamma_{[ab]}\equiv (s_1,\cdots,s_{2n+1})$ is
a path from $a$ to $b$ in $\mathbb{D}$. Recall that this means that
$a=s_1,b=s_{2n+1}$, the odd $s_{2m+1}$, $1 \leq m <n$, (if any) are
distinct vertices of the decomposition of $\mathbb{D}$ and the even
$s_{2m}$, $1 \leq m <n$, are distinct edges of the decomposition of
$\mathbb{D}$ with boundary $\{s_{2m-1},s_{2m+1}\}$. We use
$P_{(\mathbb{D},a,b)}$ to denote the probability distribution for the
interface $\gamma_{[ab]}$ from $a$ to $b$ in $\mathbb{D}$.

\begin{figure}[htbp]
\begin{center}
\includegraphics[width=.8\textwidth]{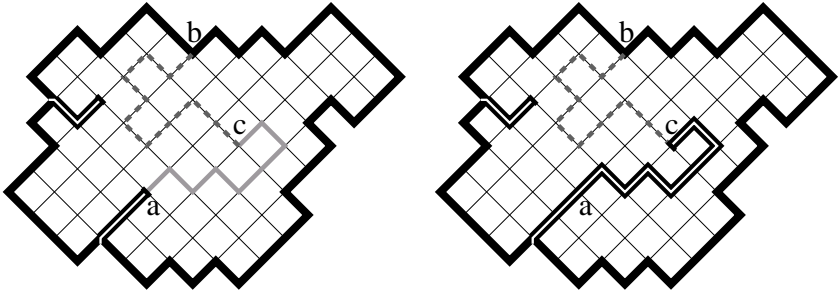}
\end{center}
\caption{An illustration of situations 1) and 2) for the case of
  loop-erased walks. What is the distribution of the dotted curve in
  both situations ?}
      \label{fig:dmp}
\end{figure}

Choose an integer $m$ such that $0 \leq m <n$ and set $s_{2m+1}\equiv
c$. Decompose $\gamma_{[ab]}=\gamma_{[ac]}\cdot \gamma_{[cb]}$, where
the $\cdot$ means concatenation. The set $\mathbb{D}'\equiv
\mathbb{D}\setminus \gamma_{]ac]}$, obtained by cutting along
$\gamma_{[ac]}$ with scissors, i.e. by removing from $\mathbb{D}$ the
sets $s_l$, $1 < l \leq s_{2m+1}$, is still a domain, and $(c,b)$ is
an admissible boundary condition for $\mathbb{D}'$. Hence we can
compare two things. \\ 
1) The probability in $(\mathbb{D},a,b)$ of
$\gamma_{[ab]}$ conditioned to start with $\gamma_{[ac]}$, that is
the ratio of the probability of $\gamma_{[ab]}$ by the probability for
the interface to start with $\gamma_{[ac]}$. \\
2) The probability of $\gamma_{[bc]}$ in
$(\mathbb{D}',c,b)$. This is illustrated on Fig.\ref{fig:dmp}.

The domain Markov property is the statement that
these two probabilities are equal. In equations
\[P_{(\mathbb{D},a,b)}(\; .\;|\gamma_{[ac]})=P_{(\mathbb{D}\setminus
  \gamma_{]ac]},c,b)}(\; .\;).\]
All the examples of interfaces introduced so far have the domain
Markov property, but for a single exception. First, it is obvious that
these two probabilities are supported on the same set, namely simple
curves along the edges of the lattice, going from $c$ to $b$ in
$\mathbb{D}\setminus \gamma_{]ac]}$. Let us however note that for
loop-erased random walks, annihilating boundary conditions are
crucial. Reflecting boundary conditions clearly do not work, if only
because the supports do not coincide in that case.

-- For percolation, the domain Markov property is seen directly by
using the definition of percolation as a growth process.

-- For the harmonic navigator, the domain Markov property rests on the
fact that the random walk can go not only on the initial boundary but
also on the beginning of the interface. This is still true of the
variants that we introduced, except the one we called the boundary
harmonic navigator, when we imposed that the initial part of the
interface be transparent and the random walk could accost only the
initial boundary.

-- For the case of the loop erased walk a little argument is needed.
Take any random walk (possibly with loops) $W_0=a,W_1,\cdots,W_l=b$
that contributes to an interface $\gamma_{[ab]}$ which is
$\gamma_{[ac]}$ followed by some $\gamma_{[cb]}$. Let $m$ be the
largest index for which the walk visits $c$. Because the interface has
to start with $\gamma_{[ac]}$, the walk $W_m=c,\cdots,W_l=b$ cannot
cross $\gamma_{[ac[}$ again, so it is in fact a walk in
$\mathbb{D}\setminus \gamma_{[ac[}$ from $c$ to $b$ leading to the
interface $\gamma_{[cb]}$. The weight for the walk
$W_0=a,W_1,\cdots,W_l=b$ is $4^{-l}$, i.e. simply the product of
weights for the walks $W_0=a,W_1,\cdots,W_m=c$ and
$W_m=c=a,\cdots,W_l=b$. Then a simple manipulation of weights leads
directly to the announced result.

-- The domain Markov property for the self avoiding walk rests (just
like the restriction property) on the fact it endows non-intersecting
walks with the uniform probability measure. Then the self avoiding
walk measure conditioned on the beginning of the interface is still
uniform, so it is the self avoiding walk measure on the cut domain.
  
-- For the statistical mechanics model, in fact more is true: we can
view $P_{(\mathbb{D},a,b)}$ not only as a probability distribution for
the interface, but as the full probability distribution for the full
configuration space and still check the identity of 1) and 2). For
orientation, first restrict attention to the O(n) model when $n$ is
an integer. The supports are the same for 1) and 2), namely any
configuration of the colors, except that the colors on both sides of
$\gamma_{[ac]}$ are fixed. The Boltzmann weight involves only nearest
neighbor interactions. The conditional probability in 1) takes into
account the interactions between the colors along the interface
$\gamma_{[ac]}$, whereas the probability in 2) does not take into
account the interactions between the colors along the cut left by the
removal of $\gamma_{]ac]}$. However, the corresponding colors are
fixed anyway, so the Boltzmann weights for the configurations that are
in the support of 1) or 2) differ by an overall multiplicative
constant, which disappears when probabilities are computed.\\
This argument extends immediately to systems with only nearest
neighbor interactions. They can be defined on any graph. If any subset
of edges is chosen and the configuration at both end of each edge is
frozen, it makes no difference for probabilities to consider the model
on a new graph in which the frozen edges have been deleted. \\
When $Q$ (Potts model) or $n$ ($O(n)$ model) are not integers, the
Boltzmann weights are not local anymore, but again the Boltzmann
weights for the configurations that are in the support of 1) or 2)
differ by an overall multiplicative constant, related to the length of
$\gamma_{[ac]}$, which disappears when probabilities are computed.

The domain Markov property --which, as should be amply evident, has nothing to
do with conformal invariance-- together with the conformal invariance
assumption is at the heart of O. Schramm's derivation of stochastic
Loewner evolutions.

We end our discussion of the domain Markov property by an illustration
of its predictive power. We have seen on the example of the $O(n)$
model that dealing with several interfaces is easy in the framework of
statistical mechanics. What about trying to define directly several
interfaces, say two, for loop erased random walks for instance? We
want that one goes from $a$ to $b$ and the second from $a'$ to $b'$.
We shall sum over pairs of random walks, but how should we restrict
the sum. Should the random walks avoid each other, or should they
simply be such that the associated loop erased walk avoid each other.
If the domain Markov property is to be preserved, the answer is
neither. The recipe can be nothing but the following: build the first
loop erased walk $\gamma_{[a,b]}$ from $a$ to $b$ in $\mathbb{D}$ and
cut the domain in two pieces, keep only the piece
$\mathbb{D}'_{\gamma_{[a,b]}}$ that contains $a'$ and $b'$ and then
build the second loop erased walk from $a'$ to $b'$ in the sub-domain.
The recipe looks asymmetric: for $\gamma_{[a,b]}$ we sum over walks in
$\mathbb{D}$, but for $\gamma_{[a',b']}$ we sum over walks in
$\mathbb{D}'_{\gamma_{[a,b]}}$. Let $S$ be this sum. Write $S=S'-S''$
where $S$ is the sum over all couples of random walks (which is
symmetric), and $S'$ is the sum of couples of random walks such that
the walk from $a'$ to $b'$ hits $\gamma_{[a,b]}$. Now split
$S''=S'''+S''''$ where $S'''$ is the sum over couples of random walks
such that the each one touches the other loop erased walk (which is
symmetric), and $S''''$ is the sum over couples of random walks such
that the one from $a$ to $b$ does not touch $\gamma_{[a',b']}$ but the
one from $a'$ to $b'$ does touch $\gamma_{[a,b]}$. Then removes the
loops that hit $\gamma_{[a,b]}$ on the walk from $a'$ to $b'$ to graft
them in the appropriate order on the walk from $a$ to $b$ and to see
that $S''''$ is in fact symmetric. This is closely related to the
general definition of multiple SLE's, either by imposing commutativity
\cite{Dubedat04c} or by imposing properties natural from the viewpoint
of statistical mechanics \cite{BBK05}.

\subsection{Other  growth processes}
\label{sec:growth}

Previous examples, either geometrical or extracted from statistical
mechanic models, are actually static. The growth dynamics arises -- or
will arise soon in the following Sections -- only via the way we
choose to described them. The fact that such dynamical description of
static objects is efficient is tided to their conformal properties.
There are however a large class of truly growth processes specifying
the dynamics of fractal domains.  The most famous is diffusion limited
aggregation (DLA) which described successive aggregations of tiny
particles. Since DLA only assumed that the growth is governed by
diffusion its domain applicability -- for instance to aggregation or
deposition phenomena -- is quite large. Of course many works,
experimental, numerical or theoretical, have been devoted to DLA, see
refs.\cite{Benetal86,Gollub99,Halsey00,Bazant05,Tanveer00,Swinney} for
alternative reviews and extra references. We shall not review all of
them but only have a glance on that field.  Another standard example,
the so-called Hele-Shaw problem, has an hydrodynamic origin
\cite{Benetal86,Saffman86,Casademunt04}. It may be viewed as
describing the invasion of an oil domain by an air bubble.  Its
dynamics leads to very interesting formation of domains with
finger-like shapes which are non-linearly selected
\cite{Shraiman86,Cambescot86}. It is one of the basics models of
non-linear pattern formations and selections. Both, DLA and Hele-Shaw,
are related to Laplacian growth (LG), see eg.\cite{Benetal86}.

\subsubsection{DLA}
\label{sec:dla}

DLA stands for diffusion limited aggregation \cite{DLA81}. It refers to
processes in which the domains grow by aggregating diffusing
particles. Namely, one imagines building up a domain by clustering
particles one by one. These particles are released from the point at
infinity, or uniformly from a large circle around infinity, and
diffuse as random walkers. They will eventually hit the domain and the
first time this happens they stick to it. By convention, time is
incremented by unity each time a particle is added to the domain. Thus
at each time step the area of the domain is increased by the physical
size of the particle. The position at which the particle is added
depends on the probability for a random walker to visit the boundary
for the first time at this position.

\begin{figure}[htbp]
\begin{center}
\includegraphics[width=0.6\textwidth]{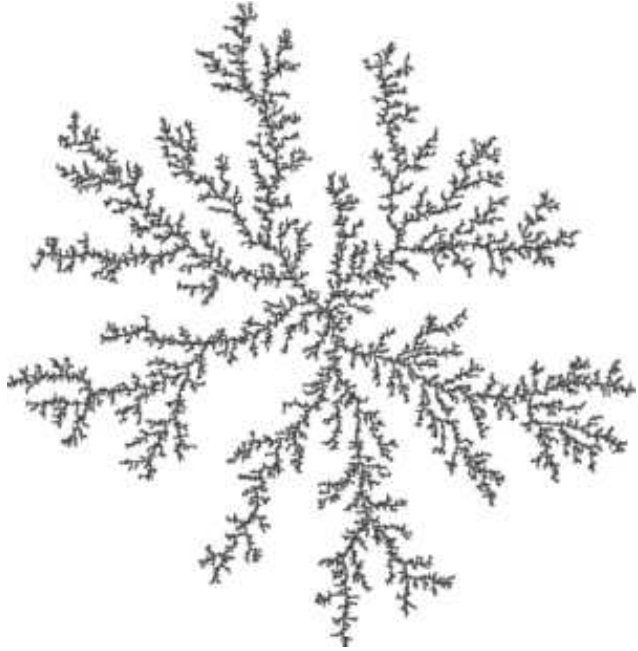}
\caption{A DLA sample.}
   \label{fig:dla}
 \end{center}
\end{figure}

In a discrete approach one may imagine that the particles are tiny
squares whose centers move on a square lattice whose edge lengths
equal that of the particles, so that particles fill the lattice when
they are glued together. The center of a particle moves as a random
walker on the square lattice. The probability $Q(x)$ that a particle
visits a given site $x$ of the lattice satisfies the lattice version
of the Laplace equation $\nabla^2Q=0$.  It vanishes on the boundary of
the domain, i.e. $Q=0$ on the boundary, because the probability for a
particle to visit a point of the lattice already occupied, i.e. a
point of the growing cluster, is zero.  The local speed at which the
domain is growing is proportional to the probability for a site next
to the interface but on the outer domain to be visited. This
probability is proportional to the discrete normal gradient of $Q$,
since the visiting probability vanishes on the interface. So the local
speed is $v_n=(\nabla Q)_n$. It is not so easy to make an unbiased
simulation of DLA on the lattice. One of the reasons is that on the
lattice there is no such simple boundary as a circle, for which the
hitting distribution from infinity is uniform. The hitting
distribution on the boundary of a square is not such a simple
function. Another reason is that despite the fact that the symmetric
random walk is recurrent is 2d, each walk takes many steps to glue to
the growing domain. The typical time to generate a single sample of
reasonable size with an acceptable bias is comparable to the time it
takes to make enough statistics on loop-erased random walks or
percolation to get the scaling exponent with two significant digits.
Still this is a modest time, but it is enough to reveal the intricacy
of the patterns that are formed. Fig.\ref{fig:dla} is such a sample.

During this process the clustering domain gets ramified and develops
branches and fjords of various scales. The probability for a particle
to stick on the cluster is much higher on the tip of the branches than
deep inside the fjords. This property, relevant at all scales, is
responsible for the fractal structure of the DLA clusters.
 
Since its original presentation \cite{DLA81}, DLA has been studied
numerically quite extensively. There is now a consensus that the
fractal dimension of 2d DLA clusters is $D_{\rm dla}\simeq 1.71$.
There is actually a debate on whether this dimension is geometry
dependent but a recent study \cite{Somfai03} seems to indicate that DLA
clusters in a radial geometry and a channel geometry have identical
fractal dimension. To add a new particle to the growing domain, a
random walk has to wander around and the position at which it finally
sticks is influenced by the whole domain. To rephrase this, for each
new particle one has to solve the outer Laplace equation, a non-local
problem, to know the sticking probability distribution. This is a
typical example when scale invariance is not expected to imply
conformal invariance.

\subsubsection{Laplacian growth and others}
\label{sec:lapla}

DLA provides a discrete analogue of Laplacian growth. The particle
size plays the role of an ultraviolet cutoff.  Laplacian growth is a
process in which the growth of a domain is governed by the solution of
Laplace equation, i.e.  by an harmonic function, in the exterior of
the domain with appropriate boundary conditions. It has many
interpretation either in terms of aggregation of particles as in DLA
but also in hydrodynamic terms (then the solution of Laplace
equation is the pressure) or electrostatic terms (then the solution is
the electrostatic potential).

To be a bit more precise \cite{Benetal86}, let $P$ be the real solution
of Laplace equation, $\nabla^2P=0$, in the complement of an inner
domain in the complex plane with the boundary behavior $P=-\log |z| +
\cdots$ at infinity and $P=0$ on the boundary curve.  The time
evolution of the domain is then defined by demanding that the normal
velocity of points on the boundary curve be equal to minus the
gradient of $P$: $v_n=-(\nabla P)_n$.

One may also formulate Laplacian growth using a language borrowed from
electrostatics by imagining that the inner domain is a perfect
conductor. Then $V=\Re{\rm e}\,\Phi_t$ is the electric potential which
vanishes on the conductor but with a charge at infinity. The electric
field $\vec{E}=\vec{\nabla}V$ is $\bar E\equiv E_x-i
E_y=\partial_z\,\Phi_t$. Its normal component $E_n=|f_t'(u)|^{-1}$ is
proportional to the surface charge density.  A slight generalization
of this model to be discussed in Section \ref{sec:LG} leads to a model
of dielectric breakdown \cite{Niemeyer84}.

In the hydrodynamic picture, one imagines that the inner domain is
filled with a non viscous fluid, say air, and the outer domain with a
viscous one, say oil. Air is supposed to be injected at the origin and
there is an oil drain at infinity. The pressure in the air domain is
constant and set to zero by convention.  In the oil domain the
pressure satisfies the Laplace equation $\nabla^2P=0$. If we neglect
the surface tension, then pressure vanishes on the boundary curve and
the model is equivalent to Laplacian growth. In presence of surface
tension then the pressure on the boundary condition is
$P=-\sigma\kappa$ with $\sigma$ the surface tension and $\kappa$ the
curvature of the boundary curve.  This is the so-called Hele-Shaw
problem. For non zero surface tension, it provides a regularization of
Laplacian growth. There are nice experiences on these systems
\cite{Swinney}.

Besides DLA, another class of discrete growth processes are
theoretically defined by iterating conformal maps.  The similarity
with the sample in Fig.\ref{fig:itermap} obtained by this method and
that obtained by aggregation, Fig.\ref{fig:dla}, is striking. But a
quantitative comparison of the two models is well out of analytic
control and belongs to the realm of extensive simulations.  We shall
described them at the end of this review, see Section
\ref{sec:disciter}.

All these models involve very nice pattern formations.  Their
relations with Loewner chains will be described in Section
\ref{sec:othergrowth},  but it is already clear that their solutions
involve analytic functions and that there are challenging physics and
mathematics behind these problems.


%% file: chap2.tex

The description of Loewner chains and SLE processes is based on coding
domain shapes in conformal maps using techniques -- especially Riemann
theorem -- from basic theory of analytical functions that we recall
briefly in this Section. More details may of course be found in
standard references \cite{Conway,Ahlfors} on analytic functions.

\subsection{Conformal mappings} 
\label{sec:confmap}

A domain is a non empty connected and simply connected open set
strictly included in the complex plane $\mathbb{C}$. Simple
connectedness is a notion of purely topological nature which in two
dimensions asserts essentially that a domain has no holes and is
contractible: the domain has the same topology as a disc. But it is a
deep theorem of Riemann that two domains are always conformally
equivalent, i.e. there is an invertible holomorphic map between them.
These maps are usually called uniformizing maps.  For instance, the
upper-half plane $\mathbb{H}$ and the unitary disc $\{z\in\mathbb{C},
|z|<1\}$ centered on the origin are two domains.  The conformal
transformation $f(z)=i\frac{1-z}{1+z}$ maps the unitary disc onto the
upper half plane with $f(0)=i$ and $f(1)=0$.

It is well known that the upper half plane has a three dimensional Lie
group of conformal automorphisms, $PSL_2(\mathbb{R})$, that also acts
on the boundary of $\mathbb{H}$. This group is made of homographic
transformations $f(z)=\frac{az+b}{cz+d}$ with $a, b, c, d$ real and
$ad-bc=1$. To specify such map we have to impose three real
conditions.  Hence, there is a unique automorphism -- possibly
followed by a transposition -- that maps any triple of boundary points
to any other triple of boundary points. Similarly there is unique
homographic transformation that maps any pair made of a bulk point and
a boundary point to another pair of bulk and boundary points.  By
Riemann's theorem, this is also true for any other domain -- at least
if the boundary is not too wild -- so that uniformizing maps are fully
specified once three conditions have been imposed.

Riemann's theorem is used repeatedly in the rest of this review. It is
the starting point of many approaches to growth phenomena in two
dimensions since it allows to code the shapes of growing domains in
their uniformizing conformal maps. To make the description precise one
has to choose a reference domain against which the growing domains are
compared. Again by Riemann theorem we may choose any domain as
reference domain -- and depending on the geometry of the problem some
choices are simpler than others. To simplify statements we use in this
Section the upper half plane $\mathbb{H}$ as the reference domain.

\begin{figure}[htbp]
\begin{center}
\includegraphics[width=\textwidth]{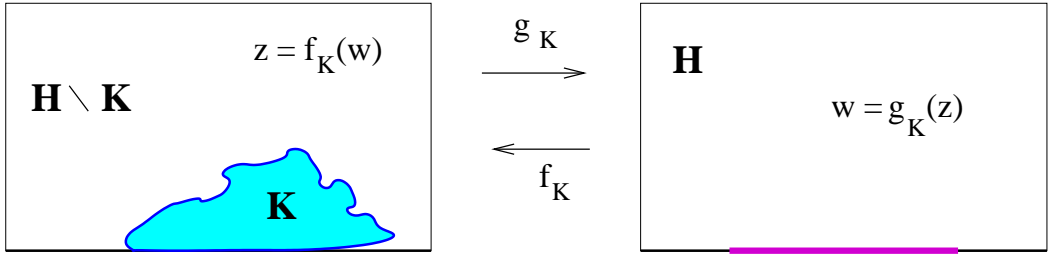}
\caption{An example of hull $K$ in the upper hall plane 
  together with the uniformizing map
  $f_K:\mathbb{H}\to\mathbb{H}\setminus\mathbb{K}$ and its invserse
  $g_K:\mathbb{H}\setminus\mathbb{K}\to\mathbb{H}$.}
     \label{fig:unifmap}
\end{center}
\end{figure}

For later use, we note that one can be a bit more explicit when the
domain $\mathbb{D}$ differs only locally from the upper half plane
$\mathbb{H}$, that is if $\mathbb{K}= \mathbb{H} \setminus \mathbb{D}$
is bounded. Such a set $\mathbb{K}$ is called a hull. See
Fig.\ref{fig:unifmap}. The real points in the closure of $\mathbb{K}$ in
$\mathbb{C}$ form a compact set which we call
$\mathbb{K}_{\mathbb{R}}$.  Let $f: \mathbb{H} \mapsto \mathbb{D}$ be
a conformal bijection and $g:\mathbb{D}\mapsto \mathbb{H}$ its
inverse. One can use the $PSL_2(\mathbb{R})$ automorphism group of
$\mathbb{H}$ to ensure that $f$ is holomorphic at $\infty$ and
$f(w)-w=O(1/w)$ there. This is called the \textit{hydrodynamic
  normalization}.  It involves three conditions, so there is no
further freedom left. We shall denote this special representative by
$f_\mathbb{K}$, which is uniquely determined by $\mathbb{K}$: any
property of $f_\mathbb{K}$ is an intrinsic property of $\mathbb{K}$.

Consider for instance the hull formed by the segment $\{z=iy,\ 
y\in[0,a]\}$, $a>0$, drawn between the origin $0$ and point $ia$ in
the upper half plane. The uniformizing map $g$ from
$\mathbb{H}\setminus[0,ia]$ to $\mathbb{H}$ and its inverse are given
by:
$$
g(z)=\sqrt{z^2+a^2}\quad,\quad f(z)\equiv g^{-1}(z)=\sqrt{z^2-a^2}
$$
The square root cut is specified by demanding that $g(z)\sim z$ at
infinity so that $g(x)$ is negative (positive) on the real axis for $x$
negative (positive). It has a cut along the segment $[0,ia]$: the
left hand of side of the segment is mapped on the real axis on the
interval $[-a,0]$ and the right hand side on $[0,+a]$.

Another simple example is for $\mathbb{K}$ a semi-disc of radius $r$
centered on the origin $\{z\in\mathbb{H},\ |z|\leq r\}$. The
uniformizing map of the upper half plane minus the semi-disc back onto
the upper half plane is $g(z) = z + \frac{r^2}{z}$ and its inverse is
$f(w)=\frac{1}{2}[w+\sqrt{w^2-4r^2}]$. It is clear that $g(z)$ is real
for $z$ real or for $z$ on the semi-circle of radius $r$, so that the
boundary of the upper half plane minus the semi-disc is mapped on the
real axis. 

A slight generalization consists in considering the infinitesimal hull
$\mathbb{K}_{\epsilon;\rho}=\{z=x+i\epsilon y,\ 0<y\leq \pi\rho(x)\}$
made of the set of point included between the real axis and the curve
$x\to i\epsilon \pi\rho(x)$, $\epsilon\ll 1$ with $x$ real. To first
order in $\epsilon$, the normalized uniformizing map of
$\mathbb{H}\setminus\mathbb{K}_{\epsilon;\rho}$ onto the upper half
plane is
\begin{eqnarray}
g(z) = z + \epsilon \int \frac{dx \rho(x)}{z-x}+ \cdots, 
\quad \epsilon\ll 1,
\label{petithull}
\end{eqnarray}
as can be seen by covering the hull by a series of semi-disc, or by
noticing that this map is real on the curve $z=x+i\epsilon\pi\rho(x)$.

Consider now again a domain $\mathbb{D}=\mathbb{H}\setminus\mathbb{K}$
with $\mathbb{K}$ generic hull and $f_\mathbb{K}$ its hydrodynamically
normalized uniformizing map.  As the boundary of $\mathbb{H}$ is
smooth, $f_\mathbb{K}$ has a continuous extension to
$\overline{\mathbb{R}}\equiv \mathbb{R}\cup \infty$, and
$f_\mathbb{K}^{-1}(\overline{\mathbb{R}}\setminus
\mathbb{K}_{\mathbb{R}})$ is a non-empty open set in
$\overline{\mathbb{R}}$ with compact complement.  We call the
complement the cut of $f_\mathbb{K}$. By the Schwarz symmetry
principle, defining $f_\mathbb{K}(z)=\overline{f_\mathbb{K}(\bar{z}})$
for $\Im{\rm m} \, z \leq 0$ gives an analytic extension of $f$ to the
whole Riemann sphere minus the cut.  Across the cut, $f$ has a purely
imaginary non-negative discontinuity which we write as a Radon-Nikodym
derivative $d\mu_{f_\mathbb{K}}/dx$. Cauchy's theorem yields
\begin{equation} \label{eq:repbydiscont}
f_\mathbb{K}(w)=w+\frac{1}{2\pi}\int_\mathbb{R} 
\frac{d\mu_{f_\mathbb{K}}(x)}{x-w}, 
\end{equation} 
Anticipating a little bit, let us note immediately that giving a
dynamical rule for the evolution of the finite positive measure
$d\mu_{f_\mathbb{K}}(x)$ is a good way to define growth processes.
A quantity that plays an important role in the sequel is
$$C_\mathbb{K}\equiv \frac{1}{2\pi}\int_\mathbb{R}
d\mu_{f_\mathbb{K}}(x),$$ a positive (unless $\mathbb{K}=\emptyset$)
number called the capacity of $\mathbb{K}$, which is such that
$f_\mathbb{K}(w)=w-C_\mathbb{K}/w+O(1/w^2)$ at infinity. The
usefulness of capacity stems from its good behavior under
compositions: if $\mathbb{K}$ and $\mathbb{K}'$ are two hulls,
$\mathbb{K}\cup f_\mathbb{K}(\mathbb{K}')$ is a hull and
\begin{equation} \label{eq:addcapa}
C_{\mathbb{K}\cup f_\mathbb{K}(\mathbb{K}')}=C_\mathbb{K}+C_{\mathbb{K}'},
\end{equation}
as seen by straightforward expansion at infinity of $f_\mathbb{K}\circ
f_{\mathbb{K}'}$, the map associated to $\mathbb{K}\cup
f_\mathbb{K}(\mathbb{K}').$ In particular capacity is a continuous
increasing function on hulls.

\subsection{Evolutions}
\label{sec:evol}

\subsubsection{Loewner chains}
\label{sec:loewner}

Evolutions of family of increasing hulls $\mathbb{K}_t$ parameterized
by some `time' $t$ are going to be coded in the evolutions of their
uniformizing map.  Let $f_t\equiv f_{\mathbb{K}_t}$ be the conformal
homeomorphism from $\mathbb{H}$ to $\mathbb{H}\setminus \mathbb{K}_t$
hydrodynamically normalized at infinity: $f_t(w)=w+O(1/w)$. Define
$g_t: \mathbb{H}\setminus \mathbb{K}_t \mapsto\mathbb{H} $ to be the
inverse of $f_t$. Then $g_t(z)=z+O(1/z)$ at infinity.

To study the evolution of the family of hulls $\mathbb{K}_t$, fix
$\varepsilon \geq 0$ and consider the hull
$\mathbb{K}_{\varepsilon,t}\equiv
g_t(\mathbb{K}_{t+\varepsilon}\setminus \mathbb{K}_t)$. Define
$f_{\varepsilon,t}\equiv f_{\mathbb{K}_{\varepsilon,t}}$. Then
$g_t=f_{\varepsilon,t}\circ g_{t+\varepsilon}$ on $\mathbb{H}\setminus
\mathbb{K}_{t+\varepsilon}$. Using the representation of
$f_{\mathbb{K}_{\varepsilon,t}}$ in terms of its discontinuity
eq.(\ref{eq:repbydiscont}), we obtain
\[ 
g_{t+\varepsilon}(z)-f_{\varepsilon,t}\circ
g_{t+\varepsilon}(z)=g_{t+\varepsilon}(z)-g_t(z)=\frac{1}{2\pi}\int_\mathbb{R}
\frac{d\mu_{f_{\varepsilon,t}}(x)}{g_{t+\varepsilon}(z)-x}
\]
For $\varepsilon$ small, the hull $\mathbb{K}_{\varepsilon,t}$ is
small so that $f_{\varepsilon,t}$ is approaching the identity map and
we may expect to be able to expand its discontinuity measure to first
order in $\varepsilon$ as $d\mu_{f_{\varepsilon,t}}(x)\simeq
\varepsilon 2\pi d\nu_t(x)$. In such cases, in the limit
$\varepsilon\to0$ we get the evolution equations:
\begin{eqnarray}
\frac{d}{dt} g_t(z) = \int_\mathbb{R}
\frac{d\nu_t(x)}{g_{t}(z)-x}
\label{evolequa}
\end{eqnarray}
These sets of equations are called ``Loewner chains''. The
Loewner measures $d\nu_t$ may depend non linearly on the map $g_t$.
They possess a simple physical interpretation. Indeed recall that the
map $f_{\varepsilon,t}$ uniformizes $\mathbb{K}_{\varepsilon,t}$ which
is the image by $g_t$ of the complement of $\mathbb{K}_t$ in
$\mathbb{K}_{t+\varepsilon}$. The hull $\mathbb{K}_{\varepsilon,t}$
may be viewed as a domain bounded by the real axis and by a curve with
height of order $\varepsilon$ and which is the image by $g_t$ of the
matter added to go from $\mathbb{K}_t$ to
$\mathbb{K}_{t+\varepsilon}$. The map $f_{\varepsilon,t}$ is then
given by equation (\ref{petithull}) to first order in $\varepsilon$ so
that $d\nu_t(x)/dx$ is proportional to the height of the curve bounding
$\mathbb{K}_{\varepsilon,t}$.

To make it more precise, let $f_t$, analytic in the upper half plane,
be the inverse of $g_t$. It satisfies Loewner equation:
$$
\frac{d}{dt}f_t(w)= -f'_t(w)\, \int_\mathbb{R}\frac{d\nu_t(x)}{w-x}
$$
This may be viewed as a Riemann-Hilbert problem for the ratio
$\partial_tf_t(w)/f_t'(w)$ since Loewner equation is equivalent to the
following boundary value problem on the real axis:
$$
\lim_{\varepsilon\to0^+}\Im{\rm m}\,
[\partial_tf_t(w)/f_t'(w)]_{w=y+i\varepsilon}  = \pi \rho_t(y)
$$
with $d\nu_t(x)=\rho_t(x)dx$. By construction the boundary curve of
the hull $\mathbb{K}_t$ is the image of the real axis, that is
$f_t(\zeta)$, $\zeta\in\mathbb{R}$. Its evolution is governed by its
normal velocity $v_n(\zeta)$ since the tangent velocity is
parameterization dependent. The normal velocity is equal to 
$|f'_t(\zeta)|\Im{\rm m}\, [\partial_tf_t/f_t'](\zeta)$, so
that it may be expressed in terms of the density $d\nu_t(x)$ as:
$$
 v_n(\zeta)\, d\zeta = \pi\, |f_t'(\zeta)|\, d\nu_t(\zeta)
 $$
 Comparing this formula with the previous heuristic interpretation
 with see that the factor $\varepsilon
 d\nu_t(\zeta)$ codes for the matter added in
 $\mathbb{K}_{\varepsilon,t}$ between time $t$ and $t+\varepsilon$ 
 while $|f'_t(\zeta)|$ is implementing the dilatation in
 going from $\mathbb{K}_{\varepsilon,t}$ to
 $\mathbb{K}_{t+\varepsilon}$.

Time parameterization has not yet been specified. In SLE context it is
very useful -- if not mandatory -- to use the capacity as time
variable so that we define the time parameter by $2t\equiv
C_{\mathbb{K}_t}$. This imposes $\int_\mathbb{R}d\nu_t(x)=2$.  The
factor $2$ is just historical. The additivity property of capacity
ensures the consistency of notation, namely $C_{\mathbb{K}_t\cup
  f_{\mathbb{K}_t}(\mathbb{K}_s)} = t+ s$.  With this time
parameterization, the maps behave as $g_t(z)=z+2t/z+O(1/z^2)$ and
$f_t(w)=w-2t/w+O(1/w^2)$ at infinity.

\subsubsection{Local growth}
\label{sec:local}

We introduce now the notion of {\it local growth} which is crucial for
interfaces. In particular it applies to the case when the hulls
$\mathbb{K}_t$ are portion of curves. Namely, let
$\gamma_{[0,\infty]}$ be a simple curve from $0$ to $\infty$ in
$\mathbb{H}$ and $\gamma_{]0,t]}$ be a portion of it with end point
$\gamma_t$.  Then $\mathbb{K}_t\equiv \gamma_{]0,t]}$ are growing
hulls with capacity $2t$ by our choice of time parameterization.  When
$\varepsilon$ is small, $\mathbb{K}_{\varepsilon,t}\equiv
g_t(\gamma_{]t,t+\varepsilon]})$ is a tiny piece of a curve and the
support of the discontinuity measure $d\mu_{f_{\varepsilon,t}}$ is
small and becomes a point when $\varepsilon$ goes to $0$. Measures
supported at a point are $\delta$ functions, so there is a point
$\xi_t$ such that, as a measure, $d\mu_{f_{\varepsilon,t}}/dx \sim
2\varepsilon \delta(x-\xi_t)$ as $\varepsilon \rightarrow 0^+$.  If
$\mathbb{K}_t$ is a more general increasing family of hulls of
capacity $2t$, we say that the condition of local growth is satisfied
if the above small $\varepsilon$ behavior holds.  At first sight, it
might seem that local growth is only true for curves, but this is not
true. We shall give an example below.

Letting $\varepsilon \rightarrow 0^+$, from the local growth
condition, we infer the existence of a real function $\xi_t$ such that 
\begin{equation}\label{eq:Loewnerevolution}
\frac{dg_t}{dt}(z)=\frac{2}{g_t(z)-\xi_t}.
\end{equation}
Had we used another parameterization of the
curve, the $2$ in the numerator would be replaced by a positive
function of the parameter along the curve.

It is useful to look at this equation from a slightly different point
of view, taking the function $\xi_t$ as the primary data. For fixed
$z$, this is a first order differential equation for $g_t$, which can
be integrated at least for $t$ small enough. The solutions $g_t(z)$ of
this equation for a given function $\xi_t$ with initial condition
$g_0(z)=z$ is called a Loewner evolution. The image of $\xi_t$ by
$g_t^{-1}$ is the tip $\gamma_t$ of the curve at time $t$. A more
proper definition is
\begin{eqnarray}
\gamma_t = \lim_{\epsilon\to 0^+} g^{-1}_t(\xi_t+i\epsilon)
\label{gammafromgt}
\end{eqnarray}
In short $g_t(\gamma_t)=\xi_t$.  It is a theorem that if $\xi_t$ is
regular enough -- namely H\"older of exponent $>1/2$ -- then
$\gamma_t=g_t^{-1}(\xi_t)$ is a curve.  In particular continuity of
$\xi_t$ is clearly a necessary condition for $\gamma_t$ to be a curve
as otherwise any jumps in time of $\xi_t$ produce branchings in
$\gamma_t$. The real function $\xi_t$ provides a
parameterization of the growing curve $\gamma_{]0,\infty]}$.

Informally, if $\mathbb{K}_t$ is a growing curve, we expect that
$g_{t+\varepsilon}(z)-g_t(z)$ describes an infinitesimal cut. This is
confirmed by the explicit solution of eq.(\ref{eq:Loewnerevolution})
for the trivial case $\xi_t\equiv 0$, which yields $g_t(z)^2=z^2+4t$,
the branch to be chosen being such that at large $z$, $g_t(z) \sim z$.
As previously explained, this describes a growing segment along the
imaginary axis. So intuitively, the simple pole in
eq.(\ref{eq:Loewnerevolution}) accounts for the existence of a cut and
different functions $\xi_t$ account for the different shapes of
curves.

\begin{figure}[htbp]
\begin{center}
\includegraphics[width=0.8\textwidth]{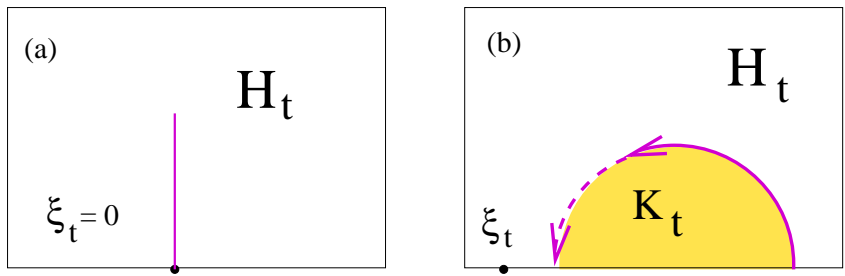}
\caption{(a) The upper half plane cuted along a vertical slit. 
(b) The upper half plane cuted along a semi-circle. At the critical
time at which the curve touches the real axis back, the hull is formed
by all points surrounded by the curve, that is by the semi-disc.}
     \label{fig:exempleloew}
\end{center}
\end{figure}

One can also solve the case when $\mathbb{K}_t$ is an arc of circle
going from $0$ to $2R$ along a circle of radius $R$. In this case the
driving function is $\xi_t=3[R-\sqrt{R^2-2t}]$. See
Fig.\ref{fig:exempleloew}. It has a square root singularity $\xi_t
\propto \sqrt{R^2-2t}$ when the arc approaches the real axis at time
$t_c=R^2/2$. The capacity remains finite, and goes to $R^2$, and the
map itself has a limit $g_{t_c}(z)=z+R^2/(z+R)$ which has swallowed
the half disk without violating the local growth condition.  One can
start the growth process again. Making strings of such maps with
various values of the radii is a simple way to construct growing
families of hulls that are not curves and that nevertheless grow
locally. Note that a square root singularity for $\xi_t$ is the
marginal behavior as if $\xi_t$ is H\"older of exponent $>1/2$,
Loewner evolution yields a simple curve.

\subsection{Miscellaneous iterations}  

Uniformizing maps are defined up to $SL(2,\mathbb{R})$ transformations
and so do the Loewner equations. These equations also take different
forms depending on which domain is used as a reference domain. Above
we chose the upper half plane as reference domain and the hydrodynamic
normalization to fix the uniformizing map uniquely.  It leads to
eq.(\ref{evolequa}) which is usually called chordal Loewner
evolutions. There are other possibilities and some of them will be
discussed below in connection with different SLE configurations.

In Section \ref{sec:radialchain}, we shall present another version of
Loewner equation, called radial Loewner evolution, which uses the unit
disk as a reference domain. The conformal map are then uniquely fixed
by imposing normalization conditions on a bulk point.

There is yet another way to generalize the previous evolutions. It
consists in discretizing the time parameter so that the evolutions are
then defined by successive iterations of conformal maps. The
elementary maps involved at each iteration code for the addition of
tiny pieces to the domain. This procedure will be described in Section
\ref{sec:disciter}.

%% file: chap3.tex

 We are interested in the continuous limit of the interfaces of 2d
statistical we just described in Section \ref{sec:examples} at
criticality.  Based on heuristic arguments of scale invariance plus
locality, it was conjectured in \cite{BPZ} about twenty years ago that
such curves should be conformally invariant (in an appropriate sense).
This statement was made really precise and powerful by Oded Schramm
who understood what are the consequences of conformal invariance for a
set of random curves and how to exploit them. This leads him to the
definition of the Loewner-Schramm stochastic evolutions (SLE).

The aim of this Section is to explain Schramm's argument in the
simplest case, called chordal SLE, describing curves joining two
boundary points of a planar domain.

Let us fix the notation.  

Consider a domain $\mathbb{D}$, with two
distinct points on its boundary, which we call $a$ and $b$. A simple
curve, denoted by $\gamma_{[ab]}$, from $a$ to $b$ in $\mathbb{D}$ is
the image of a continuous one-to-one map $\gamma$ from the interval
$[0,+\infty]$ to $\mathbb{D}\cup\{a,b\}$ such that $\gamma(0)=a$,
$\gamma(\infty)=b$ and $\gamma_{]ab[}\equiv \gamma(]0,\infty[)\subset
\mathbb{D}$.  Alternatively, a simple curve from $a$ to $b$ is an
equivalence class of such maps under increasing reparametrizations. A
point on it has no preferred coordinate but is has a past and a
future.  If $c \in \mathbb{D}$ is an interior point, we use a similar
definition for a simple curve $\gamma_{[ac]}$ from $a$ to $c$ in
$\mathbb{D}$.

For any of these, we use ${\bf P}_{(\mathbb{D},a,b)}$ to denote
the probability distribution for the interface $\gamma_{[ab]}$ from
$a$ to $b$ in $\mathbb{D}$.

\subsection{Conformal covariance}
\label{sec:confinv}

Before embarking into Schramm's argument let us point out a possible
caveat concerning conformal invariance. If a probability measure on
curves, or hulls, is defined say in the upper half plane $\mathbb{H}$,
one can always transport it to any other non empty simply connected
open sets in $\mathbb{C}$ to get new probability measures on curves or
hulls. But if we have nothing against which to compare the transported
measures, this statement is purely tautological -- and almost empty --
since it just tells us how to transport probability measures.  What is
highly nontrivial is that continuum limits of discrete 2D critical
statistical mechanics models are conformally covariant. Such models
are usually defined on a lattice, say $a\mathbb{Z}+ia\mathbb{Z}$ where
$a > 0$ is a unit of length.  Criticality is the statement that when
$a$ goes to zero certain nontrivial physical observables survive and
do not depend on any scale. The limit has to be defined carefully. If
$\mathbb{U}$ and $\mathbb{V}$ are two conformally equivalent open
subsets of $\mathbb{C}$, related by a conformal map $\varphi$, one can
consider the model in the intersection of the lattice
$a\mathbb{Z}+ia\mathbb{Z}$ with $\mathbb{U}$ or $\mathbb{V}$. When $a$
goes to $0$, scale invariance does not a priori imply that the limit
theories on $\mathbb{U}$ and $\mathbb{V}$ are related in a simple way.
Using another physical language, it was conjectured in \cite{BPZ} that
the limiting theory is well defined on the abstract Riemann surface
which is the equivalence class of all open subsets of $\mathbb{C}$
conformally equivalent to $\mathbb{U}$. For instance correlation
functions of local observables become sections of appropriate bundles,
i.e.  have transformations that involve derivatives of $\varphi$ when
going from $\mathbb{U}$ to $\mathbb{V}$.  Interfaces are directly
related by $\varphi$ and the probability law governing their
fluctuations as well.

As we shall explain, SLE curves behave geometrically as they should to
encode the statistics of critical interfaces. They have the proper
behavior under conditioning and conformal transformations.  However,
they do it in a very specific way, involving Loewner evolutions and
$1d$ Brownian motion. Schramm's argument for SLE may be decomposed in
three steps:\\
(i) a comparison of the probability distribution of curves
$\gamma_{[ab]}$ in a domain $\mathbb{D}$ conditioned on a portion
$\gamma_{[ac]}$ of the curve with the probability distribution in the
cutted domain $\mathbb{D}\setminus\gamma_{[ac]}$;\\
(ii) a formulation of conformal transport of probability distributions
of curves between two conformally equivalent domains;\\
(iii) a mixture of the two previous steps which implies a Markov and
stationarity of increments property for conformally invariant random
curves.\\
Let us make it plain.

\subsubsection{The domain Markov property}
\label{sect:condition}

We first go to the point, valid for all the discrete examples we have
described, which relates properties of conditioned probability of
curves to properties of models of statistical mechanics.

Suppose that we fix the beginning $\gamma_{[ac]}$ of a possible
interface in domain $\mathbb{D}$, up to a certain point $c$.
Then:\\
(i) we can consider the conditional distribution for the rest of
the interface and\\
(ii) we can remove the beginning of the interface from the domain to
create a new domain and consider the distribution of the
interface in this new domain.\\
This is illustrated on Fig.\ref{fig:dmp} on the lattice and in
Fig.\ref{fig:statmart} on the continuum.

We claim -- or demand -- that the distributions defined in (i) and
(ii) coincide, ie.:
\[
\gamma_{[ab]}\ {\rm in}\ \mathbb{D}\ \vert\ \gamma_{[ac]}
\equiv_{{\rm in~law}} \gamma_{[cb]}\ {\rm in}\ \mathbb{D}\setminus
\gamma_{[ac]}
\]
where the vertical line means `conditioned by'. Equivalently,  
in equations:
\begin{eqnarray}
{\bf P}_{(\mathbb{D},a,b)}[\; .\;|\gamma_{[ac]}]
={\bf P}_{(\mathbb{D}\setminus \gamma_{[ac[},c,b)}[\; .\;].
\label{eq:conditioned}
\end{eqnarray}
It is obvious that these two probabilities are supported on the same
set, namely simple curves along the edges of the lattice, going from
$c$ to $b$ in $\mathbb{D}\setminus \gamma_{[ac[}$.  

In the lattice statistical mechanics formulation, this property is a
simple consequence for instance of the locality of the interaction.  Let us however
note that for loop-erased random walks, annihilating boundary
conditions are crucial. Reflecting boundary conditions clearly do not
work, if only because the supports do not coincide in that case.

For the case of percolation and the Ising model, in fact more is true:
we can view ${\bf P}_{(\mathbb{D},a,b)}$ not only as a probability
distribution for the interface, but as the full probability
distribution for the colors of the hexagons and still check the identity
of (i) and (ii). Again, the supports are the same for (i) and (ii), namely
any configuration of the colors, except that the colors on both sides
of $\gamma_{[ac]}$ are fixed. For the case of percolation, the colors
are independent of each other so the identity of (i) and (ii) is clear.
For the Ising model, the difference is that the conditional
probabilities in (i) take into account the interactions between the
colors along the interface, whereas the probability in (ii) does not
take into account the interactions between the spins along the cut
left by the removal of the interface. However, as already mentioned
above, the corresponding colors are fixed anyway, so the Boltzmann
weights for the configurations that are in the support of (i) or (ii)
differ by a multiplicative constant, which disappears when
probabilities are computed.

This argument extends immediately to systems with only nearest
neighbor interactions. They can be defined on any graph. If any subset
of edges is chosen and the configuration at both end of each edge is
frozen, it makes no difference for probabilities to consider the model
on a new graph in which the frozen edges have been deleted.  Instead
of looking for further generalizations, we argue more heuristically
that the continuum limit for a system with short range interactions
should satisfy locality property. Its use -- which, as should be amply
evident, has nothing to do with conformal invariance -- together with
the conformal invariance assumption is at the heart of O. Schramm's
derivation of stochastic Loewner evolutions.

\subsubsection{Conformal transport}  
\label{subsubsec:cii}
 
\begin{figure}[htbp]
\begin{center}
\includegraphics[width=0.9\textwidth]{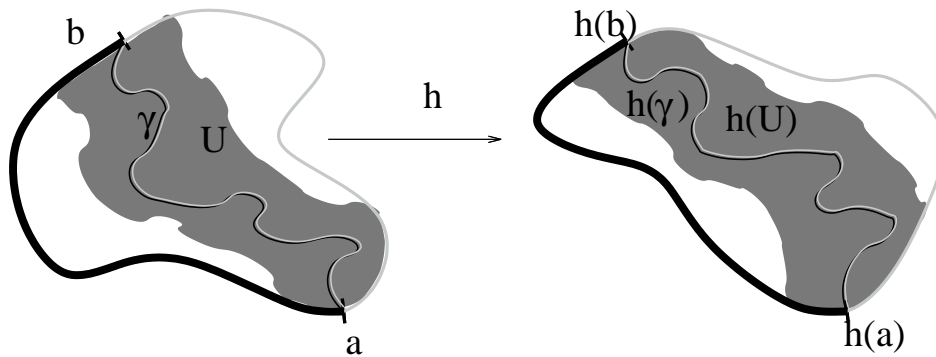}
\caption{Conformal invariance for change of domain. The measure on
  curves is simply transported by the conformal map.}
     \label{fig:conftrans}
 \end{center}
\end{figure}

For studying conformally invariant probability measures on the set of
simple curves from $a$ to $b$ in $\mathbb{D}$, there is a purely
kinematic step which demands that if $h$ is any conformal map that
sends $\mathbb{D}$ to another domain $\hat \mathbb{D}\equiv
h(\mathbb{D})$, the measure for $(h(\mathbb{D}),h(a),h(b))$ should be
the image by $h$ of the measure for $(\mathbb{D},a,b)$. Namely:
$$
h( \gamma_{[ab]}\ {\rm in}\ \mathbb{D}) \equiv_{{\rm in~law}} 
 \gamma_{[h(a)h(b)]}\ {\rm in}\ h(\mathbb{D}) 
$$
or explicitly,
\begin{eqnarray*} 
{\bf P}_{(\mathbb{D},a,b)}[\gamma_{[ab]} \subset U]=
{\bf P}_{(h(\mathbb{D}),h(a),h(b))}[\gamma_{[h(a)h(b)]} \subset h(U)],
\end{eqnarray*} 
where ${\bf P}_{(\mathbb{D},a,b)}[\gamma_{[ab]} \subset U]$ denotes the
probability for the curve $\gamma_{[ab]}$ to remain in a subset $U$
of $\mathbb{D}$. See Fig.\ref{fig:conftrans}. 

This condition is natural and it is the one that conformal field
theory suggests immediately. Let us note however that a totally
different definition of conformal invariance is understood in the
familiar statement ``two dimensional Brownian motion is conformally
invariant''.

Without any further constraint this condition is a kind of tautology
as it only tells how to transport the measure from one domain to
another. It only imposes to the measure to be invariant under the 
one parameter group of automorphisms that fixes $(\mathbb{D},a,b)$.
Indeed we could take any measure for $(\mathbb{D},a,b)$ -- well,
with the invariance under the one parameter group of automorphisms
that fixes $(\mathbb{D},a,b)$ -- and declare that the measure in
$h(\mathbb{D})$ is obtained by definition by the rule above. 

This condition becomes a non empty statement only when we start
comparing the transported measures with those obtained as continuum
limits of measures of discrete interfaces of lattice statistical
models. If this condition holds for the limit measure the interfaces
are said to be conformally invariant.

To make progress -- without going back to the discrete models -- we
need to combine conformal invariance with the domain Markov property.
This is the basic observation made by O. Schramm.

\subsubsection{Conformally invariant interfaces} 
\label{subsubsec:markovetstat}

This short Section establishes the most crucial properties of
conformally invariant interfaces: the Markov property and the
stationarity of increments of conformally invariant interfaces.
 
\begin{figure}[htbp]
\begin{center}
\includegraphics[width=0.9\textwidth]{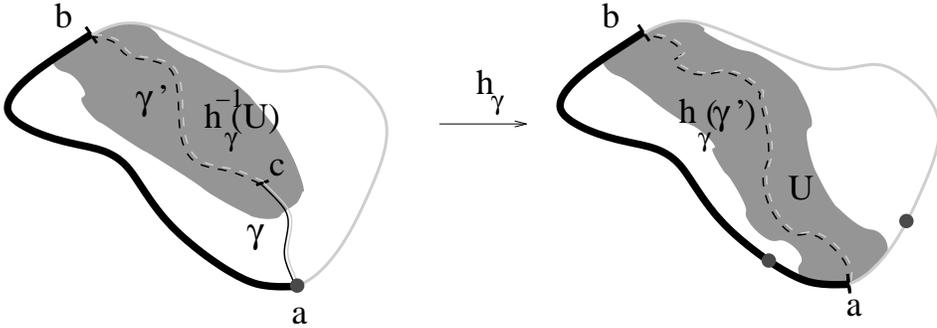}
\caption{Conformal invariance for conditional probabilities.}
     \label{fig:confinv}
\end{center}
\end{figure}

Take $c\in \mathbb{D}$ and let $\gamma_{[ac]}$ be a simple curve from
$a$ to $c$ in $\mathbb{D}$. Observe that $\mathbb{D}\setminus
\gamma_{]ac]}$ is a domain.  To answer the question ``if the beginning
of the interface is fixed to be $\gamma_{[ac]}$, what is the
distribution of the rest $\gamma'_{[cb]}$ of the interface?'' we
apply the domain Markov property to argue that this is exactly the
distribution of the interface in $\mathbb{D}\setminus \gamma_{]ac]}$.
We map this domain conformally to $\mathbb{D}$ via a map
$h_{\gamma_{[ac]}}$ sending $b$ to $b$ and $c$ to $a$:
$$
h_{\gamma_{[ac]}}(\mathbb{D}\setminus \gamma_{]ac]})= \mathbb{D}
\quad h_{\gamma_{[ac]}}(c)=a \quad h_{\gamma_{[ac]}}(b)=b
$$
so that the image by $h_{\gamma_{[ac]}}$ of the rest of the curve
$\gamma'_{[cb]}$ is curve from $a$ to $b$ in $\mathbb{D}$, see
Fig.\ref{fig:confinv}.

Conformal transport implies that the image measure is the original
measure, $ {\bf P}_{(\mathbb{D},a,b)}[\cdot\vert\gamma_{[ac]}]=
 h_{\gamma_{[ac]}}^*\circ {\bf P}_{(\mathbb{D},a,b)}[\cdot]$, so that    
$h_{\gamma_{[ac]}}(\gamma'_{[cb]})$ behaves
an interface from $a$ to $b$ that has forgotten $\gamma_{[ac]}$. 
More explicitly:
\[
{\bf P}_{(\mathbb{D},a,b)}[\gamma_{[cb]} \subset U\vert \gamma_{[ac]}]=
{\bf P}_{(\mathbb{D},a,b)}[ \gamma_{[ab]}\subset h_{\gamma_{[ac]}}(U)],
\]
for any subset $U\subset \mathbb{D}\setminus \gamma_{]ac]}$.\\
To summarize :
\begin{center}
\begin{minipage}[t]{0.9\textwidth}

$h_{\gamma_{[ac]}}(\gamma'_{[cb]})$ is independent of $\gamma_{[ac]}$ (the
Markov property) and has the same distribution as the original
interface itself from $a$ to $b$ (stationarity of increments).
\end{minipage}
\end{center}

\noindent
This property is what is meant by conformallly invariant interfaces
and is the main observation done by Schramm \cite{Schramm00}.

We end this Section with another caveat. The above statement, that SLE
is conformally covariant under domain changes, should not be confused
with the, incorrect in general, statement that SLE growth processes
are conformally invariant in the sense 2d Brownian motion is
conformally invariant. A local conformal transformation maps Brownian
motion to Brownian motion modulo a random time change. But SLE is
conformally invariant in that sense only in a special case, namely for
$\kappa=6$ which corresponds to percolation for which the central
charge and the conformal weight to be introduced later both vanish.
See Section \ref{sec:cftofsle}.

\subsection{Chordal SLE: basics} 
\label{sec:chordalsle}

The Markov and stationarity of increments property make it plain that
to understand the distribution of the full interface, it is enough to
understand the distribution of a small, or even infinitesimal, initial
segment, and then glue segments via conformal maps.

This calls for a description by differential equations, which turns
out to be conveniently described by Loewner evolutions.

For clarity let us recall basics and notations from Section
\ref{sec:anal}.  Using conformal invariance, we can restrict ourselves
to the situation when $(\mathbb{D},a,b)=(\mathbb{H},0,\infty)$.  If
$\gamma_{[0\infty]}$ is a simple curve from $0$ to $\infty$ in
$\mathbb{H}$, and $\gamma_t$ a point on it, we know that
$\mathbb{H}\setminus \gamma_{]0,t]}$ is a domain. As in previous
Section, it is convenient to use the capacity as a parameterization and
define a time parameter by $2t\equiv C_{\gamma_{]0,t]}}$.

Let $f_t$ be the conformal homeomorphism from $\mathbb{H}$ to
$\mathbb{H}\setminus \gamma_{]0,t]}$ normalized to satisfy
$f_t(w)=w-2t/w+O(1/w^2)$ at infinity. Define $g_t: \mathbb{H}\setminus
\gamma_{]0,t]} \mapsto\mathbb{H} $ to be the inverse of $f_t$. It
satisfies $g_t(z)=z+2t/z+O(1/z^2)$ at infinity.  Then, for
$\gamma_{]0,t]}$ a simple curve, there exists a real function $\xi_t$
such that the time evolution of these maps -- and therefore the growth
of the curve -- are described by the first order Loewner differential
equation (\ref{eq:Loewnerevolution}): 
$$dg_t(z)/dt=2/(g_t(z)-\xi_t).$$
The solutions of this equation for a given function $\xi_t$ with
initial condition $g_0(z)=z$ is called a Loewner evolution. By
construction, the image of $\xi_t$ by $g_t^{-1}$ is the tip of the
curve at time $t$, $\gamma_t = \lim_{\varepsilon\to 0^+}
g_t^{-1}(\xi_t+ i\varepsilon)$, so that the function $\xi_t$ provides
an alternative parameterization of the curve.

More generally, see Section \ref{sec:anal}, if $\xi_t$ is not regular
enough -- eg. if it has square root singularity -- solutions of the
Loewner equation (\ref{eq:Loewnerevolution}) may produce conformal
maps $g_t$ which uniformize the complements of hulls
$\mathbb{K}_t\equiv \mathbb{H}\setminus g_t^{-1}(\mathbb{H})$ which do
not coincide with the curves $\gamma_{]0,t]}$ but which are grown
locally from these curves.

With these tools in hands we may now conclude Schramm's argument and
present the definition of SLEs.

\subsubsection{Definition} 

If we sample locally growing hulls $\mathbb{K}_t$ with a certain
distribution, we get an associated random process $\xi_t$. In the case
of a conformally invariant distribution, we have established two
crucial properties: Markov property and stationarity of increments. To
finish Schramm's argument leading to SLE, what remains is to see the
implications of these properties on the distribution of $\xi_t$.

The argument and expressions for the meaning of Markov property and
stationarity of increments involved a map $h$ that mapped the tip of
the piece of interface to the initial marked point $a$ and the final
marked point $b$ to itself. The map $h_t(z)=g_t(z)-\xi_t$ has the
required property when the domain is the upper-half plane with $0$ and
$\infty$ as marked points: it maps the tip of the curve back to the
origin. It behaves like $h_t(z)=z-\xi_t+2t/z+O(1/z^2)$ at infinity. We
infer that for $s >t$, $h_t(\mathbb{K}_s\setminus \mathbb{K}_t)$ is
independent of $\mathbb{K}_{t'},\; t'\leq t$ (Markov property) and is
distributed like a hull of capacity $s-t=C_{h_t(\mathbb{K}_s\setminus
  \mathbb{K}_t)}$ (stationarity of increments).

The hull determines the corresponding map $h$, so this can be
rephrased as: the map $h_s\circ h_t^{-1}$, which uniformizes
$h_t(\mathbb{K}_s\setminus \mathbb{K}_t)$, is independent of
$h_{t'},\; t'\leq t,$ and distributed like an $h_{s-t}$.  As $h_s\circ
h_t^{-1}=z-(\xi_s-\xi_t)+2(s-t)/z+O(1/z^2)$ at infinity, the driving
parameter for the process $h_s\circ h_t^{-1}$ is $\xi_s-\xi_t$. To
summarize:

\begin{center}
\begin{minipage}[t]{0.9\textwidth}
the Markov property and stationarity of increments for the
interface lead to the familiar statement for the process $\xi_t$: for
$s >t$, $\xi_s-\xi_t$ is independent of $\xi_{t'},\; t'\leq t,$
(Markov property) and distributed like a $\xi_{s-t}$ (stationarity of
increments).
\end{minipage}
\end{center}

To conclude, two last physical inputs are needed. One first demands
that the interface does not branch, which means that at two nearby
times the growth is at nearby points. This implies that $\xi_t$ is a
continuous process, in the sense that it has continuous trajectories.
One also requires that the distribution of the curve is left-right
symmetric under reflexion with to respect to the imaginary axis. This
implies that $g_t(z)$ and $-\overline{g_t(-\bar z)}$ are identically
distributed which as consequence implies that $\xi_t$ and $-\xi_t$ are
alike.

One is now in position to apply a mathematical theorem: a 1d Markov
process with continuous trajectories, stationary increments and
reflexion symmetry is proportional to a Brownian motion. We conclude
that there is a real positive number $\kappa$ such that
$\xi_t=\sqrt{\kappa} B_t$ for some normalized Brownian motion $B_t$
with covariance $\mathbb{E}[B_sB_t]=\min(s,t)$. The same argument
without imposing that the time parameterization is given by the
capacity of the hull would lead to the conclusion that the driving
parameter is a continuous martingale, which is nothing but a Brownian
motion after a possibly random time change.\\

To summarize:\\
A solution of Loewner equation with a Brownian motion as driving term,
\begin{equation}
\label{eq:slec}
\frac{dg_t}{dt}(z)=\frac{2}{g_t(z)-\xi_t}
\quad,\quad \xi_t=\sqrt{\kappa} B_t,
\end{equation}
with $g_{t=0}(z)=z$, is called a chordal Schramm-Loewner evolution of
parameter $\kappa$, in short a chordal SLE, in the upper half plane
$\mathbb{H}$ with two marked points $0$ and $\infty$. Given $z\in
\mathbb{H}$, eq.(\ref{eq:slec}) may be integrated up to time $\tau_z$,
called the swallowing or exploding time, for which
$g_{\tau_z}(z)=\xi_{\tau_z}$.\\

By construction, chordal SLE enjoys the two following properties:
\begin{itemize}
\item $g_t(z)$ and $g_{\lambda^2 t}(\lambda z)$ are identically
  distributed. This is direct consequence of usual scaling property of
  the Brownian motion. Since dilatation is the only conformal map
  preserving the upper half plane with its two marked points $0$ and
  $\infty$, this ensures the coherence of the process.
\item $h_t(z)\equiv g_t(z)-\xi_t$ has the Markov property.
  Furthermore, $h_s\circ h_t^{-1}$ is independent of $h_t$ and
  distributed as $h_{s-t}$ for $s>t$. Again this is a direct
  consequence of Brownian motion properties.
\end{itemize}

The connection of this equation with interfaces relies mainly on
conformal invariance. But local growth, absence of branches, and to a
lower level locality at the interface, also play a crucial role.

\subsubsection{Basics properties}
\label{sec:basics}

The set of exact results obtained for SLE forms an impressive body of
knowledge. See for instance the physical 
\cite{GruzbergKadanov04,KagerNienhuis03,BB04z,CardyRevue05} 
or  mathematical 
\cite{Lawlerbook05,WernerFlour02,Wlectures05}
reviews. In this Section, we
list just a few ``pictorial'' properties with some comments and we
leave more detailed computations for the following Sections. They --
the properties and the comments -- should be understood with the
standard proviso ``almost surely'' or ``with probability $1$''.

We start with a surprisingly difficult result
\cite{Schramm00,LSW1,LSW2,LSW3,RohdeSchramm01,Beffara02b}.
\begin{itemize}
\item Whatever the value of $\kappa$, the pre-image of the driving
  parameter $$\gamma_t\equiv \lim_{\varepsilon\rightarrow 0^+}
  g_t^{-1}(\sqrt{\kappa}B_t+i\varepsilon)$$ is a continuous curve,
  called the SLE trace. The curve starts at $\xi_{t=0}=0$ and it
  reaches the point at infinity at infinite time, $\lim_{t\to
    \infty}\gamma_t=\infty$. The trace never crosses itself. This
  property is crucial if the trace is to be interpreted as a curve
  separating two phases.
\item For $\kappa\in [0,4]$ the SLE trace is a simple curve. For
  $\kappa \in ]4,8[$, it has double points. For $\kappa \in [8,
  \infty[$, it is space filling. A probabilistic argument for this
  property is given at the end of this Section.
\item The fractal dimension $d_{\kappa}$ of the trace is $1+\kappa /8$
  for $\kappa \leq 8$ and $2$ for $\kappa \geq 8$. See Section
  \ref{sec:dimfract}.
\end {itemize}

Using the formula for the dimension of the trace and confronting with
the numerical simulations of Section \ref{sec:examples}, it is
plausible (actually, these are among the few cases for which a
mathematical proof exists) that loop-erased random walks correspond to
$\kappa=2, d=5/4$ and percolation to $\kappa=6, d=7/4$. This is also
compatible with the general shape of the numerical samples, which
indicate that loop-erased random walks indeed lead to simple curves
and that
percolation doesn't.\\
 
The hull $\mathbb{K}_t$ is by definition $\mathbb{H}\setminus
g_t^{-1}(\mathbb{H})$. It is also the set of point which have been
swallowed by the trace at time $t$, namely 
$\mathbb{K}_t=\{z\in\mathbb{H};\ \tau_z< t\}$.
It has the following properties:
\begin{itemize}
\item The hull $\mathbb{K}_t$ is the complement of the connected
  component of $\infty$ in $\mathbb{H}\setminus \gamma_{]0,t]}$.
\item For $\kappa\in [0,4]$, the SLE hull is a simple curve coinciding
  with the trace. For $\kappa \in ]4,\infty[$, the SLE hull has a
  non empty connected and relatively dense interior.
\end{itemize}

Furthermore, there is a duality conjecture which states that the
exterior frontiers of SLE hulls for $\kappa>4$ looks locally as SLE
curves but for a dual value $\kappa_*=16/\kappa<4$. It is still
unproved.

\begin{figure}[htbp]
\begin{center}
\includegraphics[width=\textwidth]{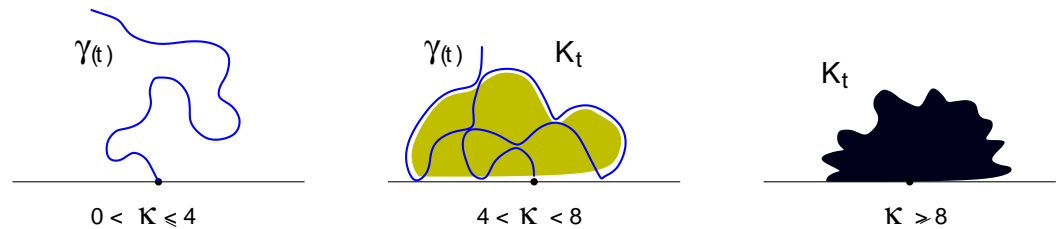}
\caption{ The phases of SLE.}
     \label{fig:sletrace}
 \end{center}
\end{figure}

These properties may seem surprising at first sight. They are the
signs that for $\kappa > 4$, the drift $\sqrt{\kappa} B_t$ goes fast
enough for the swallowing procedure to take place, as described in the
previous closing arc example, but on all scales. Although, for
$4<\kappa<8$, SLE produces non trivial hulls $\mathbb{K}_t$, the tip
of the curves $\gamma_{[0,t]}$ is always emerging from the hulls
towards infinity and never reenter into the hull. This is a
consequence of local growth.  This is summarized by
Fig.\ref{fig:sletrace}.

The existence of these different phases may be grasped by looking at
the motion of a point of the real axis. So let $x\in\mathbb{R}$ and
consider the real process $Y_t\equiv h_t(x)/\sqrt{\kappa}
=(g_t(x)-\xi_t)/\sqrt{\kappa}$. By construction it is a Bessel process
since it satisfies the stochastic equation $dY_t=
\frac{(2/\kappa)}{Y_t}dt + dB_t$. A $d$-dimensional Bessel process is
the process given by the modulus of Brownian motion in dimension $d$.
If $R_t$ denotes this modulus, it satisfies the stochastic equation
$dR_t=\frac{(d-1)/2}{R_t} + dB_t$, see Appendix \ref{app:proba}. So
the effective dimension for the process $Y_t$ is $d_{\rm eff}=
\frac{4+\kappa}{\kappa}$.  Now a Brownian motion is recurrent in
dimension less than $2$ -- meaning that it reaches the origin an
infinite number of times with probability one -- and not recurrent in
dimension bigger than $2$.  So with probability one $Y_t$ vanishes in
finite time with probability one for $\kappa>4$ and remains finite
with probability one for $\kappa<4$. Since the vanishing of $Y_t$
signals that the SLE curve has touched the real axis between $x$ and
$\infty$, this tells us that the SLE curve touches the real axis an
infinite number of time with probability one for $\kappa>4$ and does
not touch the real axis for $\kappa<4$. Markov and independent
increment properties then imply that the curve has an infinite
number of double points for $\kappa>4$ and none for $\kappa<4$.  We
shall present in a following Section an argument indicating the SLE
trace is space filling for $\kappa>8$. The phase diagram simply
expresses this behavior. 

\begin{figure}[htbp]
\begin{center}
  \[ \includegraphics[width=.4\textwidth]{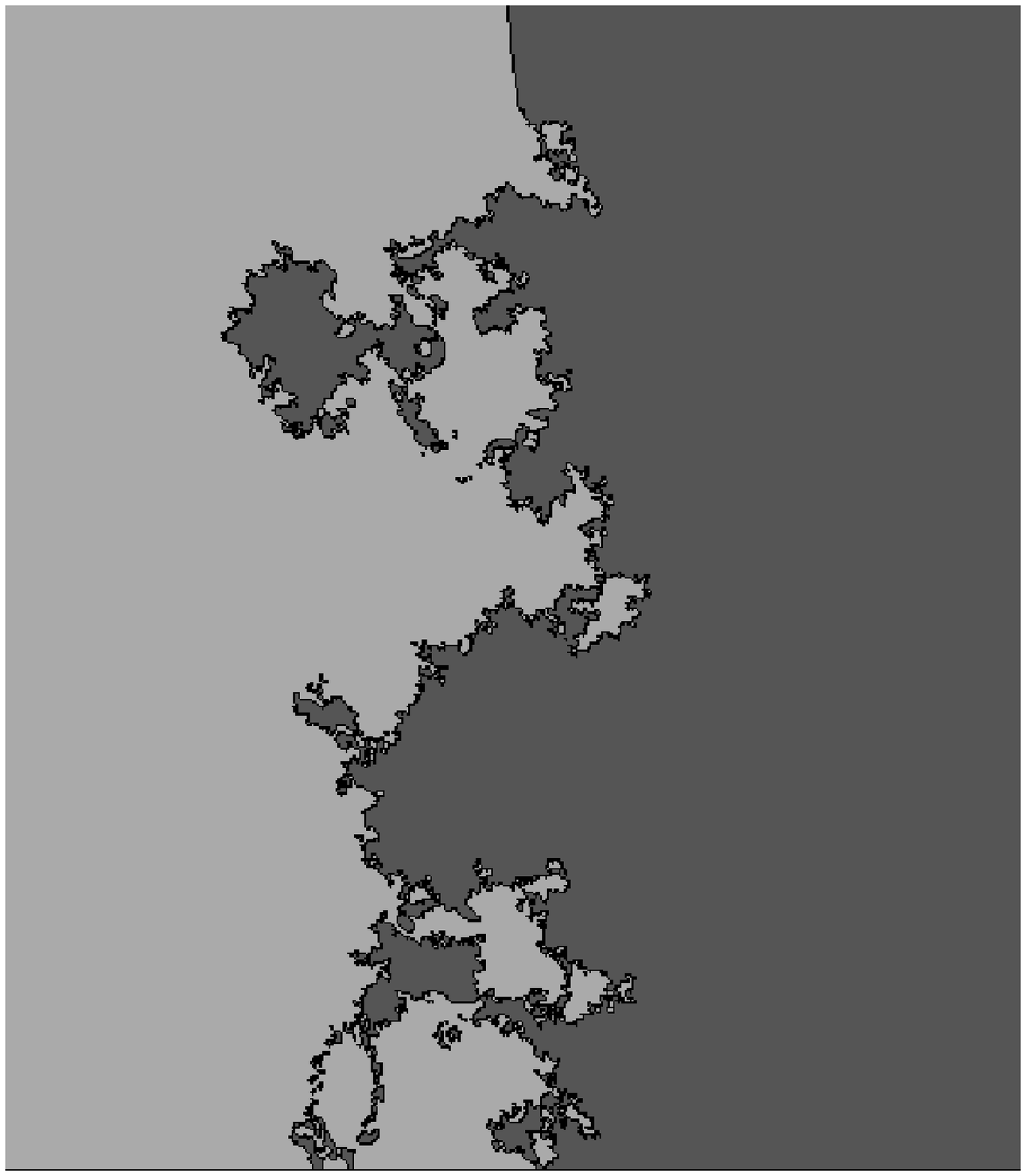} \qquad
  \includegraphics[width=.4\textwidth]{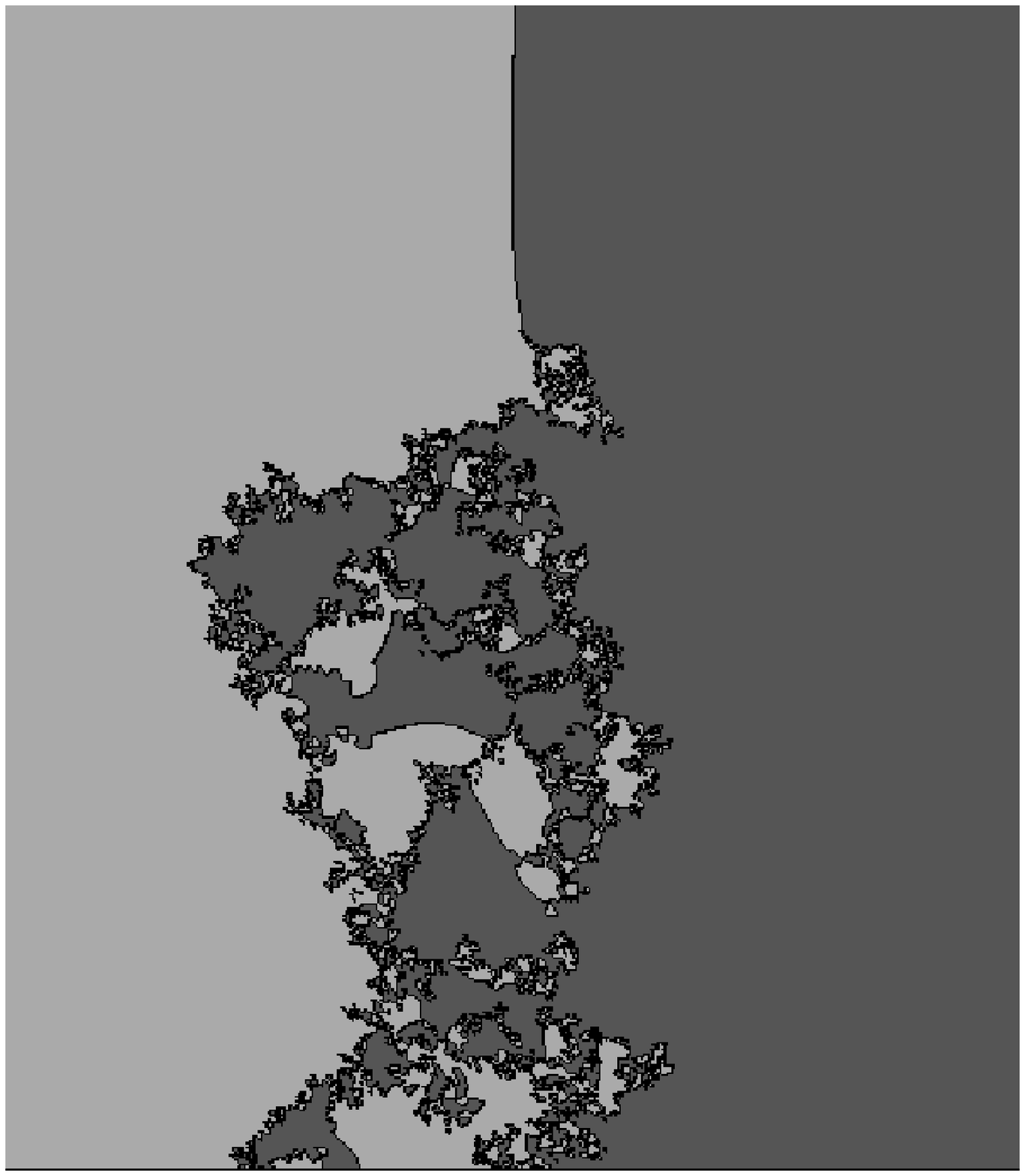}\]
\caption{ Two SLE samples for $\kappa=4$ and $\kappa=6$.}
     \label{fig:sle4et6}
 \end{center}
\end{figure}

Nice images of SLE and other growth processes can be found on Vincent
Beffara's webpage
\texttt{http://www.umpa.ens-lyon.fr/\~vbeffara/pics.php}.
We just quote two examples, $\kappa=4$ and $\kappa=6$ on Fig.\ref{fig:sle4et6}. 

\subsection{Other SLEs}
\label{sec:othersle}

Chordal SLEs yield measures $P_{(\mathbb{D},a,b)}$ on sets of curves
joining two boundary points $a$, $b$ of a domain $\mathbb{D}$.  There
are a few other versions of SLEs corresponding to different boundary
conditions. In simply connected planar domains, there are three
possible definitions (preserving reflection symmetry): chordal, radial
and dipolar SLEs.  A group theoretical explanation of this
classification will be given in the following Section \ref{sec:geosle}.
As for chordal SLE, they are defined via a version of Loewner equation
for a uniformizing map $g_t$ from which one reconstructs the curve.
They are conformally invariant in the sense that they satisfy the
Markov and identical increment properties.  They differ by their
global topological behavior but their local properties are identical.
For instance, they have three different phases depending on the value
of $\kappa$ as for chordal SLE -- with the same phase diagram -- and
their traces have identical fractal dimensions.

There has been attempts to specify SLE in two dimensional manifold of
more general topology. See Section \ref{sec:diversmisc} for an
overview.  Here we shall only present the definition of SLEs in simply
connected domains. For each SLE there is an adapted geometry -- the
upper half plane was adapted to chordal SLE -- and we shall use these
geometries to give the definitions but, of course, the processes can
be defined in any geometry by conformal transport.

\subsubsection{Radial SLE}
\label{sec:radialsle}
Radial SLE \cite{Schramm00,RohdeSchramm01,LSW1,Lawlerbook05} describes
curves joining a point on the boundary to a point in the bulk of the
domain, which have to be specified. So the radial SLE data are: the
domain $\mathbb{D}$, the starting boundary point $x_0$ and the final
bulk point $z_*$. Notice that there is no non-trivial global conformal
transformation preserving these data (contrary to the chordal case in
the upper half plane in which dilatation preserves the data).  It was
originally defined using the unit disk as reference domain
$\mathbb{D}=\{z,\ |z|<1\}$ with $1$ as the boundary point where the
SLE trace emerges and $0$ as the inside point where the SLE trace
converges. The Loewner equation then becomes:
\begin{eqnarray}
\partial_t g_t(z) = -g_t(z)\ \frac{g_t(z)+U_t}{g_t(z)-U_t},\quad
U_t=\exp(i\xi_t)
\label{radialdisc}
\end{eqnarray}
with $g_{t=0}(z)=z$.  The origin is kept fix $g_t$ and the map is
normalized so that its derivative at the origin is real. The time $t$
is defined via $g_t'(0)=e^{t}$. The radial SLE trace $\gamma_t$ is
reconstructed via the Loewner map by $\gamma_t=g_t^{-1}(U_t)$. As for
chordal SLE, the hull coincides with the trace for $0<\kappa\leq 4$
and the curve has infinitely many double points for $4\leq\kappa\leq
8$.

Radial SLE is particularly simple on a semi-infinite cylinder of
circumference $\Lambda\pi$, whose boundary is a circle. In this
geometry the defining equation simply reads:
\begin{equation}
\label{eq:sler}
\frac{dg_t}{dt}(z)
=\frac{2\Lambda^{-1}}{\tan[(g_t(z)-\xi_t)/\Lambda]},
\quad \xi_t=\sqrt{\kappa} B_t
\end{equation}
The trace $\gamma_t=g^{-1}(\xi_t)$ starts on the 
boundary of the cylinder and winds many times before
reaching infinity, which is a bulk point of the cylinder.

Without breaking conformal invariance, ie. the Markov and independent
increment property, we have the freedom in radial SLE to choose as
driving source $U_t=\sqrt{\kappa}B_t+\alpha t$, with a linear-in-time
drift term, instead of the simple choice $\xi_t=\sqrt{\kappa}B_t$. In
the disc geometry, this breaks rotation invariance but provides a way
to account for the winding number of the SLE trace around the origin.

\subsubsection{Dipolar SLE}
\label{sec:dipolsle}

To define dipolar SLE \cite{BBH04} one specifies a boundary point,
which is going to be the starting point of the trace, and a boundary
interval not including the starting point, which is going to included
the termination point of the trace.  Then dipolar SLE describes curves
starting on the specified boundary point and stopped the first instant
they hit the specified boundary interval.  So the dipolar SLE data
are: the domain $\mathbb{D}$, the starting point $x_0$ and the two
boundary points $x_\pm$ of the terminal interval $[x_-,x_+]$. As for
radial SLE, there is no non-trivial conformal map preserving this
data.  An adapted geometry is the strip of width $\pi\Delta$,
$\mathbb{S}_\Delta=\{z,\ 0<\Im{\rm m} z <\pi\Delta\}$, with two
boundary points at $\pm\infty$, in which case the equation reads:
\begin{equation}
\label{eq:sled}
\frac{dg_t}{dt}(z)=
\frac{\Delta^{-1}}{\tanh[ (g_t(z)-\xi_t)/2\Delta]},\quad
\xi_t=\sqrt{\kappa} B_t
\end{equation}
with $g_{t=0}(z)=z$. The trace $\gamma_t=g^{-1}(\xi_t)$ starts at the
origin and ends randomly on the upper boundary side. For $0<\kappa\leq
4$, the hull coincide with the trace which is then a curve joining
lower and upper boundary side, touching them only once. For
$4\leq\kappa\leq 8$, the curve has infinitely many double points so
that it does not coincide with its hull. The hulls touch the lower
boundary infinite many times but only once the upper boundary -- and
this happens at infinite time. As a consequence the hulls do not
invade the full domain but only a random subset of it, and this is one
of the main differences between dipolar SLE and chordal or radial SLE.

In the limit of infinitely large strip dipolar SLEs converge to
chordal SLEs on the upper half plane or, alternatively, chordal SLEs
describe locally dipolar SLEs in the neighborhood of their starting
points. This indirectly shows that the chordal SLE traces reach the
point at infinity with probability one.

\begin{figure}[htbp]
\begin{center}
 \includegraphics[width=.7\textwidth]{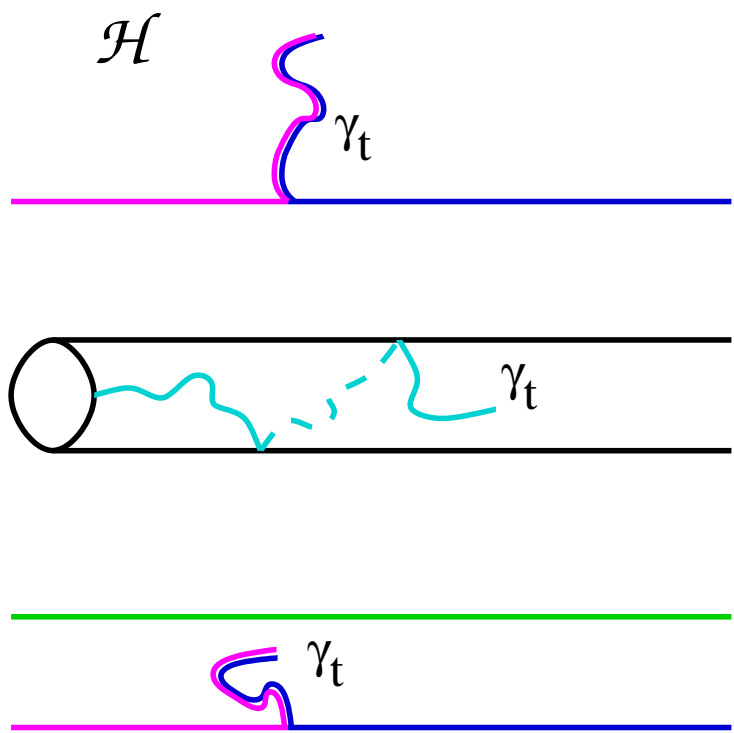} 
\caption{Three avatars of SLE: chordal, radial and dipolar.}
     \label{fig:sleavatars}
 \end{center}
\end{figure}

\subsubsection{SLE($\kappa,\rho$)}
\label{sec:slekrho}
SLE($\kappa,\rho$) involves \cite{LSWConfRest02} the same data as dipolar
SLE: the domain $\mathbb{D}$, the starting point $x_0$ and two other
boundary points $x_+$ and $x_-$.  Although not originally defined this
way, it may be viewed as a generalization of dipolar SLE in which
reflection symmetry is not imposed but conformal invariance is still
preserved. This is possible because there is no non-trivial global
conformal map preserving the data.  In the strip geometry of width
$\Delta\pi$, with the boundary points $x_\pm=\pm\infty$,
this amounts to add a linear-in-time drift to the driving term, so
that the defining equation becomes:
\begin{equation}
\label{slerhodip}
\frac{dg_t}{dt}(z)
=\frac{\Delta^{-1}}{\tanh[(g_t(z)-U_t)/2\Delta]},\quad
U_t=\sqrt{\kappa} B_t+\alpha t
\end{equation}
with $g_{t=0}(z)=z$. The effect of the drift is to push the SLE trace
preferably towards one of the two boundary points $x_\pm$. Of course
it reduces to dipolar SLE for $\alpha=0$.

SLE($\kappa,\rho$) were initially defined in the upper half plane
$\mathbb{H}$, with a marked point $x$ on the real line, via the
stochastic equation:
\begin{eqnarray}
d\hat g_t(z) &=& \frac{2dt}{\hat g_t(z)-\hat U_t},
\label{sleKRH}\\
d\hat U_t&=&\sqrt{\kappa}dB_t+\frac{\rho\, dt}{\hat U_t-\hat g_t(x)}
\nonumber
\end{eqnarray}
The trace emerges from the point $\hat U_0$, say $x<\hat U_0$, and
ends on the real axis between point $x$ and the point at infinity. For
$\rho=0$ it reduces to SLE${}_\kappa$ and SLE($\kappa,\kappa-6$) is a
chordal SLE from $U_0$ to $x$ but with a non standard normalization of
the conformal map.

The two descriptions are actually equivalent, as shown for instance in
\cite{Kytola05,SchrammWilson05}. The mapping from one definition to the other
requires mapping conformally the strip into the upper half plane, with
the appropriate normalization, and a random time change.  The
definition via eqs.(\ref{sleKRH}) treats the two marked points $x$ and
$\infty$ very asymmetrically contrary the first definition
(\ref{slerhodip}). The relation between $\alpha$ and $\rho$ is
$\rho=\frac{\kappa-6}{2}+\alpha$. In particular,
SLE($\kappa,\frac{\kappa-6}{2}$) is equivalent to dipolar SLE.

To make explicit the relation between the two descriptions is
computationally a bit long so we only give hints for it.  We start
from the first formulation in a strip of width $\pi$ and define
$k_t(z)=\exp(-t+g_t(\log z))$ and $W_t=\exp(-t+U_t)$.  The map $k_t$
is, up to a random time change and a translation, the
SLE$(\kappa,\rho)$ map in the upper half plane with marked point $x=0$
and starting point $\hat U_0=1$. So let us change time by
defining $ds=W^2_t dt$ and $Z_s=W_{t(s)}$. We set $\hat
g_s(z)=k_{t(s)}(z)+X_s$ and $\hat U_s=Z_s+X_s$ with $X_s$ solution 
$dX_s=-\frac{2ds}{Z_s}$. Then $\hat g_s$ and $\hat U_s$ satisfy the
SLE$(\kappa,\rho)$ equation (\ref{sleKRH}) with $X_s=\hat g_s(0)$ (the
marked point is $x=0$) and $\hat U_0=1$ (the starting point is $1$).

There exits a nice interplay \cite{Werner03f} between
SLE$(\kappa,\rho)$ and restriction measure to be described below in
Section \ref{sec:restrict}. Certain restriction properties and their
relations with the duality conjecture $\kappa\to16/\kappa$ have been
presented in \cite{Dubedat05a}.

\subsection{Geometry of SLE}
\label{sec:geosle}
The aim of this Section is to arrive at an alternative group
theoretical formulation of SLE processes which may later be used to
make contact with the group theoretical formulation of conformal field
theory. In our way, this will teach us what is the geometry -- in the
sense of differential geometry -- of the stochastic equations
underlying SLEs and how this geometry is linked to conformal
invariance.

\subsubsection{Conformal transport}
\label{sec:conftransp}

Our starting point is a discussion of conformal covariance for
stochastic differential equations in the following sense. It is well
known that trajectories of points on manifolds are related to vector
fields. The case of interest for us is when the manifold is a Riemann
surface $\Sigma$. Suppose $z \in \mathbb{U}\subset \mathbb{C}$ is a
coordinate system for some open subset of $\Sigma$ and $\varphi$ maps
$\mathbb{U}$ conformally to some domain $\mathbb{V} \subset
\mathbb{C}$. Suppose that an intrinsic motion of points on $\Sigma$,
when written in the local coordinate in $\mathbb{U}$, satisfies the
family of differential equations $dh_t(z)=dt\, \nu(h_t(z))$, with
initial conditions $h_0(z)=z$, where $\nu$ is holomorphic in
$\mathbb{U}$. Then, when written in the local coordinate in
$\mathbb{V}$, the corresponding map is $h^\varphi_t\equiv \varphi
\circ h_t \circ \varphi^{-1}$, which satisfies $dh^\varphi_t=dt\,
(\nu^\varphi \circ h^\varphi_t)$ with $\nu^\varphi \circ \varphi =
\varphi' \nu$.  This equation expresses that $w=\nu(z)\partial_z$ is a
holomorphic vector field on some open subset of $\Sigma$.

What happens now if the motion on $\Sigma$ is stochastic? Suppose
that $\xi_t$ a Brownian motion with covariance ${\bf
  E}[\xi_t\,\xi_s]=\kappa\, {\rm min}(t,s)$ and that the motion, 
written in local coordinate in $\mathbb{U}$ reads
\begin{eqnarray}
 \label{eqdiff}
dh_t(z)=dt\, \sigma(h_t(z))+ d\xi_t \, \rho(h_t(z)),  
\end{eqnarray} 
For each trajectory $\xi_t$ there is a random but almost surely
positive time $T$ and a non empty open domain $\mathbb{U}_T \subset
\mathbb{U}$ such that $h_t$ maps $\mathbb{U}_T$ into $\mathbb{U}$ and
solves the above differential equation for $t \in [0,T]$ and $z \in
\mathbb{U}_T$. Then, the motion in $\mathbb{V}$, obtained by transport
from $\mathbb{U}$, is given by $h^\varphi_t\equiv \varphi \circ h_t
\circ \varphi^{-1}$. By It\^o's formula it satisfies
$$
dh^\varphi_t=dt\,(\sigma^\varphi \circ h^\varphi_t) 
+ d\xi_t \,(\rho^\varphi\circ h^\varphi_t)
$$ 
with 
$\rho^\varphi \circ \varphi = \varphi' \rho$ and
$\sigma^\varphi \circ 
\varphi =\varphi' \sigma+\frac{\kappa}{2} \varphi'' \rho^2$.
By a simple rearrangement,  this means that 
\begin{equation}
\label{eq:rel}
w_{-1} \equiv \rho(z)\, \partial_z \quad \mathrm{and} 
\quad w_{-2}\equiv \frac{1}{2}\left(-\sigma(z)+
  \frac{\kappa}{2}\rho(z)\rho'(z)\right)\partial_z
\end{equation}
are holomorphic vector fields on an open subset of $\Sigma$.  Under
the motion, some points may leave this open subset before time $t$.
The corresponding random subsets of $\mathbb{U}$ and $\mathbb{V}$ are
related by $\varphi$. Eq.(\ref{eq:rel}) is another, more geometrical,
way to formulate conformal transport as it tells us how to define the
flow in one domain if we know it in another conformally equivalent
domain.

The two vector fields $w_{-1}$ and $w_{-2}$ and the Lie algebra they
generated play a peculiar role. This points towards a group
theoretical formulation which emphasizes it.

To make contact with group theory we want to view the flow $h_t$ as
element of a group. So let us assume that there is a linear space
${O}$ of holomorphic functions and a group $N$
that (anti) acts faithfully on $O$ by composition, $\mathfrak{g}_h
\cdot F \equiv F \circ h$ for $F \in {O}$ and $h \in
N$, and furthermore that $h_t \in N$ at least
up to a possibly random but strictly positive time.  In this situation
we may view $h_t$ as a random process $\mathfrak{g}_{h_t}$ on
$N$. We shall exhibit the appropriate spaces ${O}$
and groups $N$ for SLEs in the following Section.

It\^o's formula shows that $(\mathfrak{g}_{h_t}^{-1} \cdot d
\mathfrak{g}_{h_t}) \cdot F =(dt \, \sigma + d\xi_t \, \rho)F'+ dt
\,\frac{\kappa}{2}\rho^2 F''$, for any function $F\in {O}$, or
equivalently
\begin{eqnarray} 
\mathfrak{g}_{h_t}^{-1} \cdot d \mathfrak{g}_{h_t}=dt \,
(-2w_{-2}+\frac{\kappa}{2}w_{-1}^2)+ d\xi_t \, w_{-1}.
\label{SLEcov}
\end{eqnarray}
since $N$ acts faithfully on ${O}$.  This equation
may be transformed into an ordinary differential equation for the map
$g_t \equiv e^{-\xi_t w_{-1}}\cdot h_t$ which is obtained by 
transporting $h_t$ along the one parameter subgroup
generated by $w_{-1}$. The corresponding group element
is given by $\mathfrak{g}_{g_t}=\mathfrak{g}_{h_t}e^{-\xi_t w_{-1}}$
and it satisfies the ordinary differential equation
$$\mathfrak{g}_{g_t}^{-1} \cdot d \mathfrak{g}_{g_t}=-2dt \, 
(e^{\xi_t w_{-1}}w_{-2}e^{-\xi_t w_{-1}}).$$

Eq.(\ref{SLEcov}) equation involves only intrinsic geometric objects.
It clearly reveals that the vector field $w_{-1}$ drives the Brownian
motion while $w_{-2}$ specifies the drift.  It is at the heart of the
relation between SLE and conformal field theory.  The structure of the
Lie algebra generated by $w_{-1}$ and $w_{-2}$ will play an important
role and the possibility to embed this Lie algebra in the Virasoro
algebra will turn out to be crucial. As recall in Appendix 
\ref{app:cft}, it is the (essentially unique) central extension of the
Witt algebra, which is the Lie algebra of Laurent polynomial vector
fields $\ell_n$,
$$\ell_n\equiv -z^{n+1}\partial_z,$$ 
in $\mathbb{C}$ with commutation relations
$[\ell_n,\ell_m]=(n-m)\ell_{m+n}$.

\subsubsection{Group theoretical formulation}
\label{sec:groupsle}

Let us now describe the group theoretical formulation of the three
SLEs -- chordal, radial, dipolar -- and SLE$(\kappa,\rho)$. For later
use in connection with conformal field theory, we shall specify the
vector fields $w_{-1}$ and $w_{-2}$ using the standard complex
coordinate $z$ in the upper half plane $\mathbb{H}$ but, of course,
they can be transported to any domain.  
\vskip .3 truecm

$\bullet $ {\it Chordal SLE.}\\
It is useful to define $h_t(z) \equiv g_t(z) - \xi_t$ which satisfies
the stochastic (Loewner) differential equation:
$$ 
d h_t = \frac{2dt}{h_t}-d\xi_t.
$$
The germ of $h_t$ at infinity belongs to the group $N_-$ of germs of
holomorphic functions at $\infty$ of the form $z+\sum_{m \leq -1}
h_{m} z^{m+1}$ (with real coefficients) -- fixing infinity and with
derivative one there.  The group $N_-$ (anti)acts by composition on
${O}_-$, the space of germs of holomorphic functions at $\infty$
fixing $\infty$ -- but without normalized derivative their.  As above,
to $h_t$ we associate $\mathfrak{g}_{h_t}\in N_-$ which satisfies
eq.(\ref{SLEcov}) i.e.  explicitly
\begin{eqnarray}
\mathfrak{g}_{h_t}^{-1} \cdot d \mathfrak{g}_{h_t}=dt(\frac{2}{z}\partial_z
+\frac{\kappa}{2}\partial_z^2)-d\xi_t\partial_z.
\label{itoN-moi}
\end{eqnarray} 
According to our previous discussion, we identify
$w_{-1}=-\partial_z$ and $w_2=-\frac{1}{z}\partial_z$, ie.
$w_{-1}=\ell_{-1}$ and $w_{-2}=\ell_{-2}$. The first vector field is
holomorphic in $\mathbb{H}$ and tangent to the boundary, so that by
the Schwarz reflection principle it extends automatically to a
holomorphic vector field in the full complex plane. The second one is
holomorphic in $\mathbb{H}$ and tangent to the boundary except at the
origin. For the same reason it extends automatically to a holomorphic
vector field in the complex plane with the origin removed; the
extension has a simple pole with residue $2$ as its sole singularity.
Both $w_{-1}$ and $w_{-2}$ vanish at infinity, a double and triple
zero respectively, which accounts for the fact that $h_t(z)=z +O(1)$
at infinity. They have no other common zero, which is the geometric
reason why the SLE trace goes to infinity at large $t$. 

Note that to define Brownian motion along a curve, one needs
a parameterization. The fact that $\ell_{-1}$ is the infinitesimal
generator of a one parameter group of conformal automorphisms of
$\mathbb{H}$ that extend to the boundary can be viewed as providing
such a parameterization.
\vskip 0.3 truecm

$\bullet$ {\it Radial SLE.}\\
For radial SLE conformal covariance allows to choose
$\mathbb{H}$ as domain, $0$ as the boundary point where the SLE trace
emerges and $i$ as the inside point where the SLE trace converges. In
terms of geometry of vector fields, we can still use holomorphicity,
the Schwarz symmetry principle and left-right symmetry, and impose
that $-2w_{-2}$ has after extension a simple pole at the origin and
that $w_{-1}$ is holomorphic. The sole difference with the chordal
case is that this time the vector fields have to vanish at $i$, where
the SLE trace converges. This gives two real conditions, so the
situation is more rigid than in the chordal case. One finds
$w_{-2}=-\frac{(1+z^2)}{4z}\partial_z$, and
$w_{-1}=-\frac{1+z^2}{2}\partial_z$. The choice of the proportionality
factor is just a normalization.  For the space
${O}$ we choose this time the germs of holomorphic functions
at $i$ fixing $i$ and $N$ is the subspace of $O$ made of
the germs with non vanishing derivative at $i$. Hence:
$$
w_{-2}=\frac{1}{4}(\ell_{-2}+\ell_0)=-\frac{1+z^2}{4z}\partial_z
\quad,\quad 
w_{-1}=\frac{1}{2}(\ell_{-1}+\ell_1)=-\frac{1+z^2}{2}\partial_z.
$$
Observe that this time we do not use translations but another one
parameter subgroup of the group of conformal automorphisms of
$\mathbb{H}$, namely the ones fixing $i$, to parameterize the real
axis and define Brownian motion.

Radial SLE is then governed by eq.(\ref{SLEcov}).  As explained in
previous Section, the ordinary differential equation governing radial
SLE in $\mathbb{H}$ is obtained by considering
$g_t=e^{-\xi_t w_{-1}}\cdot h_t$. Since
$e^{\xi_t w_{-1}}w_{-2}e^{-\xi_t w_{-1}}
=-(\frac{1+z^2}{4})\,\frac{1+z\tan \xi_t/2}{z-\tan \xi_t/2}$, 
the map $g_t$ satisfies:
\begin{eqnarray}
\partial_t g_t(z)=\big(\frac{1+g_t(z)^2}{2}\big)\ 
\frac{1+g_t(z)\tan \xi_t/2}{g_t(z)-\tan \xi_t/2}. 
\label{radialplan}
\end{eqnarray}
This is radial SLE in the upper half plane. It is clear that the map
$g_t$ fixes the map $i$ towards which the curve converges. 

\vskip 0.3 truecm

$\bullet$ {\it Dipolar SLE.}\\
If one realizes that radial SLE is closely linked to a compact Cartan
torus of $SL_2(\mathbb{R})$, related to rigid rotations of the disk,
it is tempting to look at non compact Cartan torus of
$SL_2(\mathbb{R})$. This yields dipolar SLE. It amounts to replace the
complex fixed point $i$ by the pair of real fixed points $1$ and $-1$
and leads to
$$ w_{-2}=\frac{1}{4}(\ell_{-2}-\ell_0)
=-\frac{1-z^2}{4z}\partial_z \quad,\quad
w_{-1}=\frac{1}{2}(\ell_{-1}-\ell_1)=-\frac{1-z^2}{2}\partial_z,$$
For $O$ and $N$, one has two natural choices: germs of holomorphic
functions at $\pm 1$ fixing $\pm 1$.

One can check that the corresponding ordinary differential equation,
\begin{eqnarray}
\partial_t g_t(z)=\big(\frac{1-g_t(z)^2}{2}\big)\ 
\frac{1-g_t(z)\tanh \xi_t/2}{g_t(z)-\tanh \xi_t/2}.
\label{dipolarplan}
\end{eqnarray}
is the Loewner equation when the
Loewner map is normalized to fix $1$ and $-1$ and have the same derivative
at these two points : $g'_t(\pm 1)=e^{- t}$. 
\vskip 0.5 truecm

$\bullet$ {\it SLE($\kappa,\rho$).}\\
The group theoretical formulation of SLE($\kappa$,$\rho$) can be found
in \cite{Kytola05}. It follows by using, as in eq.(\ref{slerhodip}), the
fact that it coincides with dipolar SLE but with a driving source
$U_t=\xi_t+\alpha t$ instead of $\xi_t=\sqrt{\kappa}B_t$ with
$\rho=\frac{\kappa-6}{2}+\alpha$. This amounts to translate
$w_{-2}$ into $w_{-2}-\frac{\alpha}{2} w_{-1}$ since it simply adds a
supplementary drift term.

%% file: chap4.tex
 
In this Section we explain how stochastic processes may be
defined -- in a natural way -- in any models of statistical mechanics
such that conditioned correlation functions of the statistical models
are martingales for these processes. As a very general statement this
remark may look tautological but it is nevertheless quite instructive.
In particular it provides a key to decipher the relation between SLEs
and conformal field theories.

The main idea is very simple. Consider as above a lattice statistical
model defined on a domain $\mathbb{D}$ with boundary conditions
forcing the existence of interfaces. When computing partition or
correlation functions one has to sum over all possible configurations
of the statistical model. One may arrange this sum by first summing
over all configurations corresponding to prescribe shapes of portions
of interfaces, say of fixed total length $T$, and then summing
over all possible shapes of the portions of interfaces. By basic rules
of statistical mechanics, probabilities of occurrence of portions of
interfaces are given by ratio of conditioned partition
functions, see eq.(\ref{latticeproba}) below. Increasing the length of
these portions of interfaces amounts to add informations so that we
may view it as a process with the total length $T$ playing the role of
`time'.  The statistical sums over configurations with prescribed
portions of interfaces of total length $T$ are then `time' dependent
observables whose expectations -- with respect to the probability
distribution of the interfaces -- are time independent, because they
are equal to the statistical sums over all configurations. 

To make it plain, let us denote by $\prec \mathcal{O} \succ_{\mathbb{D}}$
the statistical sum of the observable $\mathcal{O}$ in a domain
$\mathbb{D}$ and $\prec \mathcal{O} \succ_{\vert\gamma_T}$ the
statistical sum restricted to the set of configurations corresponding
to a prescribed shape of the portions $\gamma_T$ of interfaces of
total length $T$. Then:
\begin{eqnarray}
\Expect{\, \prec \mathcal{O} \succ_{\vert\gamma_T}\,}=
\prec \mathcal{O}\succ_{\mathbb{D}}
\label{martinstatsum}
\end{eqnarray}
where the average is over all possible shapes of interfaces with the
measure $\Expect{\cdots}$ induced by the Bolztman weights, that is the
one in which the probability of occurrence of given shape of
interfaces are defined by the ratio of the partition functions
(\ref{latticeproba}). We shall named this relation the 
\textit{``statistical martingale trick''}.

The relation (\ref{martinstatsum}) applies to any statistical model,
at criticality or not. It is well defined and simple for lattice
models -- and as such it is a bit tautological. It becomes more
relevant in the continuum. There, the measure on the shapes of
interfaces is less under control, as it is difficult to control the
convergence of the ratio of the lattice partition functions. We may
however use alternative arguments or properties to specify the
measure on the interfaces, as it is done with SLEs.
Eq.(\ref{martinstatsum}) then becomes information-full as it couples
this measure on interfaces, $\Expect{\cdots}$, to the statistical
observables, $\prec\mathcal{O}\succ_\mathbb{D}$.

This observation applies to critical interfaces, and hence to
CFTs and SLEs. The remarkable observation made by O. Schramm is that
conformal invariance implies that the filtration associated to the
partial knowledge of the interfaces is that of a continuous
martingale, i.e.  that of a Brownian motion if time is chosen
cleverly. The only parameter is $\kappa$. The physical parameters of
the CFTs, for instance the central charge, can be retrieved by imposing
the condition (\ref{martinstatsum}) to the correlation functions.
This gives the relation:
$$ c = 1 -6\frac{(\kappa-4)^2}{4\kappa} $$

The relation (\ref{martinstatsum}) gives actually more. At criticality
and in the continuum the statistical averages are expected to converge
to ratio of conformal correlation functions:
$$
\prec\mathcal{O}\succ_\mathbb{D} \to 
\frac{\vev{\mathcal{O}}_{\mathbb{D},\ b.c.}}{
\vev{{\bf I}}_{\mathbb{D},\ b.c.}}
$$
The indices b.c. refers to the boundary conditions. The denominator
$\vev{{\bf I}}_{\mathbb{D},\ b.c.}$ represents the CFT partition
function. It therefore depends on the boundary conditions.  See
eq.(\ref{cftratio}) below for more detailed explanations.  As a
consequence of eq.(\ref{martinstatsum}), we learn that these ratios of
CFT correlation functions, but in the domain
$\mathbb{D}\setminus\gamma_T$ cutted along the interfaces $\gamma_T$,
are conserved in mean during the SLE evolution. More precisely, these
ratios are martingales for the SLE processes. Since martingales are
basic tools to compute probabilities, this implies that SLE
probabilities are expressible as ratio of conformal correlation
functions. This is of course in accordance with basic rules of
statistical mechanics.

We start this Section by first describing in detail the intrusion of
stochastic processes in statistical mechanics. This may sound a bit
paradoxal as statistical mechanics refers to static configurations
while stochastic processes refer to random evolutions. The resolution of
this paradox consists in viewing the `time arrow' as the increase of
informations known on the statistical system. We then apply it to
describe the SLE/CFT correspondence, starting from chordal SLE and
then moving to other SLEs. Detailed computations, mostly based on the
statistical martingale trick, are confined to the Section \ref{sec:comput}.

\subsection{Conditioning and partition functions}
\label{sec:condition}
We make the previous (tautological) argument precise.  Let $\mathcal{C}$
be the configuration space of a lattice statistical model defined on a
domain $\mathbb{D}$. For simplicity we assume $\mathcal{C}$ to be
discrete and finite but as large as desired. Let $w_c$ be the
Boltzmann weights and $Z_\mathbb{D}$ the partition function,
$Z_\mathbb{D}=\sum_{c\in\mathcal{C}}\ w_c$.

We imagine having introduced a family $\mathcal{Q}_T$ of partitions of the
configuration space whose elements $\mathcal{C}_{\gamma_T}$ 
are labeled by indices $\gamma_T$:
$$ \mathcal{C}= \bigcup_{\gamma_T}\mathcal{C}_{\gamma_T},
\qquad \mathcal{C}_{\gamma_T}\ {\rm disjoints}.
$$
The index $T$, which will be identified with `time', labels the
partitions. By convention $\mathcal{Q}_0$ is the trivial partition with
$\mathcal{C}$ as its single piece.
We assume these partitions to be finer as $T$ increases,
which means that for any $S>T$ and any element $\mathcal{C}_{\gamma_T}$
of the partition at time $T$ there exist elements of $\mathcal{Q}_S$
which form a partition of $\mathcal{C}_{\gamma_T}$. 

An example of such partitions in case of spin statistical models
consists in specifying the values of local spin variables at an
increasing number of lattice points. Block spin clustering used in
renormalization group is another way to produce such partitions.  In
the present context, we imagine that the partition is defined by
specifying the shapes and the positions of portions of interfaces of
increasing lengths -- there could be different portions of interfaces
starting at different points on the boundary of the domain.  See
Fig.\ref{fig:statmart} below. In such case, $\mathcal{C}_{\gamma_T}$
is just the set of configurations for which the portions of interfaces
coincide with the prescribed portions $\gamma_T$ of interfaces, and
indeed, specifying longer and longer portions of interfaces defines
finer and finer partitions of the configuration space.

To define a stochastic process we have to specify the probability
space and a filtration on it. By construction the probability space
should be the total configuration space $\mathcal{C}$ equipped with the
probability measure induced by the Boltzmann weights, ie. ${\bf
  P}[\{c\}]= w_c/Z_\mathbb{D}$.  To any partition $\mathcal{Q}_T$ is
associated a $\sigma$-algebra $\mathcal{F}_T$ on $\mathcal{C}$, ie. the one
generated by the elements of this partition.  Since these partitions
are finer as `time' $T$ increases, it induces a filtration $\mathcal{
  F}_T$ on $\mathcal{C}$ with $\mathcal{F}_S\subset \mathcal{F}_T$ for $T>S$.
Physically $\mathcal{F}_T$ is the set of events, observable at `time'
$T$, which are unions of the sets $\mathcal{C}_{\gamma_T}$. The fact that
we trivially get a filtration simply means that increasing `time' $T$
increases the knowledge on the system.

We define the conditioned partition function $Z_\mathbb{D}[\gamma_T]$
by the restricted sum:
$$
Z_\mathbb{D}[\gamma_T]\equiv \sum_{c\in\mathcal{C}_{\gamma_T}} w_c~.
$$
Since restricting the summation to a subset amounts to impose some
condition on the statistical configurations, $Z_\mathbb{D}[\gamma_T]$
is the partition function conditioned by the knowledge specified by
$\mathcal{C}_{\gamma_T}$.  In particular the probability of the event
$\mathcal{C}_{\gamma_T}$, ie.  the probability of occurrence of the
portions $\gamma_T$ of interfaces, is the ratio of the partition
functions
\begin{eqnarray}
{\bf P}[\mathcal{C}_{\gamma_T}]= Z_\mathbb{D}[\gamma_T]/Z_\mathbb{D}.
\label{latticeproba}
\end{eqnarray}

Now, given an observable $\mathcal{O}$ of the statistical model, ie. a
function $c\to \mathcal{O}_c$ on the configuration space, we can define
its conditional average $\prec\mathcal{O}\succ_T \equiv \Expect{ \mathcal{
  O}|\mathcal{F}_T}.$ By definition of conditioned expectations,
$\prec\mathcal{O}\succ_T$ is a function on the configuration space which
is constant on any set $\mathcal{C}_{\gamma_T}$ such that $\Expect{{\bf
  1}_{ \mathcal{C}_{\gamma_T} } \prec\mathcal{O}\succ_T}=\Expect{{\bf
  1}_{\mathcal{C}_{\gamma_T}}\mathcal{O}}$ with ${\bf 1}_{\mathcal{
    C}_{\gamma_T}}$ the characteristic function of the set $\mathcal{
  C}_{\gamma_T}\subset\mathcal{C}$.  Hence, $\prec\mathcal{
  O}\succ_T=\sum_{\gamma_T} \prec\mathcal{O}\succ_{\vert\gamma_T} \ {\bf
  1}_{\mathcal{C}_{\gamma_T}}$ with
\begin{eqnarray} 
\prec\mathcal{O}\succ_{\vert\gamma_T} \equiv
\frac{1}{Z_\mathbb{D}[\gamma_T]}
\sum_{c\in\mathcal{C}_{\gamma_T}} \mathcal{O}_c\ w_c. 
\label{statcondi}
\end{eqnarray}
This is simply the statistical average conditioned on the knowledge
specified by the set $\mathcal{C}_{\gamma_T}$.  The unconditioned
statistical average is $\prec\mathcal{
  O}\succ_\mathbb{D}=Z_\mathbb{D}^{-1}\sum_c \mathcal{O}_c w_c$.

By construction the averages of the conditioned expectation
$\prec\mathcal{O}\succ_T$ is time independent and equals to the statistical
average:
\begin{eqnarray}
\Expect{\prec\mathcal{O}\succ_T}
=\sum_{\gamma_T} {\bf P}[ \mathcal{C}_{\gamma_T}]\ 
\prec\mathcal{O}\succ_{\gamma_T} 
= \frac{1}{Z_\mathbb{D}} \sum_{c\in\mathcal{C}} \mathcal{O}_c\ w_c 
= \prec\mathcal{O}\succ_\mathbb{D}.
\label{martintrick}
\end{eqnarray}
This is a simple but a key equation.  One may be more precise and
check that $ \prec\mathcal{O}\succ_T$ is a (closed) martingale with
respect to $\mathcal{F}_T$. See Appendix \ref{app:proba} for definition.
Indeed, for $T>S$,
$$
 \Expect{\prec\mathcal{O}\succ_T |\mathcal{F}_S}=
 \Expect{ \Expect{\mathcal{O}|\mathcal{F}_T} |\mathcal{F}_S]}
=\Expect{ \mathcal{O}|\mathcal{F}_S}=\prec\mathcal{O}\succ_S,
$$
where we used standard properties of conditional expectations and
the fact that $\mathcal{F}_T\subset \mathcal{F}_S$ for $T>S$.

If the partition of the configuration space is given by specifying
portions of interfaces the restricted partition functions are simply
the partition functions of the statistical model in the domain
$\mathbb{D}_T\equiv\mathbb{D}\setminus \gamma_T$ obtained from
$\mathbb{D}$ by cutting it along the specified interfaces:
$Z_\mathbb{D}[\gamma_T]= Z_{\mathbb{D_T}}$.  Similarly the conditioned
expectation $\prec\mathcal{O}\succ_{\vert\gamma_T}$ are simply the
statistical averages in the cutted domain:
$$\prec\mathcal{O}\succ_{\vert\gamma_T}=\prec\mathcal{O}\succ_{\mathbb{D}_T}
=\prec\mathcal{O}\succ_{\mathbb{D}\setminus\gamma_T}.
$$

This observation applies to CFTs and SLEs. The CFT situation is
particularly favorable in that going from $\prec\mathcal{O}
\succ_{\gamma_T}$ to $\prec\mathcal{O}\succ_\mathbb{D}$ is pure
kinematics.

\subsection{Statistical mechanics martingales}
\label{sec:statmechmartin}
\begin{figure}[htbp]
\begin{center}
\includegraphics[width=.5\textwidth]{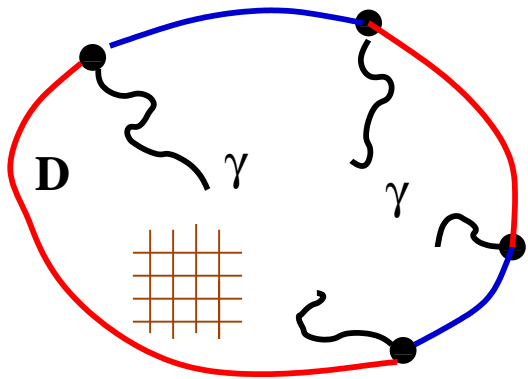}
\end{center}
\caption{A domain $\mathbb{D}$ cutted along portions of interfaces.}
      \label{fig:statmart}
\end{figure}

Our aim is now to use conformal invariance to make the statistical
martingale trick (\ref{martinstatsum},\ref{martintrick}) concrete and
powerful.  We start from the situation at the end of the previous
Section. We assume that the statistical model is defined on a domain
$\mathbb{D}$. To be able to deal with a collection of an arbitrary
number of interfaces we also assume that the boundary conditions
change at $N$ positions along the boundary of $\mathbb{D}$.  See
Fig.\ref{fig:statmart}.

Assuming the model to be at its critical point, we expect to be able
to describe its continuum limit by a conformal field theory (CFT).  At
least for a wide class of observables $\mathcal{O}$, the
partition function and statistical expectation values become CFT
correlation functions:
\begin{eqnarray*}
\prec\mathcal{O}\succ_\mathbb{D} \to
\frac{ \vev{ \mathcal{O} }_{\mathbb{D},\ b.c.} }{ 
\vev{ {\bf I} }_{\mathbb{D},\ b.c.} }
\label{cftratio}
\end{eqnarray*}
where $\vev{\cdots}_{\mathbb{D},\ b.c.}$ refer to the CFT correlation
functions in the domain $\mathbb{D}$ with specific boundary conditions
(b.c.).  We need to write the correlation function of identity ${\bf
  I}$, proportional to the partition function $Z_\mathbb{D}$, in the
denominator because the boundary conditions may already have led to
insertions of boundary changing operators \cite{Cardybc} that we have
not mentioned explicitly. This partition function depends on the
positions $x^{(j)}$ at which the boundary conditions have been changed.

With boundary condition changing at $N$ points $x^{(1)},\dots,
x^{(N)}$ on the boundary, partitioning the configuration space amounts
to specify portions of $N$ interfaces starting at these points. Denote
by $\gamma_{[x^{(j)}\gamma_T^{(j)}]}$ these interfaces -- starting at
$x^{(j)}$ and ending at $\gamma^{(j)}_T$ -- and by $\gamma_T$ their
unions.  As in previous Section, let $\mathcal{C}_{\gamma_T}$ be the set
of configuration with interfaces $\gamma_T$.  The statistical
expectation $\prec\mathcal{O}\succ_{\vert\gamma_T}$ conditioned on
$\gamma_T$ is identical to the statistical expectation on the domain
$\mathbb{D}_T$ obtained from the original domain $\mathbb{D}$ by
removing the interfaces $\gamma_T$. Thus, in the continuum and at
criticality we expect that
\begin{eqnarray*}
\prec\mathcal{O}\succ_{\vert\gamma_T} \to
\frac{\vev{\mathcal{O}}_{\mathbb{D}_T,\ b.c} }{
\vev{ {\bf I} }_{\mathbb{D}_T,\ b.c.}}
\quad,\quad
\mathbb{D}_T=\mathbb{D}\setminus\gamma_T
\end{eqnarray*}
In the continuum limit, the interfaces may have double points so that
they encircle a set of hulls $\mathbb{K}_T$. In that case
$\mathbb{D}_T$ is $\mathbb{D}$ with the hulls (and not only the traces)
removed.

For certain (but not all) observables, $\prec\mathcal{O}\succ_\mathbb{D}$
is computing a probability, which in a conformal field theory ought to
be conformally invariant.  But $\prec\mathcal{O}\succ_\mathbb{D}$ is
written as a quotient, and this means that the numerator and
denominator should transform homogeneously (and with the same factor)
under conformal transformations. In particular, the denominator should
transform homogeneously. This means that $\vev{{\bf I}}_{\mathbb{D},\ 
  b.c.}$ -- which depends on the position of the boundary condition
changes -- behaves like a product of boundary primary fields.  Then,
by locality, for any $\mathcal{O}$, the transformation of the numerator
under conformal maps will split in two pieces: one containing the
transformations of $\mathcal{O}$ and the other one canceling with the
factor in the denominator.  So we infer the existence in the CFT of a
primary boundary field, denoted by $\psi (x)$ in what follows, which
implements boundary condition changes at which interfaces anchor.
Hence we may write
$$
\vev{ {\bf I} }_{\mathbb{D},\ b.c.} = \vev{ \psi (x^{(1)})
\cdots \psi (x^{(N)}) }_{\mathbb{D}}
$$
and
$$
\vev{ \mathcal{O} }_{\mathbb{D}} = 
\vev{ \mathcal{O} \psi (x^{(1)})
\cdots \psi (x^{(N)}) }_{\mathbb{D}}.
$$

Conformal invariance relates correlation in $\mathbb{D}_T$ and
$\mathbb{D}$ provided they are conformally equivalent so that
$\vev{\mathcal{O}}_{\mathbb{D}_T}$ possesses a simple
expression in term $\vev{\mathcal{O}}_\mathbb{D}$.
Write the transformation of the observable $\mathcal{O}$ as
$g:\mathcal{O} \rightarrow \; ^g \,\mathcal{O}$ under a conformal map $g$.
Denote by $g_T$ a conformal representation of
$g_T:\mathbb{D}_T\rightarrow \mathbb{D}$ and write
$g_T(\gamma^{(j)}_{T}) \equiv X^{(j)}_{T}$. The expression for
the closed martingale $\prec\mathcal{O}\succ_T$ can then be simplified
further
\begin{equation}
\label{eq:martCFT}
\prec\mathcal{O}\succ_{\vert\gamma_T}=\prec\mathcal{O}\succ_{\mathbb{D}_T} 
\to \frac{\vev{ \, ^{g_T}\mathcal{O}\, \psi
  (X^{(1)}_T) \cdots \psi (X^{(N)}_T) }_{\mathbb{D}}} {\vev{
  \psi (X^{(N)}_T) \cdots \psi (X^{(N)}_T) }_{\mathbb{D}}}.
\end{equation}
with $^{g_T}\mathcal{O}$ the image of $\mathcal{O}$ by the map $g_T$.  The
CFT correlation functions are now defined on the original domain
$\mathbb{D}$ but the insertion point $X_T^{(j)}$ have been moved by the
uniformizing map $g_T$ which erase all the portions of the interfaces.
The Jacobians coming from the transformations of the boundary changing
primary field $\psi$ have canceled in the numerator and denominator.
Of course, we have cheated a little. For the actual map $g_T$, which is
singular at the $\gamma^{(j)}_{T}$'s, these Jacobians are infinite. A
more proper derivation would go through a regularization but locality
should ensure that the naive formula remains valid when the
regularization is removed.  Eq.(\ref{eq:martCFT}) is the starting
point of our analysis.

\subsection{CFTs of SLEs: chordal case}
\label{sec:cftofsle}
In the continuum limit and at criticality the probability measure on
interfaces (\ref{latticeproba}) induced by the Boltzmann weight is
expected to converge in an appropriate sense to a SLE measure.
However, SLEs depend only an unique parameter $\kappa$ and details on
the statistical model has been erased. The connection between SLE and
the CFT representing the statistical model at criticality is restored
by demanding that the condition expectations $\prec\mathcal{
  O}\succ_{\vert\gamma_T}$ represented in eq.(\ref{eq:martCFT}) are
SLE martingales. Imposing this relation constraints the CFT moduli as
a function of the SLE parameters. The output of the derivation we
shall explain next may be stated in simple words:

\begin{itemize}
\item SLEs with parameter $\kappa$ describe interfaces in CFTs of
  Virasoro central charge 
\begin{eqnarray}
c_\kappa= \frac{(6-\kappa)(3\kappa-8)}{2\kappa} = 1 -
  6\frac{(\kappa-4)^2}{4\kappa}.
\label{cdeSLE}
\end{eqnarray}
Notice that $c_\kappa<1$ and that it is invariant under the duality
$\kappa \leftrightarrow 16/\kappa$. Examples of correspondences are
given below.
\item The boundary conformal operator $\psi(x)$ implementing the
  change of boundary condition at the point on which the interface
  emerges has scaling dimension
\begin{eqnarray}
h_{1;2} = \frac{6-\kappa}{2\kappa}.
\label{hdeSLE}
\end{eqnarray}
It is a Virasoro primary operator degenerate at level two. In the CFT
literature this field is often denoted $\psi_{1;2}$, a notation which
makes references to the so-called Kac's labels. See Appendix
\ref{app:cft} for an explanation of this statement.
\end{itemize}

For chordal SLE there is only one interface so that there are only two
boundary changing operators, one at the starting point $x_0$ and at
the terminal point $x_\infty$ of the interface.  If, as in previous
Section, we cut the domain $\mathbb{D}$ along a portion of the
interface and remove the corresponding hull $\mathbb{K}_t$, the
boundary changing operators are then localized at the tip $\gamma_t$
of the hull and again at $x_\infty$.  The correlation function in
$\mathbb{D}_t$ are then:
$$
\vev{\mathcal{O} }_{\mathbb{D}_t, b.c.}= 
\vev{\mathcal{O}\psi(x_\infty)\psi(\gamma_t) }_{\mathbb{D}_t},
\quad \psi\equiv\psi_{1;2}
$$
Since the cutted domain $\mathbb{D}_t$ is conformally equivalent to
$\mathbb{D}$, with the Loewner map $g_t$ intertwining the two domains,
we may use conformal invariance to express these correlation functions
as correlation functions in the original domain $\mathbb{D}$. Using
the fact that $g_t(\gamma_t)=U_t$, (with $U_t=\sqrt{\kappa}B_t$ if
$\mathbb{D}=\mathbb{H}$), the statistical martingale
(\ref{eq:martCFT}) then becomes:
\begin{eqnarray}
\prec\mathcal{O}\succ_{\mathbb{D}_t} =
\frac{\vev{ ^{g_t}\mathcal{O}\psi(x_\infty)\psi(U_t)
  }_\mathbb{D} }{
\vev{\psi(x_\infty)\psi(U_t) }_\mathbb{D} }
\label{eq:martChord}
\end{eqnarray}
where $ ^{g_t}\mathcal{O}$ is the image of $\mathcal{O}$ under the conformal
map $g_t$. 

As explained below the fact that eq.(\ref{eq:martChord}) is a (local)
martingale for chordal SLEs relies on second order differential
equations satisfied by the CFT correlation functions with insertions
of the boundary operators $\psi$. The correspondence between SLE and
CFT thus relies on a identification of the Fokker-Planck like second
order differential operators driving the SLE processes with the CFT
differential equations -- which are consequences of null vector
relations satisfied by primary fields of specific representations of
the Virasoro algebra.

\subsubsection{SLE/CFT correspondence}
\label{sec:cle/cft}
We exemplify the SLE/CFT correspondence in a few specific cases. Let
us start with the example considered in original Schramm's paper: loop
erased random walk (LERW). There \cite{Schramm00,LSW01a}, it was shown
that LERW does correspond to SLE with $\kappa=2$ and so to CFT with
central charge $c=-2$ -- a very special and simple CFT.  The proof
relies first in establishing that LERW satisfies the domain Markov
property, at least in the continuum limit, and second in proving by
estimating some observables that the driving Loewner process converges
to a Brownian motion.  The dual value $\kappa=8$, which is the
marginal case in the space-filling phase, describes uniform spanning
trees \cite{Kenyon03}.

As shown by Smirnov's \cite{Smirnov}, critical percolation corresponds
to $\kappa=6$ and vanishing central charge. The dual value
$\kappa=8/3$ is expected to describe self avoiding random walk
(SAW), although there is no complete mathematical proof yet, but
see ref.\cite{LSW02}.

Boundaries of Ising spin clusters are conjectured to be described by
SLE at $\kappa=3$ and central charge $1/2$. The boundary operator
$\psi_{1;2}$, which then codes for the change of boundary conditions
from $+$ spins to $-$ spins, has dimension $1/2$.

The $Q$-state Potts models correspond to $Q=4\cos^2(4\pi/\kappa)$ with
$4\leq\kappa<8$, see ref.\cite{BB02b}. The SLE traces are expected to
describe the continuum limit of the boundary of the Fortuin-Kastelyn
\cite{FortuinKastelyn} clusters arising in the hight temperature
expansion of the Potts models. The $Q=2$ Potts model represents the
Ising model, it corresponds to the dual value $\kappa=16/3$.

The self dual value $\kappa=4$ is special in the sense that it
corresponds to a Gaussian massless free field with central charge $1$.
Assuming Dirichlet boundary conditions, the SLE curves may be thought
as discontinuity curves of the free field. The curves start from the
boundary point at which the Dirichlet boundary conditions jump by a
specific amount so that this discontinuity propagates inside the
domain. The mathematical proof that such discontinuity line exists
almost surely for each realization of the Gaussian field is not so
simple as a Gaussian free field is actually a distribution and not a
function.  This was nevertheless done is ref.\cite{SchrammSheffield05} by
taking the scaling limit of discrete model. A relation with domino
tiling is described in refs.\cite{Kenyon00,Kenyon02}.

There are two phases in the critical $O(n)$ models
\cite{Nienhuis83,Nienhuis87,CardyGas}: the dilute and the dense phase.
Parameterizing $n$ as $n=-2\cos(\pi g)$, the two values $g$ and $2-g$
give the same value of $n$ but $1\leq g \leq 2$ corresponds to the
dilute phase while $0\leq g \leq 1$ to the dense phase. The relation
with $O(n)$ loops and SLE is via $g=4/\kappa$ so that the dense phase
is for $4\leq\kappa\leq8$ and the dilute phase for $2\leq\kappa\leq4$.
This is in accordance with the SLE phases. The $O(n=1)$ model
corresponds in the dilute phase to the Ising model with $g=4/3$
($\kappa=3,\ c=1/2$) and in the dense phase to percolation with
$g=2/3$ ($\kappa=6,\ c=0$). The $O(n=0)$ model describes in the dilute
phase self avoiding walk (SAW) with $g=3/2$ ($\kappa=8/3,\ c=0)$
and in the dense phase uniform spanning trees (UST) with $g=1/2$
($\kappa=8,\ c=-2$). Loop erased random walk (LERW) corresponds to the
$O(n=-2)$ in the dilute phase with $g=2$ ($\kappa=2,\ c=-2$).  There
is no $O(n)$ models corresponding to $\kappa\in[0,2[$.

In conformal field theory these correspondences were predicted some
time ago using the coulomb gas representation of the $O(n)$ models
\cite{Nienhuis83,Nienhuis87,Kondev97,CardyGas}. The basic ingredients
of these correspondences are as follows. Recall from Section
\ref{sec:examples} that the lattice $O(n)$ models may be presented as
a loop gas model.  The first step towards the coulomb gas
representation consists in introducing a field $\phi$ at each site of
the dual lattice. There is then a one-to-one mapping between
configurations of the oriented loops and of the field. The rules are
as follows: fix $\phi=0$ on the boundary and increase (decrease)
$\phi$ by a fixed amount, say $\pi$, each time a loop is crossed which
goes to the left (right). The fact that the loops are closed makes
this procedure consistent. In the scaling limit, $\phi$ takes
continuous values and the $O(n)$ lines are identified with the level
lines of $\phi$. They form a set of non-intersecting loops.  These
level lines are also the current lines of the current
$j^\mu=\epsilon^{\mu\nu}\partial_\nu\phi$ so that the direction of the
current orientates the loops. The boundary condition ensures that the
loop cannot cross the boundary. The last but crucial step consists in
arguing that the action of $\phi$, which gives the weight of each loop
configuration, renormalizes to that of a Gaussian free field with a
background charge at infinity \cite{Nienhuis83,Nienhuis87,CardyGas}.

\subsubsection{Correlation functions}
\label{sec:correlfunct}
This Section aims at giving a proof of the SLE/CFT correspondence
using properties of CFT correlation functions.  The main point consists
in proving that for an appropriate choice of the CFTs and the boundary
condition changing operators $\psi$ the ratios of correlation
functions (\ref{eq:martChord}) are local martingales. This amounts to
check that there is no drift term in their It\^o derivative.

For simplicity we deal with chordal SLE in the upper half plane
$\mathbb{H}$ with $x_0=0$ and $x_\infty=\infty$.  The boundary
operator localized at infinity creates a state, which we denote by
$\bra{\psi_{1;2}}$, and the two point function
$\vev{\psi(x_\infty)\psi(\xi_t)}_\mathbb{H}$ becomes
trivial so that the statistical martingale (\ref{eq:martChord})
reduces to the CFT correlation function:
\begin{eqnarray}
\vev{\mathcal{O}}_{\mathbb{H}_t}= \bra{\psi_{1;2}} ^{g_t}\mathcal{O}\,
  \psi(\xi_t)\ket{0}
\label{eq:Hchord}
\end{eqnarray}
with $\ket{0}$ the conformal vacuum.
To simplify further we only sketch the computation when the
operator $\mathcal{O}$ is a product of an arbitrary number $l$ of
boundary primary fields $\mathcal{O}=\prod_{\alpha=1}^{l}
\varphi_{\delta_\alpha}(Y^{(\alpha)})$ at positions $Y^{(\alpha)}$ and
with scaling dimensions $\delta_\alpha$. The
insertion points $Y^{(\alpha)}$ are away from the boundary changing
operators $\psi$. 
This operator transform covariantly under conformal map so that:
\begin{eqnarray}
^{g_t} \mathcal{O}= \prod_{\alpha=1}^{l}\Big[\,
g'_t(Y^{(\alpha)})^{\delta_\alpha}\cdot
\varphi_{\delta_\alpha}(g_t(Y^{(\alpha)}))\, \Big]
\label{eq:multibdry}
\end{eqnarray}
The Loewner map $g_t$ is regular with positive derivative at the
insertion points $Y^{(\alpha)}$.  Substitution of $^{g_t} \mathcal{O}$ in
formula (\ref{eq:Hchord}) yields:
$$
\vev{\mathcal{O}}_{\mathbb{H}_t}=
\prod_{\alpha=1}^{l}g'_t(Y^{(\alpha)})^{\delta_\alpha}\cdot
\bra{\psi_{1;2}} \prod_{\alpha=1}^{l}
\varphi_{\delta_\alpha}(g_t(Y^{(\alpha)}))\, \psi(\xi_t)
\ket{0}
$$
We shall denote by  $J_t^\varphi=\prod_{\alpha=1}^{l}
g'_t(Y^{(\alpha)})^{\delta_\alpha}$ is the Jacobian and by
$Z_{\mathbb{H}_t}^\varphi$ the CFT correlation function.

We need to compute the It\^o derivative of $\vev{\mathcal{
    O}}_{\mathbb{H}_t}$.  Ito's formula for the $\psi$'s gives
$$d \psi(\xi_t)=\psi'(\xi_t)d \xi_t +\frac{\kappa}{2}
\psi''(\xi_t)dt. $$
Using the Loewner equation for $g_t(z)$ and its
derivative with respect to $z$, one checks that
\begin{eqnarray}
d \left(\varphi_{\delta}(g_t(Y))g'_t(Y)^\delta\right)=2g'_t(Y)^\delta
\left(\frac{\varphi'_{\delta}(g_t(Y))}{g_t(Y)-\xi_t}-\delta
  \frac{\varphi_{\delta}(g_t(Y))}{(g_t(Y)-\xi_t)^2}\right)dt.
\label{eq:Itofield}
\end{eqnarray}
The time $t$ being given, we can simplify the notation. Set $x\equiv
\xi_t$ and $y_{\alpha} \equiv g_t(Y^{(\alpha)})$ and view
$Z^\varphi_{\mathbb{H}_t}$ as a function of $y_\alpha$ and $x$. From
the chain rule we get
\begin{eqnarray*}
d (Z_{\mathbb{H}_t}^{\varphi}J^{\varphi}_t)&=& {J^{\varphi}_t}\,
\left[d\xi_t \partial_{x} + dt\left(\frac{\kappa}{2} 
\partial_{x}^2+2\sum_{\alpha}\left[
\frac{1}{y_{\alpha}-x}
\partial_{y_{\alpha}}-\frac{\delta_{\alpha}}
{(y_{\alpha}-x)^2}\right]\right)\right]Z_{\mathbb{H}_t}^{\varphi}
\end{eqnarray*}
The left hand side defines the differential operators driving the SLE
processes.  Thus, the drift term in the It\^o derivative of the
putative martingale vanishes if and only if
\begin{eqnarray}
\left(\frac{\kappa}{2} \partial_{x}^2+2\sum_{\alpha}\left[
    \frac{1}{y_\alpha-x} \partial_{y_\alpha}-\frac{\delta_{\alpha}}
    {(y_\alpha-x)^2}\right]\right)Z_{\mathbb{H}_t}^{\varphi}=0.
\label{eq:nulldiff}
\end{eqnarray}
This is a standard example of differential equation fulfills by CFT
correlation functions. It should hold true for any number of
insertions of primary operators $\varphi_{\delta_\alpha}$.  This
implies that $\psi$ is a degenerate field with a vanishing descendant
at level two and conformal weight $h_{1;2}=\frac{6-\kappa}{2\kappa}$.
The central charge is
$c_\kappa=\frac{(6-\kappa)(3\kappa-8)}{2\kappa}$.  See Appendix
\ref{app:cft} for further information on degenerate field in conformal
field theories.

\subsubsection{Operator formalism}
\label{sec:operat}
The aim of this Section is to derive the SLE/CFT correspondence using
the CFT operator formalism which was initially developed in
ref.\cite{BB02a,BB02b,BB03b}. This will reveal deep relation between
SLEs and special representations of the Virasoro algebra.

We start with the group theoretical formulation of chordal SLE as
explained in Section \ref{sec:geosle}. Recall that $h_t(z) \equiv
g_t(z) - \xi_t$ satisfies the stochastic differential equation $d h_t
= {2dt}/{h_t}-d\xi_t.$ According to Section \ref{sec:groupsle}, to
$h_t$ we can associate $\mathfrak{g}_{h_t}\in N_-$, with $N_-$ the
group of germs of holomorphic functions at $\infty$ of the form
$z+\sum_{m\leq -1} h_mz^{m+1}$. By It\^o's formula, it satisfies:
\begin{eqnarray}
\mathfrak{g}_{h_t}^{-1} \cdot d \mathfrak{g}_{h_t}
=dt(-2\ell_{-2}+\frac{\kappa}{2}\ell_{-1}^2)+d\xi_t\ell_{-1}.
\label{itoN-}
\end{eqnarray} 
with $\ell_{-2}=-z^{-1}\partial_z$ and $\ell_{-1}=-\partial_z$.
Compare with eq.(\ref{itoN-moi}). 

In conformal field theory the operators $\ell_n=-z^{n+1}\partial_z$,
with $[\ell_m,\ell_n]=(m-n)\ell_{m+n}$, are represented by operators
$L_n$ which satisfy the Virasoro algebra $\mathfrak{vir}$:
$$
[L_m,L_n]=(m-n)L_{m+n}+\frac{c}{12}m(m^2-1)\delta_{m+n;0}
$$
with $c$ the Virasoro central charge. See Appendix \ref{app:cft}
for the necessary information on the Virasoro algebra and its
representations.  We need to introduce the notation $\mathfrak{n}_-$
for the sub-algebra of $\mathfrak{vir}$ generated by the $L_n$'s with
$n<0$.  In the following we shall only deal with highest weight
representations.  These are representations of $\mathfrak{vir}$ which
possess a highest weight vector $\ket{h}$ are such that $L_n\ket{h}=0$
for $n>0$ and $L_0\ket{h}=h\ket{h}$. The parameter $h$ is called the
conformal dimension of the representation.

The representations of $\mathfrak{vir}$ are not automatically
representations of $N_-$, one of the reasons being that the Lie
algebra of $N_-$ contains infinite linear combinations of the
generators $\ell_n$'s.  However, as explained in
ref.\cite{BB03b,BB04}, highest weight representations of
$\mathfrak{vir}$ can be extended in such a way that $N_-$ get embedded
in a appropriate completion $\overline{\mathcal{U}(\mathfrak{n}_{-})}$
of the enveloping algebra of the sub-algebra $\mathfrak{n}_{-}$ of
$\mathfrak{vir}$.  This allows to associate to any $\mathfrak{g}_h\in
N_-$ an operator $G_h$ acting on appropriate representations of
$\mathfrak{vir}$ and satisfying $G_{g \circ f}=G_f\cdot G_g$ so that
the map $\mathfrak{g}_h\to G_h$ is a homomorphism.

One may think about $G_h$ as the operator implementing the
conformal map $h(z)$ of the form $z+\sum_{m\leq -1}h_mz^{m+1}$ in the
Virasoro representations and thus in the CFT Hilbert spaces.  In
particular if $\varphi_\delta(Y)$ is a boundary primary field of
scaling dimension $\delta$ acting on the representations of
$\mathfrak{vir}$ then $G_h$ acts by conjugaison as:
$$
G^{-1}_h\, \varphi_\delta(Y)\, G_h = |h'(Y)|^\delta\, 
\varphi_\delta(h(Y))
$$
More generally, the image $^h\mathcal{O}$ of an operator $\mathcal{O}$ by
the conformal map $h$ is obtained by conjugating it by the operator
$G_h$ as standard rules of quantum mechanics tells us: $^h\mathcal{O}=
G^{-1}_h\, \mathcal{O}\, G_h$.  In particular, $G_h$ acts on the CFT
stress tensor $T(z)=\sum_n L_n z^{n-2}$ as:
\begin{equation}
\label{eq:conjt}
G_h^{-1}\, T(z)\, G_h= T(h(z))\, h'(z)^2+\frac{c}{12}\, Sh(z),
\end{equation}
with $Sh(z)=(\frac{h''}{h'})'-\frac{1}{2}(\frac{h''}{h'})^2$ 
the Schwarzian derivative of $h$. This
extra term reflects the anomalous transformation law of
the CFT stress tensor under conformal transformations \cite{BPZ}.

Implementing this construction for the random Loewner map
$h_t$ yields random operators $G_{h_t}\in
\overline{\mathcal{U}(\mathfrak{n}_{-})}$ which satisfy the
stochastic It\^o equation \cite{BB02a,BB02b}:
\begin{eqnarray}
 G_{h_t}^{-1} d G_{h_t}=dt(-2L_{-2}+\frac{\kappa}{2}
L_{-1}^2)+d\xi_t L_{-1}.
\label{labelle}
\end{eqnarray}  
This follows directly from eq.(\ref{itoN-moi}) and the fact that the
correspondence $\mathfrak{g}_h\to G_h$ is a homomorphism.  This may be
viewed as defining a Markov process in the enveloping algebra
$\overline{\mathcal{U}(\mathfrak{n}_{-})}$. 

Eq.(\ref{labelle}) does not contain more information than
eq.(\ref{itoN-moi}), or than the Loewner equation, but it now makes
sense in the CFT Hilbert spaces. This immediatly leads to the
following important result:
\begin{itemize}
\item Let $\ket{\psi_{1;2}}$ be the highest weight vector in the
  irreducible highest weight representation (degenerate at level two)
  of $\mathfrak{vir}$ of central charge
  $c_{\kappa}=\frac{(6-\kappa)(3\kappa-8)}{2\kappa}$ and conformal
  weight $h_{1;2}\equiv \frac{6-\kappa}{2\kappa}$.\\
  Then $G_{h_t}\ket{\psi_{1;2}}$ is a local martingale.
  
\item Assuming appropriate boundedness conditions on $\bra{v}$, the
  scalar product $\bra{v}G_{h_t}\ket{\psi_{1;2}}$ is a martingale so
  that $\Expect{\bra{v}G_{h_t}\ket{\psi_{1;2}}|\{{G_{h_u}}\}_{u\leq
    s}}$ is time independent for $t\geq s$ and:
\begin{eqnarray}
\Expect{\,\bra{v}G_{h_t}\ket{\psi_{1;2}}\,\big|\{G_{h_u}\}_{u\leq s}}
=\bra{v}G_{h_s}\ket{\psi_{1;2}}
\label{maineq}
\end{eqnarray}
In particular, $\bra{v}G_{h_t}\ket{\psi_{1;2}}$ is conserved in mean for
any $\bra{v}$.
\end{itemize}

\noindent
This result is a direct consequence of eq.(\ref{labelle}) and the null
vector relation at level two,
$(-2L_{-2}+\frac{\kappa}{2}L_{-1}^2)\ket{\psi_{1;2}}=0$, so that
$dG_{h_t}\ket{\psi_{1;2}}=G_{h_t}L_{-1}\ket{\psi_{1;2}}d\xi_t$.  The
null vector condition is what fixes the values of the conformal weight
and of the central charge.  
\vskip .3 truecm

Since $G_{h_t}$ is the operator intertwining the
conformal field theories in $\mathbb{H}$ and in the random domain
$\mathbb{H}_t$, this result has the following important consequences.
Consider CFT correlation functions in $\mathbb{H}_t$. They can be
computed by looking at the same theory in $\mathbb{H}$ modulo the
insertion of an operator representing the deformation from
$\mathbb{H}$ to $\mathbb{H}_t$. This operator is $G_{h_t}$. 
Recall the expression (\ref{eq:Hchord}) of expectation values in the
upper half plane with the hull removed which, after translating by
$\xi_t$ to go from $g_t$ to $h_t=g_t-\xi_t$, can be written as:  
$$
\vev{\mathcal{O}}_{\mathbb{H}_t}= \bra{\psi_{1;2}}\,  ^{h_t}\mathcal{
  O}\, \psi(0)\ket{0}.
$$
The boundary operator $\psi\equiv\psi_{1;2}$ with dimension
$\frac{6-\kappa}{2\kappa}$ is the operator which create the highest
weight vector $\ket{\psi_{1;2}}$ at the tip of the hull, so that
$\psi(0)\ket{0}=\ket{\psi_{1;2}}$. Using $^{h_t}\mathcal{
  O}=G^{-1}_{h_t}\mathcal{O}G_{h_t}$ we get:
\begin{eqnarray}
\vev{\mathcal{O}}_{\mathbb{H}_t}=\bra{\psi_{1;2}} \mathcal{O}\,
  G_{h_t}\ket{\psi_{1;2}}
\label{eq:OGh}
\end{eqnarray}
where we use $\bra{\psi_{1;2}}G^{-1}_{h_t}=\bra{\psi_{1;2}}$ since
$G_{h_t}$ is the operator implementing a conformal map fixing infinity
and with derivative $1$ at infinity.

Suppose now that the central charge is
$c=(6-{\kappa})(3\kappa-8)/2\kappa$, then $\vev{\mathcal{
    O}}_{\mathbb{H}_t}$ is a local martingale because so is
$G_{h_t}\ket{\psi_{1;2}}$. In particular, the correlation functions of
the conformal field theory in the fluctuating geometry $\mathbb{H}_t$
are in average time independent:
$$
\Expect{\vev{\mathcal{O}}_{\mathbb{H}_t}}=\vev{\mathcal{O}}_\mathbb{H},
$$
a result that we also found by computing directly the CFT
correlation functions in the previous Section.

\begin{figure}[htbp]
  \begin{center}
    \includegraphics[width=\textwidth]{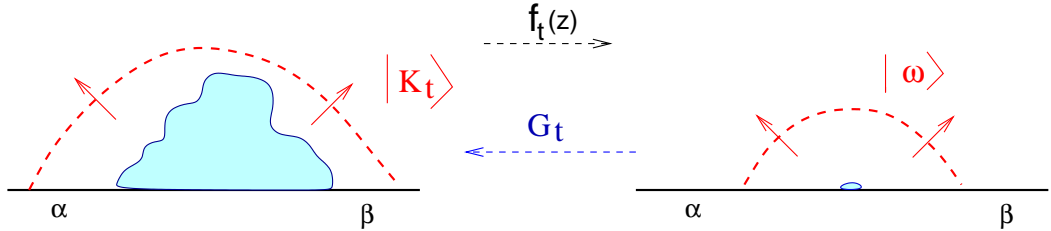}
      \caption{\em A representation of the boundary hull state 
$\ket{K_t}=G_{h_t}\ket{\psi_{1;2}}$ and of the 
        map intertwining different formulations of the CFT.}
     \label{fig:hilbert}
 \end{center}
\end{figure}

\vskip .3 truecm

The state $G_{h_t}\ket{\psi_{1;2}}$ may be interpreted as follows.
Imagine defining the conformal field theory in $\mathbb{H}_t$ via a radial
quantization, so that the conformal Hilbert spaces are defined over
curves topologically equivalent to half circles around the origin. Then, 
the SLE hulls manifest themselves as disturbances localized around
the origin, and as such they generate states in the conformal Hilbert
spaces. Since $G_{h_t}$ intertwines the CFT in $\mathbb{H}$ and in
$\mathbb{H}_t$, these states are $G_{h_t}\ket{\psi_{1;2}}$ with
$\ket{\psi_{1;2}}$ keeping track of the boundary conditions.
See Fig.\ref{fig:hilbert}.

The operator formalism shows directly that the state
$G_{h_t}\ket{\psi_{1;2}}$ is a generating function of local
martingales, since it may be expanded on any appropriate basis. This
was used in ref.\cite{BB03a} to show that the vector space of
martingale polynomials in the Taylor coefficients of the Loewner map
$h_t$ carries a representation of the Virasoro algebra. Explicit
expressions of the Virasoro generators acting on this space were given
in terms of differential operators so that all polynomial martingales
can be obtained by iterative applications of these differential
operators.

\subsection{Other SLEs}
\label{sec:sleetc}
The aim of this Section is to extend the SLE/CFT correspondence to the
other SLEs. Again the correspondence is attached to the construction
of SLE martingales using tools from CFT.  As in the chordal case we
can use either a formalism based on correlation function or an
operator formalism.

In the correlation function formalism, we shall identify the
statistical martingales as ratio of CFT correlation functions:
$$
\prec\mathcal{O}\succ_{\mathbb{D}_t}= 
\frac{\vev{ \mathcal{O} }_{\mathbb{D}_t,b.c.} }{
\vev{ {\bf I} }_{\mathbb{D}_t,b.c.} }
$$
What distinguished the different SLEs are the different boundary
conditions which depends on the marked points specific to each SLEs
and on the behavior of the SLEs map in the neighborhood of these
points.  This translates into the insertions of appropriate boundary
and/or bulk operators in the CFT correlation functions.

In the operator formalism we start from the stochastic equation
(\ref{SLEcov}) satisfied by the group element $\mathfrak{g}_{h_t}$
associated to the Loewner map. As for the chordal case, this element
is then promoted to an operator $G_{h_t}$ acting on representations of
the Virasoro algebra and thus acting on CFT Hilbert spaces.  The operator
$G_{h_t}$ is the operator which intertwines the conformal field
theories in the domain $\mathbb{D}$ and in the domain $\mathbb{D}_t$
with the hull removed so that $^{h_t}\mathcal{O}=G_{h_t}^{-1}\mathcal{
  O}G_{h_t}$ for any operator $\mathcal{O}$.  By construction it
satisfies an equation analogous to eq.(\ref{SLEcov}) of the following
form:
\begin{eqnarray} 
G_{h_t}^{-1} \cdot d G_{h_t}=dt \,
(-2W_{-2}+\frac{\kappa}{2}W_{-1}^2)+ d\xi_t \, W_{-1},
\label{eq:Gcov}
\end{eqnarray}
with $W_{-2}$ and $W_{-1}$ elements of the Virasoro algebra.  Their
precise forms depend on the type of SLE we are considering.

\subsubsection{Radial SLE}
\label{sec:slerad}
Radial SLE describes curves in a domain $\mathbb{D}$ starting from a
boundary point $x_0$ and ending at a bulk point $z_*$.  The CFT
correlation functions take into account these boundary conditions by
the insertion a boundary field $\psi$, which is going to be
$\psi_{1;2}$ as in the chordal case, at the tip of the curve and a
bulk field, which we shall soon identify as $\Phi_{0;1/2}$, at the
bulk point. Thus:
$$
\vev{ \mathcal{O}}_{\mathbb{D}_t, b.c.}
= \vev{ \mathcal{O}\, \Phi_{0;1/2}(z_*) 
\psi(\gamma_t) }_{\mathbb{D}_t}
$$
Similarly the conditioned partition function is $\vev{{\bf
    I}}_{\mathbb{D}_t, b.c.}=\vev{\Phi_{0;1/2}(z_*) \psi(\gamma_t)
}_{\mathbb{D}_t}$.  By definition of the radial SLE map $g_t$, the tip
of the curve is mapped to the driving process $U_t=g_t(\gamma_t)$ and
the terminal bulk point $z_*$ is fixed by $g_t$.  Thus for radial SLE
the statistical martingale reads:
\begin{eqnarray}
\prec\mathcal{O}\succ_{\mathbb{D}_t}=
\frac{\vev{ ^{g_t}\mathcal{O}\, \Phi_{0;1/2}(z_*) 
\psi(U_t) }_\mathbb{D} }{
\vev{ \Phi_{0;1/2}(z_*) \psi(U_t) }_\mathbb{D} }
\label{eq:radialmart}
\end{eqnarray}

It is then a matter a simple computation -- which we leave to the
reader but which is very much parallel to the one done for the chordal
case in the previous Section -- to verify that the above expectation
values is a SLE martingale provided that \cite{BBrad03}:
\begin{itemize}
\item the boundary operator $\psi$ is the boundary
  primary operator $\psi_{1;2}$ degenerate at level two with scaling
  dimension $h_{1;2}=\frac{6-\kappa}{2\kappa}$;
\item the bulk operator $\Phi_{0;1/2}$ is a spinless primary operator
  with scaling dimension
  $2h_{0;1/2}=\frac{(\kappa-2)(6-\kappa)}{8\kappa}$.
\end{itemize}
\noindent
As in the chordal case, the martingale property for $\prec\mathcal{
    O}\succ_{\mathbb{D}_t}$ essentially relies on the differential
equation satisfied by CFT correlation functions with insertion of the
degenerate operator $\psi$. The need for the insertion of the bulk
operator $\Phi_{0;1/2}$ at the point where the SLE curves terminate is
in agreement with the identification of the operators $\Phi_{0;n/2}$ as
those testing for the presence of $n$ SLE traces in the neighborhood
of a bulk point. See Section \ref{sec:comput}.

The SLE/CFT correspondence may also be done using the operator
formalism.  Let $G_{h_t}$ be the operator representing the action of
$\mathfrak{g}_{h_t}$ in the CFT Hilbert spaces for radial SLE. It
satisfies eq.(\ref{eq:Gcov}). We look at radial SLE in the upper half
plane geometry. The Loewner equation is then given in
eq.(\ref{radialplan}). We have seen in Section \ref{sec:geosle} that
$w_{-1}=\frac{1}{2}(\ell_{-1}+\ell_1)$ and
$w_{-2}=\frac{1}{4}(\ell_{-2}+\ell_0)$, so that
$$W_{-1}=\frac{1}{2}(L_{-1}+L_1),\qquad
W_{-2}=\frac{1}{4}(L_{-2}+L_0).$$
A simple rearrangement leads to
$(-2W_{-2}+\frac{\kappa}{2}W_{-1}^2)\ket{\psi_{1;2}}=
2h_{0;1/2}\ket{\psi_{1;2}}$.  From eq.(\ref{eq:Gcov}) we then deduce
that $G_{h_t}^{-1}d G_{h_t}\ket{\psi_{1;2}} =
2h_{0;1/2}\ket{\psi_{1;2}} +d\xi_t W_{-1}\ket{\psi_{1;2}}$ and thus
that
$$
e^{-2h_{0;1/2}t}G_{h_t}\ket{\psi_{1;2}}
$$
is a local martingale. The prefactor $e^{-2h_{0;1/2}t}$ accounts for
the insertion of a bulk conformal field of scaling dimension $2h_{0;1/2}$
localized at the fixed point. See ref.\cite{BBrad03} for further
details. 

The radial SLE is for instance well adapted to the $O(n)$ models with
the insertion of the operators $\psi_{1;2}$ and $\Phi_{0;1/2}$
conditioning on the presence of an $O(n)$ trace from their boundary
and bulk insertion points.

\subsubsection{Dipolar SLE}
\label{sec:sledip}
Dipolar SLE describes curves in a domain $\mathbb{D}$ starting from a
boundary point $x_0$ and stopped the first instant it hits a boundary
interval $[x_-,x_+]$ excluding the starting point. The CFT
correlation functions take into account these boundary conditions by
the insertion a boundary field $\psi$, which creates the SLE trace,
at the tip of the curve and two other boundary
fields, which we shall soon identify as $\psi_{0;1/2}$, at the
two marked points $x_\pm$. Thus we expect:
$$
\vev{ \mathcal{O}}_{\mathbb{D}_t, b.c.}
= \vev{ \mathcal{O}\, \psi_{0;1/2}(x_-) \psi_{0;1/2}(x_+)
\psi(\gamma_t) }_{\mathbb{D}_t}
$$
By definition of the radial SLE map $g_t$, the tip of the curve is
mapped to the driving process $U_t=g_t(\gamma_t)$ and the marked
boundary points $x_\pm$ fixed by $g_t$.  Thus for dipolar SLE the
statistical martingales read:
\begin{eqnarray}
\prec\mathcal{O}\succ_{\mathbb{D}_t}=
\frac{\vev{ ^{g_t}\mathcal{O}\, \psi_{0;1/2}(x_-) \psi_{0;1/2}(x_+)
\psi(U_t) }_\mathbb{D} }{
\vev{ \psi_{0;1/2}(x_-) \psi_{0;1/2}(x_+)\psi(U_t)
}_\mathbb{D} } 
\label{eq:dipolarmart}
\end{eqnarray}

Again as for radial SLE, it is then a matter of a simple computation
-- which we leave to the reader -- to verify that the above
expectation values are SLE martingales provided that \cite{BBH04}:
\begin{itemize}
\item the boundary operator $\psi$ creating the SLE trace is
  the boundary primary operator $\psi_{1;2}$
  degenerate at level two with scaling dimension
  $h_{1;2}=\frac{6-\kappa}{2\kappa}$;
\item the two boundary operators $\psi_{0;1/2}$ at the two marked
  points have each scaling dimension
  $h_{0;1/2}=\frac{(\kappa-2)(6-\kappa)}{16\kappa}$.
\end{itemize}
\noindent
As it is now usual, the martingale property for $\prec\mathcal{
    O}\succ_{\mathbb{D}_t}$ essentially relies on the differential
equation satisfied by CFT correlation functions with insertion of the
degenerate operator $\psi$.

We  now describe the operator formalism for dipolar SLE.
Let $G_{h_t}$ be the operator representing the action of
$\mathfrak{g}_{h_t}$ in the CFT Hilbert spaces for dipolar SLE. It
satisfies eq.(\ref{eq:Gcov}). We look at dipolar SLE in the upper half
plane geometry. The Loewner equation is then given in
eq.(\ref{dipolarplan}). We have seen in Section \ref{sec:geosle} that
$w_{-1}=\frac{1}{2}(\ell_{-1}-\ell_1)$ and
$w_{-2}=\frac{1}{4}(\ell_{-2}-\ell_0)$, so that 
$$W_{-1}=\frac{1}{2}(L_{-1}-L_1),\qquad
W_{-2}=\frac{1}{4}(L_{-2}-L_0).$$
A simple computation leads to
$(-2W_{-2}+\frac{\kappa}{2}W_{-1}^2)\ket{\psi_{1;2}}=
-2h_{0;1/2}\ket{\psi_{1;2}}.$ As for radial SLE, this together with
eq.(\ref{eq:Gcov}) immediately imply that
$$
e^{+2h_{0;1/2}t}G_{h_t}\ket{\psi_{1;2}}$$
is a local martingale.
The prefactor $e^{+2h_{0;1/2}t}$ accounts for the insertion of two
boundary conformal fields, each of dimension $h_{0;1/2}$,
localized at the two fixed points.

In the Ising model $(\kappa=3)$, for instance, $\psi_{1;2}$ of
dimension $1/2$ corresponds to change from $+$ to $-$ boundary
conditions while $\psi_{0;1/2}$ of dimension $1/16$ corresponds to
change from free to $+$ (or to $-$) boundary conditions. So dipolar
SLE at $\kappa=3$ describes Ising model with boundary conditions
changing from $+$ to $-$ to free and back to $+$ along the boundary.

\subsubsection{SLE($\kappa,\rho)$}
\label{sec:slekr}
As explained in Section \ref{sec:slekrho}, SLE($\kappa,\rho$) may be
viewed as dipolar SLE but with an extra drift in the driving source.
So the SLE/CFT correspondence is analogous \cite{Kytola05} to that for
dipolar SLE except that one has to change the nature of the boundary
operator inserted at the two marked point $x_\pm$ kepted fix by the
Loewner map. Thus we expect that:
$$
\vev{ \mathcal{O}}_{\mathbb{D}_t, b.c.}
= \vev{ \mathcal{O}\, \psi_{h_-}(x_-) \psi_{h_+}(x_+)
\psi(\gamma_t) }_{\mathbb{D}_t}
$$
The SLE($\kappa,\rho)$ statistical martingales similarly read:
\begin{eqnarray}
\prec\mathcal{O}\succ_{\mathbb{D}_t}=
\frac{\vev{ ^{g_t}\mathcal{O}\, \psi_{h_-}(x_-) \psi_{h_+}(x_+)
\psi(U_t) }_\mathbb{D} }{
\vev{ \psi_{h_-}(x_-) \psi_{h_+}(x_+)\psi(U_t)
}_\mathbb{D} } 
\label{slerhomart}
\end{eqnarray}
As proved in \cite{Kytola05}, the above
expectation values are SLE martingales provided that:
\begin{itemize}
\item the boundary operator $\psi$ creating the SLE trace is
  again the primary operator $\psi_{1;2}$
  degenerate at level two with scaling dimension
  $h_{1;2}=\frac{6-\kappa}{2\kappa}$;
\item the boundary operators $\psi_{h_-}$ and $\psi_{h_+}$ 
  have scaling dimensions
  $h_+=\frac{\rho(\rho+4-\kappa)}{4\kappa}$ and
  $h_-=\frac{(\rho+2)(\rho+6-\kappa)}{4\kappa}$.
\end{itemize}
\noindent
The conformal dimensions $h_\pm$ have a clear CFT interpretation.
They are the generic dimensions of the primary operators satisfying
the fusion rules with $\psi_{1;2}$ such that the three point function
$\vev{ \psi_{h_-}(x_-) \psi_{h_+}(x_+)\psi_{1;2}(x_0)}_\mathbb{D}$ is
non vanishing.  They satisfy
$$
h_+-h_-=\alpha/\kappa\quad,\quad h_++h_-=2h_{0;1/2} +\alpha^2/2\kappa
$$
with $\alpha= \frac{6-\kappa}{2}+\rho$ the drift coefficient in the
driving source $U_t=\sqrt{\kappa}B_t+\alpha\, t$, see
eq.(\ref{slerhodip}).

The operator formalism for SLE$(\kappa,\rho)$ also follows directly
from that of dipolar SLE since the vector fields $w_{-2}$ and $w_{-1}$
of SLE$(\kappa,\rho)$ are obtained from those of dipolar SLE by a
translation, see Section \ref{sec:slekrho}:
$$
W_{-2}^{{\rm SLE}(\kappa,\rho)}=W_{-2}^{\rm dip.}
-\frac{\alpha}{2}\, W_{-1}^{\rm dip.}\quad,\quad
 W_{-1}^{{\rm SLE}(\kappa,\rho)}=W_{-1}^{\rm dip.}
$$
As a consequence, the state
$$
Z^{-1}_t\, G_{h_t}\ket{\psi_{1;2}},\quad {\rm with}\quad
Z^{-1}_t= \exp{[(2h_{0;1/2}+\frac{\alpha^2}{2\kappa})t
-\frac{\alpha}{\sqrt{\kappa}}B_t]}
$$
is a local martingale. The extra term $Z^{-1}_t$ takes into account the
insertion of two operators $\psi_{h_\pm}$ in the partition function
$\vev{\psi_{h_-}(x_-) \psi_{h_+}(x_+)\psi_{1;2}(U_t)}_\mathbb{H}$.
The microscopic interpretation of $SLE(\kappa,\rho)$ in terms of
lattice statistical models is less clear but a few hints have been
given in ref.\cite{Cardy03e}.

\subsection{Multiple SLEs}
\label{sec:nsle}
\begin{figure}[htbp]
  \begin{center}
    \includegraphics[width=0.9\textwidth]{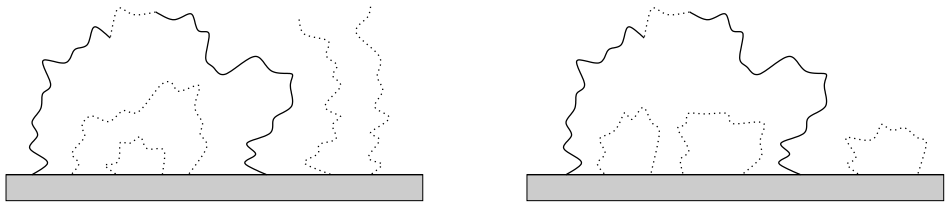}
      \caption{\em A representation of two configurations of 
        growing curves forming arches whenever they touch themself.}
     \label{fig:nsleconfig}
 \end{center}
\end{figure}

Multiple SLEs describe the local growth of $n$ interfaces in critical
models, ie. in CFT, joining fixed points on the boundary by a Loewner
chain with random driving source. See Fig.\ref{fig:nsleconfig}. The
first attempt -- however not complete -- to define them was done in
\cite{Cardy03d}. A very interesting appraoch based on commutativity of
the growths of the interfaces has then been developed in
ref.\cite{Dubedat04c,Dubedat05b}.  We shall instead follow the
approach of ref.\cite{BBK05} which is based on implementing the
statistical martingale trick to constrain the processes driving the
growth of the traces. We assume that $0 \leq \kappa <8$.

\subsubsection{The basic equations}
\label{sle:nslebasic}
We list here the set of necessary conditions and equations defining
multiple SLEs.  By conformal invariance, it is enough to give its
definition when the domain is the upper half plane $\mathbb{H}$.  The
curves, labeled by an integer $i=1,\cdots,n$ and starting at position
$X_i$, generate a hull whose complement is uniformized by a map $g_t$,
which we assume to be hydrodynamically normalized at infinity as for a
single chordal SLE.  Since the hull is generated by $n$ curves, this
map satisfies a Loewner equation with $n$ simple poles:
\begin{equation}
\label{eq:Nloewchain}
d g_t(z)=\sum_{i=1}^n \frac{2a^{(i)}_tdt}{g_t(z)-X^{(i)}_t}.
\end{equation}
The initial condition is $g_0(z)=z$.  The positive function $a^{(i)}_t$
parameterizes the speed of growth of the $i^{\rm th}$ curve. We
normalize them by $\sum_i a^{(i)}_t = 1$ so that the total capacity
of the growing hulls at time $t$ is $2t$.

The processes $X^{(i)}_t$, which are the images of the tips of the
curves by $g_t$, are solutions of the stochastic differential
equations:
\begin{equation}
\label{eq:drivepro}
d X^{(i)}_t=d\xi^{(i)}_t + \kappa 
a^{(i)}_tdt\, (\partial_{x_i}\log Z)(X_\cdot)+ \sum_{j
  \neq i} \frac{2 a^{(j)}_tdt}{X^{(i)}_t-X^{(j)}_t}.
\end{equation}
with $d\xi^{(i)}_t=\sqrt{\kappa a_t^{(i)}}\, dB_t^{(i)}$ where
$B_t^{(i)}$ are $n$ independent normalized Brownian motions.  This
choice ensures that the curves grow locally as SLE traces.  The
initial conditions are $X_0^{(i)}=X_i$ ordered in such a way that $X_1
< X_2< \cdots < X_n$.
 
The system depends on a function $Z(x_1,\cdots,x_n)$ which reflect the
interactions between the curves.
It has to fulfill the following requirements:\\
$(i)$ $Z(x_1,\cdots,x_n)$ is defined and positive for $x_1 < x_2<
\cdots< x_n$,\\
$(ii)$ $Z(x_1,\cdots,x_n)$ is translation invariant and homogeneous.
Its weight is $h_{1;n-2m+1}-nh_{1;2}$ for some nonnegative
integer $m \leq n/2$, where $2\kappa h_{1;m+1}\equiv
m(2(m+2)-\kappa).$
The number $m$ is expected to be the number of curves growing towards
infinity.\\ 
$(iii)$ $Z(x_1,\cdots,x_n)$ is annihilated by the $n$ differential
operators
\begin{eqnarray}
{\mathcal D}_i=\frac{\kappa}{2}\partial_{x_i}^2+2\sum_{j \neq
  i}\left[\frac{1}{x_j-x_i}\partial_{x_j}-
\frac{h_{1}(\kappa)}{(x_j-x_i)^2}\right].
\label{eq:eqforZ}
\end{eqnarray} 
The function $Z$ actually refers to the partition function of the
underlying statistical models. As it should be the behavior of the
curves, which are expected to represent the statistical interfaces,
depend on the partition function since it codes for the boundary
condition imposed on the statistical models.

This system of equations for $n$ curves joining together points
$X_1,\cdots,X_n$ and possibly the point at infinity has be called a
$n$SLE system \cite{BBK05}. Systems for radial and dipolar versions of
$n$SLE could probably be defined analogously. Still some mathematical
work may have to be done to make rigorous sense of this system --
these problems are still under active consideration. The problems
might be of different natures for $\kappa \leq 4$ and $4 < \kappa <8$.
Nevertheless, the $n$SLE systems is the only SLE like system
compatible with statistical mechanics in the sense that it is the only
system which admits the conditioned statistical averages as
martingales -- in the same way as chordal SLE does.


Let us sketch the argument. As for chordal SLE, we simplify the
presentation by only doing the computation when the operator $\mathcal{
  O}$ is a product of an arbitrary number $l$ of boundary primary
fields $\mathcal{O}=\prod_{\alpha=1}^{l}
\varphi_{\delta_\alpha}(Y^{(\alpha)})$ at positions $Y^{(\alpha)}$ and
with scaling dimensions $\delta_\alpha$. It transform under conformal
map as in eq.(\ref{eq:multibdry}) so that its statistical average
(\ref{eq:martCFT}) in
the upper half plane cut with the hull removed is:
$$
\prec\mathcal{O}\succ_{\mathbb{H}_t}= J_t^\varphi\
{Z_{\mathbb{H}_t}^\varphi}/{Z_{\mathbb{H}_t}}
$$
where $J_t^\varphi=\prod_{\alpha=1}^{l}
g'_t(Y^{(\alpha)})^{\delta_\alpha}$ is the Jacobian and 
$Z_{\mathbb{H}_t}^\varphi$ and $Z_{\mathbb{H}_t}$ are the CFT
correlation functions:
\begin{eqnarray}
\label{eq:marto}
Z_{\mathbb{H}_t}^\varphi &=& \vev{ \prod_{\alpha=1}^{l} 
  \varphi_{\delta_\alpha}(Y_t^{(\alpha)})\, 
  \prod_{i=1}^{n}\psi (X^{(i)}_t) }_{\mathbb{H}}\\ 
Z_{\mathbb{H}_t}&=& \vev{\prod_{i=1}^{n}
\psi (X^{(i)}_t)}_{\mathbb{H} }. 
\end{eqnarray}
where we set $Y_t^{(\alpha)}=g_t(Y^{(\alpha)})$ as in previous
Section. For proving that $\prec\mathcal{O}\succ_{\mathbb{H}_t}$ is a
(local) martingale one has to compute its It\^o derivative and check
that its drift term vanishes. So one has to write the stochastic
equation of the driving process $X^{(i)}$ as
$$
d X_t^{(i)} =d\xi^{(i)}_t + F^{(i)}_t dt 
$$
and to look under which conditions on $F^{(i)}_t$ the drift term in
$d\prec\mathcal{O}\succ_{\mathbb{H}_t}$ vanishes. The output \cite{BBK05}
is that the boundary operator $\psi$ has to be a Virasoro primary
operator degenerate at level two and that
$$
F^{(i)}_t=\kappa a ^{(i)}_t(\partial_{x_i}\log Z_{\mathbb{H}_t}) 
+ 2 \sum_{j \neq i} \frac{a^{(j)}_t}{x_i-x_j},
$$
where $Z_{\mathbb{H}_t}$ is a partition function as in the $n$SLE system.
It is under this condition that it describes the growth of $n$
interfaces in a way compatible with statistical mechanics and
conformal field theory. 

For $n=2$, the partition function $Z$ is proportional to
$(X_2-X_1)^\Delta$ with $\Delta=2/\kappa$ or
$\Delta=(\kappa-6)/\kappa$. This case, easy to study, has been
analyzed in details in \cite{BBK05}. The first choice $\Delta=2/\kappa$
selects configurations with no curve ending at infinity -- so that we
are actually describing standard chordal SLE joining to the two
initial positions of $X_1$ and $X_2$ -- while choosing
$\Delta=(6-\kappa)/\kappa$ selects configurations with two curves
emerging from the initial positions of $X_1$ and $X_2$ and ending both
at infinity.

If one demands that one of the curves is not growing, eg. by imposing
$a_2=0$ and $a_1=1$, then the case $n=2$ with
$Z=(X_2-X_1)^{\rho/\kappa}$ is equivalent to SLE$(\kappa,\rho)$,
written as in eq.(\ref{sleKRH}).

The $n$SLE system has recently been further generalized in
\cite{Graham05} by considering the possibility that each of the traces
are created by different operators. The outcome is that these
operators have to be either $\psi_{1;2}$ or $\psi_{2;1}$, which are
all degenerate at level two but for the same value of the central
charge.

\subsubsection{Arch probabilities}
\label{sec:arch}
It is known from CFT that, relaxing the positivity constraint, the
solutions to the conditions $(i),\;(ii),\;(iii)$ imposed on the
auxiliary function $Z$ of the $n$SLE system form a vector space of
dimension $d_{n,m}\equiv {n \choose m}-{n \choose
  m-1}=\frac{(n+1-2m)n!}{m!(n-m+1)!}$.  The numbers $d_{n,m}$ have
many combinatorial interpretations, but the one relevant for us is the
following. Draw $n+1$ points $X_1<X_2\cdots <X_n < \infty$ ordered
cyclically on the real line bounding the upper half plane $\mathbb H$.
Consider $n-m$ disjoint curves in $\mathbb H$ such that each $X_i$ is
an end point of exactly $1$ curve and $\infty$ is an end point of
exactly $n-2m$ curves. There are $d_{n,m}$ topologically inequivalent
configurations, called arch configurations when $n-2m=0$.
Motivated by this, it was claimed in ref.\cite{BBK05} that:\\
--- To each arch configuration $\alpha$ corresponds an extremal state
$Z_{\alpha}$ in the following sense: the solution of the $n$SLE system
with partition function $Z \propto Z_{\alpha}$ can be defined up to a
(possibly infinite) time, at which the growing curves have either
paired together or joined the point at infinity and at that time the
topology is that of the arch $\alpha$ with probability one.\\
--- One can decompose a general solution $Z$ of $(i),\;(ii),\;(iii)$
as a sum of extremal states with positive weights $p_\alpha$:
$$ 
\sum_{\alpha \,\in \, {\mathrm{arch}}}p_\alpha\, Z_{\alpha}.
$$
--- The probability that a solution of the $n$SLE system with auxiliary
function $Z$ ends in arch configuration $\alpha$ is the ratio
$$\frac{p_\alpha\, Z_{\alpha}(X_1,\cdots,X_n)}{Z(X_1,\cdots,X_n)}$$
evaluated at the initial condition $(X_1,\cdots,X_n)$.

This claim has not been yet proved in full generality but a few
examples are known \cite{BBK05}.  A first step toward a derivation of
the above results will be to explain how to construct martingales with
appropriate limiting behavior when arches are formed.  The $n$SLE
system is such that the quotient
$$
\frac{Z_{\alpha}(X^{(1)}_t,\cdots,X^{(n)}_t)}
{Z(X^{(1)}_t,\cdots,X^{(n)}_t)}
$$
are local martingales.  This can be proved directly using Ito's
formula.  They are bounded by $1$, hence they also are martingales.  A
standard argument shows that if ${\bf P}_{\alpha}$ is the probability
that the system ends in a definite arch configuration $\alpha$ (once
one has been able to make sense of it) then ${\bf
  P}_{\alpha}(X^{(1)}_t,\cdots,X^{(n)}_t)$ is a martingale.  To get a
full proof, one would then have to analyze the behavior of
$Z_{\alpha}(X^{(1)}_t,\cdots,X^{(n)}_t)$ when one arch closes, or when
one growing curve cuts the system in two, to get recursively a formula
that looks heuristically like
$$\frac{p_\alpha\, Z_{\alpha}(X^{(1)}_t,\cdots,X^{(n)}_t)}
{Z(X^{(1)}_t,\cdots,X^{(n)}_t)}\sim \delta_{\alpha,\alpha'}$$
if the
system forms asymptotically the arch system $\alpha'$ at large time
$t$.  Such a formula rests on properties of
$Z_{\alpha}(x_1,\cdots,x_n)$ when some points come close together in a
way reminiscent to the formation of arch $\alpha'$:
$Z_{\alpha'}(x_1,\cdots,x_n)$ should dominate all $Z_{\alpha}$'s,
$\alpha \neq \alpha'$ in such circumstances.

\begin{figure}[b]
\includegraphics[width=0.8\textwidth]{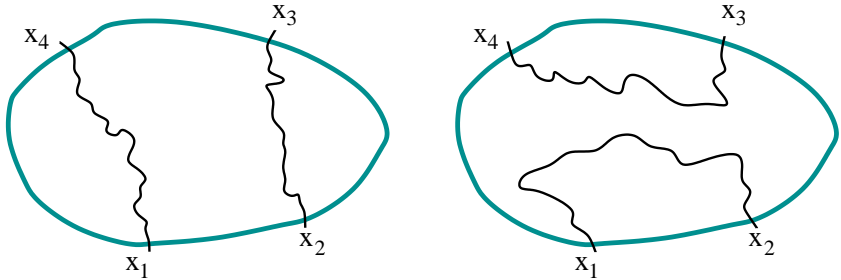}
\caption{\emph{Arch configurations for four SLE processes in 
an arbitrary domain.}}
\label{fig:4sle}
\end{figure}

To present the simplest non trivial example \cite{BBK05} we consider
critical models in the upper half plane $\mathbb{H}$ with boundary
conditions changing at $4$ points so that there is two interfaces each
joining a pair of points. See Fig.\ref{fig:4sle}. By conformal
invariance we may normalize the points so that $X_1=0$, $X_2=x$,
$X_3=1$ and $X_4=\infty$ with $0<x<1$. There are two distinct
topological configurations and therefore two pure partition functions,
which by construction may be written as correlation functions
$$
Z(x)=\langle\psi(\infty)\psi(1)\psi(x)\psi(0)\rangle_\mathbb{H}
$$
Their behavior when  points are fused are governed by CFT
fusion rules. As a consequence, $Z(x)$ behave either as
$x^{\frac{\kappa-6}{\kappa}}$ or as $x^{\frac{2}{\kappa}}$ as $x\to 0$.
The pure partition functions $Z_I$ and $Z_{II}$ are
specified by the way they behave when points are fused together:
\begin{eqnarray*}
Z_I(x) &=& x^{\frac{\kappa-6}{\kappa}}\times [1+\cdots],\quad 
~~~~~~~~~~~~ {\rm as}\ x\to0 \\ 
  &=& (1-x)^{\frac{2}{\kappa}}\times [{\rm const.}+\cdots],\quad 
~ {\rm as}\ x\to 1 \nonumber
\end{eqnarray*}
and $Z_{II}(x)=Z_I(1-x)$. 
The function $Z_I$ turns out to be the pure partition function for
configurations in which the curves join the pairs $[0x]$ and
$[1\infty]$ while $Z_{II}$ corresponds to the
configurations $[x1]$ and $[\infty 0]$.  The rationale behind these
conditions consists in imposing that the pure partition function
possesses the leading singularity, with exponent $(6-\kappa)/\kappa$,
when $x$ is approaching the point allowed by the configuration but has
sub-leading singularity, with exponent $2/\kappa$, when $x$ is
approaching the point forbidden by the configuration.
This set of conditions uniquely determines the functions $Z_I$ and
$Z_{II}$. These follows from CFT rules but may also be checked by
explicitly solving the differential equation that these functions
satisfy. Writing $Z(x)=x^{2/\kappa}(1-x)^{2/\kappa}\; G(x)$ yields,
$$ 
\kappa^2 x(1-x) G''(x) + 8\kappa (1-2x) G'(x) -
4(12-\kappa) G(x)=0 
$$
so that $G(x)$ is an hypergeometric function and
$$
Z_{II}(x) = {\rm const.} x^{2/\kappa}(1-x)^{2/\kappa}\;
F(\frac{4}{\kappa}, \frac{12-\kappa}{\kappa};\frac{8}{\kappa}|x)
$$
with the constant chosen to normalize $Z_I$ as above.
Using this explicit formula one may verify that $Z_I(x)$ is
effectively a positive number for any $x\in[0;1]$ so it has all
expected properties to be a pure partition function.
For $\kappa=4$, $Z_I(x)=\sqrt{(1-x)/x}$
and for $\kappa=2$, $Z_I(x)=(1-x^2)/x^2$.

A generic partition function $Z$ is a weighted  sum of the pure
partition functions:
$Z(x) = p_I Z_I(x) + p_{II}Z_{II}(x)$
with $p_I$ and $p_{II}$ positive. The probability of occurrence of
topological configurations $I$ and $II$ are then:
$$
{\bf P}[{\rm config}_I]= p_IZ_I(x)/Z(x)\quad,\quad
{\bf P}[{\rm config}_{II}]= p_{II}Z_{II}(x)/Z(x)
$$

--- Ising spin clusters correspond to $\kappa=3$. The boundary changing
operator $\psi$ has dimension $1/2$ and may thus be identified with a
fermion on the boundary. However the pure partition functions do not
correspond to the free fermion conformal block. By solving the
differential equation with the appropriate boundary
condition we get:
$$
Z_I(x)_{\rm spin\ Ising}= \mathrm{const.}\, 
\frac{1-x+x^2}{x(1-x)}\int_x^1 dy\frac{(y(1-y))^{2/3}}{(1-y+y^2)^2}
$$
The total partition function $Z_I(x)+Z_I(1-x)$ is proportional to
$\frac{1-x+x^2}{x(1-x)}$, which is the free fermion result. Hence, 
the Ising configuration probabilities, which are nothing but 
crossing formula for spin clusters, are:
$$
{\bf P} [{\rm config}_I]_{\rm spin\ Ising}= \int_x^1 
dy\frac{(y(1-y))^{2/3}}{(1-y+y^2)^2}\; \Big/ \int_0^1
dy\frac{(y(1-y))^{2/3}}{(1-y+y^2)^2}
$$

--- FK Ising clusters correspond to $\kappa=16/3$. The operator $\psi$ has
then dimension $1/16$. The pure partition function are given by:
$$
Z_I(x)_{\rm FK\ Ising} = \frac{(1-x)^{3/8}}{x^{1/8}(1+\sqrt{x})^{1/2}}
$$
and the crossing probabilities by:
\begin{eqnarray*}
{\bf P} [{\rm config}_I]_{\rm FK\ Ising} =
    \frac{\sqrt{(1-x) + (1-x)^{3/2}}}{\sqrt{x + x^{3/2}}
    + \sqrt{(1-x) + (1-x)^{3/2}}}
\end{eqnarray*}

--- The other critical random cluster (or Potts) models with $0 \leq Q
\leq 4$ have $Q = 4 \cos^2 \big( \frac{4 \pi}{\kappa} \big)$,
$4 \leq \kappa \leq 8$, and it is straightforward to obtain
explicit crossing formulas involving only hypergeometric functions.
The case $\kappa=6$ reproduces Cardy's crossing formula for
percolation.

\subsection{Miscellaneous}
\label{sec:diversmisc}

Another approach to the CFT/SLE correspondence has been proposed in
\cite{FW02,WernerFried03}. It uses the restriction properties to be
discussed in Section \ref{sec:restrict}. The restriction property
holds for SLE only for $\kappa=8/3$, which corresponds to the central
charge $c=0$. However, the restriction property can be recovered for
$\kappa<8/3$ be dressing the SLE curves with Brownian bubbles
associated to the Brownian loop soup to be described in Section
\ref{sec:browsoup}. This approach was later generalized in
\cite{Doyon05} to identify the CFT stress-tensor within the SLE
framework.

Generalizations of SLEs to a larger class of CFTs, with central
charges bigger than one and with more degrees of freedom than those
described above, have of course been considered. These include
supersymmetric \cite{Rasmussen03a,Lesage03} or logarithmic
\cite{Rasmussen03b,Rouhani04} generalizations. The cases corresponding
to CFT with Lie-group symmetry, described by the Wess-Zumino-Witten
models, have been discussed in ref.\cite{Bettelheim05}. The outcome of
these studies is that for describing interfaces in CFT with higher
central charges one needs to add extra degrees of freedom along the
interfaces but as a geometrical objects, ie. as curves, these
interfaces are still described by the standard SLE with some effective
parameter $\kappa$ corresponding to some effective central charge less
than one. This had to be expected because Schramm's argument implies
that conformally invariant curves have to be described by the standard
SLEs.

SLEs in different topologies than that of a disc have been considered:
refs.\cite{Dubedat03b,Zhan03,BB04} discussed SLEs in annuli and
ref.\cite{BauerFried04a,BauerFried04b,BauerFried05} defined SLEs in
multiply connected domains. These explicit constructions show that the
moduli parameters of the domains change in time while the SLE curves
are growing. This is clear in the case of the annuli in which case the
only moduli is the ratio of radii of the inner and outer circles. An
annulus cutted along a piece of curve grown say from the outer circle
is topologically equivalent to another annulus but with different
ratio of radii.  In the annulus case the SLE process stops at the
instant the curve touches the inner circle. An attempt to formalize
SLEs as motions in the moduli spaces of Riemann surfaces has been
written in \cite{FriedKalk03}.

%% file: chap5.tex

 This Section aims at presenting some of the remarkable -- and quite
beautiful -- structures and properties of conformally invariant curves
or hulls. Some of them are pure SLE properties, others involve extra
conformally invariant measures on hulls.  The first is a locality
property of SLE$_6$ which is inherited from the locality property of
percolation. The second refers to the notion of restriction measures
which concerns measures on hulls such that the measures for hulls on a
domain $\mathbb{D}$ conditioned not to touch some fixed hull
$\mathbb{A}$ is identical to the measure for hulls on
$\mathbb{D}\setminus\mathbb{A}$. This properties holds for SLE only
for $\kappa=8/3$. It also holds for Brownian excursion and this
provides a way to prove \cite{LSW00a,LSWConfRest02} Mandelbrot's
conjecture that the fractal dimension of the exterior perimeter of
Brownian excursion is $4/3$. We also presents Duplantier's predictions
concerning the fractal spectrum of harmonic measures of conformally
invariant hulls. To prove this prediction is one of the remaining big
challenge of the SLE project. Finally we describe the Brownian loop
soup which was introduced in ref.\cite{LWsoup03} in connection with
restriction measures.

\subsection{Locality of $SLE_6$}
\label{sec:locality6}
In percolation there is no interaction in the sense that the color of
the sites are chosen with given probability independently to those of
the other sites of the system. This implies that, on the lattice, the
exploration process used to recursively construct an interface is
defined by local rules. In the continuum, and at criticality, this
translates into the locality property of SLE$_6$ which signifies that
its hull does not feel the boundary of the domain, or portion the
domain, before it visits it.

To be more concrete consider chordal SLE in the upper half plane
$\mathbb{H}$.  Let us deform it by removing a hull $\mathbb{A}$.  Let
$\phi_\mathbb{A}$ be the map uniformizing
$\mathbb{H}\setminus\mathbb{A}$ onto $\mathbb{H}$ normalized by
$\phi_\mathbb{A}(0)=0$, $\phi_\mathbb{A}(\infty)=\infty$ and
$\phi_\mathbb{A}'(\infty)=1$. We want to compare SLE in $\mathbb{H}$
and in $\mathbb{H}\setminus\mathbb{A}$. By conformal transport, this
amounts to compare the SLE trace $\gamma_{[0,t[}$ in $\mathbb{H}$ and
its image $\phi_\mathbb{A}(\gamma_{[0,t[})$, again in $\mathbb{H}$,
with the hope of proving that they have identical law,
\begin{eqnarray}
\gamma_{[0,t[}\
\equiv_{\rm in~law} \phi_\mathbb{A}(\gamma_{[0,t[})
\label{localprop}
\end{eqnarray}
up to a possible random time change. This turns out to be true only at
$\kappa=6$.

\begin{figure}[htbp]
\begin{center}
\includegraphics[width=0.9\textwidth]{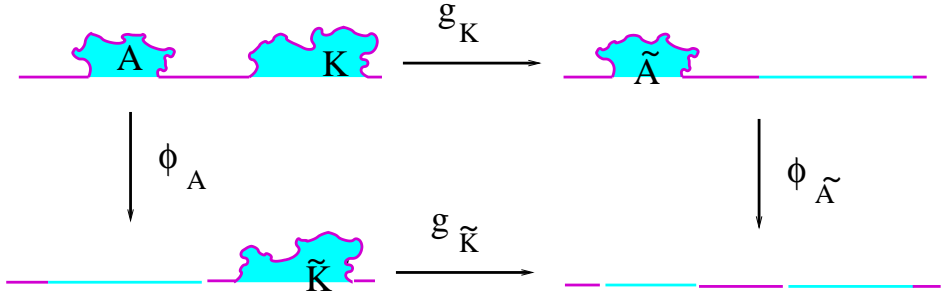}
\caption{The commutative diagram associated to the two alternative ways
  of uniformizing the complement of two hulls. It codes for the
  relation $\phi_{\tilde A}\circ g_K= g_{\tilde K}\circ \phi_A$.}
     \label{fig:2hullcom}
 \end{center}
\end{figure}

This proof is an interesting exercise \cite{LSW1,Lawlerbook05}. It is
based on the fact that we can uniformize the complement of the union
of the hull and of the trace in two different ways: either erasing
first the trace $\gamma_{[0,t[}$ using the SLE Loewner map $g_t$ and
then the modified hull $\mathbb{A}_t$ using a map
$\phi_{\mathbb{A}_t}$, or erasing first the hull $\mathbb{A}$ using
$\phi_\mathbb{A}$ and then the modified curve $\phi_A(\gamma_{[0,t[})$
using another Loewner map $k_t$. See Fig.\ref{fig:2hullcom}. This translates
into the commutative diagram \cite{LSW1,LSWConfRest02}:
\begin{eqnarray}
\phi_{\mathbb{A}_t}\circ g_t = k_t \circ \phi_\mathbb{A}.
\label{comdiag}
\end{eqnarray}
Here $\phi_{\mathbb{A}_t}$ is normalized the same way as
$\phi_\mathbb{A}$.  Since $k_t$ is a map (properly normalized)
uniformizing of the complement of a curve, it satisfies the Loewner
equation $\partial_t k_t(z)= a_t/(z-W_t)$. The source $W_t$ is the
image of the modified curve, $k_t(\phi_\mathbb{A}(\gamma_t))$, which
is equal to $W_t=\phi_{\mathbb{A}_t}(\xi_t)$ thanks to the commutative
diagram.  The evolution equation for $\phi_{\mathbb{A}_t}$ directly
follows from the commutative diagram:
$$
\partial_t\phi_{\mathbb{A}_t}(w) =
\frac{a_t}{\phi_{\mathbb{A}_t}(w)-W_t} 
- \frac{2\phi_{\mathbb{A}_t}'(\xi_t)}{w-\xi_t} 
$$
The map $\phi_{\mathbb{A}_t}$ is regular at point $\xi_t$ because the
modified hull $\mathbb{A}_t$ is away from it. Demanding that its
time derivative is also regular at $\xi_t$ fixes
$a_t=2\phi_{\mathbb{A}_t}'(\xi_t)^2$. The limit is then 
$\partial_t\phi_{\mathbb{A}_t}(\xi_t)=-3\phi_{\mathbb{A}_t}''(\xi_t)$.

To prove locality we have to prove that $k_t$ defines a SLE process up
to a random time change which amounts to prove that $W_t$ is a
Brownian motion up to the random time change specified by $a_t$.
It\^o's formula gives $dW_t= \partial_t\phi_{\mathbb{A}_t}(\xi_t) dt +
\phi_{\mathbb{A}_t}'(\xi_t)d\xi_t
+\frac{\kappa}{2}\phi_{\mathbb{A}_t}''(\xi_t)dt$. Thanks to the
previous evaluation of $\partial_t\phi_{\mathbb{A}_t}(\xi_t)$ the
first and  last terms are similar so that:
$$
dW_t=(\frac{\kappa-6}{2})\phi_{\mathbb{A}_t}''(\xi_t)dt  +
\phi_{\mathbb{A}_t}'(\xi_t)d\xi_t
$$
The drift term vanishes for $\kappa=6$ so that $W_{t(s)}$ is a
Brownian motion up to the change $ds=\phi_{\mathbb{A}_t}'(\xi_t)^2dt$.
This proves locality at $\kappa=6$.

\subsection{Restrictions}
\label{sec:restrict}
\begin{figure}[htbp]
\begin{center}
\includegraphics[width=0.6\textwidth]{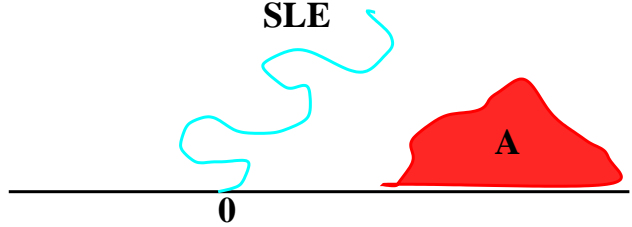}
\caption{A SLE curve growing in the upper half plane minus a hull
  $\mathbb{A}$ or conditioned not to touch this hull.}
     \label{fig:restrict}
 \end{center}
\end{figure}

Restriction measures \cite{LSWConfRest02} are measures for conformally
invariant random hulls with the property that the law of the hulls in
a domain $\mathbb{D}$ conditioned not to visit a fixed hull
$\mathbb{A}$ is identical to that of the random hulls in the domain
$\mathbb{D}\setminus \mathbb{A}$ with the hull removed, see
Fig.\ref{fig:restrict}.

These laws of been studied and classified in \cite{LSWConfRest02}:
there is a one parameter family of such measures.  By conformal
invariance it is enough to look at them for hulls in the upper half
plane.  Let $\mathbb{K}\subset \mathbb{H}$ be the random hulls.  Their
law is characterized by the set of probabilities ${\bf
  P}[\mathbb{K}\cap \mathbb{A}=\emptyset]$ that they do not intersect
fixed hulls $\mathbb{A}$. As proved in \cite{LSWConfRest02}, the
restriction property imposes that these probabilities are of the
following form:
\begin{eqnarray}
{\bf P}[\mathbb{K}\cap \mathbb{A}=\emptyset]
= \phi_{\mathbb{A}}'(0)^\alpha, \quad {\rm for}\ 
\mathbb{K}\subset\mathbb{H}
\label{resmeasure}
\end{eqnarray}
with $\phi_{\mathbb{A}}$ the conformal map uniformizing
$\mathbb{H}\setminus \mathbb{A}$ back onto $\mathbb{H}$, normalized
by $\phi_{\mathbb{A}}(0)=0$, $\phi_{\mathbb{A}}(\infty)=\infty$
and $\phi_{\mathbb{A}}'(\infty)=1$. The exponent $\alpha$ specifies
the measure. It has to be bigger than $5/8$. The greater is $\alpha$
the thicker are the hulls: $\alpha=5/8$ corresponds to random simple
curves -- actually to SLE$_{8/3}$ as we shall see below.

It is an instructive exercise to check that the measures
(\ref{resmeasure}) indeed satisfy the restriction property, ie. the
law in $\mathbb{H}$ of the hulls $\mathbb{K}$ conditioned not to touch
a given hull $\mathbb{A}$ is identical to the law in
$\mathbb{H}\setminus \mathbb{A}$:
$$
\mathbb{K}\subset\mathbb{H}\ \Big\vert
(\mathbb{K}\cap\mathbb{A}=\emptyset) 
\equiv_{\rm in~law} 
\mathbb{K}\subset\mathbb{H}\setminus \mathbb{A}
$$
To check it we pick another arbitrary hulls $\mathbb{B}$ disjoint
from $\mathbb{A}$, we look at the probabilities that the hulls viewed
as conditioned hulls in $\mathbb{H}$ or as hulls in
$\mathbb{H}\setminus\mathbb{A}$ do not intersect $\mathbb{B}$ and we
prove that:
$$
{\bf P}_{\rm in~\mathbb{H}}[\mathbb{K}\cap\mathbb{B}=\emptyset\ 
\Big\vert \mathbb{K}\cap\mathbb{A}=\emptyset]= {\bf P}_{\rm
  in~\mathbb{H}\setminus\mathbb{A}}[\mathbb{K}\cap\mathbb{B}=\emptyset]
$$
The conditioned probability in the left hand side is the ratio of
the probability that $\mathbb{K}$ does not touch
$\mathbb{A}\cup\mathbb{B}$ by the probability that it does not touch
$\mathbb{A}$, so it is
$\phi_{\mathbb{A}\cup\mathbb{B}}'(0)^\alpha/\phi_\mathbb{A}'(0)^\alpha$
where $\phi_{\mathbb{A}\cup\mathbb{B}}$ and $\phi_\mathbb{A}$ are the
respective uniformizing conformal maps properly normalized. The
complement of the union $\mathbb{A}\cup\mathbb{B}$ can be uniformized
in two steps: first erasing the hull $\mathbb{A}$ using
$\phi_\mathbb{A}$ and then erasing the image $\hat\mathbb{B} \equiv
\phi_\mathbb{A}(\mathbb{B})$ of the remaining hull using the
appropriate normalized conformal map $\phi_{\hat\mathbb{B}}$. This
yields $\phi_{\mathbb{A}\cup\mathbb{B}}=\phi_{\hat
  \mathbb{B}}\circ\phi_\mathbb{A}$ so that:
$$
{\bf P}_{\rm in~\mathbb{H}}[\mathbb{K}\cap\mathbb{B}=\emptyset\ 
\Big\vert \mathbb{K}\cap\mathbb{A}=\emptyset]= 
\phi_{\hat \mathbb{B}}'(0)^\alpha,\quad
\hat\mathbb{B}=\phi_\mathbb{A}(\mathbb{B})
$$
This is also ${\bf P}_{\rm in~\mathbb{H}\setminus\mathbb{A}}
[\mathbb{K}\cap\mathbb{B}=\emptyset]$ because the law in
$\mathbb{H}\setminus \mathbb{A}$ is defined from that in the upper
half plane by conformal transport.

As explained in \cite{Werner03f}, there exists a nice interplay
between SLE$(\kappa,\rho)$ and restriction measures. Indeed an
SLE$(\kappa,\rho)$ can be viewed as an SLE$_\kappa$ conditioned not to
touch a restriction sample with exponent
$\alpha=\rho(\rho+4-\kappa)/4\kappa$. Furthermore, an
SLE$(\kappa,\rho)$ conditioned not touch a restriction sample with
exponent $\alpha$ is another SLE$(\kappa,\bar \rho)$ with
$2\bar\rho=\kappa-4+\sqrt{16\kappa\alpha+(2\rho+4-\kappa)^2}$.  A
relation between reflected Brownian motions and restriction measures
is given in \cite{LSWConfRest02}.

\subsubsection{The restriction property of SLE${}_{8/3}$}
\label{sec:rest8/3}
Here we present the arguments \cite{LSWConfRest02} showing that
chordal SLE${}_{8/3}$ satisfies the restriction property with
$\alpha=5/8$.  Note that this value is that of the dimension of the
operator $\psi_{1;2}$ creating the SLE trace: $h_{1;2}=5/8$ at
$\kappa=8/3$. So we have to compute the probability that the
SLE$_{8/3}$ curve does not touch a bounded hull $\mathbb{A}$ away from
the origin. If it is equal to $\phi_\mathbb{A}'(0)^{5/8}$, then, by
the usual Markov and independent increment properties of SLE, the
process (up to the hitting time of $\mathbb{A}$),
$$
M_t(\mathbb{A})=\phi_{\mathbb{A}_t}'(\xi_t)^{5/8},
\quad \kappa=8/3
$$
is a martingale with $\mathbb{A}_t\equiv g_t(\mathbb{A})$ with
$g_t$ the SLE Loewner map and $\phi_{\mathbb{A}_t}$ the corresponding
uniformizing map fixing $0$ and $\infty$ with
$\phi_{\mathbb{A}_t}'(\infty)=1$. We shall prove it a bit later.

Reciprocally, if $M_t(\mathbb{A})$ is a martingale then
$$
{\bf P}[\gamma_{[0,\infty[}\cap \mathbb{A}=\emptyset]=
\phi_{\mathbb{A}}'(0)^{5/8}, \quad \kappa=8/3
$$
Indeed, choose a very large semi circle ${\mathcal C }_R$ of radius
$R$ in $\mathbb{H}$ centered at the origin. Let $\tau_R$ be the first
time when the trace touches either $\mathbb{A}$ or ${\mathcal C }_R$.
Then $\tau_R$ is a stopping time and we claim that
$\phi_{\hat\mathbb{A}_{\tau_R}}'(\xi_{\tau_R})$ is $0$ if the SLE hull
hits $\mathbb{A}$ at $\tau_R$ and goes to $1$ for large $R$ if the SLE
hull hits ${\mathcal C }_R$ at $\tau_R$. Indeed, when the hull
approaches $\mathbb{A}$, one or more points on $\hat\mathbb{A}_{t}$
approach $\xi_t$, and at the hitting time, a bounded connected
component is swallowing $\xi_t$ indicating that the derivative has to
vanish there.  On the other hand, if ${\mathcal C }_R$ is hit first,
then $\hat\mathbb{A}_{\tau_R}$ is dwarfed so that
$\phi_{\hat\mathbb{A}_{\tau_R}}$ is close to the identity map away
from $\hat\mathbb{A}_{\tau_R}$ and in particular at the point
$\xi_{\tau_R}$. (The normalization of the conformal maps are crucial
for proving these statements.)  Hence, the martingale
$\phi_{\hat\mathbb{A}_t}'(\xi_t)^{5/8}$, at $t=\tau_R$, is $0$ if
$\mathbb{A}$ is hit before ${\mathcal C }_R$ and close to $1$ if the
opposite is true.  But the expectation of a martingale is time
independent, so that the probability that the trace does not hit
$\mathbb{A}$ is $\phi_\mathbb{A}'(0)^{5/8}$. This shows that
SLE$_{8/3}$ satisfies the restriction property.

The proof that $M_t(\mathbb{A})$ is a martingale is similar to the
proof of the locality property \cite{LSWConfRest02}. It is again based
on the commutative diagram (\ref{comdiag}): $\phi_{\mathbb{A}_t}\circ
g_t = k_t \circ \phi_\mathbb{A}$. Let us recall that $k_t$ is the map
uniformizing the complement of the image curve
$\phi_\mathbb{A}(\gamma_{[0,t[})$. It satisfies the Loewner equation
$\partial_t k_t(z)= a_t/(z-W_t)$ with $W_t=\phi_{\mathbb{A}_t}(\xi_t)$
and $a_t=2\phi_{\mathbb{A}_t}'(\xi_t)^2$.  The evolution equation for
$\phi_{\mathbb{A}_t}$ follows from the commutative diagram:
$$
\partial_t\phi_{\mathbb{A}_t}(w) =
\frac{2\phi_{\mathbb{A}_t}'(\xi_t)^2}{\phi_{\mathbb{A}_t}(w)-W_t} 
- \frac{2\phi_{\mathbb{A}_t}'(\xi_t)}{w-\xi_t} 
$$
The map $\phi_{\mathbb{A}_t}$ is regular at point $\xi_t$ because
the modified hull $\mathbb{A}_t$ is away from $\xi_t$.  Taking the
derivative with respect to $w$ and then the limit $w\to\xi_t$ gives:
$$
\partial_t\phi_{\mathbb{A}_t}'(\xi_t) = 
\phi_{\mathbb{A}_t}''(\xi_t)^2/2\phi_{\mathbb{A}_t}'(\xi_t)
- 4\phi_{\mathbb{A}_t}'''(\xi_t)/3
$$
This allows to compute the It\^o derivative of
$\phi_{\mathbb{A}_t}'(\xi_t)^\alpha$ :
\begin{eqnarray*}
d \phi_{\mathbb{A}_t}'(\xi_t)^\alpha
= \alpha \phi_{\mathbb{A}_t}'(\xi_t)^\alpha[\,(3\kappa-8)
\frac{\phi_{\mathbb{A}_t}'''(\xi_t)}{6\phi_{\mathbb{A}_t}'(\xi_t)}
+(\kappa\alpha-\kappa+1)
\frac{\phi_{\mathbb{A}_t}''(\xi_t)^2}{2\phi_{\mathbb{A}_t}'(\xi_t)^2}\,]
dt + [\cdots] d\xi_t
\end{eqnarray*}
The drift term vanishes for $\kappa=8/3$ and $\alpha=5/8$, which
proves that $M_t(\mathbb{A})$ is a local martingale.

For $\kappa\not=8/3$, the process $\phi_{\mathbb{A}_t}'(\xi_t)^{5/8}$ is
not a martingale but the following one~\cite{LSWConfRest02},
\begin{eqnarray}
\phi_{\mathbb{A}_t}'(\xi_t)^{h_{1;2}}\, \exp-\frac{c}{6}\int_0^t ds\,
(S\cdot\phi_{\mathbb{A}_s})(\xi_s), 
\label{martinres}
\end{eqnarray}
with $S\cdot\phi$ the Schwarzian derivative of $\phi$,
$c=(6-\kappa)(8\kappa-3)/2\kappa$ the central charge and
$h_{1;2}=(6-\kappa)/2\kappa$, is a martingale. The fact that it is a
martingale can be proved as above using the commutative diagram
\cite{LSWConfRest02} or using the algebraic formulation of SLE
\cite{BB03b}.  The extra term is present for non zero central charge
because this martingale codes the way SLEs react to deformations of
the domain.  Similar formula exist for other SLEs, eg. for radial SLE
\cite{LSWConfRest02,BBrad03}.

\subsubsection{The Brownian exterior perimeter}
\label{sec:browext}
Naively, a two dimensional Brownian excursion $\mathfrak{B}_t$ from
say $0$ to $\infty$ is a Brownian walk started at the origin
conditioned to remain in the upper half plane and to escape to
$\infty$.  This has to be made more precise since such events happen
with probability zero!  The cleaner mathematical definition consists
in setting $\mathfrak{B}_t=B_t+iY_t$ with $B_t$ a standard Brownian
motion and $Y_t$ a Bessel process. Instead, we choose the most
physical one, and we consider a Brownian walk started at point
$i\epsilon$ ($\epsilon>0$) conditioned to escape first through line
$i\Lambda+\mathbb{R}$ $(\Lambda>0)$ before touching the real axis
$\mathbb{R}$. This happens with probability $\epsilon/\Lambda$. We
then take the limit $\epsilon\to0$ and $\Lambda\to\infty$ and we look
at events for which the conditioned probabilities are finite in this
limit.

Let us show that Brownian excursion satisfies the restriction
property. Let $\mathbb{A}$ be a hull in the upper half plane.  We have
to compute the probability ${\bf
  P}(\mathfrak{B}_{]0,\infty[}\cap\mathbb{A}=\emptyset)$ that the
Brownian excursion does not touch this hull. By the above definition,
this is the probability for a 2d Brownian motion started at
$i\epsilon$ not to touch $\mathbb{A}$ conditioned to escape first on
$i\Lambda+\mathbb{R}$.  So it is the ratio of the probability that the
2d Brownian motion started at $i\epsilon$ does not touch $\mathbb{A}$
and escapes through $i\Lambda+\mathbb{R}$ by the probability -- equal
to $\epsilon/\Lambda$ -- that it escapes through
$i\Lambda+\mathbb{R}$.  To compute it we use conformal invariance of
the 2d Brownian motion.  Let $\phi_\mathbb{A}$ be the conformal map
uniformizing $\mathbb{H}\setminus \mathbb{A}$ onto $\mathbb{H}$ fixing
$0$ and $\infty$ and such that $\phi_\mathbb{A}'(\infty)=1$. The image
of the starting point under this map is
$\phi_\mathbb{A}(i\epsilon)\simeq i\epsilon\phi_\mathbb{A}'(0)$ for
$\epsilon\ll 1$ and the straight line $i\Lambda+\mathbb{R}$ at
infinity is mapped under a line at infinity slightly waving around
$i\Lambda+\mathbb{R}$ since $\phi_\mathbb{A}(\infty)=\infty$ and
$\phi_\mathbb{A}'(\infty)=1$. Thus, the probability that the 2d
Brownian motion started at $i\epsilon$ does not touch $\mathbb{A}$ and
escapes through $i\Lambda+\mathbb{R}$ is, by conformal invariance, the
probability that the 2d Brownian motion started at
$i\epsilon\phi_\mathbb{A}'(0)$ escapes through the line
$i\Lambda+\mathbb{R}$. This is equal to
$\epsilon\phi_\mathbb{A}'(0)/\Lambda$. Finally, the probability that
the Brownian excursion does not touch the hull $\mathbb{A}$ is:
\begin{eqnarray*}
{\bf P}[\mathfrak{B}_{]0,\infty[}\cap\mathbb{A}=\emptyset]
= \frac{ \epsilon\phi_\mathbb{A}'(0)/\Lambda}
{\epsilon/\Lambda}= \phi_\mathbb{A}'(0)
\end{eqnarray*}
Since this is valid for any hull, it proves that Brownian
excursions in the upper half plane -- more precisely hulls obtained
by filling the domain surrounded by the Brownian excursions -- form a
restriction measure with exponent $1$.

We can now compare Brownian excursions and self avoiding walks.
Consider $5$ independent Brownian excursions. By construction they
form a restriction measure with exponent $5$ -- that is the
probability that none of these excursions hit a hull $\mathbb{A}$ is
$\phi_\mathbb{A}'(0)^5$. Similarly $8$ independent SLE$_{8/3}$ --
ie. $8$ self avoiding walks -- also form a restriction measure with
exponent $5$ since each of them form a restriction measure with
exponent $5/8$. As a consequence we have an identification of the
distribution of these hulls \cite{LSWConfRest02}:
$$
5\ {\rm Brownian~excursions} \equiv_{\rm in~law}  
8\ {\rm SLE}_{8/3}
$$
This identification yields to a poor (wo)man argument for
Mandelbrot's conjecture on the fractal dimension of the Brownian
exterior perimeter. Indeed, the exterior perimeter of the hull formed
by filling the space surrounded by the five Brownian excursions is
locally the exterior perimeter of one of these excursions. Similarly,
the exterior perimeter of the hull formed by the eight SLE$_{8/3}$ is
locally one of these SLEs since they are simple curve. Thus the fractal
dimension of the exterior perimeter of a Brownian excursion equals
that of $SLE_{8/3}$. Namely
$$ 
{\rm dim.~Brownian~perimeter} = d_{8/3} = 4/3
$$
The mathematical proof of this statement has been one of the main
achievement of the SLE project \cite{LSW00a,LSWConfRest02}. It was
originally based on the observation that the outer boundary of
Brownian excursion is the same as that of SLE$_6$. SLE also gives the
dimension $3/4$, which is the dimension of the set of double points in
SLE$_6$, for the set of cut points of Brownian paths.

\subsection{Multifractal harmonic measure}
\label{sec:multifract}

\subsubsection{Harmonic measure}
\label{sec:harmo}

The harmonic measure of a planar domain $\mathbb{U}$ is linked to the
hitting probability distribution of a random walker when she/he escapes
from $\mathbb{U}$. To be more precise let us consider a planar domain
$\mathbb{U}$ with the topology of a disc and point $z_0$ in the
interior of $\mathbb{U}$. Consider a two dimensional Brownian motion
$X_t$ started at point $z_0$ and stopped at the first instant
$\tau_\mathbb{U}$ it exits from $\mathbb{U}$. Then the harmonic
measure $\mu_{z_0}$, which is a measure on the boundary, is the
probability that the Brownian motion escapes from the domain through a
subset of its boundary:
$$
\mu_{z_0}({F})\equiv {\bf P}_{z_0}[X_{\tau_\mathbb{U}}\in {F}],
\quad {F}\subset \partial\mathbb{U}
$$
By construction, it is harmonic as a function of the starting
position $z_0$ with boundary condition $\mu_{z_0}({F})=1$ if $z_0\in
F$ and $0$ if $z_0$ is approaching the complement of $F$ on $\partial
\mathbb{U}$. The harmonic measure has many applications to generalized
Dirichlet problems \cite{Oksendal}.

The simplest example is for $\mathbb{U}$ the unit disc and $z_0$ the
origin. The hitting distribution is then uniform on the unit circle so
that in this case $d\mu_0=d\theta/2\pi$ with $\theta\in[0,2\pi[$ the
angle parameterizing the unit circle. This gives a way to compute the
harmonic measure (for sufficiently regular boundary). Let $w$ be a
conformal map uniformizing $\mathbb{U}$ onto the unit disc with
$w(z_0)=0$, then $\mu_{z_0}(F)=\int_F |dw(z)|$ by conformal invariance
of the two dimensional Brownian.

The harmonic measure is sensible to the local geometry of the boundary
and linked to the behavior of the (derivative of the) uniformizing map
close the boundary.  Consider for instance the wedge $\mathcal{
  W}_\theta\equiv \{z\in\mathbb{C}, 0<{\rm arg}\, z<\theta\}$ of angle
$\theta$. A uniformizing map of $\mathcal{W}_\theta$ onto the unit disc
is $w(z)=\frac{z^{\pi/\theta}+i}{z^{\pi/\theta}-i}$. For $z$ at a
distance $\varepsilon$ from the boundary $\varepsilon\, |w'(z)|$
behaves as $\varepsilon^{\pi/\theta}$. The harmonic measure
$\mu_*(\partial W_\theta \cap B_\varepsilon)$ of the portion of the
boundary of the wedge contained in the ball of size $\varepsilon$
centered at the tip of the wedge scales the same way as
$\varepsilon^{\pi/\theta}$.

We shall be interested in cases in which a portion of the boundary
of the domain is the boundary of critical clusters so that it looks
locally as a SLE curve with parameter $\kappa<4$. We may for instance
have in mind a domain minus SLE hulls (run during sufficiently enough
time).

\subsubsection{Multifractal spectrum}
\label{sec:multispec}
The multifractal spectrum of harmonic measures of boundaries of
critical clusters has been predicted by B. Duplantier in a serie of
papers
\cite{Duplantier99a,Duplantier99b,Duplantier00,DuplantierBinder02}
using arguments based on applications of 2D gravity and of the KPZ
formula \cite{KPZ} for gravitationally dressed dimensions. See the
reviews \cite{Duplantier03,Duplantier03a}.  It is defined by
considering the expectation values of moments of the harmonic measure
of the boundary curve. More precisely, let us define
$$
\mathcal{Z}_n = {\bf E}\left[{ \sum_{B_\varepsilon} 
[\mu_*(\gamma\cap B_\varepsilon)]^n }\right] 
$$
where the sum is over a set of boxes $B_\varepsilon$ of size
$\varepsilon$ covering the boundary curve $\gamma$ and $\mu_*$ is
the harmonic measure.  For $n=0$ the sum is simply the number of boxes
needed to cover the curve. So by definition of the fractal dimension
it scales as $\mathcal{Z}_0\approx \varepsilon^{-d_\kappa}$. For $n=1$
the sum reproduces the total harmonic measure $\mu_*(\gamma)$ of
the curve which we assumed to be normalized to one, so that $\mathcal{
  Z}_1=1$. More generally, scaling behavior of $\mathcal{Z}_n$ defines
the multifractal exponents $\tau_n$ via $\mathcal{Z}_n\approx
\varepsilon^{\tau_n}$.  Duplantier's prediction is:
\begin{eqnarray}
\tau_n=\frac{n-1}{2} + \frac{\kappa+4}{16\kappa}\,
\big[{\sqrt{16n\kappa +(\kappa-4)^2}-(\kappa+4)}\big]
\label{multiexpo}
\end{eqnarray}
As it should be, it is invariant under the duality $\kappa \to
16/\kappa$ and $-\tau_0$ reduces to the fractal dimension of the
exterior perimeter.  The harmonic measure is only sensible to the
exterior perimeter, so we shall choose the branch $\kappa<4$.

By standard multifractal arguments \cite{Mandel74,Procaccia}, the
multifractal dimensions $f_\alpha$ are obtained from $\tau_n$ by a
Legendre transform.  Let us recall it. Consider the set $\mathcal{
  C}_\alpha$ of points in which the harmonic measure, evaluated on a
ball $B_\varepsilon$ intersecting the curve at a point of $\mathcal{
  C}_\alpha$, scales as $\varepsilon^\alpha$. By definition,
$f_\alpha$ is the fractal dimension of $\mathcal{C}_\alpha$.  We may
replace the sum defining $\mathcal{Z}_n$ by a sum over all boxes of sizes
$\varepsilon$ covering the full domain -- there are of order $\approx
\varepsilon^{-2}$ of such boxes -- and over the sets $\mathcal{C}_\alpha$
weighted by the probability that they intersect one of these balls --
this probability scales as $\varepsilon^{2-f_\alpha}$ by definition of
the fractal dimension. Over the set $\mathcal{C}_\alpha$ the $n^{\rm th}$
moment of the harmonic measure behaves as $\varepsilon^{n\alpha}$.
Hence, $\mathcal{Z}_n$ scales as $\int d\nu_\alpha\,
\varepsilon^{n\alpha-f_\alpha}$. Saddle point approximation valid for
$\varepsilon\to 0$ gives
$$ \tau_n= n\alpha -f_\alpha,\quad df_\alpha/d\alpha =n $$
This is a Legendre transform so that 
$f_\alpha$ may be computed by inverting it:
$f_\alpha = \tau_n-n\alpha$ with $d\tau_n/dn=\alpha$.
The result is:
\begin{eqnarray}
f_\alpha = \frac{(\kappa+4)^2}{16\kappa}\,
\frac{3\alpha-2}{2\alpha-1} -\frac{(\kappa-4)^2}{16\kappa}\, \alpha 
\label{multidim}
\end{eqnarray}
with $\alpha\in]1/2,+\infty[$. It satisfies Makarov's theorem
\cite{Makarov} which states that $\tau'(1)=1$ or alternatively
$f_{\alpha=1}=1$.  By construction, the maximum of $f_\alpha$ is for
$df_\alpha/d\alpha=0$ so that $f_\alpha\vert_{\rm max}=-\tau_0$ is the
fractal dimension.

The sets $\mathcal{C}_\alpha$ are sets of points in which the exterior
perimeter has locally the geometry of a wedge with angle $\pi/\alpha$,
so that the harmonic measure scales locally as $\varepsilon^\alpha$.
The typical geometry is a wedge of angle $\theta^*=\pi/\alpha^*=
\pi(1-\kappa/4)$, we choose the branch $\kappa<4$. It corresponds to
the value $\alpha^*$ at which $f_\alpha$ reaches its maximum.  The
minimum value $\alpha_{\rm min}=1/2$ corresponds to the maximum angle
$2\pi$ so that the curve looks locally like a needle. This minimum
value governs the large $n$ behavior of the multifractal exponents
$\tau_n\simeq n/2$ for $n\gg 1$. Large values of $\alpha$ correspond
to small angles and therefore to local fjords, and $f_\alpha\simeq
n^*\alpha$ asymptotically for $\alpha \gg 1$.  They govern the
behavior of the lowest moments as $n\searrow n^*$. Moments of order
less that $n^*$ do not exist.

No mathematically complete proof of Duplantier's predictions have yet
been published. Besides original Duplantier's arguments there are
however hints for these formulas based on conformal field theory.  The
first hint comes by observing \cite{BB02b} that KPZ formulas of 2D
gravity \cite{KPZ} naturally arise in SLE because they are linked to
operator product expansion with the conformal operator $\psi_{1;2}$
creating a SLE curve. This implies that moments of the SLE
uniformizing map should naturally be expressed in terms of the
gravitationally dressed dimensions.  The second set of hints
\cite{Wiegmann05} is based on the observation that the harmonic
measure evaluated on a ball of size $\varepsilon$ and the derivative
of the uniformizing map $w'(z)$ evaluated at a distance $\varepsilon$
from the boundary scale the same way:
$$
\mu_*(\gamma\cap B_\varepsilon)\approx \varepsilon\, |w'(z)|, \quad
{\rm dist}(z,\gamma)\approx \varepsilon
$$
Hence their $n^{\rm th}$ moments behave the same way and the
multifractal exponents $\tau_n$ may be expressed as
$$\tau_n=x_n+n-d_\kappa,$$
where $x_n$ denote the exponents of the conformal map, ${\bf
  E}[|w'(z)|^n]\approx \varepsilon^{x_n}$ for $z$ at distance
$\varepsilon$ from the boundary, and
where the extra term $-d_\kappa$ comes from the sum over the balls
in the definition of $\mathcal{Z}_n$. The exponents $x_n$ are estimated
\cite{Wiegmann05} by using the trick of statistical martingales starting
from the CFT correlation function 
$$
\langle\cdots\Phi_h(z,\bar z) \Phi_{0;1}(z_0,\bar z_0)
\psi_{1;2}(x_0)\rangle_\mathbb{D}
$$
Computing this correlation function in the fixed domain
$\mathbb{D}$ and in its average in the deformed domain $\mathbb{D}_t$
with the hull removed yields informations on expectations of
(derivative of) the uniformizing maps.  The boundary operator
$\psi_{1;2}$ creates the boundary curve at point $x_0$, the bulk
operator $\Phi_{0;1}(z_0,\bar z_0)$ conditions on the presence of the
curve in the neighborhood of point $z_0$ and the operator
$\Phi_h(z,\bar z)$ of dimension $2h$ is there to test scaling
properties of the uniformizing map at point $z$. It is clear that the
properties of the uniformizing map for $z$ close to the boundary are
then coded into the operator product expansions of these operators.
Details involve the Coulomb gas technique and lead to the formula
$$
x_n = \frac{n}{2} - \frac{\kappa+4}{8}\, \Delta_n $$
with
$\Delta_n$ given by the KPZ formula:
$\Delta_n\left({\Delta_n-\frac{\kappa-4}{\kappa}}\right)=
\frac{4}{\kappa} n$. This coincides with eq.(\ref{multiexpo}). It
would be very interesting to have a more complete proof.

\subsection{The Brownian loop soup}
\label{sec:browsoup}

The Brownian loop soup has been introduced in \cite{LWsoup03}, see
also the book \cite{Lawlerbook05}. It is a Poisson realization from a
measure on unrooted loops, so that it describes an ensemble -- ie. a
soup -- of possibly overlapping closed curves -- ie. loops. This
measure, which is constructed using the 2d Brownian motion, satisfies
both conformal invariance and the restriction property. The Brownian
loop soup has deep relations with SLEs and with restriction measures.
Boundaries of clusters made by the loops of the soup have recently
been conjectured to be of the SLE type \cite{Wernersoup03}.

We first start by describing loops, random walk loops, and the loop
soup in a discrete setting. This was considered in
\cite{LawlerTrujillo04}.  Our approach is slightly different as we
emphasize the connection with statistical mechanics. We then go to the
continuum and present elements of the Brownian loop soup.

\subsubsection{Discrete considerations on loops}
\label{sec:discretloops}

The discrete counterpart of the Brownian loop soup is interesting in
its own sake, and leads naturally to a statistical mechanics
framework. So we present this case in relative detail. The (formal)
adaptation to the Brownian loop soup is then straightforward.

If $V$ is a set, which can be assumed for simplicity reasons to be
finite or countable, a weighted graph on $V$ is a map $A$ from $V
\times V$ to ${\mathbb R}^+$. We use a matrix notation $A_{v,v'}$
instead of $A(v,v')$ because matrix products appear naturally in what
follows. If $A_{v,v'}>0$ we say that there is an edge from $v$ to
$v'$, and that this edge carries weight $A_{v,v'}$. This defines an
oriented graph $G_A$ associated to $A$ and explains the name weighted
graph.

Two canonical examples arise starting from a simple graph $G$ with
vertex set $V$. The first one is when $A$ is the adjacency matrix of
$G$, with $A_{v,v'}=1$ or $A_{v,v'}=0$ depending whether $\{v,v'\}$ is
an edge of $G$ or not. The simplicity of $G$ means that $A$ has zeroes
on the diagonal. This case is related to path counting with the
uniform measure.  The second one is when $2A-I$ is the discrete
Laplace operator on $G$, i.e. $A_{v,v'}=1/d$ if $v$ has $d$ neighbors
in $G$ and $v'$ is one of those, but $A_{v,v'}=0$ in all other cases.
This case is related to random walks on $G$. In the applications we
have in mind, $G$ will be the graph of a regular lattice, and the two
cases differ only by a normalization.

Fix a weighted graph $A$ with associated graph $G_A$. A loop in $G$ of
length $n\geq 1$ is a sequence $(v_0,v_1,\cdots,v_{n-1})$ of vertices
such that 
\[ A_{v_0,v_1}A_{v_1,v_2} \cdots
A_{v_{n-2},v_{n-1}}A_{v_{n-1},v_0}>0,\] 
i.e. such that the $n$
edges are present in $G_A$. The space of loops of length $n$ is
denoted by ${\mathcal P}_n$, and ${\mathcal P}\equiv \cup_{n\geq 1}
{\mathcal P}_n$ (a disjoint union) is the space of (rooted) loops. 

The cyclic group of order $n$, ${\mathbb Z}/n{\mathbb Z}$ acts
naturally on loops of length $n$, permuting cyclically
$(v_0,v_1,\cdots,v_{n-1})$. An equivalence class of loops of length
$n$ under this action is called an unrooted loop of length $n$. The
space of unrooted loops of length $n$ is denoted by ${\mathcal L}_n
\equiv {\mathcal P}_n/({\mathbb Z}/n{\mathbb Z})$, and $ {\mathcal
  L}\equiv \cup_{n\geq 1} {\mathcal L}_n$ (a disjoint union) is the
space of unrooted. We denote by $||L||$ the dimension of the class
$L$.  By construction $||L||=n/|{\rm Aut}L|$ with $|{\rm Aut}L|$ the
order of the subgroup ${\rm Aut}L$ of $\mathbb{Z}/n\mathbb{Z}$ fixing
any representative $P$ of the class $L$. We call ${\rm Aut}L$ the
automorphism group of $L$.

Fix two positive numbers $\alpha$, $\lambda$ and,
if $P\in {\mathcal P}_n$ is the rooted loop $(v_0,v_1,\cdots,v_{n-1})$,
define the weight of $P$ to be 
$$
w(P)= \lambda \alpha^n\, A_{v_0,v_1}\cdots A_{v_{n-1},v_0}.
$$
By averaging this induces a measure on unrooted loops $L\in \mathcal{L}_n$ via
$w(L)\equiv \frac{1}{n}\sum_{P\in L} w(P)$ where the sum is on loops
belonging to the equivalence class $L$. Explicitly, if $L\in {\mathcal
  L}_n$ is the class of the loop $(v_0,v_1,\cdots,v_{n-1})$ its weight
is
\begin{eqnarray}
w(L)\equiv  \frac{\lambda\alpha^n}{|{\rm Aut}L|}\ 
A_{v_0,v_1}A_{v_1,v_2} \cdots A_{v_{n-2},v_{n-1}}A_{v_{n-1},v_0}
\label{wpourL}
\end{eqnarray}
It is indeed independent of the loop representing $L$. We use this
weight to define a (positive) measures $\nu$ on ${\mathcal L}_n$,
hence on the disjoint union $ {\mathcal L}\equiv \cup_{n\geq 1}
{\mathcal L}_n$ in an obvious way as follows. Namely, if $B\equiv
(B_1,B_2,\cdots) \subset {\mathcal L}$ then
$$
\nu(B)\equiv \sum_{n\geq 1} \sum_{L_n\in B_n} w(L_n),
$$
a possibly infinite number.  Observe that 
$$
\sum_{L\in{\mathcal L}_n}w(L)
= \frac{\lambda\alpha^n}{n}\, {\rm Tr}\, A^n,
$$
as can be seen by writing the sum over $L\in{\mathcal L}_n$ as a
sum over $P\in{\mathcal P}_n$ and noting that the $|{\rm Aut}L|$ factor is
just what is needed to get a free sum over sequences of vertices
$(v_0,\cdots,v_{n-1})$.  In the language of statistical mechanics,
$\alpha$ is a fugacity.

\subsubsection{Discrete considerations on the loop soup}
\label{sec:disloopsoup}
We can go one step further, from a measure on unrooted loops to a
measure on the unrooted loop soup, see ref.\cite{LawlerTrujillo04}. The
unrooted loop soup is simply the set of maps from ${\mathcal L}$ to
$\mathbb N$, or equivalently the set of formal linear combinations
${\mathbf m} \equiv \sum_{L\in{\mathcal L}} m_L L$ of unrooted loops
with (non-negative) integer coefficients. We extend the definition of
the weight $w$ to the loop soup by
$$
w({\mathbf m})\equiv \prod_{L\in{\mathcal L}}
\frac{w(L)^{m_L}}{m_L!}.
$$
Indeed, if all $m_L$'s vanish but for one which is equal to one,
the weight of the corresponding soup is the weight of its single
component.

A formal manipulation which could be
made rigorous for instance by taking $V$ finite and $\alpha$ small
enough shows that the partition function $Z\equiv\sum_{\mathbf
  m}w({\mathbf m})=e^{\sum_{L\in{\mathcal L}}w(L)}$. But
$\sum_{L\in{\mathcal L}}w(L)=\sum_{n\geq 1} \frac{\lambda}{n}\alpha^n Tr
\; A^n$ so that 
$$
Z=\left[\ {\rm det} \; (1-\alpha A)\ \right]^{-\lambda}.
$$

In the same spirit, if $z_L$ are formal variables, an immediate
computation shows that 
$$
{\bf E}[\prod_{L\in{\mathcal L}}z_L^{m_L}]
=\prod_{L\in{\mathcal L}}e^{(z_L-1)w(L)},
$$
saying that the $m_L$'s are independent Poisson variables with mean
$w(L)$. It is immediate that the random variables $N_B\equiv
\sum_{L\in B}m_L $ for $B \subset {\mathcal L}$ such that $\nu
(B)<+\infty$ define the Poisson random measure on ${\mathcal L}$
associated to the measure $\nu$. Namely, $N_B$ counts the number of
loops in the set $B$ and their characteristic functions
$$
{\bf E}[ z^{N_B}] = \exp[(z-1)\nu(B)]
$$
are those of Poisson variables with mean $\nu(B)$. This explains
the name Poisson soup of loops, as defined in the Appendix
\ref{app:proba}. Furthermore, given any function $L\to C_L$ on the
unrooted loop space $\mathcal{L}$, we have:
$$
{\bf E}[\sum_{L\in\mathcal{L}}m_L\, C_L] =\sum_n \frac{1}{n}
\sum_{P\in\mathcal{P}_n} w(P)\, C_P
$$
where the function $C_P= C_L$ if $P$ belongs to the class
$L$. As before, this can be checked by writing the sum over
$L\in{\mathcal L}_n$ as a sum over $P\in{\mathcal P}_n$ and using
that the $|{\rm Aut}L|$ factor is just what is needed to get a free
sum over rooted loops.

 The operator $\Gamma \equiv (1-\alpha A)^{-1}$ plays an important
role. It deserves the name of Green function or propagator: think of
the case when $2A-I$ is the discrete Laplace operator on the
intersection of a domain and a regular lattice of mesh $\varepsilon$,
and adjust $\alpha(\varepsilon)-2 =O(\varepsilon^2)$ to recover the
free (massive) Euclidean propagator of field theory. 

Further interesting observables can be evaluated. If $L\in{\mathcal
  L}$ has a representative $(v_0,v_1,\cdots,v_{n-1})$, and $v\in V$
define $c_L(v)=\delta_{v,v_0}+\cdots+\delta_{v,v_{n-1}}$, which is
simply the number of occurrences of $v$ in $L$. If $v_1,\cdots,v_k$,
$k\geq2$, are distinct vertices, the observable $N(v_1,\cdots,v_k)\equiv
\sum_{L\in{\mathcal L}} m_L \prod_{j=1}^k c_L(v_j)$ counts the
occupation of vertices $v_1,\cdots,v_k$ in the soup.  A tedious
computation shows that
\[{\bf E}[N(v_1,\cdots,v_k) ] =\lambda
\sum_{v'_1,\cdots,v'_k} \Gamma_{v'_1,v'_2}\cdots  \Gamma_{v'_{k-1},v'_k}
 \Gamma_{v'_k,v'_1},\] 
where the sum is over all cyclic orderings of
$v'_1,\cdots,v'_k$ of $v_1,\cdots,v_k$.  
These sample computations should be enough to convince the reader that
the Poisson loop soup is a probabilistic construction which shows some
analogies with free field theory without being identical to it.

Let us conclude these elementary constructions by mentioning another
important loop soup. We can start just from a graph $G$ and replace
the data of a weighted graph given by $A$ by something else. We
construct as before the space of unrooted self avoiding loops on $G$
(loops with $v_0,v_1,\cdots,v_{n-1}$ all distinct), and define the
self avoiding unrooted loop measure $\nu^{sa}$ and the corresponding
self avoiding loop soup by taking each unrooted loop of length $n$ to
have weight $\lambda \alpha^l$. Making explicit computations with the
self avoiding loop soup is a very difficult open problem.

\subsubsection{Brownian loops}
\label{sec:loopbrown}

We want to describe measures on continuous closed curves in the plane,
see ref.\cite{Lawlerbook05}.  We want these curves to look locally like
Brownian motions, and these come with a preferred time
parameterization that can be recovered by the quadratic
variation\footnote{It may be useful at this point and for the rest of
  this Section to have a glance at the paragraphs on Brownian motion
  in Appendix \ref{app:proba}}. So we can view each sample of the
closed curves we are interested in as parameterized and having a well
defined time duration. This leads to represent closed curves as
periodic functions from $\mathbb R$ to $\mathbb C$, the period
depending on the curve. Remember that a non constant periodic function
$\gamma$ has a well defined minimal period $t_{\gamma} \in \;
]0,+\infty[$, which we call ``the'' period of the function.

With these intuitive considerations in mind, let $\mathcal P$ denote
the space of continuous non constant periodic functions from $\mathbb
R$ to $\mathbb C$ (we hope that the use of the same notation as in the
discrete case will help the reader and not confuse him). If $\mathbb
D$ is a domain in $\mathbb C$, let ${\mathcal P}_{\mathbb D}$ denote
the subspace $\{\gamma \in {\mathcal P}\; | \; \gamma ({\mathbb
  R})\subset {\mathbb D}\}$ of continuous non constant periodic
functions from $\mathbb R$ to $\mathbb D$.

We define a measure on $\mathcal P$ by giving finite dimensional
distributions. Let $K(z,t)\equiv \frac{1}{2\pi t}
\exp({-\frac{|z|^2}{2t}})$ denote the heat kernel in $\mathbb C$. Fix
$n\geq 1$.  If $0\leq u_1< \cdots < u_n<1$ and $z_1,\cdots,z_n \in
{\mathbb C}$ set $u_{n+1}\equiv 1+u_1$, $z_{n+1}\equiv z_1$ and define
\[ 
K_n(t,z_1,u_1,\cdots,z_n,u_n)\equiv K(z_2-z_1,t(u_2-u_1))\cdots 
K(z_{n+1}-z_{n},t(u_{n+1}-u_{n})).\]
Let $A$ be a Borel subset on $]0,+\infty[$, $B_1,\cdots,B_n$ be
Borel subsets of $\mathbb C$ and define the measure of the (cylinder)
subset $C(A,u_1,B_1,\cdots,u_n,B_n)$ of
$\mathcal P$
consisting of curves $\gamma$ whose period $t_{\gamma}$ belongs to $A$
and such that $\gamma(tu_i)$ belongs to $B_i$ for $i=1,\cdots,n$ to be
\begin{eqnarray*}
\mu(C(A,u_1,B_1,\cdots,u_n,B_n))&\equiv & \\
& & \hskip -2.0 truecm
\int_A \frac{dt}{t} \int_{B_1}
d^2 z_1 \cdots \int_{B_n} d^2 z_n\, K_n(t,z_1,u_1,\cdots,z_n,u_n).
\end{eqnarray*}
Here, no constraint has been imposed to the root $z_0$ of loops so
that we have integrated $z_0$ over the complex plane.
Note that $\mu$ on cylinder sets takes values in $[0,+\infty]$ for
$n\geq 0$.  As for the standard definition of Brownian motion via
cylinder sets, the Kolmogorov consistency condition is satisfied (i.e.
if $n\geq 2$ and $B_{i}={\mathbb R}$ for some $i$, the measure of
$C(A,u_1,B_1,\cdots,u_n,B_n)$ where $u_i$ and $B_i$ are omitted is
recovered) so that $\mu$ on $\mathcal P$ extends to the $\sigma$-algebra
generated by all cylinder sets.  We can restrict $\mu$ to a measure
$\mu_{\mathbb D}$ on ${\mathcal P}_{\mathbb D}$. For an appropriate
topology, a curve close to a curve contained in $\mathbb D$ is itself
in $\mathbb D$, so ${\mathcal P}_{\mathbb D}$ is large enough to
ensure that $\mu_{\mathbb D}$ is nontrivial. By construction,
$\mu_{\mathbb D}$ has the restriction property i.e. if ${\mathbb D}'
\subset {\mathbb D}$, $\mu_{{\mathbb D}'}$ is the restriction of
$\mu_{\mathbb D}$ to those loops that are in ${\mathbb D}'$.

As our aim is to produce locally Brownian loops, the formula for
$\mu$ needs little explanation. It mimics closely the definition of
the measure of cylinder sets for Brownian motion, and the measure on
periods $dt/t$ is reminiscent of what has been done in the discrete
setting above : we are defining a measure on rooted loops. Note that
the loops with a nontrivial automorphism group can be shown to have
measure $0$ and no symmetry factor correction needs to be taken into
account. One can decompose $\mathcal P$ according to periods as a
disjoint union ${\mathcal P} = \cup_{t>0} {\mathcal P}_t$. There is a
circle action on ${\mathcal P}_t$ : if $r\in S^1 \equiv {\mathbb
  R}/{\mathbb Z}$ and $\gamma \in {\mathcal P}_t$ the translated loop
$ \gamma$ is defined by $^r \gamma(tu)\equiv \gamma(t(u+r)$. We can
consider ${\mathcal L}_t \equiv {\mathcal P}_t/ S^1$ and the space of
unrooted loops ${\mathcal L} \equiv \cup_{t>0} {\mathcal L}_t$. The
image of the measure $\mu$ on $\mathcal P$ is a measure $\nu$ on
${\mathcal L}$ for which the periods are integrated with the (uniform,
counting) measure $dt$.  This amounts to average uniformly on the
starting point of the loop. For instance the measure of the set
of unrooted loops with period $t$ belonging to $A$ and visiting the
balls $B_i$ and spending time $tu_i$ between the visit of $B_{i}$
and $B_{i+1}$ is:
$$
\int_A\frac{dt}{t} \int_{B_1} d^2z_1\cdots \int_{B_n}d^2z_n\
K(z_2-z_1,tu_1)\cdots K(z_{n+1}-z_n,tu_{n})
$$
with $z_{n+1}=z_1$ and $\sum_ku_k=1$.  The measure $\nu$ restricts
to measures $\nu_{\mathbb D}$ which again have the restriction
property.

The factor $1/t$ has another nice interpretation. It
ensures that if $p \neq 0$ and $q$ are two complex numbers
\[\mu(C(|p|^2A,u_1,pB_1+q,\cdots,u_n,pB_n+q))= 
\mu(C(A,u_1,B_1,\cdots,u_n,B_n))\] 
as a direct substitution in the definition shows. Hence the measure on
$\mathcal P$ is invariant under similarities, and so is its
image $\nu$ on $\mathcal L$. But there is more.

In fact, the measure $\nu$ on ${\mathcal L}$ is even conformally
invariant but the measure $\mu$ on $\mathcal P$ is not. Inspired by
the case of Brownian motion, we define an action of conformal
transformations on ${\mathcal P}$. If $\gamma \in {\mathcal
  P}_{\mathbb D}$ and $f$ is a conformal (one to one) map from
${\mathbb D}$ to $\tilde{\mathbb D}$, set $\tilde{s}\equiv \int_{0}^s
|f'(\gamma(u)|^2du$ (so that $\tilde{s}$ is an increasing continuous
function of $s$) and $\tilde{\gamma}(\tilde{s})=f(\gamma(s))$. This
gives a bijection from ${\mathcal P}_{\mathbb D}$ to ${\mathcal
  P}_{\tilde{\mathbb D}}$ which changes the period from $t$ to
$\tilde{t}=\int_{0}^t |f'(\gamma(u)|^2du$. To see the effect on $\mu$,
split it as $\mu=({dt}/{t})\,\mu_t$ where $\mu_t$ is a measure on
loops of period $t$.  As recalled in Appendix \ref{app:proba}, $2d$
Brownian motion is conformally invariant, and the same argument
applies to loops. If $\gamma$ of period $t$ is a $\mu_t$ sample in
${\mathcal P}_{\mathbb D}$, $\tilde{\gamma}$ is a $\mu_{\tilde{t}}$
sample in ${\mathcal P}_{\tilde{\mathbb D}}$. Hence the sole
discrepancy between $\mu$ on ${\mathcal P}_{\tilde{\mathbb D}}$ and
the image of $\mu$ by $f$ from ${\mathcal P}_{\mathbb D}$ comes from
the variation of $({dt}/{t})$.  Now $({d\tilde{t}}/{\tilde{t}})=
|f'(\gamma(t)|^2({dt}/{\tilde{t}})$ which is not $({dt}/{t})$.
Observe that $({dt}/{\tilde{t}})$ does not depends on the starting
point of $\gamma$, but $|f'(\gamma(t)|^2=|f'(\gamma(0)|^2$ does: for
$\gamma$ rotated to start at $u$, we would get $|f'(\gamma(u)|^2$. But
the measure $\nu$ on the unrooted loop space ${\mathcal L}$ is
unchanged because we have to average over the starting point $u\in
[0,t]$ with the uniform measure $({du}/{t})$, and this reconstructs
$({dt}/{t})$.  Hence $\nu$ is conformally invariant despite the fact
that $\mu$ is not.

One remark on notations is in order : fixing $t$, taking $u_1=0$ and
setting $z\equiv z_1$, the factor $\int_{B_2} d^2 z_2 \cdots
\int_{B_n} d^2 z_n K_n(t,z_1,u_1,\cdots,z_n,u_n)$ can be used to
define a measure on loops starting at $z$. The mass of this measure is
seen to be $1/ 2\pi t$ by taking $n=1$. Normalized to become a
probability, this defines the Brownian bridge starting at $z$. If the
Brownian bridge probability is used instead of our unnormalized
measure in the definition of the measure of Brownian loops, an
additional $1/ 2\pi t$ appears in the formulae, as in most
presentations by probabilist \cite{LWsoup03,Lawlerbook05}.

\subsubsection{The Poisson soup of Brownian loops}
\label{sec:souploops}
Now that the measure $\nu$ on ${\mathcal L}$ is defined, we can
introduce \cite{LWsoup03} a new non-negative parameter $\lambda$ and
define the Poisson soup of Brownian loop (in short the loop soup $LS$)
of intensity $\lambda$ as the random Poisson measure associated to the
measure $\lambda\nu$.  Recall that this amounts to demand that the
number of unrooted loops belonging to some given subset of
$B\subset{\mathcal L}$ are Poisson variables with mean
$\lambda\nu(B)$. See Appendix \ref{app:proba} for more details.  The
loop soup inherits naturally the restriction property and conformal
invariance from the analogous properties of $\nu$. These properties
make it a very important tool in the study of SLE. Indeed, SLE$_\kappa$
for $\kappa \leq 8/3$ has the loop soup of intensity $\lambda(\kappa)
=\frac{(8-3\kappa)(6-\kappa)}{2\kappa}$ as a natural companion since
the intensity compensate the central charge
$\lambda(\kappa)+c(\kappa)=0$.  We refer the reader to the literature
\cite{LWsoup03} for details and mention only a few examples.  A
complete understanding of the loop soup from a CFT view point is still
missing, but the identity $\lambda+c=0$ suggests a deep relationship.

As explained in Section \ref{sec:examples}, (chordal) loop-erased
random walks are obtained by erasing loop as they appear on a $2d$
random walk which is just the simple symmetric random walk along the
horizontal axis, but is an excursion along the vertical axis (i.e. a
simple symmetric random walk conditioned to reach an arbitrarily high
altitude before going back to the origin, this is also a discrete
analog of the $3d$ Bessel process). In the continuum, such a
systematic removal of loops is impossible, because Brownian motion
makes loops at all scales. But one can work the other way round
\cite{LWsoup03}.  The loop soup companion of SLE$_2$, known to describe
the continuum limit of loop erased random walks, is LS$_2$. One can
show that if one takes an LS$_2$ sample and attaches its loops to a
growing SLE$_2$ sample when they are hit, one gets a sample of a $2d$
Brownian motion conditioned to reach an arbitrarily high altitude
before coming back to the horizontal axis.

For a general $\kappa \leq 8/3$, one can do the analogous
construction. Take LS$_{\lambda(\kappa)}$ sample, fill in the loops and
attach them to a growing SLE$_\kappa$ sample if they touch it. One
can show \cite{LSWConfRest02} that the resulting hull is a sample of
the restriction measure with parameter
$\alpha(\kappa)=(6-\kappa)/2\kappa$.

%% file: chap6.tex

This Section is devoted to illustrate possible computations with SLE
-- as such it is probably the most technical part of this review.
They deal with boundary or bulk properties giving informations on
geometrical properties of the SLE hulls or traces.  We shall only
present samples of such computations -- as there is almost no limits
to possible computations. We have selected those which yield to
concrete results and which enlighten the relation between SLE and CFT.
Miscellaneous results obtained via SLE are described at the end of
this Section.

In CFT literature the conformal operators conditioning on the presence
of SLE curves in the neighborhood of their insertion point have been
identified some time ago using Coulomb gas technique
\cite{Nienhuis83,Nienhuis87}.  These computations give confirmations
of these identifications. The nature of these operators depends
whether the point is on the boundary or in the bulk of the domain:

\begin{itemize}
\item The operators `creating' $n$ SLE traces at a boundary point are
  the boundary operators $\psi_{1;n+1}$ with dimension
  $$h_{1;n+1}=n(4+2n-\kappa)/2\kappa$$
  
\item The operators `creating' $n$ SLE curves at a bulk point are the
  bulk operators $\Phi_{0;n/2}$ with dimension
  $$2h_{0;n/2}=[4n^2-(\kappa-4)^2]/16\kappa$$ 
\end{itemize}

In Section \ref{sec:cftofsle}, we already identified the boundary
operator $\psi_{1;2}$ as creating one SLE curve on the boundary. This
is the operator used in constructing the statistical mechanics
martingales.  To check that the operator $\psi_{1;3}$ creates two
curves from the boundary we shall compute the density probability for
the SLE curves to touch the real axis, since one curve touching the
real axis looks locally as two curves merging from it.  This
probability vanishes for $\kappa<4$. For $4<\kappa<8$, it is
proportional to $(dx/|x|)^{(8-\kappa)/\kappa}$. Its scaling is indeed
compatible with the dimension $h_{1;3}=(8-\kappa)/\kappa$. For
$\kappa>8$ the curve is space filling so that it covers the real axis.
More generally we shall compute the hitting probability that the SLE
curves visit an interval on the real axis and show how it is related
to CFT correlation functions. Identification of the other operators
follow by recursively fusing them as CFT fusions correspond to the
merging of the SLE traces.

In Section \ref{sec:sleetc} on radial SLE we already identified the
bulk operator $\Phi_{0;1/2}$ as conditioning on the presence of a SLE
trace at its insertion point. Since by cutting it an SLE trace passing
through a neighborhood of a point may be viewed as two SLE traces
emerging from it, the operator $\Phi_{0;1}$ may be viewed either as
creating two SLE curves or as forcing one SLE curve to pass in the
neighborhood of a point.  The computation of the fractal dimension of
the SLE curves, which is based on computing the probability for a SLE
curve to pass in the neighborhood of a bulk point (see below), will
confirm this identification.  The fractal dimension of the SLE curve
is linked to the dimension of this operator via $d_\kappa=2-2h_{0;1}$.
More generally, since the operator $\Phi_{0;n/2}$ are those
conditioning on the presence of $n$ SLE traces emerging from the
neighborhood of their bulk insertion point, the dimension of the set
of points with $n$ curves emerging from their neighborhood is
$d_\kappa(n)=2-2h_{0;n/2}$ that is:
\begin{eqnarray}
d_\kappa(n)=[(\kappa+4)^2-4n^2]/8\kappa.
\label{fracdims}
\end{eqnarray}
By Kolmogorov's $0/1$ law, these points exist almost surely if
$d_\kappa(n)>0$ but do not if $d_\kappa(n)<0$.  These two cases
correspond to whether the operator $\Phi_{0;n/2}$ is relevant or not.
The dimension $d_\kappa(2)=1+\kappa/8$, which is the fractal dimension
of the curve, is of course positive -- so the curve exists.  The
dimension $d_\kappa(4)=[(\kappa+4)^2-64]/8\kappa$ is the dimension of
the set of double points of the SLE curves -- since four traces emerge
from a neighborhood. It is negative for $\kappa<4$ but positive for
$\kappa>4$, so that with probability one double points exist for
$\kappa>4$ but do not for $\kappa<4$. This is in accordance with the
different phases of the SLE traces discussed in Section
\ref{sec:basics}.  The dimension $d_\kappa(6)$ is positive for
$\kappa>8$, that is in the phase in which the SLE trace is
space-filling.

Another set of important results are crossing formulas.  The most
famous is that of Cardy \cite{Cardy92} which gives the probability that
there exists a percolating cluster in critical percolation connecting
to opposite sides of a rectangle. Its original derivation involves the
relation between percolation and the Q-states Potts models in the
limit $Q\to 0$ and Cardy's intuition on boundary conformal field
theories. It was motivated by numerical computations of crossing
probabilities in critical percolation done by Langlands et al
\cite{Langlands94}. The latter did play an important role because they
exhibit explicit and concrete manifestations of conformal invariance
in two dimensional critical percolation.

In its original formulation \cite{Cardy92}, Cardy's formula was written
in terms of hypergeometric function, because it was related to CFT
correlation functions in the upper half plane and then transported to
the rectangle by conformal invariance. It gives the probability
$\pi_v$ that there is a percolating cluster from the top to the bottom
of a rectangle with aspect ratio $r$ -- equal to the height over the
width:
\begin{eqnarray}
\pi_v = \frac{3\Gamma(2/3)}{\Gamma(1/3)^2}\, \eta^{1/3}
\, _2F_1(1/3,2/3,4/3;\eta)
\label{cardyrect}
\end{eqnarray}
with ${}_2F_1$ the hypergeometric functions and $\eta=[(1-k)/(1+k)]^2$
for an aspect ratio $r=K(1-k^2)/[2K(k^2)]$ where $K(u)$ is the
complete elliptic integral of the first kind. This formula did agree
very well the numerical data of \cite{Langlands94}.

\begin{figure}[htbp]
\begin{center}
\includegraphics[width=0.5\textwidth]{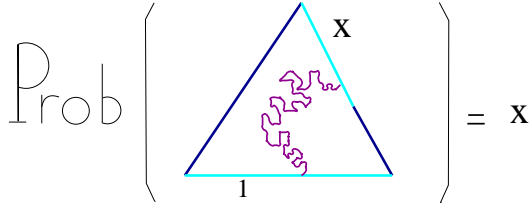}
\caption{Crossing probability in a equilateral triangle.}
     \label{fig:slesmirnov}
 \end{center}
\end{figure}

It was later realized by Carleson (unpublished) that the formula
becomes simpler if we look at it in a triangle instead in a rectangle.
So let us consider site percolation in a triangular lattice with sites
colored in black/white with probability 1/2 to ensure criticality.
SLE and CFT deal with universal properties valid in the continuum
limit in which the mesh of the lattice goes to zero. Formulas simplify
if we look at percolation in an equilateral triangle with sides of
length $1$ and with corners at positions $0$, $1$ and $e^{i\pi/3}$.
Assume boundary conditions such that all sites on the bottom side
$[0,1]$ of the triangle are black. Cardy's formula is for the
probability for the existence of clusters of black sites connecting
the bottom side to a segment attached to the opposite corner
$e^{i\pi/3}$ and of length $x<1$, see Fig.\ref{fig:slesmirnov}. In
this setting it becomes particularly simple as it is linear in $x$:
$$
{\bf P}[\ {\rm crossing~ in~a~triangle}\ ] = x 
$$
This was proved by Smirnov \cite{Smirnov} by actually considering a
generalization of this probability involving a point in the bulk and
not only on the boundary. Namely, with identical boundary conditions
and given a point $z$ inside the triangle, he looked for the
probability that there exists a path, included in a cluster of black
sites, connecting the sides $[0,1]$ and $[1,e^{i\pi/3}]$ and leaving
the point $z$ on its right. Smirnov's proof is based on the
observation that this probability is harmonic as a function of $z$.
Namely:
$$
{\bf P}[\ {\rm cluster~ on~the~left~of}~z\ ]= -\frac{2}{\sqrt{3}}\,
\Im{\rm m}(z\,e^{-i\pi/3})
$$
With $z$ approaching the boundary it reduces to Cardy's formula.

Below, we present the derivation of Cardy's formula using chordal SLE
and that of Smirnov formula using dipolar SLE. However, we start with
computations of hitting probabilities in order to exemplify the
techniques. All these results, which require computing specific
probabilities, can be found either using probabilistic or conformal
field theory arguments.  As usual with Markov processes, most of these
probabilities may be computed by identifying the appropriate
martingales. Since the SLE/CFT correspondence of Section
\ref{sec:cftofsle} shows that SLE martingales are provided by ratios
of CFT correlation functions, all these SLE probabilities can be
expressed in terms ratio of CFT correlation functions.  We shall
alternatively use chordal, radial or dipolar SLEs to illustrate these
facts.

\subsection{Boundary hitting probabilities}
\label{sec:hitproba}
Boundary hitting probabilities are the probabilities that the SLE
curve visits a set of intervals on the real axis in a given order.  We
are going to show on a simple example how these probabilities are
related to particular CFT correlation functions of boundary primary
fields. These relations reveal connections between topological
properties of SLE paths and fusion algebras and operator production
expansions in conformal field theory.

We consider chordal SLE and we assume $4<\kappa<8$ so that the SLE
trace touches the real axis infinitely many times with probability
one. Consider the probability, first computed in \cite{RohdeSchramm01},
that the SLE curve touches an interval $[x,X]$ on the positive real
axis. We shall compute its complement ${\bf P}[\gamma_{[0;\infty[}\cap
[x,X]=\emptyset]$ that the curve does not touch the interval which is
also equal to the probability that the two points $x$ and $X$ are
swallowed at the same instant. The result is the following expression
with $s=x/X$:
\begin{eqnarray}
{\bf P}[\gamma_{[0;\infty[}\cap [x,X]=\emptyset]=
\frac{s^{\frac{\kappa -4}{\kappa}}\Gamma\left(\frac{4}{\kappa}
\right)}{\Gamma\left(\frac{\kappa-4}{\kappa}\right)\Gamma
\left(\frac{8-\kappa} {\kappa}\right)}\int_0^1 d\sigma \sigma
^{-\frac{4}{\kappa}} (1-s\sigma)^{2\frac{4-\kappa}{\kappa}}.
\label{eq:visit1}
\end{eqnarray}
Its behavior as $X\to x$ gives the probability density for the SLE
trace to touch the real axis in the neighborhood of the point $x$:
\begin{eqnarray}
{\bf P}[\gamma_{[0;\infty[}\cap[x,x+dx]\not=\emptyset] 
\propto \ (dx/|x|)^{{(8-\kappa)}/{\kappa}} \nonumber
\end{eqnarray}
It agrees with the dimension $h_{1;3}=(8-\kappa)/\kappa$ of the
operator coding for two SLE paths emerging from the real axis.

To simplify notation, let $p(x,X)\equiv {\bf
  P}[\gamma_{[0;\infty[}\cap [x,X]=\emptyset]$. By dimensional
analysis it only depends on the ratio $x/X$ so that $p(x,X)=p(x/X,1)$.
This probability is $0$ if $x\to 0$, or $X\to \infty$, and it is $1$
in the limit of coinciding points $X\to x$.

Let us start with the probabilistic argument. It is based on
constructing an appropriate martingale using the Markov property of
chordal SLE. Thus consider the probability that the SLE trace touches
the interval $[x,X]$ conditioned on the knowledge of the SLE trace
$\gamma_{[0;t[}$ up to time $t$ for $t<\tau_x$. As a conditional
probability -- and thus conditional expectation value -- this is
a martingale. Indeed, if we average it over $\gamma_{[0,t[}$ we
reproduce the  probability that the SLE trace touches
the interval $[x,X]$. By Markov property we may start again the SLE
process at time $t$ by erasing the SLE trace using the Loewner
conformal map $h_t=g_t-\xi_t$. By conformal invariance the image of
the rest of the SLE trace is then distributed as the original SLE
trace but the point $x$ and $X$ have been moved to position $h_t(x)$
and $h_t(X)$. Hence this conditional probability is the probability
that the SLE trace touches the interval $[h_t(x),h_t(X)]$:
$$
{\bf P}[\gamma_{[0;\infty[}\cap [x,X]=\emptyset\Big\vert
\gamma_{[0;t[}] =
{\bf P}[\gamma_{[0;\infty[}\cap [h_t(x),h_t(X)]=\emptyset]
$$
Because this conditional probability is a martingale,
the drift term in its It\^o derivative vanishes. It thus satisfies the
second order differential equation:
\begin{eqnarray}  
\left(\frac{2}{x}\partial_x+\frac{2}{X}\partial_X +
\frac{\kappa}{2}(\partial_x+ \partial_X)^2   \right)p(x,X)=0. 
\label{eq:diffP1}
\end{eqnarray}
Since it only depends on $s=x/X$ this translates into:
$$
\left(\frac{d^2}{ds^2}+\left(\frac{4}{\kappa s} +
\frac{2(4-\kappa)}{\kappa(1-s)}\right)\frac{d}{ds}
\right)p(s,1)=0.
$$
The integration is then straightforward with the normalization
$p(s=0,1)=0$ and $p(s=1,1)=1$. It gives the formula
(\ref{eq:visit1}) quoted above. 

To illustrate the SLE/CFT correspondence we now re-derive this
expression using CFT techniques. This is again based on exhibiting the
appropriate martingale but now the latter is expressed in terms of CFT
data~\cite{BB02b}. To prepare for this computation, we study the CFT
correlation function
$$
\bra{\psi_{1;2}} \varphi_{0}(X)\varphi_{0}(x)
\ket{\psi_{1;2}}_\mathbb{H}.
$$
For $4<\kappa <8$ we may choose it such that it vanishes as $x\to
0$ and takes value $1$ at $X\to x$.  Indeed, if $x$ comes close to
$0$, we can expand this function by computing the operator product
expansion of $\varphi_{0}(x)\ket{\psi_{1;2}}$.  This is constrained by
the fusion rules which arise from the null vector
$(4L_{-2}-{\kappa}L_{-1}^2)\ket{\psi_{1;2}}=0$.  It can involve at
most two conformal families of dimension
$h_{1;2}=\frac{6-\kappa}{2\kappa}$ or
$h_{1;0}=\frac{\kappa-2}{\kappa}$.  We demand that only the conformal
family of dimension $h_{1;0}$ appears in the operator product
expansion. Then, $\varphi_{0}(x)\ket{\psi_{1;2}} \sim x ^{\frac{\kappa
    -4}{\kappa}} \ket{\psi_{1;0}},$ with $\ket{\psi_{1;0}}$ the state
created by $\psi_{1;0}(0)$. This goes to $0$ for $\kappa > 4$.  If the
points $x$ and $X$ come close together, the operator product expansion
$\varphi_{0}(X)\varphi_{0}(x)$ is more involved. General rules of
conformal field theory ensure that the identity operator contributes,
but apart from that, there is no a priori restrictions on the
conformal families $\varphi_{\delta}$ that may appear.  However, only
those for which $\bra{\psi_{1;2}} \varphi_{\delta}\ket{\psi_{1;2}}\neq
0$ remain, and this restricts to two conformal families, the identity
and $\psi_{1;3}$. Namely, when $x$ and $X$ come close together,
$$
\bra{\psi_{1;2}} \varphi_{0}(X)\varphi_{0}(x)
\ket{\psi_{1;2}}_\mathbb{H}\simeq
1+ \cdots +\hat C\, (X-x)^{h_{1;3}}\,
\bra{\psi_{1;2}}\psi_{1;3}(x)
\ket{\psi_{1;2}}_\mathbb{H} +\cdots 
$$
with $h_{1;3}=\frac{8-\kappa}{\kappa}$ and $\hat C$ some fusion
coefficient.  The dominant contribution to
$\bra{\psi_{1;2}}\varphi_{0}(X)\varphi_{0}(x)
\ket{\psi_{1;2}}_\mathbb{H}$ is either $1$ or $(X-x)^{h_{1;3}}$,
depending on whether $\kappa<8$ or $\kappa>8$.  Hence, if $4<\kappa
<8$, the correlation function vanishes as $x\to 0$ and takes value $1$
at $X\to x$.

Now, for nonzero $t$, we consider the CFT correlation function which
is a martingale thanks to the statistical martinagle trick:
\begin{eqnarray*}
\bra{\psi_{1;2}} \varphi_{0}(h_t(X))
\varphi_{0}(h_t(x)) \ket{\psi_{1;2}}_\mathbb{H} 
\end{eqnarray*}
If the position $a_x$ of the SLE trace at $t=\tau_x$ satisfies $x< a_x
< X$, then $h_{\tau_x}(X)$ remains away from the origin but $\lim _{t
  \nearrow \tau_x}h_t(x)=0$ and the correlation function vanishes.  On
the other hand, if $X\leq a_x$, it is a general property of hulls that
$\lim _{t \nearrow \tau_x}h_t(x)/h_t(X)=1$ and the correlation
function is unity.  Thus
$$
\lim _{t \nearrow \tau_x} 
\bra{\psi_{1;2}} \varphi_{0}(h_t(X))
\varphi_{0}(h_t(x)) \ket{\psi_{1;2}}_\mathbb{H} 
=\mathbf{1}_{\{\tau_x=\tau_X\}}.
$$
with $\mathbf{1}_{\{\tau_x=\tau_X\}}$ the characteristic function
of the events with $\tau_x=\tau_X$. From the martingale property
extended to the stopping time $\tau_x$, we infer that the expectation
values of this martingale is equal to its values at initial time.
Since ${\bf E}[\mathbf{1}_{\{\tau_x=\tau_X\}}]$ is the probability
that the curve does not touch the interval $[x,X]$ we get:
\begin{eqnarray}
{\bf P}[\gamma_{[0;\infty[}\cap [x,X]=\emptyset]=
\bra{\psi_{1;2}} \varphi_{0}(X)\varphi_{0}(x)
\ket{\psi_{1;2}}_\mathbb{H}.
\label{excur0}
\end{eqnarray}
Furthermore, the fact that
$\bra{\psi_{1;2}}\varphi_{0}(X)\varphi_{0}(x)(-4L_{-2}+
{\kappa}L_{-1}^2)\ket{\psi_{1;2}}=0$ translates into a differential
equation for the correlation function which coincides with
eq.(\ref{eq:diffP1}). See Appendix \ref{app:cft}.  The differential
operator annihilates the constants, a remnant of the fact that the
identity operator has weight 0. With the chosen normalization for
$\varphi_0(x)$, the relevant solution vanishes at the origin. The
integration is then straightforward and it gives the formula
(\ref{eq:visit1}).

This example is instructive, because it clearly shows how the CFT
correlation functions are selected according to the topological
behavior specified by the probabilities one computes. It shows in a
fairly simple case that the thresholds $\kappa=4,8$ for topological
properties of SLE appear in the CFT framework as thresholds at which
divergences emerge in operator product expansions. Probability for
visiting, in a given order, collections of intervals of the real axis
are similarly related to CFT correlation functions with insertion a
boundary operator of dimension zero at each end points of these
intervals. Some of them have been explicitly computed in ref.\cite{BB04}.
However the general rules relating these probabilities to the specific
conformal correlation functions, and thus to the specific intermediate
families, have not been given yet.

\subsection{Cardy's crossing formulas}
\label{sec:cardy}
\begin{figure}[htbp]
\begin{center}
\includegraphics[width=0.8\textwidth]{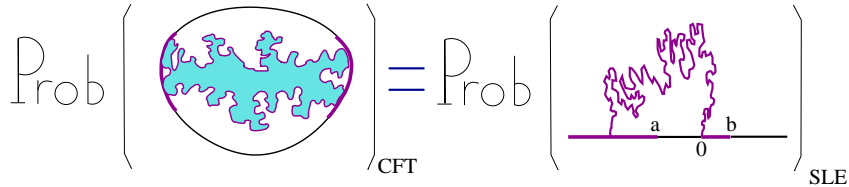}
\caption{Crossing probability is identical 
to SLE swallowing probabilities.}
     \label{fig:crossingcardy}
 \end{center}
\end{figure}

Cardy's formula for critical percolation applies to SLE$_6$.  It may
be extended \cite{LSW1,LSW2,LSW3} to a formula valid for arbitrary 
$\kappa>4$. The problem is then formulated as follows.  Consider
chordal SLE.  Let $-\infty<a<0<b<\infty$ and define the stopping times
$\tau_a$ and $\tau_b$ as the first times at which the SLE trace
touches the interval $(-\infty,a]$ and $[b,+\infty)$ respectively.  By
definition, the crossing probability is the probability that the trace
hits first the interval $(-\infty,a]$ before it hits the interval
$[b,+\infty)$, that is ${\bf P}[\tau_a<\tau_b].$ 
See Fig.\ref{fig:crossingcardy}.

The probabilistic argument is the same as in previous Section.  Let
$p_c(a,b)\equiv {\bf P}[\tau_a<\tau_b]$ be the crossing probability.
By the Markov and identical increment properties of SLE, the process
$t\to p_c(h_t(a),h_t(b))$ is a martingale. Hence the drift term of its
It\^o derivative vanishes. This yields the second order differential
equation:
$$
\left(\frac{2}{a}\partial_a+\frac{2}{b}\partial_b +
\frac{\kappa}{2}(\partial_a+ \partial_b)^2   \right)p(a,b)=0. 
$$
By scaling argument, $p_c(a,b)$ only depends on the ratio $r=-a/b>0$ with
boundary condition $p_c(a=0,b)=1$, since then $a$ is clearly swallowed
first, and $p_c(a,b=0)=0$, since then $b$ is swallowed first. Direct
integration gives:
\begin{eqnarray}
p_c(a,b)=
\frac{\Gamma(2(\kappa-4)/\kappa)}{\Gamma((\kappa-4)/\kappa)^2}\
 \int^\infty_r d\sigma\,
\sigma^{-4/\kappa}(1+\sigma)^{2(4-\kappa)/\kappa}, 
\label{cardyint}
\end{eqnarray}
with $r=-a/b$. This is  proportional to the hypergeometric function:
$$
p_c(a,b)=1-\frac{\kappa r^{1-4/\kappa}}{\kappa-4}\,
_2F_1(\frac{4}{\kappa},1-\frac{4}{\kappa},2-\frac{4}{\kappa},
\frac{r}{1+r})
$$
For $\kappa=6$ it reduces to Cardy's formula \cite{Cardy92}.

The CFT derivation is simpler than the one of previous Section.  As
it starts to become usual it consists in exhibiting the appropriate
CFT martingale. So let us consider the following correlation function
which, by the SLE/CFT correspondence, is a martingale:
$$ 
F(h_t(a)/h_t(b))\equiv 
\bra{\psi_{1;2}}\varphi_{0}(h_t(a))
\varphi_{0}(h_t(b))\ket{\psi_{1;2}}_\mathbb{H}  
$$
with $\varphi_{0}$ a boundary conformal field of scaling
dimension zero. By dimensional analysis it is only a function of the
ratio $h_t(a)/h_t(b)$. There exist actually two linearly 
independent correlators, one of them being constant, but we shall not
specify yet which non constant correlation function we pick.  
$F(a/b)$ can be computed explicitly, in terms of hypergeometric
function, thanks to the null vector equation $(4L_{-2}-\kappa
L_{-1}^2)\ket{\psi_{1;2}}=0$. 

The basic observation is that the ratio $h_t(a)/h_t(b)$ takes two
simple non-random values depending whether $\tau_a<\tau_b$ or not, ie.
depending whether the point $a$ have been surrounded by the SLE trace
first or not. Let $\tau={\rm min}(\tau_a,\tau_b)$. If $\tau_a<\tau_b$,
the point $a$ has been swallowed first so that $h_\tau(a)\simeq 0$
while $h_\tau(b)$ remains finite and thus $\lim_{t \nearrow
  \tau}\,h_t(a)/h_t(b)=0$.  Similarly if $\tau_b<\tau_a$ then
$\lim_{t\nearrow \tau}\,h_t(a)/h_t(b)=\infty$. Thus we have
$$
\lim_{t \nearrow \tau}\, F(h_t(a)/h_t(b))= 
{\bf 1}_{\{\tau_a<\tau_b\}}\, F(0)
+{\bf 1}_{\{\tau_a>\tau_b\}}\, F(\infty).
$$
Making the argument precise require checking that $F(0)$ and
$F(\infty)$ are finite. As in previous Section, this follows from CFT
fusion rules and it is true for $4<\kappa<8$.  We can then compute
$\lim_{t \nearrow \tau}\, {\bf E}[F(h_t(a)/h_t(b)]$ in terms of the
swallowing probability ${\bf P}[\tau_a<\tau_b]$:
$$
\lim_{t \nearrow \tau}\,  {\bf E}[F(h_t(a)/h_t(b)]=
 {\bf P}[\,\tau_a<\tau_b\,]\, F(0)
+ (1-{\bf P}[\,\tau_a<\tau_b]\,)\, F(\infty)
$$
where we used that ${\bf P}[\,\tau_a>\tau_b\,]=1-{\bf
  P}[\,\tau_a<\tau_b\,]$.  Applying the martingale property so that
$\lim_{t \nearrow \tau}\, {\bf E}[\, F(h_t(a)/h_t(b))\,] = F(a/b)$, we
get:
\begin{eqnarray}
{\bf P}[\,\tau_a<\tau_b\,] = 
\frac{F(a/b)-F(\infty)}{F(0)-F(\infty)}
\label{cardyform}
\end{eqnarray}
For $\kappa=6$, this is Cardy's formula \cite{Cardy92}. In this form
the formula is independent of the chosen normalization for $F$, but
it can be further simplified by choosing boundary conditions on $F$
such that $F(0)=1$ and $F(\infty)=0$, which is possible for
$4<\kappa<8$.  It is very reminiscent of the probability for a one
dimension Markov process driven by a stochastic differential equation
to escape on a prescribed side of an interval containing its starting
point.

\subsection{Harmonic probabilities and Smirnov's formula}
\label{sec:smirnov}
Analogues of Smirnov's formula give informations on bulk properties.
For $\kappa>4$ they cannot be found using chordal or radial SLE because
the hulls then invade the full domain. So we have to deal with dipolar
SLE defined say in the strip $\mathbb{S}=\{z,\ 0<\Im{\rm m} z<\pi\}$
with marked point $x_\pm=\pm\infty$ and $x_0=0$.

We shall evaluate the probabilities $P_l(z,\bar z)$ -- resp.
$P_r(z,\bar z)$ -- for a bulk point $z$ not to be swallowed by the SLE
trace and to be on the left -- resp. the right -- of the trace. This
is the probability for the point $z$ to be on the left -- resp. the
right -- of the exterior frontier of the SLE hull viewed from the
boundary point $x_-$ -- resp. $x_+$. It is therefore the probability
for the existence of a path joining $x_-$ -- resp. $x_+$ -- to the
boundary interval $[x_+,x_-]$ leaving the point $z$ on its right --
resp.  left -- and included into one cluster of the underlying model
of statistical mechanics. The result is \cite{BBH04}:
\begin{eqnarray}
P_l(z,\bar z)= 1 - \frac{\Im{\rm m}\, \hat F(z)}{\Im{\rm m}\,
  \hat F(\infty)},\quad
\hat F(z)\equiv \int_{-\infty}^z \frac{du}{(\sinh u/2)^{4/\kappa}}.
 \label{probleft}
\end{eqnarray}
A noticeable property is that it is a harmonic function.
For $\kappa=6$ it reproduces Smirnov's formula \cite{Smirnov}.
At the end of this Section we give another formula for the 
probability for the point $z$ to be in the SLE hull.

The function $\hat F(z)$ has a nice geometrical interpretation: it
uniformizes the strip onto a triangle with corners $\hat
F(-\infty)=0$, $\hat F(+\infty)=e^{-i2\pi/\kappa}I$ and $\hat
F(0)=e^{-i4\pi/\kappa}J$ with $I=\int_{-\infty}^{+\infty} dy(\cosh
y/2)^{-4/\kappa}$ and $J= \int_0^\infty dy (\sinh y/2)^{-4/\kappa}$.
We have $I=2J\cos(2\pi/\kappa)$. The geometry becomes even simpler in
the case $\kappa=6$ because  the triangle is then equilateral. This
explains why Smirnov's formula is simply linear in an equilateral
triangle. More on the relation between SLE$(\kappa,\rho)$ and triangle
may be found in ref.\cite{Dubedat02}.

As usual, a way to compute these probabilities is to notice that the
process $t\to P(h_t(z),\overline{h_t(z)})$ is a local martingale.
Indeed, since $h_s\circ h_t^{-1}$, $s>t$, is independent of $h_t$ and
distributed as $h_{s-t}$, the function $P(h_t(z), \overline{h_t(z)})$
is the wanted probability conditioned on the process up to time $t$
and, as such, it is a martingale.  As a consequence, the drift term in
its It\^o derivative vanishes which implies that $P(z,\bar z)$
satisfies the following differential equation:
\begin{eqnarray}
\kappa \partial_z\bar\partial_{\bar z} P +
(\coth\frac{z}{2} +\frac{\kappa}{2}\partial_z)\partial_z P
+ (\coth\frac{\bar z}{2} +\frac{\kappa}{2}\partial_{\bar z})
\partial_{\bar z} P =0. \nonumber
\end{eqnarray}
Quite remarkably this equation has enough harmonic solutions
to compute $P_l$ and $P_r$.  For the probability to be on
the left of the hull, the boundary conditions are: $P_l(-\infty)=1$,
$P_l(+\infty)=0$ and $P_l(0)=0$.  Similar conditions hold for $P_r$.
That these boundary conditions are enough to specify a unique solution
is due to the fact that this equation is of second order in $\Re {\rm
  e} \; z$ but only first order in $\Im {\rm m} \; z$, so that
boundary conditions at three points are enough to fix the function on
the whole boundary.  They follow by noticing that if point $z$ is
swallowed at time $\tau_z$ then $\lim_{t\nearrow\tau_z}h_t(z)=0$, if
it is not swallowed but is on the left of the trace then $\lim_{t
  \nearrow\infty}h_t(z)=-\infty$, and if it is not swallowed but is on
the right of the trace then $\lim_{t \nearrow\infty}h_t(z)=+\infty$.
These conditions are such that at the stopping $\widehat \tau_z={\rm
  min}(\tau_z,\infty)$ the martingale $P(h_t(z),\overline{h_t(z)})$
projects on the appropriate events.

The solution of the martingale equation satisfying the appropriate
boundary conditions is clearly given by the harmonic function
(\ref{probleft}).  The function $\hat F(z)$ is well-defined and
analytic on the strip $\mathbb{S}$ for all $\kappa$'s. For $\kappa>4$,
$\hat F(z)$ is bounded and has a continuous extension to the closure
of the strip.  As a check one may verify that $P_l(z,\bar z)$ behaves
as expected on the boundary. On the positive real axis, $(\sinh z/2)$
is real and positive so that $P_l(x)=0$ for $x\in \mathbb{R}_+$, in
accordance with the fact that no point on the real axis can be on the
left of the trace. On the negative real axis, $(\sinh x/2)^{4/\kappa}=
e^{i4\pi/\kappa}(\sinh|x|/2)^{4/\kappa}$ and
\begin{eqnarray*}
P_l(x)= 1 - \frac{1}{J}\int_{|x|}^{+\infty}
\frac{dy}{(\sinh y/2)^{4/\kappa}},\quad x\in \mathbb{R}_-,
\end{eqnarray*}
It gives the probability of the hull not to spread further than $x$ on
the negative real axis.  On the upper boundary,
\begin{eqnarray}
P_l(z=i\pi+x)= 1 - \frac{1}{I}\int_{-\infty}^x
\frac{dy}{(\cosh y/2)^{4/\kappa}}, \quad z\in i\pi+\mathbb{R},
\label{eq:dipupper}
\end{eqnarray}
since there $(\sinh z/2)^{4/\kappa}=e^{2i\pi/\kappa}(\cosh
x/2)^{4/\kappa}$.  This yields the density probability for the trace
to stops on an interval $[x,x+dx]$ on the upper boundary.  

For $\kappa=4$ the SLE trace is a simple curve so that no point
away from the curve is swallowed in a finite time.  This case is
marginal in the sense that the integral defining $\hat F(z)$ is only
logarithmically divergent. By extension, we have:
\begin{eqnarray}
P_l(z,\bar z)= \frac{1}{\pi} \Im {\rm m}\, 
\left[\log(\tanh \frac{z}{4})\right].
\label{probk=4}
\end{eqnarray}
This satisfies the martingale equation for $\kappa=4$ and the
appropriate boundary conditions: $P_l(x\in\mathbb{R}_+)=0$ and
$P_l(x\in\mathbb{R}_-)=1$. Contrary to the cases $\kappa >4$, it is
discontinuous at the origin.  On the upper boundary the distribution
of the trace is given by:
$$ P_l(i\pi +x)= 1 -\frac{2}{\pi} \arctan(e^{x/2}),\quad
x\in\mathbb{R}.$$

The probability (\ref{probk=4}) possesses a nice free field CFT
interpretation. For $\kappa=4$, the Virasoro central charge is $c=1$
and $h_{1;2}=1/4$ and $h_{0;1/2}=1/16$. Central charge $c=1$
corresponds to bosonic free field.  Let us denote by $X$ this field.
$h_{1;2}=1/4$ is the conformal weight of the boundary vertex operator
$V_{1;2}=\cos X/\sqrt{2}$ which may be thought of as the boundary
condition changing operator intertwining two boundary intervals on
which two different Dirichlet boundary conditions are imposed.
$h_{0;1/2}=1/16$ is the dimension of the twist field $\sigma$ which is
the boundary condition changing operator intertwining between
Dirichlet and Neumann boundary conditions. Thus the probability
$P_l(z,\bar z)$ is proportional to the free field expectation value
$$
\vev{ X(z,\bar z)}_{\mathbb{S},\, D;D;N}=
\vev{\sigma(x_+)\sigma(x_-) X(z,\bar z) V_{1;2}(x_0)}_{\mathbb{S}},
$$
where 'D;D;N' refers to Dirichlet boundary conditions on the lower
boundary $[x_-,x_0]$ and $[x_0,x_+]$, but with a discontinuity at
$x_0$ and Neumann boundary condition on the upper boundary
$[x_-,x_+]$.  The fact $P_l(z,\bar z)$ satisfies the Dirichlet
boundary conditions on the lower boundary is clear by construction but
one may verify that it actually satisfies the Neumann boundary
condition on the upper boundary. The fact that it is a harmonic
function is then a consequence of the free field equation of $X$.

For $\kappa>4$, these probabilities are proportional to the CFT
correlation functions
\begin{eqnarray*}
\vev{\psi_{0;1/2}(x_-)\psi_{0;1/2}(x_+)\Phi_0(z,\bar
  z)\psi_{1;2}(x_0)}_\mathbb{S}
\end{eqnarray*}
involving a weight zero bulk primary field $\Phi_0$.  This operator
$\Phi_0$ has a simple interpretation in the Coulomb gas formulation
\cite{FF84,DMS:CFT} of CFT: it a linear combination of the primitive
of the screening current $Q_-$ and the identity operator, ie:
$$
\Phi_0(z,\bar z) = {\rm const'.}\,{\bf 1}
+ \Re {\rm e}\, [\, {\rm const.}\, \int^z dw\, Q_-(w)\, ]. 
$$
Indeed this operator has conformal weight zero, satisfies the
appropriate fusion rules and fulfills the charge conservation
requirement which demands that the sum of the coulomb charges of the
operators involve in the correlation function minus the background
charge belongs to the lattice generated by the screening charges.

Similarly, we may compute the probability $P_{in}(z,\bar z)$ for the
point $z$ to be in the SLE hull for $\kappa>4$. We do not distinguish
the events in which the point has been swallowed from the right or
from the left. It is solution of the same second order differential
equation as $P_l$ but with different boundary conditions:
$P_{in}(\pm\infty)=0$ and $P_{in}(0)=1$. The result is:
\begin{eqnarray}
P_{in}(z,\bar z)= \frac{\Im {\rm m}[e^{i2\pi/\kappa}\,\hat F(z)\,]}{
\Im {\rm m}[e^{i2\pi/\kappa}\,\hat F(0)\,]}
\label{probin}
\end{eqnarray}
with same function $\hat F(z)$ as above. We have:
$\Im{\rm m} [e^{i\frac{2\pi}{\kappa}}\hat F(0)] 
= -\sin(\frac{2\pi}{\kappa})\,J$.
Again, $P_{in}$ has the expected behavior on
the boundary.  Since $e^{i2\pi/\kappa}\, F(z)$ is real on the upper
boundary, we have $P_{in}(z,\bar z)=0$ for $z\in i\pi +\mathbb{R},$ in
agreement with the fact that no point on the upper boundary can be
swallowed. $P_{in}$ is even on the real axis and
$$ P_{in}(x)=  \frac{1}{J}\int_{|x|}^{+\infty}
\frac{dy}{(\sinh y/2)^{4/\kappa}},\quad x\in \mathbb{R}.$$
This is of course complementary to $P_l(x)$ for $x$ negative.

For $\kappa<4$ the SLE trace is a simple curve so that the probability
to be inside the hull has no meaning but one may still try to evaluate
the probability to be on the left, or on the right, of the trace. It
turns out that these probabilities are not any more harmonic
functions. However, the probability to hit the upper boundary is still
given by formula (\ref{eq:dipupper}), as proved in \cite{BBH04}.  It
is has been numerically check in the case of the Ising model in
\cite{BBH04}.

\subsection{Fractal dimensions}
\label{sec:dimfract}

The fractal dimension of a set may be defined via box counting. Let
$N_\varepsilon$ be the number of boxes of typical linear size
$\varepsilon$ needed to cover the set. This number increases as
$\varepsilon\to 0$ and -- if it exists -- we may define the fractal
dimension $d_\kappa$ by $d_\kappa=-\lim_{\varepsilon\to 0} \log
N_\varepsilon/\log\varepsilon$, ie. $N_\varepsilon$ follows the power
law $N_\varepsilon\approx \varepsilon^{-d_\kappa}$.  In our case, the
set is a curve. Imagine that this curve is inscribed in a domain of
typical size $L$ which may be covered by $(L/\varepsilon)^D$ boxes of
size $\varepsilon$ -- in two dimensions $D=2$. To cover the curve we
need $N_\varepsilon\approx (L/\varepsilon)^{d_\kappa}$ so that the
probability to find the curve in one of the infinitesimal box scales
as $\varepsilon^{D-d_\kappa}$.  This is the strategy we follow to
compute the fractal dimension of the SLE curves. Namely we evaluate
the probability that the SLE path approaches a bulk point $z_0$ at a
distance less than $\varepsilon$:
$${\bf P}[\gamma_{[0,+\infty)}\cap
\mathbb{B}_\varepsilon(z_0)\not=\emptyset] 
\approx \varepsilon^{2-d_\kappa} ,\quad \varepsilon\to 0
$$ 
with $\mathbb{B}_\varepsilon(z_0)$ the ball of radius $\varepsilon$
centered in $z_0$. As already mentioned, this yields:
$$
d_\kappa= 1 + \kappa/8,\ {\rm for}\ \kappa<8,\quad\quad d_\kappa=2,\ 
{\rm for}\ \kappa>8
$$
This formula was predicted by B. Duplantier
\cite{Duplantier03,Duplantier03a} and rigorously proven by V. Beffara
\cite{Beffara02a,Beffara02b}. See also \cite{DuplantierSaleur87} in
the case of percolation ($\kappa=6$ and $8/3$). The complete
determination of $d_\kappa$ requires also establishing a two point
estimate, which is much harder to obtain but which may be found in the
nice reference \cite{Beffara02b}.

Computing this probability is done, as usual, by identifying the
appropriate martingale using either probabilistic or CFT arguments. It
leads to the identification of the conformal primary field
$\Phi_{0;1}(z_0,\bar z_0)$ of bulk scaling dimension
$2h_{0;1}=\frac{8-\kappa}{8}$ as the field conditioning on the presence
of a SLE curve in the neighborhood of the point $z_0$.

We use chordal SLE in the upper half plane.
So let $z_0\in \mathbb{H}$, $\Im {\rm m}z_0>0$, be a point in the upper
half plane and $\delta_t(z_0)$ its distance to the SLE curve
$\gamma_{[0,t]}$ stopped at time $t$. We shall evaluate $\delta_t(z_0)$
using the conformal radius of $\gamma_{[0,t]}$ seen from $z_0$.
To compute it, let $k_t(z)$, defined by
$$
k_t(z)= \frac{g_t(z)-g_t(z_0)}{g_t(z)-\overline{g_t(z_0)}},
$$
be a uniformizing map of $\mathbb{H}\setminus\mathbb{K}_t$ onto the
unit disk with $k_t(z_0)=0$, $k_t(\infty)=1$. The conformal radius of
$\gamma_{[0,t]}$ viewed from $z_0$ is defined as $\rho_t(z_0)\equiv
|k_t'(z_0)|^{-1}$.  An explicit computation gives $\rho_t(z_0)=|2\Im
{\rm m} h_t(z_0)/h'_t(z_0)|$ with $h_t=g_t-\xi_t$.  K\"obe
$1/4$-theorem states that $(1/4)\rho_t(z_0)\leq \delta_t(z_0)\leq
\rho_t(z_0)$ so that $\delta_t(z_0)$ and $\rho_t(z_0)$ scale the same
way.  One may check that $\rho_t(z_0)$ is always decreasing as time
goes by.  So instead of estimating the distance between the SLE path
and $z_0$, we shall estimate its conformal radius,
$\rho(z_0,\gamma)=\lim_{t\to \tau_{z_0}}\rho_t(z_0)$, and the
probability ${\bf P}[\rho(z_0,\gamma)\leq \varepsilon ]$.

The image $U_t\equiv k_t(\gamma(t))$ of the tip of the curve by $k_t$
is on the unit circle. Setting $U_t\equiv e^{i\alpha_t}$ defines a
process $\alpha_t$ on the unit circle with $\alpha_t\to 0$ or $2\pi$
as $t\to \tau_{z_0}$ depending whether $z_0$ has been swallowed
clockwise, or counterclockwise, by the SLE trace. Actually, up to a
random time reparametrization, $ds=(\frac{2\Im {\rm
    m}h_t(z_0)}{|h_t(z_0)|^2})^2dt$, this process is driven by
$d\alpha_s= \frac{\kappa-4}{4}\cot (\alpha_s/2) + d\xi_s$.  Estimating
${\bf P}[\rho(z_0,\gamma)\leq \varepsilon ]$ can be formulated
\cite{Beffara02b} as a survival probability problem for the process
$\alpha_s$ but, in order to understand its CFT origin, we shall
compute it using a CFT martingale.  For $\kappa<8$, let us
consider the expectation value
$$ 
\hat M_t(z_0)\equiv |h_t'(z)|^{2h_{0;1}}\,
\bra{\psi_{1;2}}\Phi_{0;1}(h_t(z_0),\bar{h_t(z_0)})
\ket{\psi_{1;2}}_\mathbb{H}
$$
with $\Phi_{0;1}$ the bulk conformal field of weight
$2h_{0;1}=(8-\kappa)/8$. By construction this is well defined up to
time $t<\tau_{z_0}$.  The correlation function
$\bra{\psi_{1;2}}\Phi_{0;1}(z_0,\bar z_0)\ket{\psi_{1;2}}$ may be
computed exactly using the level two null vector.  It is equal to
$|2\Im{\rm m}z_0|^{-2h_{0;1}} (\sin\alpha_0/2)^{8/\kappa-1}$
with $z_0/\bar z_0= e^{i\alpha_0}$ so that
$$
\hat M_t(z_0)= \Big\vert \frac{h'_t(z_0)}{2\Im{\rm m}h_t(z_0)}
\Big\vert^{2h_{0;1}}\, (\sin\alpha_t/2)^{\kappa/8-1}
= \rho_t(z_0)^{-2h_{0;1}}\, (\sin\alpha_t/2)^{8/\kappa-1}
$$
Let $\tau_{z_0}^\varepsilon$ be either the time at which the
conformal radius $\rho_t(z_0)$ reaches the value $\varepsilon$,
if $\rho(z_0,\gamma)\leq \varepsilon$, or the
swallowing time $\tau_{z_0}$ if the point $z_0$ is swallowed before 
the conformal radius reaches this value, 
i.e. if $\rho(z_0,\gamma)> \varepsilon$. 
The time $\tau_{z_0}^\varepsilon$ is a stopping time.
Since $h'_t(z_0)$ vanishes faster than $h_t(z_0)$ as $t\to
\tau_{z_0}$, the martingale
$\hat M_t(z_0)$ vanishes as $t\to \tau_{z_0}$ for $\kappa<8$.
Therefore as time $t$ approaches $\tau_{z_0}^\varepsilon$, the
martingales $\hat M_t(z_0)$ projects on configuration with the
curve at a distance from $z_0$ less than $\varepsilon$, ie.:
$$ 
\hat M_{\tau_{z_0}^\varepsilon}(z_0)= \varepsilon^{-2h_{0;1}}\, 
(\sin\alpha_{\tau_{z_0}^\varepsilon}/2)^{8/\kappa-1}\
{\bf 1}_{\{\rho(z_0,\gamma)\leq \varepsilon\}}
$$
Up to the angular dependence
$(\sin\alpha_{\tau_{z_0}^\varepsilon}/2)^{8/\kappa-1}$, which does not
play any role in the scaling analysis, $\hat
M_{\tau_{z_0}^\varepsilon}(z_0)$ is proportional to the characteristic
function of the set of curve passing at distance $\varepsilon$ from
$z_0$ so that its expectation behaves as the probability ${\bf
  P}[\rho(z_0,\gamma)<\varepsilon]$.  By construction $\hat M_t(z_0)$
is a martingale so that ${\bf E}[\hat
M_{\tau_{z_0}^\varepsilon}(z_0)]=\hat M_{t=0}(z_0)$ and
$$
{\bf P}[\rho(z_0,\gamma)\leq \varepsilon]\approx
\varepsilon^{2h_{0;1}} 
$$
as $\varepsilon\to 0$.  This one point estimate
yields to the fractal dimension $d_\kappa= 2 - 2h_{0;1}$
or equivalently $d_\kappa=1+\kappa/8$.

More generally, one may look for the zig-zag density probabilities that
the SLE paths visit balls $\mathbb{B}_\varepsilon(z_p)$ centered in
points $z_p$.  This is clearly proportional to CFT correlation
functions $\bra{\psi_{1;2}} \Phi_{0;1}(z_1,\bar z_1)\cdots
\Phi_{0;1}(z_n,\bar z_n) \ket{\psi_{1;2}}$.  Different orders of
visiting the points $z_p$ corresponds to different correlation
functions alias conformal blocks.  If no order among the visited balls
is specified, these correlation functions have no monodromy and they
thus correspond to the complete CFT correlation functions.  Zig-zag
probabilities with specified ordering in the visits would be exchanged
as one moves the points $z_p$ around. In other words, there is
probably a quite direct relation between CFT monodromies, quantum
groups, and braiding properties of samples of SLE traces.

\subsection{Miscellaneous}

An important output of SLE was the mathematical proof
\cite{LSW1,LSW2,LSW2} for the values of the Brownian intersection
exponents. See the comprehensible reviews \cite{Werner00b,Lawler03b}.
Many of these exponents have been predicted using techniques from
conformal field theory, see for instance
\cite{DuplantierSaleur86,DuplantierK88,AizenmanDuplantier99}.  These
exponents describe how the probabilities for a set of Brownian paths
not to intersect decreases with time. Let $B_t^j$, $j=1,\cdots, n$ be
$n$ independent planar Brownian motions started from $n$ different
points in the plane and run during a time $t$. The probability that
they do not intersect decreases as $t^{-\zeta_n}$ with
$\zeta_n=(4n^2-1)/24$. The same probability but for Brownian motions
confined in the upper half plane decreases with time as $t^{-\tilde
  \zeta_n}$ with $\tilde \zeta_n=n(2n+1)/6$. More generally, one may
look at the non intersecting probabilities of $k$ packs of Brownian
motions, each of them made of $n_j$, $j=1,\cdots,k$, independent
Brownian motions started from distinct points. These probabilities
decrease with time with exponents $\zeta_{n_1,\cdots,n_k}$ and
$\tilde \zeta_{n_1,\cdots,n_k}$ depending whether the Brownian paths
are in the plane or in the upper half plane. Although many properties
of these exponents \cite{Werner00b,Lawler03b} were known -- such that
the cascade relations they satisfy \cite{Lawler96} and more
\cite{LawlerWerner99,LawlerWerner00,SmirnovWerner01} -- the exactness
of CFT predictions
\cite{DuplantierSaleur86,DuplantierK88,AizenmanDuplantier99} was
proved only recently by Lawler, Schramm and Werner
\cite{LSW1,LSW2,LSW3} using SLE. The proof of these results relies on
a universality argument \cite{LawlerWerner00} which states that any
conformally invariant process satisfying the restriction property has
crossing and intersection exponents that are intimately related to the
Brownian intersection exponents.  As the boundary of SLE$_6$ is
conformally invariant and satisfies the restriction property, the
computation of its exponents yields the Brownian intersection
exponents. The relation between SLE$_6$ and Brownian motion is even
more precise in the sense that the hull generated by SLE$_6$ is the
same as the hull generated by a Brownian motion with oblique
reflection \cite{WernerEdin03}.

Cardy's crossing formula \cite{Cardy92} for percolation, its proof
by Smirnov \cite{Smirnov} and the numerical simulations of
\cite{Langlands94} did play an important in the birth of SLEs.
Another crossing formula for percolation has been predicted by Watts
using CFT argument \cite{Watts}. This formula, which was also
motivated by the numerical analysis of \cite{Langlands94}, gives the
probability in critical percolation that there exists a percolating
clusters crossing simultaneously from the right to left and from top
to bottom of rectangle. Although Watts's formula remained conjectural
for a while it has now been proved in \cite{Dubedat04} using excursion
decompositions of SLEs. Crossing formula for critical percolation in
an annulus have also been predicted by Cardy using CFT
\cite{Cardy02c}. Their SLE companions have been derived in
\cite{Dubedat03b}. Finally, another excursion formula for
percolation has been proved using SLE in \cite{Schrammperc}.

The one-arm exponent which governs the decrease of the probability
that the critical percolating cluster has diameter of order $R$ --
this probability behaves as $R^{-5/48}$ -- has been SLE proved in
\cite{LSW01b}. A two-arm exponent, also called backbone exponent,
which describes the decrease of the probability that there are two
disjoint open crossings from a circle of radius $r$ to a circle of
radius $R$ has been estimated in \cite{LSW3}. This prediction is yet
out of reach of conformal field theory.

The proof that the scaling limit of loop erased random walks and
uniform spanning trees are respectively SLE$_2$ and $SLE_8$ are given
in refs.\cite{Schramm00,LSW01a}. Although there is no much doubt,
there is yet no mathematical proof that self-avoiding walks converge
to SLE$_{8/3}$ in the continuum limit, but see ref.\cite{LSW02}.  More
informations on spanning trees and related domino tilling may be found
in ref.\cite{Kenyon03}.

%% file: chap7.tex

This Section deals with more general 2D growth processes than SLEs.
Although, they do not fulfill the local growth and conformal
invariance properties of SLEs, they are nevertheless described by
dynamical conformal maps. We first present systems whose conformal
maps have a time continuous evolution and give examples. We then go on
by presenting a discrete version thereof in terms of iterated
conformal maps. This field of research is much less developed and
understood, at least mathematically, than SLEs. There are many
questions still unanswered today.

Examples of models -- including DLA, dielectric breakdown, Hele-Shaw
problems, etc. -- have been introduced in Section \ref{sec:growth}.
They have a large domain of applicability
\cite{Benetal86,Gollub99,Halsey00,Bazant05,Tanveer00} and many
examples have been given in Section \ref{sec:growth}.  We shall
elaborate more on them. They are all linked to Laplacian growth which
is one of the simplest examples of such growth processes.  We shall
spend more times on Laplacian growth since it is a rich system which
possesses an underlying integrable structure but which simultaneously
produces singularities leading to dendritic growth.

In this part, the exterior of the unit disk,
$\mathbb{U}=\{w\in\mathbb{C},\ |w|>1\}$, is used as the reference
geometry. So the growth dynamics are going to be described by radial
Loewner chains, simple variants of the chordal Loewner chains. 

We end this Section by a brief discussion of discrete Loewner chains
defined by iterations of conformal maps.

\subsection{Radial Loewner chains}
\label{sec:radialchain}
\begin{figure}[htbp]
\begin{center}
\includegraphics[width=0.9\textwidth]{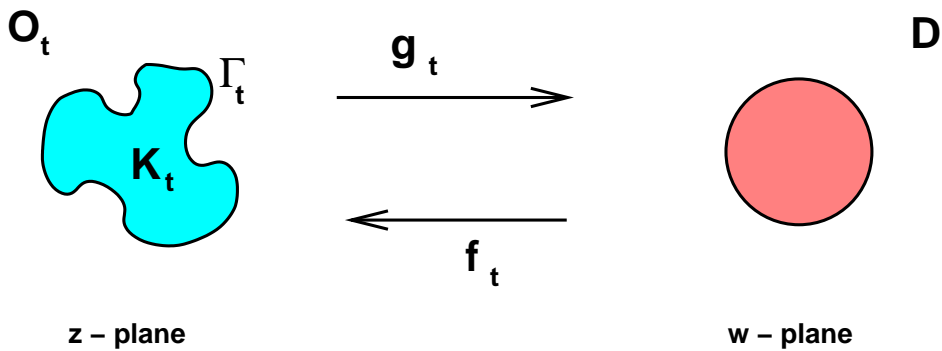}
\caption{Uniformizing maps intertwining the `physical' $z$-plane and
 the `mathematical' $w$-plane.}
     \label{fig:radialgrowth}
\end{center}
\end{figure}

Let $\mathbb{K}_t$ be a family of growing closed planar sets such that
their complement in the complex plane $\mathbb{O}_t\equiv
\mathbb{C}\setminus\mathbb{K}_t$ also have the topology of a disk. See
Fig.\ref{fig:radialgrowth}.  To fix part of translation
invariance we assume that the origin belongs to $\mathbb{K}_t$ and the
point at infinity to $\mathbb{O}_t$.

Loewner chains describe the evolution of family of conformal maps
$f_t$ uniformizing $\mathbb{U}=\{w\in \mathbb{C};\ |w|>1\}$ onto
$\mathbb{O}_t$. It thus describes the evolution of the physical
domains $\mathbb{O}_t$. We normalize the maps $f_t:\mathbb{U}\to
\mathbb{O}_t$ by demanding that they fix the point at infinity,
$f_t(\infty)=\infty$ and that $f_t'(\infty)>0$.  With $t$
parameterizing time, Loewner equation for the evolution of $f_t$
reads:
\begin{eqnarray}
\frac{\partial}{\partial t} f_t(w) = wf_t'(w)\,
\oint \frac{\rho_t(u)du}{2i\pi u} \,
\Big({\frac{w+u}{w-u}}\Big)
\label{loew}
\end{eqnarray}
The integration is over the unit circle 
$\{u\in\mathbb{C}, |u|=1\}$. The Loewner density $\rho_t(u)$ codes
for the time evolution. It may depends on the map $f_t$ in which case
the growth process in non-linear.
For the inverse maps $g_t\equiv f_t^{-1}:\mathbb{O}_t\to \mathbb{U}$, 
Loewner equation reads:
\begin{eqnarray}
\frac{\partial}{\partial t} g_t(z) = -g_t(z)
\oint\frac{\rho_t(u)du}{2i\pi u} 
\Big({\frac{g_t(z)+u}{g_t(z)-u}}\Big)
\label{loewbis}
\end{eqnarray} 
Compare with the equation governing radial SLE.

The behavior of $f_t$ at infinity fixes a scale since at infinity,
$f_t(w)\simeq R_tw+O(1)$ where $R_t>0$, with the dimension of a {\tt
  [length]}, is called the conformal radius of $\mathbb{K}_t$ viewed
from infinity.  $R_t$ may be used to analyze scaling behaviors, since
Kobe 1/4-theorem (see e.g. \cite{Conway,Ahlfors}) ensures that $R_t$
scales as the size of the domain.  In particular, the (fractal)
dimension $D$ of the domains $\mathbb{K}_t$ may be estimated by
comparing their area $\mathcal{A}_t$ with their linear size measured
by $R_t$: $\mathcal{A}_t\asymp R_t^{D}$ for large $t$ -- the
proportionality factor contains a cutoff dependence which restores
naive dimensional analysis.


The boundary curve $\Gamma_t\equiv \partial \mathbb{O}_t$ is the image of
the unit circle by $f_t$.  We may parameterize the
boundary points by $\gamma_{t;\alpha}=f_t(u)$ with $u=e^{i\alpha}$.
The Loewner equation codes for the evolution of the shape of $\mathbb{O}_t$
and thus for the normal velocity of the boundary points.  Only the
normal velocity is relevant as the tangent velocity is
parameterization dependent.  The tangent to the curve is $\tau = iu
f_t'(u)/|f_t'(u)|$ and the outward normal is $n=-i\tau$ so that the
normal velocity at $\gamma_t$ is $v_n=\Re{\rm e}[\bar n\,
\partial_tf_t(u)]$, or
$$ v_n = |f_t'(u)|\, \Re{\rm e}[\partial_tf_t(u)/ uf_t'(u)].$$

The r.h.s. is determined by the Loewner equation (\ref{loew}) because
this equation may be viewed as providing the solution of a boundary
value problem.  Indeed, recall that for $\hat h(u)$ a real function on
the unit circle, $h(w)=\oint\frac{du}{2i\pi u}
\Big(\frac{w+u}{w-u}\Big)\hat h(u)$ is the unique function analytic
outside the unit disk whose real part on the unit circle is $\hat h$.
Thus, since $\partial_tf_t(w)/wf_t'(w)$ is analytic in $\mathbb{U}$,
the Loewner equation (\ref{loew}) is equivalent to:
$$v_n = |f_t'(u)|\, \rho_t(u)$$
or more explicitly:
\begin{eqnarray}
(\partial_\alpha \gamma_{t;\alpha})\,
(\partial_t \overline{\gamma_{t;\alpha}}) -
(\partial_\alpha \overline{\gamma_{t;\alpha}})\,
(\partial_t \gamma_{t;\alpha}) = 2i\, 
|f_t'(u)|^2\, \rho_t(u)
\label{bdry}
\end{eqnarray}

Hence, the evolution of the domain may be encoded either in the
evolution law for its uniformizing conformal map as in eq.(\ref{loew}) 
or in the boundary normal velocity as in eq.(\ref{bdry}).
The two equations are equivalent.

\subsection{Laplacian growth as a Loewner chain} 
\label{sec:LG}

\subsubsection{Basics} 
\label{sec:LGbasic}
Laplacian growth (LG) is a process in which the growth of the domain
is governed by the solution of Laplace equation, i.e. by an harmonic
function, in the exterior of the domain with appropriate boundary
conditions. It originates from the hydrodynamic Hele-Shaw problem to
be described below, see eg. \cite{Benetal86}.

To be precise, let $P$ be the real solution of Laplace equation,
$\nabla^2P=0$, in $\mathbb{O}_t$ with the boundary behavior $P=-\log
|z| + \cdots$ at infinity and $P=0$ on the boundary curve
$\Gamma_t=\partial \mathbb{O}_t$. The time evolution of the domain is
then defined by demanding that the normal velocity of points on the
boundary curve be equal to minus the gradient of $P$: $v_n=-(\nabla
P)_n$.

This problem may be written as a Loewner chain since, as is well
known, Laplace equation is solved via complex analysis by writing $P$
as the real part of an analytic function.  One first solves Laplace
equation in the complement of the unit disk with the appropriate
boundary conditions and then transports it back to the physical domain
$\mathbb{O}_t$ using the map $f_t$. This gives:
\begin{eqnarray}
P= -\Re{\rm e}\ \Phi_t\quad {\rm with}\quad 
\Phi_t(z) = \log g_t(z) 
\nonumber
\end{eqnarray}

The evolution equation for the map $f_t$ is derived using 
that the boundary normal velocity is $v_n=-(\nabla P)_n$. 
The above expression for $P$ gives:
$$ 
v_n = -(\nabla P)_n = |f_t'(u)|^{-1}
$$
at point $\gamma_t=f_t(u)$ on the boundary curve.  As explained in the
previous Section, this is enough to determine $\partial_t f_t(w)$ for
any $|w|>1$ since this data specifies the real part on the unit circle
of the analytic function $\partial_tf_t(w)/wf_t'(w)$ on the complement
of the unit disk. The result is:
\begin{eqnarray}
\partial_t f_t(w) = wf_t'(w)\,\oint_{|u|=1}
\frac{du}{2i\pi u\, |f_t'(u)|^2} 
\Big({\frac{w+u}{w-u}}\Big)
\label{lg}
\end{eqnarray}
It is a Loewner chain with $\rho_t(u)=|f_t'(u)|^{-2}$.  

As we shall see below, Laplacian growth is an integrable system, which
may be solved exactly, but it is ill-posed as the domain develops
singularities (cusps $y^2\simeq x^3$) in finite time.  It thus needs
to be regularized. There exist different ways of regularizing it.

A larger class of problems generalizing Laplacian growth have been
introduced. Their Loewner measures are as in Laplacian growth but with a
different exponent:
$$\rho_t(u)=|f_t'(u)|^{-\alpha}\quad,\quad 0\leq\alpha\leq2.$$
Using an
electrostatic interpretation of the harmonic potential, one usually
refers to the case $\alpha=1$ as a model of dielectric breakdown
because the measure is then proportional to the local electric field
$E_n=|f_t'(u)|^{-1}$. This is a phenomenological description. Just as
the naive Laplacian growth these models are certainly ill-posed.  They
also require ultraviolet regularization.

\subsubsection{Singularities}
\label{sec:LGsing} 
The occurrence of singularities in Laplacian growth may be grasped by
looking for the evolution of domains with a $Z_n$ symmetry uniformized
by the maps
$$
f_t(w)=R_tw(1+\frac{\beta_t}{n-1}w^{-n})
$$
for some $n>2$ and with $|\beta_t|\leq 1$. This form of 
conformal maps is preserved by the dynamics. The conformal radius
$R_t$ and the coefficients $\beta_t$ evolve with time according to
$\partial_t R^2_t= 2/(1-\beta_t^2)$ and $\beta_t = (R_t/R_c)^{n-2}$ with
$R_c$ some integration constant.  The singularity appears when
$\beta_t$ touches the unit circle which arises at a finite time $t_c$.
At that time the conformal radius is $R_c$.

\begin{figure}[htbp]
\begin{center}
\includegraphics[width=0.8\textwidth]{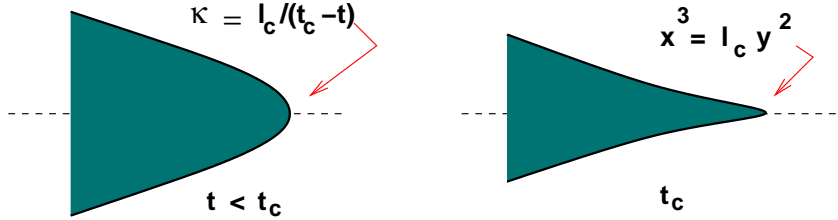}
\caption{Cups formation in Laplacian growth.}
     \label{fig:cusp}
 \end{center}
\end{figure}

At $t_c$ the boundary curve $\Gamma_{t_c}$ has
cusp singularities of the generic local form 
$$
\ell_c\,(\delta y)^2\simeq (\delta x)^3
$$
with $\ell_c$ a characteristic local length scale.
In the present simple case $\ell_c\simeq R_c$. 
At time $t\nearrow t_c$, the dynamics is regular in the
dimensionless parameter $\ell_c^{-1}\sqrt{t_c-t}$. 
The maximum curvature of the boundary curve scales
as $\kappa_{\rm max} \simeq {\ell_c}/{(t_c-t)}$ near $t_c$ and it is
localized at a distance $\sqrt{t_c-t}$ away from the would be cusp
tip. See Fig.\ref{fig:cusp}.

This behavior is quite generic. Conformal maps $f_t(w)$ such that
their derivatives are polynomials in $w^{-1}$ are stable by the
Laplacian growth dynamics. By construction, their zeroes are localized
inside the unit disk. A singularity in the boundary curve occurs if
one of these zeroes converges to the unit circle.  The singularity is
then a cusp $\ell_c\,y^2\simeq x^3$ as can be seen by expanding
locally the conformal map around its singular point.

Once regularized with an explicit ultraviolet cut-off, the processes
are expected to be well defined for all time. The curvature of of the
boundary curve is then expected to remain finite at all time.  Using
scaling theory, a crude estimate of its maximum around the would be
singularities may be obtained by interchanging the short distance
scale $\sqrt{(t_c-t)}$ near the singularity in the unregularized
theory with the UV cutoff of the regularized theory. For the Hele-Shaw
problem to be defined below, the maximum curvature is then expected to
scale as $\kappa_{\rm max}\simeq \ell_c/\sigma^2$ as the surface
tension $\sigma\to 0$.  The effect of the regularization procedure on
the domain properties is presently unclear. The domain structures may
a priori depend on how the problem has been regularized. For the
Hele-Shaw problem, the cusp production is expected to be replaced by
unlimited ramifications leading to dendritic growth.

\subsubsection{Regularization}
\label{sec:LGreg}
We now describe a few of the possible regularization of Laplacian
growth which have been considered in the literature.
Of course DLA provides a regularization of Laplacian growth. 
Another one is a hydrodynamic regularization, called the Hele-Shaw
problem, which actually was studied well before Laplacian growth.  The
differences with Laplacian growth are in the boundary conditions which
now involve a term proportional to the surface tension.  

Let us recall that it may be formulated as follows \cite{Benetal86}.
One imagines that the domain $\mathbb{K}_t$ is filled with a non
viscous fluid, say air, and the domain $\mathbb{O}_t$ with a viscous
one, say oil. Air is supposed to be injected at the origin and there
is an oil drain at infinity. The pressure in the air domain
$\mathbb{K}_t$ is constant and set to zero by convention.  In
$\mathbb{O}_t$ the pressure satisfies the Laplace equation
$\nabla^2P=0$ with boundary behavior $P=-\phi_\infty \log |z| +
\cdots$ at infinity reflecting the presence of the oil drain.  The
boundary conditions on the boundary curve are now $P=-\sigma\kappa_t$
with $\sigma$ the surface tension and $\kappa_t$ the curvature of the
boundary curve~\footnote{The curvature is defined by $\kappa\equiv
  -{\vec{n}.\partial_s\vec{\tau}}/{\vec{\tau}^2} = \Im{\rm
    m}[{\bar\tau\partial_s\tau}/{|\tau|^3}]$ with $\vec{\tau}$ the
  tangent and $\vec{n}$ the normal vectors. An alternative formula is:
  $\kappa=|f_t'(u)|^{-1}\Re{\rm e} [1+\frac{uf_t''(u)}{f_t'(u)}]$. For
  a disk of radius $R$, the curvature is $+1/R$.}.  The fluid velocity
in the oil domain $\mathbb{O}_t$ is $\vec{v}=-\vec{\nabla}P$. Laplace
equation for $P$ is just a consequence of incompressibility. The
evolution of the shape of the domain is specified by imposing that
this relation holds on the boundary so that the boundary normal
velocity is $v_n=-(\nabla P)_n$ as in Laplacian growth.

Compared to Laplacian growth, the only modification is the boundary
condition on the boundary curve. This term prevents the formation of
cusps with infinite curvature singularities. The parameter
$\phi_\infty$ sets the scale of the velocity at infinity.
In the following we set $\phi_\infty=1$. By dimensional analysis
this implies that $[{\tt time}]$ scales as $[{\tt length}^2]$
and the surface tension $\sigma$ has dimension of a $[{\tt length}]$.
It plays the role of an ultraviolet cut-off.

A standard procedure \cite{Benetal86} to solve the equations for the
Hele-Shaw problem is by first determining the pressure using complex
analysis and then computing the boundary normal velocity.  By Laplace
equation, the pressure is the real part of an analytic function, $P=
-\Re{\rm e} \Phi_t$. The complex velocity $v=v_x+iv_y$ is $\bar v=
\partial_z\Phi_t$. At infinity $\Phi_t(z)\simeq \log z+\cdots$ and
$\bar v\simeq 1/z+\cdots$.  The boundary conditions on $P$ demand that
$$
(\Phi_t\circ f_t)(w) = \log w + \sigma \vartheta_t(w)
$$
where $\vartheta_t(w)$ is analytic in $\mathbb{U}$, the exterior of
the unit disk, with boundary value $\Re{\rm e}[\vartheta_t(u)]=
\kappa_t(f_t(u))$ with $\kappa_t$ the curvature.  Explicitly
$$
\vartheta_t(w)=\oint\frac{du}{2i\pi u}\,
\Big(\frac{w+u}{w-u}\Big)\,\kappa_t(f_t(u))
$$  
The evolution of $f_t$ is then found by evaluating the boundary normal 
velocity $v_n=\Re{\rm e}(\nabla\Phi)_n$ at point $\gamma_t=f_t(u)$: 
$$ 
v_n = \Re{\rm e}[\, n\partial_z\Phi_t\,]
= |f_t'(u)|^{-1}\, \Re{\rm e}[1+\sigma u \partial_u\theta_t(u)]
$$
As above, this determines uniquely
$\partial_tf_t(u)$ and it leads to a Loewner chain (\ref{loew})
with density:
\begin{eqnarray}
\rho_t(u) = |f_t'(u)|^{-2}\, \Big( 1 + \sigma \epsilon_t(u)\Big)
\quad,\quad \epsilon_t(u) = \Re{\rm e}[ u\partial_u \vartheta_t(u)]
\label{meshLG}
\end{eqnarray}
The difference with Laplacian growth is in the extra term
proportional to $\sigma$. It is highly non-linear and non-local. 
This problem is believed to be well defined at all times for $\sigma$
positive.

Another way to regularize Laplacian growth amounts to introduce an
ultra-violet (UV) cutoff $\delta$ in the physical space by evaluating
$|f_t'|$ at a finite distance away from $\partial\mathbb{O}_t$. A
possible choice \cite{MakCarl} is $\rho_t(u)^{1/2}=\delta^{-1}{\rm
  inf}\{\varepsilon:\, {\rm dist}[f_t(u+\varepsilon
u);\partial\mathbb{O}_t]=\delta\}$.  An estimation gives
$\rho_t(u)\asymp |f_t'(u+\hat\varepsilon_u u)|^{-2}$ where
$\hat\varepsilon_u$ goes to $0$ with $\delta$, so that it naively
approaches $|f_t'(u)|^{-2}$ as $\delta\to 0$.
Another possible, but less physical, regularization consists in
introducing an UV cutoff $\nu$ in the mathematical space so that
$\rho_t(u)=|f_t'(u+\nu u)|^{-2}$.

\subsection{Integrability of Laplacian growth} 
\label{sec:LGint}
Laplacian growth is an integrable system, at least up to the cusp
formation. Let us recall that it corresponds to a Loewner chain with a 
density $\rho_t(u)=|f_t(u)|^{-2}$, or equivalently to the quadratic
equation
\begin{eqnarray}
(\partial_\alpha \gamma_{t;\alpha})\,
(\partial_t \overline{\gamma_{t;\alpha}}) -
(\partial_\alpha \overline{\gamma_{t;\alpha}})\,
(\partial_t \gamma_{t;\alpha}) = 2i\,
\label{lgbdry}
\end{eqnarray}
for the dynamics of the boundary points $\gamma_{t;\alpha}=f_t(u)$,
$u=e^{i\alpha}$. What makes the model integrable is the fact that the
r.h.s of eq.(\ref{lgbdry}) is constant. Eq.(\ref{lgbdry}) is then
similar to a quadratic Hirota equation.  Hints on the integrable
structure were found in \cite{shraiben} and much further developed in
\cite{richards,integr}. Laplacian growth is also deeply related to
random matrix models \cite{LGmatrix}.

\subsubsection{Conserved quantities}
\label{sec:LGcons}
We now define an infinite set of quantities which are conserved in the
naive unregularized LG problem. They reflect its integrability. We
follow ref.\cite{richards,integr}.  These quantities may be defined via a
Riemann-Hilbert problem on $\Gamma_t$ specified by,
 \begin{eqnarray}
S_+(\gamma)-S_-(\gamma) = \bar \gamma \quad,\quad \gamma\in \Gamma_t
\label{lgrh}
\end{eqnarray} 
for functions $S_-$ and $S_+$ respectively analytic in the outer
domain $\mathbb{O}_t$ and in the inner domain $\mathbb{K}_t$.
We fix normalization by demanding $S_-(\infty)=0$.
We assume $\Gamma_t$ regular enough for this
Riemann-Hilbert problem to be well defined. 
As usual, $S_\pm$ may be presented as contour integrals:
$$
S_\pm(z)=-\oint_{\Gamma_t}\frac{d\gamma}{2i\pi} \frac{\bar
  \gamma}{z-\gamma}.
$$ 

The conserved quantities are going to be expressed in terms of
$S_\pm$. We thus need their time evolution.  Differentiation of
eq.(\ref{lgrh}) with respect to time and use of the evolution
equation (\ref{lgbdry}) gives:
$$
 \partial_t S_+(\gamma)-\partial_t S_-(\gamma) = 
2{g_t'(\gamma)}/{g_t(\gamma)} 
$$
Notice now that ${g_t'(\gamma)}/{g_t(\gamma)}$ is the boundary
value of $(\log g_t)'$ which by construction is analytic in
$\mathbb{O}_t$. We may thus rewrite this equation as a trivial
Riemann-Hilbert problem, $ \partial_t S_+(\gamma)-(\partial_t S_- +
2(\log g_t)')(\gamma) =0$, so that both terms vanish:
\begin{eqnarray}
\partial_tS_+(z)=0
\quad {\rm and}\quad
(\partial_t S_- + 2(\log g_t)')(z)=0
\label{scw+-}
\end{eqnarray}

Since $S_+$ is analytic around the origin, we may expand it in
power of $z$.  Equation $\partial_t S_+(z)=0$ then tells us that
$S_+(z)$ is a generating function of conserved quantities:
$S_+(z)=\sum_{k\geq 0}z^{k} I_k$ with 
\begin{eqnarray}
I_k=\oint_{\Gamma_t} \frac{d\gamma}{2i\pi} \bar\gamma \gamma^{-k-1}
\quad,\quad \partial_t I_k=0. 
\label{lgik}
\end{eqnarray}
This provides an infinite set of conserved quantities.

Since $S_-$ is analytic around infinity, it may be expanded in power
of $1/z$: $S_-(z)=-\mathcal{A}_t/\pi z+\cdots$ with
$\mathcal{A}_t=-\frac{i}{2}\oint_{\Gamma_t}d\gamma \bar \gamma$ the
area of the inner domain $\mathbb{K}_t$.  The second equation
$(\partial_t S_- + 2(\log g_t)')(z)=0$ with $g_t(z)=R_t^{-1}z+O(1)$
then implies $\partial_t\mathcal{A}_t=2\pi.$ The area of the domain
grows linearly with time, up to the time at which the first cusp
singularity appears.  This is actually a direct consequence of the
fluid incompressibility.

\subsubsection{Simple solutions}
\label{sec:LGsol}

A particularly simple class of conformal maps, solutions of the
Laplacian growth equation, are those such that their derivatives are
polynomials in $w^{-1}$. They may be expanded as:
\begin{eqnarray}
f_t(w)=\sum_{n=0}^N f_nw^{1-n},\quad f_0=R_t>0
\label{poly}
\end{eqnarray}
with $N$ finite but arbitrary. The dynamical variables are the $N+1$
coefficient $f_0,\cdots, f_N$. 
They are all complex except $f_0$ which is real.
It will be convenient to define the function $\bar f_t$ by
$\bar f_t(w)=\overline{f_t(\overline{w})}$.

The fact that this class is stable under the dynamics follows from the
Loewner equation (\ref{lg}). The trick consists in using the fact that
the integration contour is on the unit circle so that 
$|f_t'(u)|^2=f_t'(u)\bar f_t'(1/u)$. The contour integral then
involves a meromorphic function of $u$ so that it can be evaluated by
deforming the contour to pick the residues. This is enough to prove
that $\partial_tf_t(w)$ possesses the same structure as $f_t(w)$
itself so that the class of functions (\ref{poly}) is stable under the
dynamics.

Alternatively one may expand the quadratic equation
(\ref{lgbdry}) to get a hierarchy of equations:
$$
\sum_{n\geq 0} (1-n)[f_n\dot{\bar f}_{j+n} + \bar f_n\dot f_{-j+n}]
=2\delta_{j;0}
$$
For $j=0$, this equation tells us again that the area of the domain grows
linearly with time. Besides this relation there are only $N$
independent complex equations for $j=1,\cdots, N$ which actually 
code for the conserved quantities.

To really have an integrable system we need to have as many
independent integrals of motion as dynamical variables. Thus we need
to have $N$ conserved quantities. These are given by the $I_k$'s
defined above which may be rewritten as
$$
I_k=\oint_{|u|=1} \frac{du}{2i\pi} 
\frac{f_t'(u)\bar f_t(1/u)}{f_t(u)^{k+1}}
$$
Only the first $N$ quantities, $I_0,\cdots, I_{N-1}=R^{1-N}\bar f_N $
are non-vanishing. They are independent.
They can be used to express algebraically all $f_n$'s, $n\geq 0$, 
in terms of the real parameter $f_0=R_t$. The area law,
$$ 
\mathcal{A}_t=\pi[R_t^2+\sum_{n\geq 1}(1-n)|f_n|^2]=2\pi t,
$$
with the $f_n$'s expressed in terms of $R_t$, then reintroduces
the time variable by giving its relation with the conformal radius.

\subsubsection{Algebraic curves}
\label{sec:LGcurve}
As was pointed out in \cite{integr}, solutions of Laplacian growth and
their cusp formations have an elegant geometrical interpretation
involving Riemann surfaces.

Recall that given a sufficiently smooth real curve $\Gamma_t$ drawn on
the complex plane one may define a function $S(z)$, called the
Schwarz function, analytic in a ribbon enveloping the curve such that
$$ 
S(\gamma) = \overline{\gamma},\quad \gamma\in \Gamma_t
$$
By construction, the Schwarz function may be expressed in terms of
uniformizing maps of the domain bounded by the curves as $S(z)=\bar
f_t(1/g_t(z))$.

The Riemann-Hilbert problem (\ref{lgrh}) defining the conserved
charges then possesses a very simple interpretation: $S_\pm$ are the
polar part of the Schwarz function respectively analytic inside or
outside $\Gamma_t$, i.e. $S(z) = S_+(z) - S_-(z).$ Thus the polar part
$S_+$, analytic in the inner domain, is conserved.  The polar part
$S_-$, analytic in the outer domain, evolves according to
eqs.(\ref{scw+-}). Since $\log g_t(z)$ is analytic in the outer
domain, these equations are equivalent to the single equation:
\begin{eqnarray}
\partial_t S(z) = -2 (\log g_t(z))'
\label{eqSZ}
\end{eqnarray}

Now the physical curve $\Gamma_t$ may be viewed as a real slice of a
complex curve, alias a Riemann surface. The latter is constructed
using the Schwarz function as follows. Recall that $s=S(z)$ is
implicitly defined by the relations $z=f_t(w)$, $s=\bar f_t(1/w)$. In
the case of polynomial uniformizing maps we get the pair of equations
\begin{eqnarray}
z&=& f_0w+f_1+f_2w^{-1}+\cdots +f_Nw^{1-N}\nonumber\\
s&=& \bar f_0w^{-1}+\bar f_1 +\bar f_2 w+\cdots +\bar f_Nw^{N-1}
\nonumber 
\end{eqnarray}
Eliminating $w$ yields an algebraic equation for $z$ and $s$ only:
\begin{eqnarray}
{\bf R}:\quad R(z,s)=0
\label{lgrie}
\end{eqnarray}
with $R$ a polynomial of degree $N$ in both variables, $R(z,s)=\bar
f_N z^N+ f_N s^N+\cdots$.  Eq.(\ref{lgrie}) defines an algebraic curve
${\bf R}$. It is of genus zero since by construction it is uniformized
by points $w$ of the complex sphere. It has many singularities which
have to be resolved to recover a smooth complex manifold.

The Riemann surface ${\bf R}$ may be viewed as a $N$-sheeted covering
of the complex $z$ plane: each sheet corresponds to a determination of
$s$ above point $z$. At infinity, the physical sheet corresponds to
$z\simeq f_0w$ with $w\to\infty$ so that $s\simeq (z/f_0)^{N-1}\,\bar
f_N$, the other $N-1$ sheets are ramified and correspond to $z\simeq
f_N/w^{N-1}$ and $s\simeq \bar f_0/w$ with $w\to 0$ so that $z\simeq
(s/\bar f_0)^{N-1}\, f_N$. Hence infinity is a branch point of order
$N-1$.

By the Riemann-Hurwitz formula the genus $g$ is $2g-2=-2N+\nu$ with
$\nu$ the branching index of the covering. Since the point at infinity
counts for $\nu_\infty=N-2$, there should be $N$ other branch
points generically of order two.  By definition they are determined by
solving the equations $R(z,s)=0$ and $\partial_s R(z,s)=0$. Since the
curve is uniformized by $w\in\mathbb{C}$, these two equations imply
that $z'(w)\partial_zR(z(w),s(w))=0$. Hence either $z'(w)=0$,
$\partial_zR\not=0$, and the point is a branch point, or
$z'(w)\not=0$, $\partial_zR=0=\partial_sR$, and the point is
actually a singular point which needs to be desingularized. So the $N$
branch points at finite distance are the critical points of
the uniformizing map $z=f_t(w)$.

The curve ${\bf R}$ possesses an involution
$(z,s)\to (\bar s,\bar z)$ since $R(\bar s, \bar
z)=\overline{R(z,s)}$ by construction. The set of points fixed by this
involution has two components: (i) a continuous one parametrized by
points $w=u$, $|u|=1$ --this is the real curve $\Gamma_t$ that we
started with-- and (ii) a set of $N$ isolated points which are actually
singular points.

The cusp singularity of the real curve $\Gamma_t$ arises when 
a isolated real point merges with the continuous real slice
$\Gamma_t$. Locally the behavior is as for the curve
$u^2=\varepsilon\, v^2 + v^3$ with $\varepsilon\to 0$.

\begin{figure}[htbp]
\begin{center}
\includegraphics[width=0.8\textwidth]{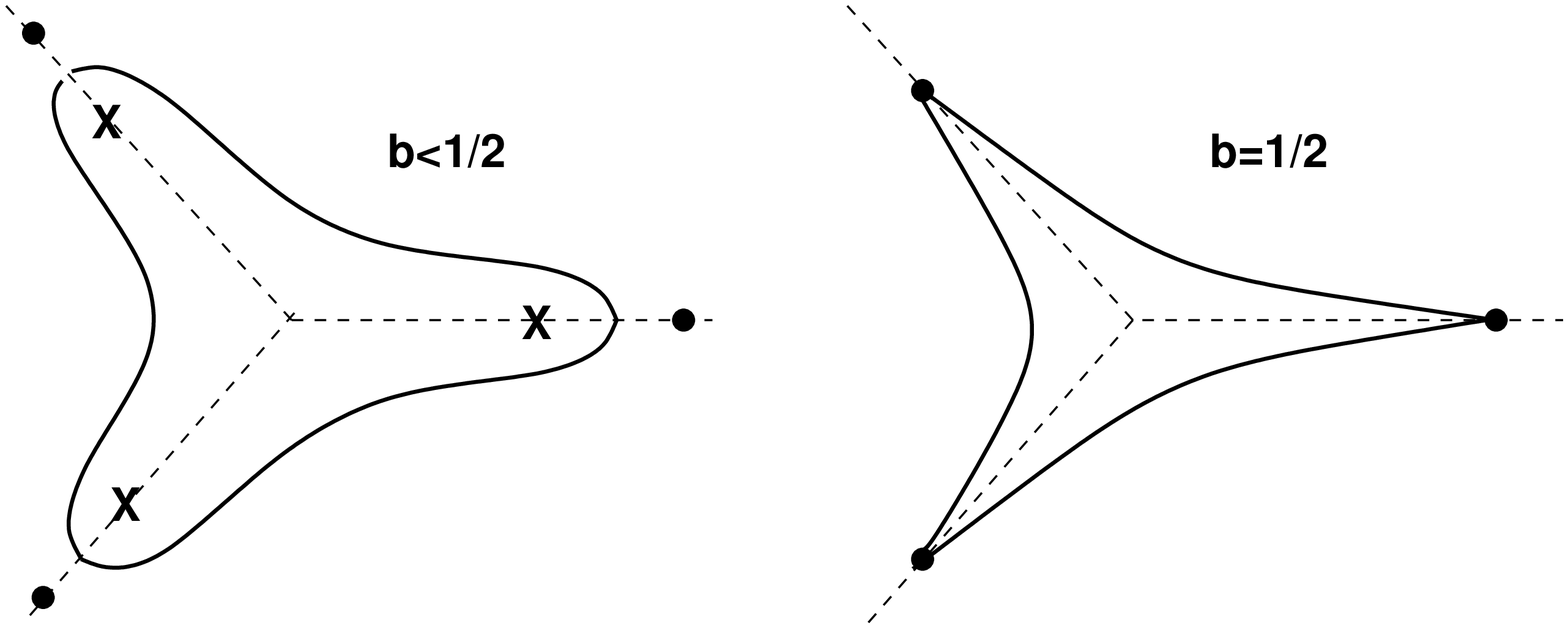}
\caption{Subcritical and critical algebraic curves. Black circles are
  singular points. Crosses are branch points.}
     \label{fig:lgcurve}
 \end{center}
\end{figure}

The simplest example is for $N=3$ with $\mathbb{Z}_3$ symmetry so that
$f_t(w)=w+b/w^2$ and
$$ 
w^2\, z= w^3 +b\quad,\quad w\, s=1+bw^3
$$
We set $f_0=1$ and $f_3=b$.
Without lost of generality we assume $b$ real.
The algebraic curve is then
$$
R(z,s)\equiv bz^3+bs^3 -b^2 s^2z^2 + (b^2-1)(2b^2+1)sz -(b^2-1)^3=0
$$
Infinity is a branch point of order two.  The three other
branch points are at $z=3\omega\,(b/4)^{1/3}$,
$s=\omega^2\,(2b^2+1)(2b)^{-1/3}$ corresponding to
$w=\omega(2b)^{1/3}$ with $\omega$ a third root of unity. They are
critical points of $z(w)$.  There are three singular points at
$z=\omega\,(1-b^2)/b$, $s=\omega^2\, (1-b^2)/b$ corresponding to
$w=\omega(1\pm\sqrt{1-4b^2})/2b$.  The physical regime is for $b<1/2$
in which case the real slice $\Gamma_t=\{z(u),\, |u|=1\}$ is a simple
curve.  The singular points are then in the outer domain and the
branch points in the inner domain. The cusp singularities arise for
$b=1/2$.  For $b>1/2$ there are no isolated singular points, they are
all localized on the real slice so that $\Gamma_t$ possesses double
points. See Fig.\ref{fig:lgcurve}.


\subsection{Discrete iterations}
\label{sec:disciter}
\begin{figure}[htbp]
\begin{center}
\includegraphics[width=0.8\textwidth]{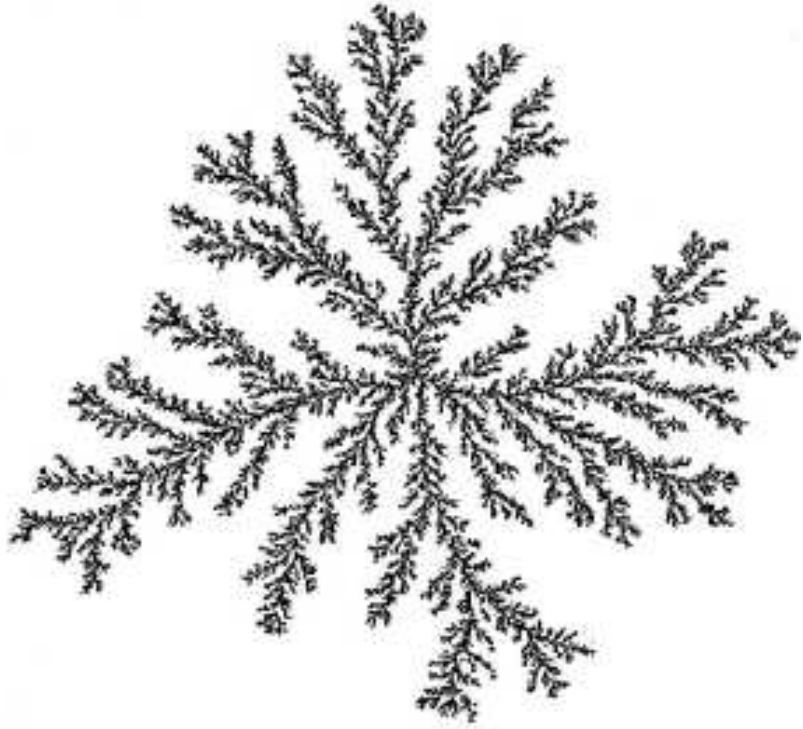}
\caption{An exemple of cluster obtained by iterating conformal
  maps. Notice the similarity with the DLA sample of 
Fig.\ref{fig:dla}.}
     \label{fig:itermap}
 \end{center}
\end{figure}

As proposed in \cite{lgdiscret}, an alternative way to mimic the
gluing of elementary particles as in DLA processes consists in
composing elementary conformal maps, each of which corresponds to
adding an elementary particle to the domain. See Fig\ref{fig:itermap}.
This provides yet another regularization of Laplacian growth.

One starts with an elementary map corresponding to the gluing of a
tiny bump, of linear size $\lambda$, to the unit disk. A large variety
of choices is possible, whose influence on the final structure of the
domain is unclear. An example is given by the following formul\ae\
($g_\lambda$ is the inverse map of $f_\lambda$):
\begin{eqnarray}
g_\lambda (z) &=& z\  \frac{ z\cos\lambda-1}{z-\cos\lambda}\nonumber\\
f_\lambda (w)= &=&
(2\cos\lambda)^{-1}\left[{w+1+\sqrt{w^2-2w\cos2\lambda
      +1}}\right]\nonumber 
\end{eqnarray}
where $f_\lambda$ correspond to the deformation of the unit disk
obtained by gluing a semi-disk centered at point $1$ and whose two
intersecting points with the unit circle define a cone of angle
$2\lambda$. For $\lambda\ll 1$, the area of the added bump is of order
$\lambda^2$.  But other choices are possible and have been used
\cite{discret,Jensen}.

Gluing a bump around point $e^{i\theta}$ on the unit circle is
obtained by rotating  these maps. The uniformizing maps are then
$$ 
f_{\lambda;\theta}(w)= e^{i\theta}\,f_{\lambda}(we^{-i\theta})
$$

\begin{figure}[htbp]
\begin{center}
\includegraphics[width=0.9\textwidth]{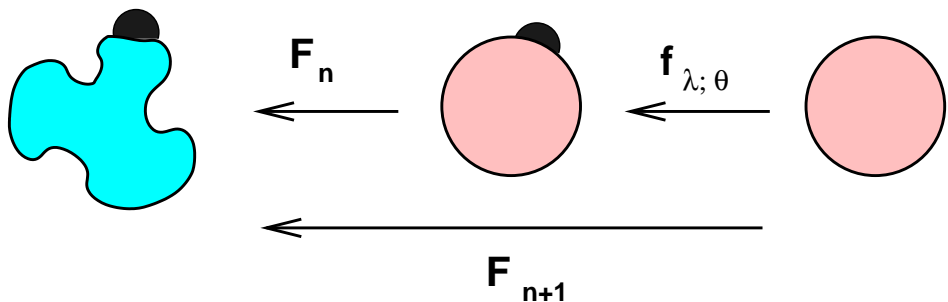}
\caption{Iteration of conformal maps.}
     \label{fig:iteration}
 \end{center}
\end{figure}

The growth of the domain is obtained by successively iterating the maps
$f_{\lambda_n;\theta_n}$ with various values for the size
$\lambda_n$ and the position $\theta_n$ of the bumps. 
See Fig.\ref{fig:iteration}.
Namely, if after $n$ iterations the complement of the unit disk is
uniformized into the complement of  the domain by the map
$F_{(n)}(w)$, then at the next $(n+1)^{\rm th}$ iteration the
uniformizing map is given by:
\begin{eqnarray}
F_{(n+1)}(w)= F_{(n)}(\, f_{\lambda_{n+1};\theta_{n+1}}(w)\, )
\label{iter}
\end{eqnarray}
For the inverse maps, this becomes
$G_{(n+1)}=g_{\lambda_{n+1};\theta_{n+1}}\circ G_{(n)}$.


To fully define the model one has to specify the choice of the
parameter $\lambda_n$ and $\theta_n$ at each iteration. 
Since $\lambda_n$ codes for the linear size of the added bump and since
locally conformal maps act as dilatations, the usual choice is to
rescale $\lambda_{n+1}$ by a power of $|F_{(n)}'(e^{i\theta_n})|$ as:
$$ 
\lambda_{n+1} = \lambda_0\ |F_{(n)}'(e^{i\theta_n})|^{-\alpha/2},
\quad 0\leq \alpha \leq 2
$$
The case $\alpha=2$ corresponds to DLA as the physical area of the
added bump are approximatively constant and equal to $\lambda_0$ at
each iterations.  In the other case, the area of the added
bump scales as $|F_{(n)}'(e^{i\theta_n})|^{2-\alpha}$.

The positions of the added bump are usually taken uniformly
distributed on the mathematical unit circle with a measure
$d\theta/2\pi$.

It is clear that this discrete model with $\alpha=2$ provides a
regularization of Laplacian growth with $\lambda_0$ playing the role
of an ultraviolet cutoff. This may also be seen by looking at the {\it
  naive} limit of a small cutoff. Indeed, a naive expansion as
$\lambda_n\ll 1$ gives that $F_{(n+1)}=F_{(n)}+\delta F_{(n)}$ with
$$
\delta F_{(n)}(w)\simeq \lambda_n\, wF_{(n)}' (w)\,
\left(\frac{w+e^{i\theta_n}}{w-e^{i\theta_n}}\right)
$$
where we used the expression of $f_{\lambda}$ for $\lambda\ll 1$.
Using the recursive expression for $\lambda_n$ and averaging over
$\theta$ with a uniform distribution yields:
$$
\langle \delta F(w) \rangle = \lambda_0\, w F'(w)\,
\oint \frac{d\theta}{2\pi} |F'(e^{i\theta})|^{-\alpha}\,
\frac{w+e^{i\theta}}{w-e^{i\theta}}
$$
For $\alpha=2$ this reproduces the Loewner chain for Laplacian
growth.  But this computation is too naive as the small cutoff limit
is not smooth, a fact which is at the origin of the non trivial
fractal dimensions of the growing domains.

There are only very few mathematical results on these discrete models.
The most recent one \cite{rhode} deals with the simplest (yet
interesting but not very physical) model with $\alpha=0$. It proves the
convergence of the iteration to well-defined random maps uniformizing
domains of Hausdorff dimension $1$. However, these models have been
studied numerically extensively. There exists a huge literature on
this subject but see ref.\cite{discret,Jensen} for instance. These studies
confirm that the fractal dimension of DLA clusters with $\alpha=2$ is
$D_{\rm dla}\simeq 1.71$ but they also provide further informations on
the harmonic measure multi-fractal spectrum.  Results on the $\alpha$
dependence of the fractal dimension may be found eg. in
ref.\cite{hastbis}.

Various generalizations have been introduced. For instance, in
ref.\cite{gener} a model of iterated conformal maps has been defined
in which particles are not added one by one but by layers.  These
models have one control parameter coding for the degree of coverage of
the layer at each iterative step. By varying this parameter the
model interpolates between discrete DLA and a discrete version of the
Hele-Shaw problem. The fractal dimension of the resulting clusters varies
with this parameter \cite{Barra04}.

\subsection{Miscellaneous}

DLA and Laplacian growth have a large domain of applicability since
they are based on simple Brownian diffusion, and a large number of
works have devoted to them. See
\cite{Benetal86,Gollub99,Halsey00,Bazant05,Tanveer00} for reviews. But
precise -- and mathematically clean -- descriptions are unknown. In
particular it is yet not clear whether these descriptions are
universal or not, and numerical simulations point in different
directions. Indeed, Laplacian growth has to be regularized and, as we
discussed, there are different ways to regularize it -- DLA, the
Hele-Shaw problem, the discrete iterations provide different
regularizations. Each of them involve an ultraviolet cut-off, and it
is yet not clear whether the singular limits of vanishingly small
cut-off are identical. If it is, then universality holds. It is
tempting to suppose that a renormalization group inspired approach
should provide a way to answer this question.

Finally, it will be interesting to extend the previous considerations
to fracture related problems, see eg. \cite{fracture}.

%% file: app1.tex

We summarize here some of the basics tools of probability theory with
the aim of filling part of the gap between the mathematic and physics
languages. 

\subsection{Tribes and measurable spaces}

If $\Omega$ is any set, a tribe (or $\sigma$-algebra) on $\Omega$ is
a subset of $2^{\Omega}$ which contains $\Omega$, is stable by
complementation and by denumerable unions. Thus ${\mathcal F} \subset
2^{\Omega}$ is a tribe on $\Omega$ if 

-- $\textit{i}$) $\Omega \in
{\mathcal F}$, \\
-- $\textit{ii}$) if $A \in {\mathcal F}$ its complement
$A^c \in {\mathcal F}$,\\ 
-- $\textit{iii}$) if $A_n\in {\mathcal F}$ for
$n \in {\mathbb N}$ then $\cup_n A_n \in {\mathcal F}$.\\
Two trivial examples are ${\mathcal
  F}=\{\Omega,\emptyset\}$, the trivial tribe, and ${\mathcal
  F}=2^{\Omega}$, the total tribe. Because of $\textit{ii}$),
$\textit{iii}$) can be replaced by \\
-- $\textit{iii}'$) if $A_n\in
{\mathcal F}$ for $n \in {\mathbb N}$ then $\cap_n A_n \in {\mathcal
  F}$,\\ 
i.e. by stability under denumerable intersections.  

An arbitrary intersection of tribes is still a tribe. An arbitrary
subset ${\mathcal S}\subset 2^{\Omega}$ is contained in a smallest
tribe denoted by $\sigma({\mathcal S})$, the intersection of all
tribes containing ${\mathcal S}$. If $\Omega$ is a topological space,
the smallest tribe containing the open sets is called the Borel tribe,
usually denoted by ${\mathcal B}$.

The pair $(\Omega,{\mathcal F})$ is usually refered as a measurable
space. A map $f$ from a measurable space $(\Omega,{\mathcal F})$ to
another measurable space $(\Omega',{\mathcal F}')$ is called measurable
if $f^{-1}(A')\in {\mathcal F}$ whenever $A' \in {\mathcal F}'$.

\vspace{.5cm}

\begin{ex}\label{ex:ct}: coin tossing.\end{ex}

\vspace{-.3cm}

\noindent Think for example of a simple model of a coin tossing game.
Games of length $n\geq 1$ can be modeled to take place in the space
$\Omega_n\equiv \{-1,1\}^{[1,n]}$, where $1$ stands for ``head'' say,
and $-1$ for ``tail''. Thus $\Omega_n$ is made of finite sequences
$X_1,\cdots,X_n$ with values in $\pm 1$.  Infinite games take place in
$\Omega \equiv \{-1,1\}^{{\mathbb N}^*}$, which is made of infinite
sequences $X_1,X_2,\cdots$. Troncation after the $n^{th}$ term gives a
map $\pi_n$ from $\Omega$ onto $\Omega_n$ for each $n \geq 1$. For
each $n$ $2^{\Omega_n}$ is a tribe on $\Omega_n$ and ${\mathcal
  F}_n=\{\pi_n^{-1}(A), A \in 2^{\Omega_n}\}$ is a tribe for $\Omega$,
which conveys intuitively the knowledge of what happens in the first
$n$ toses of the coin. Clearly ${\mathcal F}_1 \subset {\mathcal F}_2
\subset \cdots$ is an increasing sequence of tribes.  The smallest
tribe containing all of them, denoted by ${\mathcal F}$, is larger
than the union $\cup_n {\mathcal F}_n$ (which is not a tribe !). The
subset of $\Omega$ made of sequences in which $1$ appears at least
once is in ${\mathcal F}$, but in no ${\mathcal F}_n$. The same is
true of $\{ \omega \in \Omega, S_n(\omega)/n \; \mathrm {converges}
\}$ where $S_n$ is the sum of the first $n$ steps, $S_n\equiv
X_1+\cdots+X_n$. One way to see it is to write this set as $\cap
_{k=1}^{\infty} \cup_{l=1}^{\infty}\cap_{n>m>l} A_{k,m,n}$ where
$$A_{k,m,n}\equiv \left\{ \omega \in \Omega, 
\left| \frac{S_n(\omega)}{n}-\frac{S_m(\omega)}{m} \right| <
\frac{1}{k}\right\} \in {\mathcal F}_n.$$ 

\vspace{.5cm}

In probability theory, the members of ${\mathcal F}$ are assigned
probabilities in a consistent way (see below), and consistency is one
among many of the reasons to consider other tribes than the total
tribe (see the coin tossing example to find other reasons).

\vspace{.5cm}

\noindent \textit{The statistical mechanics viewpoint.} 

\noindent Let us explain now why tribes remain most of the time behind
the scene in statistical mechanics.

In statistical mechanics, the configuration space $\Omega$ is often
finite (think for example of Ising variables on a finite number of
sites) or sometimes denumerable (as in height models) to start with. 
Then ${\mathcal F}=2^{\Omega}$ will turn out to be a consistent
choice. Taking the thermodynamic limit is in general a nontrivial step
from the point of view of probability theory, but the difficulties are
most of the time of technical nature, and do not need to be adressed
carefull by physicists to get the correct answer to the questions they
are interested in.

If $\Omega$ is finite or countable and ${\mathcal F}$ is a tribe, it
is not difficult to show that there is a finite or countable index set
$I$ and a partition of $\Omega=\cup_{i\in I}\Omega_i$ such that the
members of ${\mathcal F}$ are the unions $\cup_{j\in J}\Omega_j$ when
$J$ runs over the subsets of $I$, i.e. ${\mathcal F}$ is the smallest
tribe containing all the sets of the partition. The $\Omega_i$'s are
just the minimal elements of ${\mathcal F}$ for the inclusion. We say
that $\Omega=\cup_{i\in I}\Omega_i$ is the partition associated to
${\mathcal F}$.


Hence in the context of finite or countable configuration spaces,
there is an equivalence between tribes and partitions. Partitions are the
standard approach of statistical mechanics. An archetypal example in
these notes is to partition $\Omega$ according to the position of (the
beginning of) an interface.

\subsection{Probability spaces and random variables}

A measure space is a triple $(\Omega,{\mathcal F},\mu)$ where
${\mathcal F}$ is a tribe on $\Omega$ and $\mu$ a map from ${\mathcal
  F}$ to $[0,+\infty]$ such that if $A_n$, $n\in \mathbb{N}$ is a 
sequence of disjoint members of
${\mathcal F}$ and $A=\cup_n A_n$ then $\mu(A)=\sum_n \mu(A_n)$ ($\mu$
is said to be countably additive).

Among measure spaces, probability spaces are most important in these
notes, a notable exception being Poisson random measures to be
defined below. 

A measure space $(\Omega,{\mathcal F},p)$ is a probability space if 
$p(\Omega)=1$. 

If $(\Omega,{\mathcal F},p)$ is a probability space and
$(\Omega',{\mathcal F}')$ a measurable space, a random variable
$X$ on $(\Omega,{\mathcal F},p)$ with values $(\Omega',{\mathcal F}')$
is simply a measurable map from $(\Omega,{\mathcal F})$ to
$(\Omega',{\mathcal F}')$. Quite often, random variables take values
in ${\mathbb R}$ endowed with the Borel tribe.

An arbitrary collection of random variables on $(\Omega,{\mathcal
  F},p)$ (with possibly different target spaces) generates a subtribe
of ${\mathcal F}$, namely the smallest tribe for which all random
variables in the collection are measurable.

A random variable $X$ induces a probability $p_X$ on its target space
$(\Omega',{\mathcal F}')$ by $p_X(A')\equiv p(X^{-1}(A'))$, for which
the notation $p(X\in A')$ is also intuitively appealing. This induced
probability is called the probability distribution of $X$.   

\vspace{.5cm}

\begin{ex}\label{ex:pd}: Poisson distribution.\end{ex} 

\vspace{-.3cm}

\noindent Fix $\lambda \in [0,+\infty[$, take $\Omega=\{0,1,\cdots\}$,
${\mathcal F}=2^{\Omega}$ and, for $A\in \Omega$, $p(A)=e^{-\lambda}
\sum_{n\in A} \frac{\lambda^n}{n!}$. It is immediate that
$(\Omega,{\mathcal F},p)$ is a probability space. A slight extension
is when $X$ is a random variable on a probability space
$(\Omega,{\mathcal F},p)$ with values in $\{0,1,\cdots\}$ and
probability distribution as above. This distribution is called the
Poisson distribution of parameter $\lambda$.

\vspace{.5cm}

Whenever $\Omega$ is not countable, probability distributions are
usually defined by specifying probabilities of a simple subclass
${\mathcal S}$ of ${\mathcal F}$ such that ${\mathcal
  F}=\sigma({\mathcal S})$, and using an extension theorem to get a
probability defined on all of ${\mathcal F}$. Doing it by hand for a
special case is painful. The extension theorems work when some
consistency condition holds for the probabilities specified on
${\mathcal S}$. The reader can look at example \ref{ex:fct} and 
refer to the litterature for more details.

\vspace{.5cm}

\begin{ex}\label{ex:fct}: Fair coin tossing.
\footnote{With notations as in example \ref{ex:ct}.}\end{ex}

\vspace{-.3cm}

Take $\Omega \equiv \{-1,1\}^{{\mathbb N}^*}$ with tribe ${\mathcal
  F}=\sigma(\cup_n {\mathcal F}_n)$. Each element $\omega$ of $\Omega$
is an infinite sequence $X_1,X_2,\cdots$, which we can write in a
tautological way $X_1(\omega),X_2(\omega),\cdots$ and the coordinate
maps $\omega \mapsto X_n(\omega)$ are measurable for $n=1,2,\cdots$. By
construction, ${\mathcal F}_n$ is the smallest tribe making
$X_1,\cdots,X_n$ measurable.

Define a probability $p_n$ on ${\mathcal F}_n$ by
$p_n(A)=|\pi_n(A)|/2^n$ for $A \in {\mathcal F}_n$ (recall that
$\pi_n(A)$ is a subset of $\Omega _n$ and $|\Omega _n|=2^n$).

The probabilities $p_n$ are consistent in the
following way : if $A \in {\mathcal F}_m$ and $n\geq m$ then $A \in
{\mathcal F}_m$ and $p_n(A)=p_m(A)$. So we can assemble the $p_n$'s
into a function $p$ on ${\mathcal S}\equiv \cup_n {\mathcal F}_n$.

If $A_k$ is a sequence of disjoint elements of ${\mathcal S}$ such
that $\cup_k A_k$ is again in $ {\mathcal S}$, then $p(\cup_k
A_k)=\sum_k p(A_k)$. The proof is simple if tedious. This condition is
enough to garanty a consistent extension of $p$ to a probability on
${\mathcal F}=\sigma({\mathcal S})$. This is one of the useful
extension theorems in the field. 

\vspace{.5cm}

\begin{ex}\label{ex:ud}: The uniform distribution.\end{ex}

\vspace{-.3cm}

\noindent Take $\Omega=[0,1]$ with the Borel tribe ${\mathcal B}$. Le length
$b-a$ of an open interval $]a,b[$, $0 \leq a < b \leq 1$, can be
extended in a unique way to a probability measure on
$(\Omega,{\mathcal B})$, called the uniform distribution, which is nothing
but the well-known Lebesgue measure.

\vspace{.5cm}

Consider the map $f$ from $\{-1,1\}^{{\mathbb N}^*}$ to $[0,1]$
defined by $f(X_1,X_2,\cdots)=\sum_{n>0} b_n/2^{n},$ where $b_n\equiv
(X_n+1)/2\in \{0,1\}$ The sequence $(b_1,b_2,\cdots)$ is simply the
binary expansion\footnote{Recall that the dyadic rationals have in
  fact two binary expansions, but this in not a problem from the
  probabilistic viewpoint because they form a set of probability $0$.}
of the real number $f(X_1,X_2,\cdots)$. As a consequence, this map is
such that if $A \in \cup_n {\mathcal F}_n$, the image $f(A)$ is a
finite union of closed intervals and the Lebesgue measure of $f(A)$
coincides with $p(A)$. This indicates that from a probabilistic
viewpoint $([0,1],{\mathcal B},dx)$ and $(\{-1,1\}^{{\mathbb
    N}^*},{\mathcal F},p)$ are essentially indistiguishable. In fact,
one can show that $(\{-1,1\}^{{\mathbb N}^*},{\mathcal F},p)$ is in
some precise sense equivalent as a probability space to any
nondenumerable probability space. Let us give two modest
illustrations. If $d\geq 2$ is and integer, on can by split a sequence
$X=(X_1,X_2,\cdots)$ in $d$ sequences,
$X^{(1)}=(X_1,X_{d+1},X_{2d+1},\cdots),\cdots,
X^{(d)}=(X_d,X_{2d},X_{3d},\cdots)$ to show quickly that
$([0,1],{\mathcal B},dx)$ and $([0,1]^d,{\mathcal B},d^dx)$ are one
and the same probability space. One can also split a sequence
$X=(X_1,X_2,\cdots)$ into a denumerable family of sequences
$X^{(1)}=(X_1,X_3,X_5,\cdots),X^{(2)}=(X_2,X_6,X_{10},\cdots),
X^{(3)}=(X_4,X_{12},X_{20},\cdots),\cdots$, a fact Wiener used for its
original definition of Brownian motion (see below).

\vspace{.5cm}
 
\begin{ex}\label{ex:gd}: The Gaussian distribution.\end{ex}

\vspace{-.3cm}

\noindent Take $\Omega={\mathbb R}$ with the Borel tribe ${\mathcal
  B}$, and define $p(A)= \int_A \frac{dx}{\sqrt{2\pi}} e^{-x^2/2}
\equiv \int {\mathbf 1}_A \frac{dx}{\sqrt{2\pi}} e^{-x^2/2}$ for $A\in
{\mathcal B}$.  This is the standard Gaussian distribution. If $X$ is
a random variable which is either constant or such that $(X-b)/a$
follows the standard Gaussian distribution for some real numbers $a
\neq 0$ and $b$, $X$ is called a Gaussian random variable. When $b=0$,
$X$ is called a centered Gaussian random variable.

If $X_1,\cdots,X_n$ are real random variables, the vector
$(X_1,\cdots,X_n)$ is called Gaussian if any linear combination
$\sum_i c_iX_i$ is a Gaussian random variable.

For instance, if $(U,V)$ is uniformly distributed in the unit square
$[0,1]^2$, one can check that $(\log U \cos 2\pi V, \log U \sin 2\pi
V)$ is a Gaussian vector. In fact the two components are
independent\footnote{The general notion of independence is recalled
  below.}  standard Gaussian random variables. Combining this with our
remarks on the ''size'' of the probability space of fair coin tossing,
one sees that it can accomodate a countable family of independent
standard Gaussian random variables.

\vspace{.5cm}

\noindent \textit{The statistical mechanics viewpoint.} 

\noindent In statistical mechanics, an energy function $E$ on
the (finite or denumerable) space $\Omega$ is given, and there is a simple
formula for the relative probability of $\omega$ and $\omega'$ at
temperature $T=1/\beta$: 
\[p(\omega)/p(\omega')= e^{\beta(E(\omega')-E(\omega))}.\]
The partition function $Z=\sum_{\omega \in \Omega}e^{-\beta
  E(\omega)}$ gives the normalization of the probability.  If
$Z<+\infty$, this defines a probability on $(\Omega,2^{\Omega})$.

A real random variable is any
function from $\Omega$ to ${\mathbb R}$, also called an observable.
If we look at another tribe ${\mathcal F}$ with associated
partition $\Omega=\cup_{i\in I}\Omega_i$, a
random variable for $(\Omega,{\mathcal F})$ is a function from
$\Omega$ to ${\mathbb R}$ constant on each $\Omega_i$.

\subsection{(Conditional) expectations}

Let $(\Omega,{\mathcal F},p)$ be a probability space.

Finite sums and products of random variables with values in
$(\mathbb{R},{\mathcal B})$ are again random variables. A useful
quantity associated to a real random variable is its average, usually
called expectation in the probabilistic context. It is defined at
first only for so called simple random variables, those which can be
written as $X=\sum_{i=1}^n x_i{\mathbf 1}_{A_i}$ for some integer $n$,
real numbers $x_i$ and measurable sets $A_i \in {\mathcal F}$ for
$i=1,\cdots,n$. This decomposition is in general not unique, but the
expectation, defined by $\Expect{X} \equiv \sum_{i=1}^n x_ip(A_i)$
can be shown to be well defined. If $\Omega$ is finite, every random
variables is simple. In the other cases, one tries to approximate more
general random variables by simple ones, and define the expectation by
a limiting procedure. For instance, if $\Omega$ is countable, then any
tribe ${\mathcal F}$ is the smallest tribe containing all sets of a
certain partition $\Omega=\cup_{i\in I}\Omega_i$ into a finite or
countable number of pieces. The most general random variable can
uniquely be written $X=\sum_{i\in I} x_i{\mathbf 1}_{\Omega_i}$. The
limiting procedure allows to define the expectation of $X$ under the
condition $\sum_{i\in I} |x_i|p(\Omega_i) <\infty$ by the formula
$\Expect{X} \equiv \sum_{i\in I} x_ip(\Omega_i)$, a formula wich
could also be taken as a definition in this simple case. In the
general case, an expectation with values in $[0,+\infty]$ can be
defined for any positive random variable, and $\Expect{X}$ can be
defined if $\Expect{|X|}<+\infty$.

\vspace{.5cm}

\noindent \textit{The statistical mechanics viewpoint.} 

\noindent If $\Omega$ is countable, we can consider the tribe
${\mathcal F}=2^{\Omega}$. A real random variable, or observable, is a
function $X$ from $\Omega$ to ${\mathbb R}$ and 
\[\Expect{X} \equiv \left< X \right> \equiv \frac{1}{Z}
\sum_{\omega \in \Omega}X(\omega) e^{-\beta E(\omega)},\]
whenever the sum is absolutely convergent.

\vspace{1cm}

The reader is probably familiar with the notion of conditional
probability : if $(\Omega,{\mathcal F},p)$ is a probability space,
$A,B \in {\mathcal F}$ and $p(B)\neq 0$ the probability of $A$ given
that $B$ occurs (or simply the probability of $A$ given $B$) is
defined to be $p(A|B)\equiv p(A\cap B)/p(B)$. The events $A$ and $B$
are called independent if $p(A\cap B)=p(A)p(B)$ and then
$p(A|B)=p(A)$. Hence conditional probabilities and independence convey
the correct intuitive meaning. 

Independence can be formulated at different levels.  The events of a
family $\{A_{\alpha}, \alpha \in I\}$ are called independent if
$p(\cap_J A_{\alpha})=\prod_J p(A_{\alpha})$ for any finite subset $J$
of $I$.  The tribes $\{{\mathcal F}_{\alpha}, \alpha \in I\}$ are
called independent if the events $\{A_{\alpha}, \alpha \in I\}$ are
independent whenever $A_{\alpha}\in {\mathcal F}_{\alpha}$ for all
$\alpha$'s in $I$. The random variables $\{X_{\alpha}, \alpha \in I\}$
are called independent if the tribes $\sigma(X_{\alpha})$ they
generate are independent. If moreover the functions $\{f_{\alpha},
\alpha \in I\}$ are measurable functions from $\mathbb R$ to $\mathbb
R$ such that $\Expect{|f_{\alpha}(X_{\alpha})|}<+\infty$ for $\alpha \in I$
and $J$ is a finite subset of $I$, then
$\Expect{(\prod_Jf_{\alpha}(X_{\alpha}))}=\prod_J
\Expect{(f_{\alpha}(X_{\alpha}))}$.  
Conversely, this multiplicative property for all measurable functions
from $\mathbb R$ to $\mathbb R$ such that
$\Expect{|f_{\alpha}(X_{\alpha})|}<+\infty$ for $\alpha \in I$ ensures that
the random variables $\{X_{\alpha}, \alpha \in I\}$ are independant. This
is easy to check for simple random variables\footnote{Variables of the
  type $X=\sum_{i=1}^n x_i{\mathbf 1}_{A_i}$ where one can assume
  without loss of generality that the the $x_i$'s are distinct. Then
  $\sigma(X)$ is simply the finite tribe generated by the $A_i$'s.}.

\vspace{.5cm}
 
\begin{ex}\label{ex:kslln}: The Kolmogorov strong law of large numbers.\end{ex}

\vspace{-.3cm}

Let $X_n$, $n=1,2,\cdots$ is a sequence of real independant
identically distributed random variables on $(\Omega,{\mathcal F},p)$
with partial sums $S_n=X_1+\cdots,X_n$, $n\geq 1$.

-- If $\Expect{|X_n|}<+\infty$ and $\Expect{X_n}=\mu$, the sequence
$S_n/n$ converges to $\mu$ almost surely (i.e. the subset of $\Omega$
such that $S_n/n$ does not converges to $\mu$ fits into elements of
${\mathcal F}$ of arbitrary small probability). 

-- If $\Expect{|X_n|}=+\infty$, the sequence $S_n/n$ diverges almost
surely.

\vspace{.5cm}

\begin{ex}\label{ex:prm}: Poisson random measures.\end{ex}

\vspace{-.3cm}

\noindent If $(\Omega,{\mathcal F},\mu)$ is a measure space and
${\mathcal F}_0\equiv \{B \in {\mathcal F} \; | \; \mu(B) < \infty \}$
there exits a collection of integer valued random variables $\{N_B\; |
B\in {\mathcal F}_0\}$ such that

-- $\textit{i}$) $N_B$ is a Poisson random variable with mean $\mu(B)$,

-- $\textit{ii}$) if $B_1,\cdots,B_n \in {\mathcal F}_0$ are disjoint,
the variables $N_{B_1},\cdots,N_{B_n}$ are independent,

-- $\textit{iii}$) if $B,B' \in {\mathcal F}_0$,
$Cov(N_B,N_{B'})\equiv
\Expect{N_BN_{B'}}-\Expect{N_B}\Expect{N_{B'}}=\mu(B\cap B')$.\\
This collection is called the Poisson random measure on
$(\Omega,{\mathcal F},\mu)$. Intuitively, a sample is a collection of
points in $\Omega$, the random variables $N_B$ describe the number of
points in region $B$, which follows a Poisson distribution. Disjoint
regions are idependent. Conditions $\textit{i}$), $\textit{ii}$),
$\textit{iii}$) ensure that the number of points in a disjoint union
is (almost surely) the sum of the numbers of points in each separate
region.

\vspace{.5cm}

This notions of expectations and conditional probabilities are
combined in a very useful way in the concept of conditional
expectation.

Let $X$ be an ${\mathcal F}$ random variable with
$\Expect{|X|}<+\infty$ and ${\mathcal F}'$ be a subtribe of ${\mathcal
  F}$. A conditional expectation of $X$ given ${\mathcal F}'$ is an
${\mathcal F}'$ measurable random variable $Y$such that
$\Expect{|Y|}<+\infty$ and
\[\Expect{X{\mathbf 1}_{A}} =\Expect{Y{\mathbf 1}_{A}}\] 
for any $A\in{\mathcal F}'$. The notation $Y=\Expect{X|{\mathcal F}'}$
is standard. Let us stress that $\Expect{X|{\mathcal F}'}$ is by definition
${\mathcal F}'$ measurable. The above definition is not a
constructive, but it turns out to be a characterization which is most
useful to work with conditional expectations. The general abstract
argument for the existence of a conditional expectation
$\Expect{X|{\mathcal F}'}$ relies on the Radon Nykodim theorem or on
projections in Hilbert spaces of square integrable random variables,
i.e. on cornerstones of measure theory (see \cite{dudley}).

Note that if $X$ is ${\mathcal F}'$ measurable, then $X$ itself
satisfies the properties of $\Expect{X|{\mathcal F}'}$.  One can
also see that if ${\mathcal F}''$ is a subtribe of ${\mathcal F}'$,
\[\Expect{\Expect{X|{\mathcal F}'}|{\mathcal F}''}
=\Expect{\Expect{X|{\mathcal F}''}|{\mathcal F}'}=\Expect{X|{\mathcal
    F}''}.\] Thus, when conditional expectations are nested, the
smallest tribe wins.

More modestly, we construct conditional expectations in the case when
$\Omega$ is finite or countable, so that ${\mathcal F}$ is associated
to a finite or countable partition $\Omega=\cup_{i\in I}\Omega_i$.
Suppose that ${\mathcal F}'$ is a subtribe of ${\mathcal F}$.  Then
$I$ can be partitionned into a finite or countable number of pieces
$I=\cup_{j\in J} I_j$ in such a way that ${\mathcal F}'$ is the
smallest tribe containing all $\Omega'_j \equiv \cup_{i\in
  I_j}\Omega_i$. If $X=\sum_{i\in I} x_i{\mathbf 1}_{\Omega_i}$ is a
random variable with expectation (i.e.  $\sum_{i\in I}
|x_i|p(\Omega_i) <\infty$ as above), and $X'= \sum_{j\in
  J}x'_j{\mathbf 1}_{\Omega'_j}$ is an ${\mathcal F}'$ measurable
random variable, $\Expect{X{\mathbf 1}_{\Omega'_j}}=
\Expect{X'{\mathbf 1}_{\Omega'_j}}$ says that $p(\Omega'_j)x'_j\equiv
\sum_{i\in I_j} x_ip(\Omega_i)$.  As $p(\Omega'_j)=\sum_{i\in
  I_j}p(\Omega_i)$, this formula fixes $x'_j$ if $p(\Omega'_j)\neq 0$
but leaves the value of $x'_j$ undetermined if $p(\Omega'_j)=0$. Then,
for any choice of the $x'_j$ satisfying the above conditions and , for
$A\in{\mathcal F}'$, $\Expect{X{\mathbf 1}_{A'}}= \Expect{X'{\mathbf
    1}_{A'}}$, i.e.  $X'$ is a conditional expectation of $X$ given
${\mathcal F}'$. So conditional expectations exist, in general they
form a class of ${\mathcal F}'$-measurable random variables that
coincide except on a set of probability $0$.

\vspace{.5cm}

\noindent \textit{The statistical mechanics viewpoint.} 

\noindent In statistical mechanics, the starting point is not absolute
probabilities, but relative probabilities. This implies the use of
partition functions, and trivializes the notion of conditional
expectations, in that physicists manipulate them all the time without
ever giving them a special name.

If $\Omega$ is countable, we can consider the tribe ${\mathcal
  F}=2^{\Omega}$ associated with a partition of $\Omega$ into
singletons. The context may dictate to split $\Omega$ in larger
pieces. For instance, in the Ising model, we may compare different
possible boundary conditions, and partition $\Omega$ accordingly.  Or
as in these notes, fix boundary conditions that imply the presence of
an interface and partition the configuration space according to (part
of) the position of the interface. As a last illustration in the
context of the renormalization group, we may split $\Omega$ according
to the value of the magnetization of blocks of spin. All these
contexts lead to a partition $\Omega=\cup_{j\in J}\Omega'_j$ with
associated tribe ${\mathcal F}'$, partial partition functions
$Z_j=\sum_{\omega \in \Omega'_j} e^{-\beta E(\omega)}$, and, if $X$ is
any observable, partial averages $$\left< X \right>_j\equiv
\frac{1}{Z_j}\sum_{\omega \in \Omega'_j}X(\omega) e^{-\beta
  E(\omega)}.$$ The random variable taking the constant value $\left< X
\right>_j$ on $\Omega'_j$ is nothing but the conditional expectation
of $X$ given ${\mathcal F}'$.

It should be clear from the examples that conditional expectations
are a general framework for all situations when one want to
concentrate on certain degrees of freedom and average over the others. 
In some sense, the statistical mechanics framework is the most
symmetrical, in that absolute probabilities are only a secondary concept,
so that conditioning is transparent. Except for the special role
played by the temperature which in principle appears in the Boltmann
weight but not in the energy function, nothing indicates
that $\Omega$ itself and the associated Boltzmann weights have not been
obtained by a previous conditioning. 

\subsection{Martingales and stopping times : discrete setting}

If $(\Omega,{\mathcal F},p)$ is a probability space, an increasing
sequence ${\mathcal F}_n,n=0,1,2,\cdots$ of subtribes of ${\mathcal F}$
is called a filtration. A sequence of random variables is also called
a (random or stochastic) process. Most often, the target is the same,
for all terms in the sequence. If the target is ${\mathbb R}$ one
talks of a real process. 

\noindent Given such a filtration, 

-- a sequence of random variables $A_n,n=0,1,2,\cdots$ is adapted if
$A_n$ is ${\mathcal F}_n$ measurable for each $n$.

-- a real adapted process $M_n,n=0,1,2,\cdots$ is a martingale if
$\Expect{|M_n|}<\infty$ for each $n$ and $\Expect{M_n|{\mathcal
  F}_m}=M_m$ for $0< m <n$. Note that this condition by itself implies
that the sequence $M_n$ is adapted, but for the problem that
conditional expectations have ambiguities (on sets of measure zero).

-- a real random variable $T$ with values in $0,1,2,\cdots,+\infty$ is said to
be a stopping time if the event $T\leq n$ is in ${\mathcal F}_n$ for
each $n$, or equivalently (in this discrete setting) if the event $T=
n$ is in ${\mathcal F}_n$ for each $n$. It is an exercise to show that
${\mathcal F}_T \equiv\{A\in {\mathcal F} : A\cap \{T \leq n\} \in
{\mathcal F}_n \; \mathrm{for} \; \mathrm{each} \; n\} $ is a tribe that
summarizes the information ''collected up to $T$''.

\vspace{.2cm}

One should view the parameter $n$ as a discrete time, ${\mathcal F}_n$
as the knowlege accumulated up to time $n$. An adapted sequence is one
whose description at time $n$ does not require knowledge of the
future. A martingale is such that its expectation in the future given
the knowledge accumulated up to now is its present value. A stopping
time is a random time for which the present knowledge is enough to
decide if it has occurred in the past. Note that if $h$ is an
increasing map from $\mathbb N$ to $\mathbb N$ and $X_n$ is adapted
(resp. a martingale) for ${\mathcal F}_n$, then $X_{h(n)}$ is adapted
(resp. a martingale) for the filtration ${\mathcal F}_{h(n)}$.

 From the general rule of nesting of expectations, ${\mathbb
  E}M_n=\Expect{\Expect{M_n|{\mathcal F}_0}}$, so if the
sequence $M_n$ is a martingale,  $\Expect{M_n}=\Expect{M_0}$ :
martingales are time independant in average. 

If $X_n$ is a sequence of random variables and $N$ is a real random
variable with values in $0,1,2,\cdots$, one can construct a new random
variable $X_N$ by setting $X_N(\omega)\equiv X_{N(\omega)}(\omega)$ for
$\omega \in \Omega$ , or equivalently, $X_N=\sum_n X_n{\mathbf
  1}_{N=n}$.

Part of the usefulness of martingales comes from the following : if
$M_n,n=0,1,2,\cdots$ is a martingale, and $T$ is a bounded stopping
time (i.e. there is an integer $m$ such that $T$ takes values in
$0,1,\cdots,m$), $\Expect{M_T}=\Expect{M_0}$. The proof is simple
and instructive. If $T\leq m$
\begin{eqnarray*} \Expect{M_T} & = & \Expect{
  \sum_{n=0}^{m}M_n{\mathbf 1}_{T=n}}
  =  \sum_{n=0}^{m}\Expect{M_n{\mathbf 1}_{T=n}}\\
  & = & \sum_{n=0}^{m}\Expect{\Expect{M_m|{\mathcal F}_n}
  {\mathbf 1}_{T=n}}\\
  & = & \sum_{n=0}^{m}\Expect{M_m{\mathbf 1}_{T=n}}
   =  \Expect{M_m\sum_{n=0}^{m}{\mathbf 1}_{T=n}}\\
  & = & \Expect{M_m}
   =  \Expect{M_0}.
\end{eqnarray*}
The first equality is the definition of $M_T$, the third is the
martingale property of the sequence 
$M_n$, the fourth is the definiting property of conditional
expectations because ${\mathbf 1}_{T=n}$ is ${\mathcal F}_n$
measurable. The other equalities are obvious. 

This result can be adapted to deal with unbounded stopping times, as
we shall see in applications. 

Another use of martingales is that they allow to define new
probability distributions. Suppose $M_n$ is a martingale such that
$M_0=1$ and $M_n >0$ (with probability $1$) for $n\geq 0$. If $X$ is
an ${\mathcal F}_n$ measurable random variable for some $n$, define
$\tilde\Expect{X}\equiv \Expect{XM_n}$. This is a consistent
definition because the martingale property ensures that
$\tilde\Expect{XM_n}$ is the same for all $n$'s such that $X$ is
${\mathcal F}_n$ measurable. In the same spirit, If $A \in {\mathcal
  F}_n$, define $\tilde{p}(A)\equiv \Expect{{\mathbf 1}_AM_n}$.
This defines a consistent family of probability distribution on
$(\Omega,{\mathcal F}_n)$. Under technical growth conditions for $M_n$
in $n$ that we shall not try to make precise, $\tilde{p}$ extends to
a probability measure on $(\Omega,\sigma(\{{\mathcal F}_n\})$. Note
that this tribe may be strictly smaller than ${\mathcal F}$.

We start by illustrating these concepts for the
simple random walk and then turn to statistical mechanics.

\vspace{.5cm}

\begin{ex}\label{ex:mfct}: Martingales and fair coin
  tossing.\footnote{With notations as in examples
    \ref{ex:ct},\ref{ex:fct}.}\end{ex}

\vspace{-.3cm}

Recall $\Omega \equiv \{-1,1\}^{{\mathbb N}^*}$ is a space of infinite
sequences $X_1,X_2,\cdots$. Set $S_0=0$ and $S_n=S_{n-1}+
X_n=X_1+\cdots+X_n$ for $n \geq 1$. The tribe ${\mathcal F}_n$ is the
smallest tribe making $X_1,\cdots,X_n$ all measurable, and an
${\mathcal F}_n$ measurable random variable is simply an arbitrary
function of $X_1,\cdots,X_n$, defined on $\{-1,1\}^n$.

With the probability defined in \ref{ex:fct}, chosen to model
intuitively independent tosses of a fair coin, the $X_n's$ are easily
checked mathematically to be independent random variables, and
$\Expect{X_n|{\mathcal F}_m}=0$ for $m<n$.

An adapted process is simply a sequence $A_n=f_n(X_1,\cdots,X_n)$
where $f_n$ is a function on $\{-1,1\}^n$.  As $X_{n+1}^2=1$, the most
general function of $X_{n+1}$ can be written in a unique way as
$aX_{n+1}+b$. Hence,
$f_{n+1}(X_1,\cdots,X_n,X_{n+1})-f_n(X_1,\cdots,X_n)$ can be written
in a unique way as
\begin{eqnarray*}
f_{n+1}(X_1,\cdots,X_n,X_{n+1})-f_n(X_1,\cdots,X_n) &=&\\
 & & \hspace{-3cm} g_n(X_1,\cdots,X_n)X_{n+1}+h_n(X_1,\cdots,X_n)
\end{eqnarray*}
More abstractly, any adapted process $A_n$ can be defined recursively
in a unique way by a formula $A_{n+1}-A_n=B_nX_{n+1}+C_n$ where $B_n$
and $C_n$ are adapted processes. This leads us to the topic of
stochastic difference equations.

Introducing the notation $\Delta U_n\equiv U_{n+1}-U_n$ for finite
differences, we note that $\Delta S_n=X_{n+1}$ and $\Delta n=1$, so
that the above equation can be rewritten $\Delta A_n = B_n\Delta S_n +
C_n\Delta n$, which is equivalent to
\[A_n=A_0+\sum_0^{n-1}B_m\Delta S_m+\sum_0^{n-1}C_m\Delta m.\]
The second sum looks very much like a
Riemann-Stiljes sum, but the first one is of another nature because
$\Delta S_m$ oscillates. In the discrete setting, this is harmless,
but a good version for continuous time requires the construction of a
new integral, the It\^o integral. Integrals are amenable objects mostly
due to the change of variable formula and integration by parts. Let us
look at their discrete counterparts. Suppose $k_n$ is a sequence of
functions from $\mathbb R$ to $\mathbb R$ and look at the process
$k_n(A_n)$. The outcome is 
\[\Delta k_n(A_n)=B^{(k)}_n\Delta S_n +
C^{(k)}_n\Delta n\] where
\[B^{(k)}_n=\frac{k_{n+1}(A_n+B_n+C_n)-k_{n+1}(A_n-B_n+C_n)}{2},\]
 which
looks like a first order derivative, and
\begin{eqnarray*} C^{(k)}_n & = & (k_{n+1}(A_n+C_n)-k_n(A_n))\\
  & + & \frac{k_{n+1}(A_n+B_n+C_n)
    -2k_{n+1}(A_n+C_n)+k_{n+1}(A_n-B_n+C_n)}{2} 
\end{eqnarray*}
which looks like the sum of a first derivative due to $\Delta n$ in
the original equation and the explicit $n$ dependence in $k_n$, and a
second derivative due to the oscillating nature of $\Delta S_n$ tamed
by the fact that $(\Delta S_n)^2=1$. We could  pedantically call this
the discrete It\^o formula. The serious It\^o formula relies heavily on a
continuous time counterpart of $(\Delta S_n)^2=1$. We shall come back
to this later.  For instance, if $\Delta A_n=\alpha A_n \Delta S_n $,
one finds $\Delta \log A_n=\frac{1}{2}\log \frac{1+\alpha}{1-\alpha}
\Delta S_n +\frac{1}{2}\log (1-\alpha^2) \Delta n$, leading to $\log
\frac{A_n}{A_0}=\frac{1}{2}\log \frac{1+\alpha}{1-\alpha} S_n
+\frac{1}{2}\log (1-\alpha^2) n$.  In the same way, if $ A^{(i)}_n$,
$i=1,2$ are adapted processes, and $A_n=A^{(1)}_nA^{(2)}_n$, one finds
that $\Delta A_n=B_n\Delta S_n + C_n\Delta n$ with $B_n=(A^{(1)}_n+
C^{(1)}_n)B^{(2)}_n+B^{(1)}_n(A^{(2)}_n+ C^{(2)}_n)$ and
$C_n=A^{(1)}_nC^{(2)}_n+C^{(1)}_nA^{(2)}_n+B^{(1)}_nB^{(2)}_n$. For
instance $\Delta S_n^2=2 S_n\Delta S_n +\Delta n$.

 Stochastic difference equations can be used in several ways. On the
one hand, one can take $B_n$ and $C_n$ in full generality as given
function of $X_1,\cdots,X_n$. On the other hand, it causes no harm and
can be very useful to add a dependence in $A_1,\cdots,A_n$ in $B_n$
and $C_n$, because then the recursion relation itself ensures that
indeed $B_n$, $C_n$ and $A_n$ are adapted. We shall give illustrations
below. An important example is when $B_n$ and $C_n$ are given functions
$b_n$ and $c_n$ of $A_n$, and $A_{n+1}-A_n=b_n(A_n)X_{n+1}+c_n(A_n)$.
This defines an adapted Markov process which is called a diffusion.

In particular, ${\mathbb
  E}(A_{n+1}|{\mathcal F}_n)=A_n+C_n$. So the process $A_n$ cannot be
a martingale unless $C_n=0$, i.e. the process $C_n$ vanishes. The law
for nesting conditional expectations (the smallest tribe wins) ensures
recursively that this is also sufficient condition.

Hence, we have obtained a characterization of martingales in fair coin
tossing : the sequence $M_n$ is a martingale if and only if $M_0$ is a
constant and there is an adapted process $B_n$ such that
$M_{n+1}-M_n=B_nX_{n+1}$ for $n \geq 0$. Again $B_n$ can be viewed
either as an explicit function of $X_1,\cdots,X_n$ or as defined
implicitely via the recursion. 

A few examples will illustrate this flexibility better than words.

If we take $B_n=1$ for $n \geq 0$, and $M_0=0$ we find that $M_n=
S_n\equiv \sum_{m=1}^n X_m$ is a martingale. If we take $B_n=2S_n$ for
$n \geq 0$, and $M_0=0$, we find that $M_n= S_n^2-n$ is a martingale.
In these two examples, $B_n$ was given apriori as a function of
$X_1,\cdots,X_n$. Now fix a constant $\alpha$, set $M_0=1$ and take
$B_n=\alpha M_n$ for $n \geq 0$. Then, it is the difference equation
itself that ensures that $B_n$ is adapted and $M_n$ is a martingale.
The solution to $M_{n+1}-M_n=\alpha M_nX_{n+1}$ is $M_n=\prod_{m=1}^n
(1+\alpha X_n)$. Write $\alpha =\tanh H$ ($H$ may be complex) to get
$M_n=e^{HS_n-n\log \cosh H}$. Observe that $M_n=1+H
S_n+\frac{H^2}{2}(S_n^2-n)+O(H^3)$ at small $H$, so that the last
example contains the previous two.  In these three examples, we have
not proved that $\Expect{|M_n|} < +\infty$ but this is obvious because
$M_n$ is bounded on $\Omega$ for fixed $n$.

\vspace{.5cm}

\begin{ex}\label{ex:rpmfct}: The ruin problem, martingales and fair coin
  tossing.\footnote{With notations as in examples
    \ref{ex:ct},\ref{ex:fct},\ref{ex:mfct}.}\end{ex}

\vspace{-.3cm}

One of the standard problems in fair coin tossing is the ruin problem.
Fix two integers $a,b$ such that $a < 0 <b$. If $-a$ and $b$ are the
initial fortunes of two gamblers and $S_n$ is the gain of the first
gambler, then he is ruined if $S_n$ reaches $a$ before it reaches $b$.
Let $T$ be the first time at which $S_n$ reaches $a$ or $b$.  Because
$\{T\leq n\}=\cup_{m=1}^n \{S_m=a\}\cup \{S_m=b\}$ and $\{S_m=a\}\cup
\{S_m=b\}\in {\mathcal F}_m \subset {\mathcal F}_n$ for $m\leq n$, $T$
is a stopping time. Obviously, $T$ is not a bounded stopping time.
However, $p(T\geq n)$ can be shown to decreases at least exponentially
in $n$. Indeed, set $c=b-a$ and choose an integer $m$ such that $m\geq
c/2$. If $I$ is any interval of length $c$ and one starts the random
walk somewhere in $I$, making $m$ steps in the appropriate direction
takes the walk to the boundary or out of $I$, so if
$k$ is an integer and $n\geq km$, $p(T\geq n)\leq (1-2^{-m})^k$. In
particular, $p(T=+\infty)=0$.

If we forget about the fact that $T$ is unbounded, we get
\[\Expect{S_T}=\Expect{S_T^2-T}=0,\]
 i.e.
\[ ap(S_T=a)+bp(S_T=b)=a^2p(S_T=a)+b^2p(S_T=b)-\Expect{T}=0\] 
which combined with $p(S_T=a)+p(S_T=b)=1$ yields 
\[p(S_T=a)=\frac{b}{b-a} \quad p(S_T=b)=\frac{-a}{b-a} \quad 
\Expect{T}=-ab.\]
These results are indeed correct, but we have not justified them, and
indeed, they indicate that some care is needed. Let
$T'$ be the first time at which $S_n$ reaches $b$. Again, $T'$ is a
stopping time. Moreover, $p(T'=+\infty)\leq p(S_T=a)$ for any $a$ so
$p(T'=+\infty)=0$ : the walk $S_n$ goes through $b$ with probability
$1$. But $S_{T'}=b$ so obviously $E(S_T')=b \neq 0=S_0$. One can
analyze the details of this failure by looking carefully at what
happens when $a \rightarrow -\infty$. 

Let us instead illustrate why $\Expect{M_T}=M_0$ holds for certain
martingales despite the fact that $T$ is an unbounded stopping time.
The basic trick is to define, for integer $m$, $T_m=\min (m,T)$. Then
for each $m$, $T_m$ is a bounded stopping time and ${\mathbb
  E}M_{T_m}=M_0$ for any martingale, showing that it is enough to
prove that $\lim_{m\rightarrow +\infty} \Expect{M_T-M_{T_m}}=0$
for the martingale at hand to conclude. For instance, $ a \leq S_n
\leq b$ for $0 \leq n \leq T$ and $T_m \leq T$ for all $m$'s. So
$|S_T-S_{T_m}|$ is $0$ for $m\leq T$ and $\leq b-a$ for $m > T$. Hence
$\Expect{|S_T-S_{T_m}|}\leq (b-a)p(T>m)$ which goes to $0$ when
$m\rightarrow +\infty$. Hence $\Expect{S_T}=0$.  We get analogously
that $\Expect{|S^2_T-S^2_{T_m}|}\leq \max(-a,b)^2p(T>m)$ which goes
to $0$ when $m\rightarrow +\infty$. On the other hand,
$\Expect{T-T_m}= \sum_{n>m}(n-m)p(T=n)$ which again goes to $0$ when
$m\rightarrow +\infty$. Hence $\lim_{m\rightarrow
  \infty}\Expect{|(S^2_T-T)-(S^2_{T_m}-T_m)|}=0$ and
$\Expect{S^2_T-T}=0$ as 
announced.

As a last example, choose $M_n=e^{HS_n-n\log \cosh H}$ for
real $H$ as martingale.  For $0 \leq n \leq T$, $0 < M_n \leq
e^{|H|\max (b,-a)}$, and  $\Expect{|M_T-M_{T_m}|}\leq
p(T>m)e^{|H|\max (b,-a)}$ which goes to $0$ when $m\rightarrow +\infty$.
 Hence $\Expect{M_T}=1$, which gives enough information to compute the
ditribution of $T$. To keep formul\ae\ simple, assume that $a+b=0$. 
The martingale property gives $\cosh (b H)\Expect{(\cosh
H)^{-T}}=1$ leading for instance to $\log p(T\geq n)\sim n \log \cos\pi/(2b)$,
improving significantly the naive bound obtained above for the
exponential decay of $p(T\geq n)$. 

\vspace{.5cm}

\begin{ex}\label{ex:msm}: Martingales in statistical mechanics.
\end{ex}

\vspace{-.3cm}
The construction has been explained in full details in Section
\ref{sec:statmech}.

It is now time to turn to the continuous time  setting. 

\subsection{Brownian motion}

\subsubsection{Random processes in the large}

A random process on a probability space 
$(\Omega,{\mathcal F},p)$ is a familly $\{X_t\}_{t \in I}$ of
random variables with values in the same measurable space, where $I$
is a totally ordered set, in concrete examples either $\{0,1,\cdots,
N\}$, $\{0,1,\cdots\}$ (random process in discrete time),
$[0,T]$ or $[0,+\infty[$ (random process in continuous time). 

It can be useful to put more structure in the definition of a random
process and add a filtration ${\mathcal F}_t$, $t\in I$, i.e. an
increasing family of tribes, all included in ${\mathcal F}$, such that
$\{X_t\}$ is ${\mathcal F}_t$-measurable. Then, it is $\{X_t,{\mathcal
  F}_t\}_{t \in I}$ which is referred to as a random process. When no
such filtration is assumed, ${\mathcal F}_t$ can be taken to be the
smallest tribe making $\{X_{t'}\}$ measurable for all $t'\leq t$.

The notion of adapted process, martingale and stopping time can be
easily recopied from the discrete definitions for continuous time processes.

\subsubsection{The definition of Brownian motion}

We describe Brownian motion in $d$ dimensions, starting at the origin
in ${\mathbb R}^d$. Stochastic Loewner evolutions involve one
dimensionnal Brownian motion (d=1) but the Brownian loop soup (see
section \ref{sec:loopbrown}) is in  ${\mathbb R}^2=\mathbb C$.

Our aim is to put a probability measure on $\Omega={\mathcal
  C}_0([0,+\infty[,{\mathbb R}^d)$, the space of continuous functions
$\gamma$ from $[0,+\infty[$ to ${\mathbb R}^d$ such that
$\gamma(0)=0$. If $n \geq 1$, $0<t_1<\cdots<t_n$ and $A_1,\cdots,A_n$
are Borel subsets of ${\mathbb R}^d$, the subspace
$C(\{t_m,A_m\}_{m=1,\cdots,n})$ of $\Omega$ consisting of the
$\gamma$'s such that $\gamma(t_i)\in A_i$ for $i=1,\cdots,n$ is called
a cylinder set. We define $\mathcal F$ to be the smallest tribe
containing cylinder sets and ${\mathcal F}_t$ the smallest tribe
containing cylinder sets with $C(\{t_m,A_m\}_{m=1,\cdots,n})$ with
$t_n \leq t$.  

The basic object to define the probability measure is the heat kernel
in $d$ dimensions $K({\mathbf x},t)\equiv \frac{1}{(2\pi t)^{d/2}}
\exp{-\frac{||{\mathbf x}||^2}{2t}}$.  The measure of the cylinder set
$C(\{t_m,A_m\}_{m=1,\cdots,n})$ is defined to be
\begin{eqnarray*}
  \mu(C(\{t_m,A_m\}_{m=1,\cdots,n})) & \equiv & \\ 
  & &  \hspace{-5cm} \int_{A_1} d^d {\mathbf x}_1
  \cdots \int_{A_n} d^d {\mathbf x} _n K({\mathbf x}_1,t_1) K({\mathbf
    x}_2-{\mathbf x}_1,t_2-t_1)\cdots K({\mathbf x}_n-{\mathbf
    x}_{n-1},t_n-t_{n-1}).
\end{eqnarray*}

If $n\geq 1$ and $A_{l}$ is the whole real line for some $l$, the
integral over ${\mathbf x}_l$ can be performed explicitely, and
$\mu(C(\{t_m,A_m\}_{m=1,\cdots,n;\, m\neq l}))$ is recovered. This is
an obvious consistency condition if $\mu$ is to extend to a
probability measure on
$\mathcal F$. An
extension theorem by Kolmogorov ensures this is also a sufficient
condition. 

It turns out to be important to extend the ${\mathcal
  F}_t$'s and $\mathcal F$ with sets of measure zero, but we shall 
remain dumb\footnote{In all acceptations of the word, maybe.} on that.

Note that for  $\lambda \in {\mathbb R}\backslash \{0\}$,
\[\mu(C(\{\lambda^2t_m,\lambda A_m\}_{m=1,\cdots,n}))=
\mu(C(\{t_m,A_m\}_{m=1,\cdots,n})),\]
i.e. $\mu$ is scale invariant on cylinder sets. Hence
if $\lambda \in {\mathbb R}\backslash \{0\}$, and $B_t$ is a Brownian
motion, then $\frac{1}{\lambda}B_{\lambda^2t}$ is also a Brownian
motion. 
\vspace{.5cm}

Let us turn for a while to dimension $d=1$. A point $\omega \in
\Omega$ is a continuous function from $[0,+\infty[$ to ${\mathbb R}$,
and the Brownian motion process is denoted by $B_t$, where $B_t(\omega)
\equiv \omega(t)$. From the definition, 
\begin{description} 
\item[BM] For $0<t_1<\cdots<t_n$, the vector
$(B_{t_1},B_{t_2}-B_{t_1},\cdots,B_{t_n}-B_{t_{n-1}})$ is centered
Gaussian with independant components of variance
$(t_1,t_2-t_1,\cdots,t_n-t_{n-1})$
\end{description}

One can generalize the notion of Brownian motion as any process $B_t$
on a probability space such that $B_t$ is almost surely continuous and
starting at $0$, and moreover satisfies \textbf{BM}.

This extension is useful for instance to construct Brownian samples.
Take $0 \leq t' <t''$. First, $B_{t'}$ and $B_{t''}$ being fixed, the
distribution of $B_t$ for $t\in [t',t'']$ is independant of what has
happended before $t'$ or what will happen after $t''$. Second, setting
\[ t=\frac{t'+t''}{2} \qquad X=
\frac{2B_t-B_{t''}-B_{t'}}{(t''-t')^{1/2}}\]
$X$ is a standard Gaussian random variable
because the numerator is the difference of two independent centered
Gaussians of variance $(t''-t')/2$. One
can iterate to construct $2^n$ independent standard Gaussian random
variables from $B_{1/2^n},\cdots,B_{k/2^n},\cdots,B_1$. In the passage
from $n$ to $n+1$, $2^n$ new standard Gaussian random variables come
into play. Going the other way round, one can construct Brownian
samples on $t\in [0,1]$ by iterating as follows. Let $h$ be the
''tent'' function, $h(t)=2t$ on $[0,1/2]$, $h(t)=1-2t$ on $[1/2,1]$
and $h(t)=0$ elsewhere. Define, for $n\geq 0$ and $k=0,\cdots,2^n-1$,
$h_{n,k}(t)\equiv h(2^nt-k)$ and suppose that $Y$ and $X_{n,k}$ for
$n\geq 0$ and $k=0,\cdots,2^n-1$ form a family of independent
normalized centered Gaussian random variable on some probability
space. Then one can show that $tY+\sum_{n,k}
\frac{1}{2^{1+n/2}}X_{n,k}h_{n,k}(t)$ is almost surely convergent to a
random continous function of $t\in [0,1]$, call it $W_t$, and $W_t$ is a
Brownian process in the generalized sense. If the random variables $Y$
and $X_{n,k}$ are sampled from a Brownian sample $B_u, u\in [0,1]$
then $W_t$ and $B_t$ coincide at the dyadic rationals $k/2^n$.
By cutting at some $n$ one gets nice simulations of Brownian motion.
One can show that if $B_t$ is a Brownian motion, $tB_{1/t}$ is a
Brownian motion as well. So
gluing in the appropriate way two independent Brownian motions on
$[0,1]$ one gets a Brownian motion on $[0,+\infty[$.

The Brownian bridge ($d=1$) can be defined via a measure on ${\mathcal
  C}_{0,0}([0,1],{\mathbb R})$, the space of continuous functions
$\gamma$ from $[0,1]$ to ${\mathbb R}$ such that
$\gamma(0)=\gamma(1)=0$. The formula for the measure of a cylinder set
$C(\{t_m,A_m\}_{m=1,\cdots,n})$ with $0<t_1<\cdots<t_n<1$ is defined to be
\begin{eqnarray*} 
\mu(C(\{t_m,A_m\}_{m=1,\cdots,n})) & \equiv &
\sqrt{2\pi}\int_{A_1} dx_1 \cdots \int_{A_n} dx_n \\ & & \hspace{-4cm}
K(x_1,t_1)K(x_2-x_1,t_2-t_1)\cdots K(x_n- x_{n-1},t_n-t_{n-1})K(-x_n,1-t_n).
\end{eqnarray*}

The Brownian bridge is simply Brownian motion $B_t$ on $[0,1]$
conditionned 
to $B_1=0$ (via a limiting procedure). If $B_t$ is a Brownian motion,
$B_t-tB_1$ is a Brownian bridge. The Brownian bridge can be simulated
simply by forgetting about $Y$ and  keeping only the $X_{n,k}$'s.

\subsubsection{Some martingales}

Our starting point is Brownian motion on $\Omega={\mathcal
  C}_0([0,+\infty[,{\mathbb R})$. Remember that $\mathcal F$ is the
smallest tribe containing cylinder sets and ${\mathcal F}_t$ the
smallest tribe containing cylinder sets
$C(\{t_m,A_m\}_{m=1,\cdots,n})$ with $t_n \leq t$. As the increments
of Brownian motion are independent, $B_{t+s}-B_t$ is independent of
${\mathcal F}_t$ for $s\geq 0$, i.e. $\Expect{f(B_{t+s}-B_t)X
|{\mathcal F}_t}=X\Expect{f(B_{t+s}-B_t)}$ if the random variable $X$
is ${\mathcal F}_t$-measurable. For instance
\begin{eqnarray*} \Expect{B_{t+s}|{\mathcal F}_t}& = & 
\Expect{B_t|{\mathcal F}_t}+ \Expect{B_{t+s}-B_t|{\mathcal F}_t} \\
  & = & B_t+\Expect{B_{t+s}-B_t}\\ & = & B_t
\end{eqnarray*}
 so $B_t$ is a martingale (note
      that $\Expect{|B_t|}=\sqrt{\frac{2t}{\pi}} < +\infty$). 
In the same spirit, writing 
$B^2_{t+s}=B_t^2+2B_t(B_{t+s}-B_t)+(B_{t+s}-B_t)^2$ and taking the
conditionnal expectation with respect to ${\mathcal F}_t$ yields
$\Expect{B^2_{t+s}|{\mathcal F}_t}=B^2_t+0+s$
As $\Expect{B_t^2}=t<+\infty$, we conclude that $B^2_t-t$ is a
martingale.
Finally, writing $e^{HB_{t+s}}=e^{HB_t}e^{H(B_{t+s}-B_t)}$ and taking
 the conditionnal expectation with respect to ${\mathcal F}_t$ yields
$\Expect{e^{HB_{t+s}}|{\mathcal F}_t}=e^{HB_t}e^{sH^2/2}$.  As
$\Expect{|e^{HB_t}|}=e^{t(\Re e\, H)^2/2}< +\infty$ for complex $H$ we
conclude that $e^{HB_t-tH^2/2}$ is a martingale. So we have retreived
continuous time analogs for the simplest random walk martingales
defined above. However, the need of a continuous analog of stochastic
difference equations and stochastic sums is plain. 

\subsection{The quadratic variation of Brownian motion}

In the theory of Rieman-Stiljes integrals, one defines
$\int_0^t f(u)dg(u)$ by limits of Riemann sums. 

If $t_0=0< t_1 < \cdots t_n=t$ is a subdivision,define
$V \equiv \sum_{m=0}^{n-1} |g_{t_{m+1}}-g_{t_m}|$,
$S_{ret} \equiv \sum_{m=0}^{n-1} f(t_m)(g(t_{m+1})-g(t_m))$,
$S_{adv} \equiv \sum_{m=0}^{n-1} f(t_{m+1})(g(t_{m+1})-g(t_m))$,
and $2S\equiv S_{adv}+S_{ret}$. The function $g(t)$ is said to have
bounded variation if $V$ is bounded on the set of all subdivisions. It
can be shown that if $g(t)$ has bounded variation and (say) $f(t)$,
$g(t)$ are continuous, $S_{ret}$, $S_{adv}$ and $S$ have a common
limit when the mesh $\sup_{m} (t_{m+1}-t_m)$ of the subdivision goes
to $0$.

Suppose we want to make sense of $\int_{0}^t 2B_sdB-s$ in the same
way. So 
$S_{ret} = 2\sum_{m=0}^{n-1}B_{t_m}(B_{t_{m+1}}-B_{t_m})$,
$S_{adv} = 2\sum_{m=0}^{n-1}B_{t_{m+1}}(B_{t_{m+1}}-B_{t_m})$
and $ S = \sum_{m=0}^{n-1} (B_{t_{m+1}}+B_{t_m})(B_{t_{m+1}}-B_{t_m})$.
 Now $S$ is a telescopic
sum whose value is always $B_t^2$. On the other hand
$S_{adv}-S_{ret}=2Q$ where $Q(t_1,\cdots,t_n)\equiv \sum_{m=0}^{n-1}
(B_{t_{m+1}}-B_{t_m})^2$ a quadratic sum to be contrasted with the
linear sum $V=\sum_{m=0}^{n-1} |B_{t_{m+1}}-B_{t_m}|$.

First suppose that the subdivision is regular, i.e.  $t_k=tk/n$. By
the scale invariance of Brownian motion, $Q(t/n,2t/n,\cdots,nt/n)$ is
distributed as $Q(t,2t,\cdots,nt)/n$ and $Q(t,2t,\cdots,nt)$ is a sum
of $n$ independant identically distributed random variables with
average $t$ : the strong law of large numbers (see example \ref{ex:kslln})
implies that $Q(t/n,2t/n,\cdots,nt/n)$ converges almost surely to $t$.

Central to It\^o's theory is the following combinatorial identity.

\vspace{.3cm}

Assume that for $m=0,\cdots,n-1$, $X_m$ and $\Delta_m$ are random
variables on some probability space, with the property that $\Delta_l$
and $\Delta_m$ are independant for $l\neq m$ and $\Delta_m$ is
independant of $X_l$ for $l<m$. Define $\mathbb{E}\Delta_m^2 \equiv
\delta_m$ and assume that $\mathbb{E}\Delta_m^4=3\delta_m^2$.  Then
\[\mathbb{E}(\sum_{m=0}^{n-1} X_m \Delta_m^2 -\sum_{m=0}^{n-1} X_m
\delta_m)^2=2\sum_{m=0}^{n-1} \delta_m^2 \mathbb{E}X_m^2.\]

\vspace{.3cm}

 Note that
the relation between the second and fourth moment of $\Delta_m$ is
true for a centered Gaussian.

As a first application, take $X_m=1$ (a constant random variable) and
$\Delta_m=B_{t_{m+1}}-B_{t_m}$. Then 
$\mathbb{E}Q(t_1,\cdots,t_n)=t$ and the combinatorial identity yields
$\mathbb{E}(Q(t_1,\cdots,t_n)-t)^2=2\sum_{m=0}^{n-1} (t_{m+1}-t_m)^2
\leq 2t\sup_{m} (t_{m+1}-t_m)$ which goes to $0$ if the mesh of the
subdivision goes to $0$, so that $Q(t_1,\cdots,t_n)$ converges to $t$
in the $\mathbb{L}^2$ topology.

More generally, if $X_t$ is a random process such that the function
$\mathbb{E}X_s^2$ is (Riemann) integrable on $s\in [0,t]$
$\sum_{m=0}^{n-1}(t_{m+1}-t_m)^2\mathbb{E}X_{t_m}^2\leq \sup_{m}
(t_{m+1}-t_m) \sum_{m=0}^{n-1}(t_{m+1}-t_m)\mathbb{E}X_{t_m}^2$ goes
to $0$ if the mesh of the subdivision goes to $0$. If moreover $X_s$
is independant $B_t-B_s$ for $t>s$ and the samples of $X_s$ are
(Riemann) integrable on $[0,t]$, we infer from the combinatorial lemma
that $\sum_{m=0}^{n-1}X_{t_m}(B_{t_{m+1}}-B_{t_m})^2$ converges to
$\int_0^t X_sds$ in the $\mathbb{L}^2$ topology when the mesh of the
subdivision goes to $0$. This leads to the suggestive infinitesimal
notation $(dB_t)^2=dt$ which is the ''miraculous'' rigid analog to the
discrete $X_n^2=1$ lying at the origin of It\^o's calculus.

One could extend these results in several directions, but the point
is that the Brownian motion has, for more than enough definitions of
convergence, a well defined quadratic variation
which is deterministic and equal to $t$. 

The discretization $S_{ret}$ leads to $\int_{0}^t
2B_sdB_s=B_t^2-t$, while the discretization $S_{adv}$ would lead to
$\int_{0}^t 2B_sdB_s=B_t^2+t$ and the discretization $S$ to
$\int_{0}^t 2B_sdB_s=B_t^2$. This discrepency shows that $B_t$ has
infinite variation with probability one and some consistent convention
has to be taken to define stochastic integrals. Not all conventions
are equally convenient. The symmetric choice $S$
(Stratanovich's convention) is popular in physics but there are good
reasons to prefer the discretization $S_{ret}$ (It\^o's convention)
because it leads to martingales, as the above special case exemplifies.

\subsubsection{Stochastic integrals and It\^o's formula}

As usual, integrals are first defined for a special class of
integrants, and then extended by a limiting procedure about which we
shall say almost nothing. If $(B_t,{\mathcal F}_t)$ is a Brownian motion on a
space $(\Omega,{\mathcal F},p)$, a simple process $U(t)$ is a random
function for which there exists an increasing deterministic sequence
$0=t_0 < t_1 < \cdots < t_n$ and a sequence of random variables
$U_1,\cdots,U_n$ such that $U_i$ is ${\mathcal F}_{t_i}$ measurable,
$U(t)=U_m$ for $t\in [t_m,t_{m+1}[$ and $U(t)=0$ for $t\geq t_{n}$.
Then $\int U(s) dB_s \equiv \sum_{m=0}^{n-1}
U_m(B_{t_{m+1}}-B_{t_m})$. If $T\geq 0$ and $U(t)$ is a simple
process, then so is $U(t){\mathbf 1}_{t\in[0,T]}$ Then $\int_0^T U(s)
dB_s \equiv \int U(s){\mathbf 1}_{s\in[0,T]} dB_s$. There is a deep
relationship with Hilbert space theory here, and it is natural to
assume that $\Expect{U_m^2} < +\infty$ for each $m=0,\cdots,n-1$.
Then $\Expect{(\int U(s) dB_s)^2}= \Expect{\int U(s)^2ds}$, a
formula at the heart of the extension of the stochastic integral to
more complicated processes. This has to be done with care to avoid a
wild (non measurable) behavior of the stochastic integral as a
function of $\omega$. It is easy to check that if $X(t)$ is a simple
stochastic process, $\int_0^T U(s) dB_s$ is a martingale. If we take
for $U(t)$ a piecewise constant interpolation of Brownian motion, we
recover the definition of $S_{ret}$. In general $\int_0^T U(s) dB_s$,
even if defined, needs not be a martingale. It is a local martingale,
which is almost as useful, because local martingales can be shown to
be martingales when stopped at appropriate sequences of stopping
times. The reader is refered to the litterature for precise
definitions. We shall almost surely make no  distinction between local
martingles and martingales in these notes.

The exemple of $B_t^2$ shows that differentials cannot be computed in the
classical way for stochastic integrals. Indeed, we have
$B_t^2=\int_0^t 2B_s dB_s+\int_0^t ds$, where the first integral is an
It\^o integral and the second one an ordinary (say Riemann) integral.
More generally, suppose that some process $X_t$ can be written as
$X_t=X_0+\int_0^t U_sdB_s+\int_0^tV_sds$ where  $X_0$ is a constant
random variable and $U_t, V_t$ are adapted
processes (then so is $X_t$). A short-hand notation (and nothing more)
is $dX_t= U_tdB_t+V_tdt$. If $f(t,x)$ is smooth enough (three
times continuously differentiable is more than enough), $f(t,X_t)$ can
also be represented as an integral $f(t,X_t)=f(0,X_0)+\int_0^t
P_sdB_s+\int_0^tQ_sds$ given by It\^o's formula :
\begin{eqnarray*}
  P_t & = & U_t\frac{\partial f}{\partial x}(t,X_t)\\
  Q_t & = & \frac{\partial f}{\partial t}(t,X_t)+ V_t\frac{\partial
    f}{\partial x}(t,X_t)+\frac{U_t^2}{2}\frac{\partial^2 f}{\partial
    x^2}(t,X_t).
\end{eqnarray*}

Our handwaving argument goes as follows : first, we can use simple
processes as approximations in the integrals defining $X_t$. The
resulting integrals converge to $X_t$, and as $f(t,x)$ is continuous
in $x$, it is enough to prove It\^o's formula when $U_t$ and $V_t$ are
simple processes. Take a small $\varepsilon >0$ and a subdivision
$0=t_0 < t_1 < \cdots < t_n=t$ such that on each interval
$[t_m,t_{m+1}[$ the processes $U_s$ and $V_s$ are constant but
$\sup_{m} (t_{m+1}-t_m) < \varepsilon$. Set $\delta_m=t_{m+1}-t_m$,
$\Delta_m=B_{t_{m+1}}-B_{t_m}$ and expand
\[f(t_{m+1},X_{t_{m+1}})-f(t_m,X_{t_m})=f(t_m+\delta_m,X_{t_m}+
U_{t_m}\Delta_m+V_{t_m}\delta_m)\] in powers of $\delta_m$ and
$\Delta_m$. The term involving $\Delta_m^1$ is an It\^o sum, the terms
involving $\delta_m^1$ are Riemann sums. In the limit $\varepsilon
\rightarrow 0$, their sum over $m$ have a limit as an It\^o or a Riemann
integral. The sum over $m$ of term involving $\Delta_m^2$ is converted
via the combinatorial identity to the same sum with $\Delta_m^2$
replaced by $\delta_m$ plus a correction term which is small in the
$\mathbb{L}^2$ topology. These terms account for It\^o's formula. The
error that arises from keeping only these contribution is small (even
after the sum over $m$) because it involves sums over $m$ of
$O(\delta_m\Delta_m,\Delta_m^3,\delta_m^2)$ : though $\Delta_m$ and
its powers have fluctuations, the sum over $m$ tames these -- as for
$\Delta_m^2$ via the combinatorial identity -- yielding a total error
of order at most $\sqrt{\varepsilon}$ (this is a time for some
energetic handwaving on our side !).

Once the notion of It\^o derivative is around, one can define stochastic
differential equations. There are subtleties between so-called weak and
strong solutions of strochastic diffferential equations, but the basic
tool for existence of solutions is Picard iteration as for standard
differential equations. A useful particular class is formed by
It\^o diffusions, i.e. processes which solve an equation of the form
$X_t=X_0+\int_0^t a(s,X_s)dB_s +\int_0^t b(s,X_s)ds$. 

To conclude this section, we extend the notion of quadratic variation.
If the sum $\sum_{m=0}^{n-1} (X_{t_{m+1}}-X_{t_m})^2$ defined for
$0=t_0 < t_1 < \cdots < t_n=t$ has a (in general random) limit when
$\sup_{m} (t_{m+1}-t_m)$ goes to $0$, this limit is called the
quadratic variation of $X_t$, usually denoted by $<X_t>$, a notation
chosen almost surely to confuse physicists.  Brownian motion has a
deterministic quadratic variation $t$, but this is more the exception
than the rule. If it exists, $<X_t>$ is a non-decreasing process. If
$X_t=X_0+\int_0^t U_sdB_s+\int_0^tV_sds$, the process $X_t$ has a
well-defined quadratic variation and $<X_t>=\int_0^t U_s^2 ds$, as
suggested by a naive formal computation. It\^o's formula can be recast
in differential notation as
\[df(t,X_t)=\frac{\partial f}{\partial x}(t,X_t)dX_t+ \frac{\partial
  f}{\partial x}(t,X_t)dt+\frac{1}{2}\frac{\partial^2 f}{\partial
  x^2}(t,X_t)d<X_t>.\]

If $c$ is a constant and $X_t$,$Y_t$ have finite quadratic
variation, then so does $Z_t=X_t+cY_t$, and
\[ \frac{<X_t+cY_t>-<X_t>-c^2<Y_t>}{2c}\equiv <X_t,Y_t>\] is independant
of  $c$ and called the cross-variation of $X_t$ and $Y_t$.

 From It\^o's change of variable formula, it is easy to obtain a
formula for the It\^o derivative of a product : if $X_t=X_0+\int_0^t
U_sdB_s+\int_0^tV_sds$ and $Y_t=X_0+\int_0^t R_sdB_s+\int_0^tW_sds$,
and $Z_t=X_t+cY_t$ for some constant $c$, the It\^o derivative of
$Z_t^2$ is quadratic in $c$ and the linear term yields
$d(X_tY_t)=X_tdY_t+Y_tdX_t+d<X_t,Y_t>$ and $<X_t,Y_t>=\int_0^t
U_sR_sds$.

On the other hand, if $X_t$ and $Y_t$ are independant
Brownian motions, their cross-variation is easily checked to vanish.
We leave to the reader the straightforward extension of It\^o's formula
when $X_t=X_0+\sum_{i=1}^d\int_0^t U^{(i)}_sdB_s^{(i)}+\int_0^tV_sds$
where $(B^{(1)}_t,\cdots,B^{(d)}_t)$ is a $d$-dimensional Brownian motion,
i.e. $B^{(1)}_t,\cdots,B^{(d)}_t$ are $d$ independant Brownian
motions.

\subsubsection{A few applications of It\^o's formula}

Among the things that make Brownian motion such an important and
ubiquituous probabilistic object are the following characterizations :

-- If $X_t$ is a continuous process with independant stationary
increments (not assumed to be Gaussian !), there are constants $\kappa
\geq 0$ and $c$ such that $X_t=\sqrt{\kappa} B_t+ct$ for some Brownian
$B_t$. 

Note that Brownian motion is a special member in a famous class of
processes, those with independent stationary increments. There is a
classification of these processes, and it turns out that ''continuity
implies gaussianity''. This result is one of the ingredients in
Schramm's proof that conformaly invariant interfaces are related to
Brownian motion.

-- If $X_t$ is a continuous (local) martingale with quadratic
variation $t$, then $X_t$ is a Brownian motion.

\vspace{.5cm}

\begin{ex}\label{ex:bp}: Bessel processes.\end{ex}

\vspace{-.3cm}

If $(B^{(1)}_t,\cdots,B^{(d)}_t)$ is a $d$-dimensional Brownian motion,
let $R_t$ be the ''distance to the origin process'',also called
''Bessel process in dimension $d$'' : 
$R_t=\sqrt{(B^{(1)}_t)^2+\cdots+(B^{(d)}_t)^2}$. It is known that
(almost surely) $d$
dimensionnal Brownian motion is recurrent (i.e visits all points an
infinite number of times) for $d<2$, dense for $d=2$ and escapes to
infinity for $d >2$.

It\^o's formula yields
\[dR_t=\sum_{i=1}^d  \frac{\partial R_t}{\partial
  B^{(i)}_t}dB^{(i)}_t+\frac{1}{2} \sum_{i,j=1}^d \frac{\partial^2 R_t}{\partial
  B^{(i)}_t \partial B^{(j)}_t}d<B^{(i)}_t,B^{(j)}_t>.\] But
$<B^{(i)}_t,B^{(j)}_t>=\delta^{i,j}t,$ leading to
\[dR_t=\frac{1}{R_t}\sum_{i=1}^d B^{(i)}_tdB^{(i)}_t+\frac{d-1}{2R_t}dt.\]
The stochastic contribution look complicated, but one checks easily
that the quadratic variation of the (local) martingale $B_t\equiv \int_0^t
\frac{1}{R_s}\sum_{i=1}^d B^{(i)}_sdB^{(i)}_s$ is $t$. Hence $B_t$ is
a Brownian motion and we arrive to the conclusion that $R_t$ satisfies
the stochastic differential equation
\[dR_t=dB_t+\frac{d-1}{2R_t}dt.\]
Setting $\sqrt{d-1}X_t\equiv 2(R_t-B_t)$ yields
$dX_t=\frac{2dt}{X_t-\xi_t}$ where $\sqrt{d-1}\xi_t=-2B_t$. Hence
$X_t$ satisfies the restriction of Loewner's radial equation to the
positive real axis, for $\kappa=4/(d-1)$. This leads immediately to
the transition between hulls which are simple curves 
which do neither hit the real axis nor have self contacts for $\kappa \leq 4$
(i.e. $d\geq 2$, when $R_t$ does not return to the origin), and and
thick hull for $\kappa > 4$.

\vspace{.5cm}

\begin{ex}\label{ex:tc}: Time change.\end{ex}

\vspace{-.3cm}

Suppose $\tau_t(\omega)$ is an adapted continuous real non-negative
non-decreasing random process with $\tau_0=0$ almost surely.  Then,
for fixed $\tau$, $T_{\tau}=\inf\{t, \tau_t=\tau\}$ is a stopping
time, the first time at which $\tau_t$ reaches $\tau$. From the
definition of martingales and their good behavior with respect to
stopping times we infer the following. If $M_t,{\mathcal F}_t$ is a
martingale, then $M_{T_{\tau}},{\mathcal F}_{T_{\tau}}$ is also a
martingale ($\tau$ is the time parameter, which may not cover the
whole positive real axis).

Suppose $M_t=\int_0^t U_sdB_s$ is a (local) martingale and set
$\tau_t\equiv \int_0^t U_s^2ds$, the quadratic variation of $M_t$.
Then $M_{T_{\tau}},{\mathcal F}_{T_{\tau}}$ is a (local) martingale
too, with quadratic variation is $\tau$. Hence $M_{T_{\tau}},{\mathcal
  F}_{T_{\tau}}$ is a Brownian motion, though  possibly defined only on a
finite interval. This is sometimes loosely rephrased as : Brownian
motion is the only continuous local martingale modulo time changes.

For fixed $t$, the distribution of $M_t$ is not Gaussian at all in
general. However, when looked at the stopping time $T_{\tau}$ it is
Gaussian. Note in passing the remarkable Skorohod theorem, which goes
in the opposite direction somehow : any 
distribution function with $0$ mean is the distribution function of
Brownian motion at an appropriate stopping time. 

\vspace{.5cm}

\begin{ex}\label{ex:cipbm}: Conformal invariance of planar Brownian motion.\end{ex}

\vspace{-.3cm}

 From the definition of Brownian motion, if $R$ is a rotation in $d$
dimensions, and $T,\lambda$ are positive reals, the map $f$ from
${\mathcal C}_0([0,+\infty[,{\mathbb R}^d)$ to itself defined by
\[
(fB)_t=\left\{ 
  \begin{array}{l} B_t \quad 0\leq t\leq T \\
    B_T+R(B_{T+\lambda^2(t-T)}-B_T)/\lambda \quad T\leq t
  \end{array}\right. 
\]
preserves the Brownian probability measure. This transformation can be
iterated for different parameters, and vaguely speaking Brownian
motion is invariant under ''local'' dilations and rotations when time
is rescaled appropriately. When $d=2$, conformal transformations have
no shear : they preserve angles and look locally like a
dilation-rotation. Hence we expect that $2d$ Brownian motion is
conformally invariant. The simplest way to state this precisely is
It\^o's formula. Suppose $X_t$ and $Y_t$ are two independant Brownian
motions, set $Z_t=X_t+iY_t$ and consider a conformal transformation
$f:{\mathbb D}\subset {\mathbb C} \rightarrow {\mathbb D}' \subset
{\mathbb C}$ fixing the origin. The multidimensionnal It\^o formula
combined with the Cauchy equations yields
\[df(Z_t)=f'(Z_t)dZ_t+\frac{1}{2}f''(Z_t)d(<X_t>-<Y_t>+2i<X_t,Y_t>).\]
Observe that $<X_t>-<Y_t>+2i<X_t,Y_t>$ could be seen as the quadratic
variation of the complex process $Z_t$ if we would accept to put $c=i$
in our definition of the cross variation, but this would be nothing
but a convention. Now $<X_t>=<Y_t>=t$ and $<X_t,Y_t>=0$ if $X_t$ and
$Y_t$ are independant Brownian motions. One infers first that
$<X_t>-<Y_t>+2i<X_t,Y_t>=0$ so that $f(Z_t)$ is a (local) martingale.
Second one infers that $d<\Re e \, f(Z_t)>=d<\Im m \, f(Z_t)>=|f'(Z_t)|^2dt$ and
$d<\Re e \, f(Z_t),\Im m \, f(Z_t)> =0$. Thus, the same time change
$\tau=\int_0^t |f'(Z_s)|^2ds$ turns the real and imaginary parts of
$f(Z_t)$ into Brownian motions, which are gaussian processes, so that
the vanishing of the cross variation ensures independance. Hence
$f(Z_t)$ is a two dimensionnal Brownian motion after a time chance,
proving the conformal invariance of the two dimensionnal Brownian
motion.

\vspace{.5cm}

\begin{ex}\label{ex:gt}: Girsanov's theorem.\end{ex}

\vspace{-.3cm}

We have already seen in the discrete setting that martingales can be
used to deform probability laws. Let us illustrate the great flexibility
gained in the continuous setting.

Let $M_t,{\mathcal F}_t$ be a nonnegative martingale on
$(\Omega,{\mathcal F},p)$ such that $M_0=1$. If $X$ is ${\mathcal
  F}_s$-measurable and $t\geq s$ then basic rules of conditional
expectations yield $\Expect{XM_t}=\Expect{XM_s}$ so that we can
make a consistent definition $\tilde\Expect{X}\equiv \Expect{XM_t}$ 
whenever $X$ is ${\mathcal F}_t$ measurable. Then
$\tilde\Expect{\cdots}$ is easily seen to be a positive linear
functionnal with $\tilde\Expect{1}=1$. Hence the definition
$\tilde{p}_t(A)\equiv \tilde\Expect{{\mathbf 1}_A}$ for $A \in
{\mathcal F}_t$ makes $(\Omega,{\mathcal F}_t,\tilde{p}_t)$ a
probability space. Under some technical growth conditions on $M_t$,
$\tilde{p}_t$ extends to a probability law on $\sigma\{\cup_t
{\mathcal F}_t\}$

Now suppose that $$M_t=e^{\int_0^t V_sdB_s-\frac{2}{2} V_s^2ds}$$ for
some adapted process $V_s$. It\^o's formula shows that $M_t$ satisfyes
the stochastic integral equation $M_t=1+\int_0^t M_sV_sdB_s$ and is a
(local) martingale. Let $X_t$ be a process satisfying
$X_t=B_t-\int_0^t V_sds$. Girsanov's theorem states that for each $T
>0$, $X_t$ is a Brownian motion on $[0,T]$ for $(\Omega,{\mathcal
  F}_T,\tilde{p}_T)$. Here are elements of a heuristic proof.

A simple special case is $M_t\equiv e^{HB_t-tH^2/2}$, which we know is
a martingale on the Brownian motion space satisfying the conditions
above. Conversely, suppose that $W_t$ is a continuous process such
that $e^{HW_t-tH^2/2}$ is a martingale on some probability space with
a filtration ${\mathcal F}_t$. If $t_1 < \cdots < t_n <t$, 
$e^{\sum_{m=1}^{n}H_mW_{t_m}}$ is
${\mathcal F}_{t_{n}}$ measurable and
\[
\Expect{e^{\sum_{m=1}^{n}H_mW_{t_m}+HW_t-tH^2/2}} =
\Expect{e^{\sum_{m=1}^{n}H_mW_{t_m}+HW_{t_n}-t_nH^2/2}}. 
\] 
This leads to a recursive formula 
\[ \Expect{e^{\sum_{m=1}^{n}H_mW_{t_m}+WB_t}}=
e^{(t-t_n)H^2/2}\Expect{e^{\sum_{m=1}^{n}(H_m+H\delta_{m,n})W_{t_m}}
}\]
from which it follows that the finite dimensionnal distributions of
the continuous process $W_t$ are those of a Brownian motion, so that
$W_t$ is a Brownian motion. 

We can now go to the case of a general $M_t$ again. The lesson of the
previous computation is that to show that $X_t$ is a Brownian motion for
$(\Omega,{\mathcal F}_T,\tilde{p}_T)$, it is enough to show that $X_t$
is continuous in $t$ and that $e^{HX_t-tH^2/2}$ is a martingale for
$(\Omega,{\mathcal F}_T,\tilde{p}_T)$ i.e. that $e^{HX_t-tH^2/2}M_t$
is a (local) martingale for the original probability law. We compute
its It\^o derivative using It\^o's change of variable and product
formul\ae. First $de^{HX_t-tH^2/2}=e^{HX_t-tH^2/2}HdX_t$, then
$d(e^{HX_t-tH^2/2}M_t)=e^{HX_t-tH^2/2}M_tHdB_t$ as announced.

Path integrals trivialize (for good or bad) this argument : one writes
the Wiener measure as ${\mathcal D}x(s) \exp {-\frac{1}{2}\int_0^t
  \dot{x}(s)^2ds}$ and in the same notation, $$M_t=\exp[ {\int_0^t
  V(s)\dot{x}(s)ds-\frac{1}{2} \int_0^t V(s)^2ds}].$$
This is
misleading because in general $V(s)$ may depend on $x(s')$ for all
$s'<s$. The full measure becomes ${\mathcal D}x(s) \exp
{-\frac{1}{2}\int_0^t (\dot{x}(s)-V(s))^2ds}$ and a formal triangular
change of variables $y(s)=x(s)-\int_0^tV(s)ds$ combined with
translation invariance of the (non existing) Lebesgue measure
${\mathcal D}x(s)$ yields Girsanov's result.

As a simple application, take again $M_t\equiv e^{HB_t-tH^2/2}$. Then
$X_t=B_t-Ht$ is a Brownian motion with constant drift, which looks
like a Brownian motion again when the original measure is multiplied
by $e^{HB_t-tH^2/2}$. 

Note that the Bessel process $R_t$ in $d$ dimensions also becomes a
Brownian motion when the original measure is multiplied by 
\[ M_t=\exp[ {-\frac{d-1}{2}\int_0^t\frac{dB_s}{R_s}-
  \frac{(d-1)^2}{8}\frac{ds}{R_s^2}}].\]

This is an appropriate point to end this appendix. 






%% file: app2.tex
We restrict this presentation to a bare minimum, referring the newcomer
to the many articles, reviews and books on the subject
(\cite{DMS:CFT,BPZ}). The reader who knows too little or too much
about CFT can profitably skip this Section.

Observables in CFT can be classified according to their behavior under
conformal maps. Local observables in quantum field theory are called
fields. For instance, in the Ising model, on an arbitrary (discrete)
domain, the average value of a product of spins on different (well
separated) sites can be considered. Taking the continuum limit at the
critical point, we expect that on arbitrary domains $\mathbb{D}$ there
is a local observable, the spin. The product of two spins at nearest
neighbor points corresponds to the energy operator. In the continuum
limit, this will also lead to a local operator. In this limit, the
lattice spacing has disappeared and one can expect a definite (but
nontrivial) relationship between the energy operator and the product
of two spin fields close to each other. As on the lattice the product
of two spins at the same point is $1$, we can expect that the identity
observable also appears in such a product at short distances. Local
fields come in two types, bulk fields whose argument runs over
$\mathbb{D}$ and boundary fields whose argument runs over
$\partial\mathbb{D}$. 

The simplest conformal transformations in the upper-half plane are
real dilatations and boundary fields can be classified accordingly. It
is customary to write $\varphi_{\delta}(x)$ to indicate that in a real
dilatation by a factor $\lambda$ the field $\varphi_{\delta}(x)$ picks
a factor $\lambda^{\delta}$. By a locality argument, boundary fields
in a general domain $\mathbb{D}$ (not invariant under dilatations) can
still be classified by the same quantum number. The number $\delta$ is
called the conformal weight of $\varphi_{\delta}$. Similarly, bulk
fields are classified by their scaling dimension $\delta=h+\bar h$ and
their spin $s=h-\bar h$ with $(h,\bar h)$ their chiral conformal weights.

There are interesting situations in which (due to degeneracies) the
action of dilatations cannot be diagonalized, leading to so called
logarithmic CFT. While this more general setting is likely to be
relevant for several aspects of SLE, we do not need it.

Under general conformal transformations, the simplest objects in CFT
are so called primary fields. Their behavior is dictated by the
simplest generalization of what happens under dilatations: for a bulk
primary field $\Phi_{h,\overline{h}}$ of weights $(h,\overline{h})$,
$\Phi_{h,\overline{h}}(z,\overline{z})dz^hd\overline{z}^{\overline{h}}$ is
invariant, and for a boundary conformal field $\varphi_\delta$ of
weight $\delta$, $\varphi_\delta(x)|dx|^{\delta}$ is invariant under
conformal transformations.
  
The basic principles of conformal field theory state that correlation
functions in a domain $\mathbb{D}$ are known once they are known in a
domain $\mathbb{D} _0$ and an explicit conformal map from
$\mathbb{D}$ to $\mathbb{D}_0$ preserving boundary conditions is
given.  Suppose $\varphi_{\delta_1},\cdots \varphi_{\delta_n}$ are
boundary primary fields of weights $\delta_1,\cdots,\delta_n$ (bulk
fields may be considered similarly). If $f$ is a conformal map from
domain $\mathbb{D}$ to a domain $\mathbb{D}_0$, CFT postulates that
$$ 
\vev{\prod_{j=1}^n \varphi_{\delta_j}(x_j)}_\mathbb{D} =  
\prod_{j=1}^n |f'(x_j)|^{\delta_j}\
\vev{\prod_{j=1}^n \varphi_{\delta_j}(f(x_j))}_{\mathbb{D}_0}. 
$$
Symbolically, this can be written as
$\varphi_{\delta}(x) \rightarrow
\varphi_{\delta}(f(x))|f'(x)|^{\delta}.$
\medskip

As usual in quantum field theory, to a symmetry corresponds an
observable implementing it. Infinitesimal deformations of the
underlying geometry are implemented in local field theories by
insertions of the stress-tensor.  In conformal field theories, the
stress-tensor is traceless so that it has only two independent
components, one of which, $T(z)$, is holomorphic (except for possible
singularities when its argument approaches the arguments of other
inserted operators).  The field $T(z)$ itself is not a primary field
but a projective connection so that it behaves under conformal
transformation as
$$
\vev{\cdots T(z) \cdots}_\mathbb{D}= \statav{\cdots
  T(f(z))f'(z)^2+\frac{c}{12}\mathrm{S}f(z)\cdots}_{\mathbb{D}_0},
$$
with $c$ the CFT central charge and $\mathrm{S}f(z)=
\left(\frac{f''(z)}{f'(z)}\right)'-\frac{1}{2}
\left(\frac{f''(z)}{f'(z)}\right)^2$ the Schwarzian derivative of $f$
at $z$.  When $c=0$, $T$ is be a $(2,0)$ primary field i.e. an
holomorphic quadratic differential.

This applies to infinitesimal deformations of the upper half plane.
Consider an infinitesimal hull $\mathbb{K}_{\epsilon;\mu}$, whose boundary
is the curve $x\to\epsilon\,\pi\mu(x)$, $x$ real and $\epsilon\ll1$,
so that $\mathbb{K}_{\epsilon;\mu}=\{z=x+iy\in \mathbb{H},\ 
0<y<\epsilon\,\pi\mu(x)\}$.  Assume for simplicity that
$\mathbb{K}_{\epsilon;\mu}$ is bounded away from $0$ and $\infty$.
Let $\mathbb{H}_{\epsilon;\mu}\equiv
\mathbb{H}\setminus\mathbb{K}_{\epsilon;\mu}$.  To first order in
$\epsilon$, the uniformizing map onto $\mathbb{H}$ is
$$
z+ \epsilon\,\int_{\mathbb{R}}\frac{\mu(y)dy}{z-y} + o(\epsilon).
$$ 
To first order in $\epsilon$ again,
correlation functions in $\mathbb{H}_{\epsilon; \mu}$
 are related to those in $\mathbb{H}$ by insertion of $T$:
\begin{eqnarray}
&& \frac{d}{d\epsilon}
\vev{(\cdots \Phi_{h,\overline{h}}(z,\overline{z})\cdots
  \varphi_\delta(x)\cdots)}_{\mathbb{H}_{\epsilon; \mu}}
\Big\vert_{\epsilon=0^+}  
\nonumber \\
&& \hskip -1.5 truecm
=-\int_{\mathbb{R}} dy \mu(y)\, 
\vev{ T(y)(\cdots \Phi_{h,\overline{h}}(z,\overline{z})\cdots
  \varphi_\delta(x)\cdots)}_{\mathbb{H}}
\label{Tdeform}
\end{eqnarray}
Clearly, the stress tensor $T$ is the operator implementing
infinitesimal conformal deformations.

Finite conformal transformations are implemented in conformal field
theories by insertion of operators, representing some appropriate
exponentiation of insertions of the stress tensor. Let $\mathbb{D}$ be
conformally equivalent to the upper half plane $\mathbb{H}$ and $f$
the corresponding uniformizing map.  Then, following \cite{BB03b},
the finite conformal deformations that leads from the conformal field
theory on $\mathbb{D}$ to that on $\mathbb{H}$ can be represented by
an operator $G_f$:
\begin{eqnarray}
\vev{(\cdots \Phi_{h,\overline{h}}(z,\overline{z}) \cdots
  \varphi_\delta(x) \cdots) }_\mathbb{D}= 
 \statav{ G_f ^{-1}\left(\cdots
     {\Phi}_{h,\overline{h}}(z,\overline{z})
\cdots {\varphi_\delta}(x) \cdots \right) G_f}_{\mathbb{H}}.
\nonumber 
\end{eqnarray}
This relates correlation functions in $\mathbb{D}$ to correlation
functions in $\mathbb{H}$ where the field arguments are taken at the
same point but conjugated by $G_f$.  Here, radial quantization is
implicitly assumed.
\medskip

Let us now describe the action of the stress tensor on local fields
and the associated action of the Virasoro algebra.  When a (smooth)
boundary is present, the Schwarz reflection principle allows to extend
$T$ by holomorphicity.  Holomorphicity also implies that if $O$ is any
local (bulk or boundary) observable at point $z \in \mathbb{D}$ and
$v$ is vector field meromorphic close to $z$, the contour integral
$L_vO \equiv\oint_z dw v(w)T(w)O$ along an infinitesimal contour
around $z$ oriented counterclockwise is again a local field at $z$,
corresponding to the infinitesimal variation of $O$ under the map
$f(w)=w+\varepsilon v(w)$. It is customary to write $L_n$ for
$v(w)=w^{n+1}$. They statisfy the Virasoro commutation relations:
$$
[L_n,L_m]= (n-m)L_{n+m} + \frac{c}{12}n(n^2-1)\delta_{n+m;0}
$$

It is one of the postulates of CFT that all local fields can be
obtained as descendants of primaries, i.e. by applying this
construction recursively starting from primaries. The correlation
functions of descendant fields are obtained in a routine way from
correlations of the primaries. But descendant fields do not transform
homogeneously.  When $v$ is holomorphic at $x$, $L_vO$ is a familiar
object. For instance, if $\varphi_{\delta}$ is a primary boundary
field, one checks readily that $L_n\varphi_{\delta}=0$ for $n\geq 1$,
$L_0\varphi_{\delta}=\delta \varphi_{\delta}$ and
$L_{-1}\varphi_{\delta}=\Re e \; [\partial_x\varphi_{\delta}]$. The
other descendants are in general more involved, but by definition the
stress tensor $T=L_{-2}{\bf 1}$ is the simplest descendant of the
identity ${\bf 1}$. It does indeed not transform homogeneously.

A primary field and its descendants form what is called a conformal
family.  Not all linear combinations of primaries and descendants need
to be independent. The simplest example is the identity observable,
which is primary with weight $0$ and whose derivative along the boundary
vanishes identically\footnote{For other primary fields with the same
  weight if any, this does not have to be true.}. 

The next example in order of complexity is of utmost importance for
the rest of this paper. If the weight and the central charge satisfy
$(2h+1)c=2h(5-8h)$, the field
$$
-2(2h+1)L_{-2}\varphi_{h}+3L_{-1}^2\varphi_{h}
$$
is again a primary, i.e. it transforms homogeneously under
conformal maps. Parametrized the central charge as
$c=(6-\kappa)(3\kappa-8)/2\kappa$ and the weight by
$h=(6-\kappa)/2\kappa$ the above field is proportional to $-2
L_{-1}^2\varphi +\frac{\kappa}{2}L_{-2}\varphi$, and we recognize the
key operator involved in the SLE/CFT correspondance.  In this case,
consistent CFTs can be constructed for which it vanishes identically.
The above field is then called a null-vector. This puts further
constraints on correlators.

For example, when $\mathbb{D}$ is the upper half plane, so that the
Schwarz principle extends the stress tensor $T$ to the full plane, the
contour for $L_{-2}$ can be deformed and shrunken at infinity.  Then,
for an arbitrary boundary primary correlator one has the differential
equation:
\begin{eqnarray}
\label{eq:sing}
\left(\frac{3}{2(2h+1)}\partial_x^2 +\sum_{\alpha=1}^{l}\left[
  \frac{1}{y_{\alpha}-x}\partial_{y_{\alpha}}-
  \frac{\delta_{\alpha}}{(y_{\alpha}-x)^2} \right]\right) & & \nonumber \\
& & \hspace{-3.5cm}  \vev{\varphi_{\delta}(\infty) \prod_{\alpha=1}^l
\varphi_{\delta_{\alpha}}(y_{\alpha}) \varphi_{h}(x)}_\mathbb{H}=0. 
\end{eqnarray}  
It is customary to call this type of equation a null-vector equation.

Note that the primary field of weight $\delta$ sitting at $\infty$ has
led to no contribution in this differential equation. Working the
other way round, this equation valid for an arbitrary number of
boundary primary fields with arbitrary weights characterizes the field
$\varphi_{h}$ and the relation between $h$ and the central charge $c$.

The case of three points correlators is instructive.  Global conformal
invariance implies that
$$
\vev{\varphi_{\delta}(y)\varphi_{\delta'}(y') \varphi_{h}(x)}_\mathbb{H}
\propto\ |y-y'|^{h-\delta-\delta'}
|x-y|^{\delta'-h-\delta}|y'-x|^{\delta-\delta'-h}.
$$
The proportionality constant might depend on the cyclic ordering of
the three points. But if the differential equation for $\varphi_{h}$
is used, a further constraint appears. The three point function can be
non vanishing only if
$$
3(\delta-\delta')^2-(2h+1)(\delta+\delta')= h(h-1).$$
This
computation has a dual interpretation : consider a correlation
function with any number of fields, among them a $\varphi_{\delta}(y)$
and a $\varphi_{h}(x)$. If $x$ and $y$ come very close to each other
they can be replaced by an expansion in terms of local fields. This is
called fusion. Several conformal families can appear in such an
expansion, but within a conformal family, the most singular
contribution is always from a primary. This argument applies even if
$c$ and $h$ are arbitrary.  But suppose they are related as above and
the differential equation eq.(\ref{eq:sing}) is valid. This equation
is singular at $x=y$ and at leading order the dominant balance leads
to an equation where the other points are spectators. One finds that
the only conformal families that can appear are the ones whose
conformal weight $\delta'$ satisfies the fusion rule.

This is enough CFT background for the rest of this paper. 

\vskip 1.5 truecm